\documentclass[a4paper,onecolumn,twoside,12pt,final]{book}

\usepackage{graphicx}
\usepackage{citesort}
\usepackage{amsmath}

\begin{document}

\begin{titlepage}

\title{Variational Principles in General Relativity}

\author{Brian D.~Baker \thanks{bbaker@phys.ufl.edu}\\
          Department of Physics\\
          University of Florida\\
	  Gainesville, FL 32611
       }


\date{\today}
\frontmatter
\maketitle

\end{titlepage}
\chapter*{Acknowledgements}
When faced with the realization of compiling a list of individuals and groups whom I would wish to thank for their help, inspiration, and support over the years, I quickly realized this would be a daunting task. It is a sad truth that I will be sure to forget some people at the time of writing this acknowledgment. For those people, I apologize profusely, but know that I am, and have been, greatly appreciative of their support.

I would like to begin by thanking my parents, Phil and Cherie Baker. They have always been supportive of me, and encouraged my inquisitive mind at an early age. My successes in life are due in large part to the values they instilled in me at an early age: lessons which have contributed to my success as a graduate student, and lessons which I know will allow me to succeed in life after graduate school.

A special thank you also goes to my sister Carrie Klippenstein, her husband Stacy, and my two favorite nephews Steven and Ty. They were always very supportive of me and actively inquired about my work, even when we lived on opposite sides of the country.

I owe a tremendous debt of gratitude to my advisor, Dr.~Steven Detweiler. It has been an extreme honor and pleasure to work with him. Steve has always been very supportive of my work and my life, and has always provided invaluable advice in both arenas. He is a trusted friend, and he will be greatly missed when I leave Gainesville.

Very special thanks go to my best friend Preeti Jois-Bilowich. She gave me the strength and support I needed when I felt like giving up on physics and on life. I look forward to continuing the special bond we share as I enter the next chapter of my life.

Many thanks go to my cell-mate and good friend, Eirini Messaritaki. Not only was Eirini an invaluable resource for new physics ideas, but she would also patiently sit and listen to me rant on endlessly about a litany of topics. She was always a voice of reason, and often lifted my spirits when the daily grind of teaching would wear me down.

Dr.~Richard Haas deserves a place on my acknowledgments list. I can't begin to count the late evenings in which Rich and I would (seemingly) be the only ones working in the building. His enlightening insight, keen sense of humor, side-project business pursuits, and love of fine cognac provided much enjoyment.

I would also like to thank Charles Parks for his continuing support and friendship. He has been encouragement personified---urging me forward and supporting me in a variety of ways at the time I needed it most.

Thanks to Dr.~Ilya Pogorelov for his friendship, useful conversations, encouragement, and endless help when I was having difficulty with Unix or Matlab. His enlightened perspective on life and physics always made for enjoyable company.

A very special ``heh!!'' to Dr.~Richard Woodard, for fighting the good fight and helping to keep the humans at bay. He is an inspiration to us all for his dedication to the truth and to his students. May he never forget our rallying cry:\\ \indent ``$\Sigma\Pi{\text{A}}\Sigma{\text{IK}}\Lambda{\text{E}}\Sigma$ ~~${\text{T}}{\text{O}}\Upsilon$ ~~${\text{K}} {\text{O}}\Sigma{\text{M}}{\text{O}}\Upsilon$ ~${\text{E}}{\text{N}}\Omega\Theta{\text{E}}{\text{I}}{\text{T}}{\text{E}}$! ~~${\text{A}}\Sigma$ ~~$\Xi{\text{E}}\Sigma{\text{H}}{\text{K}}\Omega\Theta{\text{O}}\Upsilon{\text{M}}{\text{E}}$ ~~${\text{N}}{\text{A}}$ ~~$\Sigma\Upsilon{\text{N}}{\text{T}}{\text{P}}{\text{I}}\Psi{\text{O}}\Upsilon{\text{M}}{\text{E}}$ ~~${\text{T}}{\text{O}}\Upsilon\Sigma$  ~${\text{A}}{\text{N}}\Theta{\text{P}}\Omega\Pi{\text{O}}\Upsilon\Sigma$ ~~${\text{K}}{\text{A}}{\text{T}}{\text{A}}\Pi{\text{I}}{\text{E}}\Sigma{\text{T}}{\text{E}}\Sigma$ ~~${\text{M}}{\text{A}}\Sigma$!''

Very special thanks to Dr.~Darin Acosta, his wife Janis, and their children Angela and Alex. Their unending kindness and understanding will not be forgotten, and they have left me with some pleasant memories of the old neighborhood.

Thank you to John and Sylvia Raisbeck for occasionally pulling me away from the long hours at the office, and reminding me what it is like to socialize. Their kindness and company were a welcome addition to the monotony of graduate school.

Special thanks go to my good friend Kungun Mathur, who was always willing to visit with me despite her incredibly difficult schedule. She always lifted my spirits, and our visits will remain one of the bright memories of my stay at Gainesville.

Innumerable thanks go to Susan Rizzo and Darlene Latimer, who make the lives of countless graduate students and faculty much more tolerable, thanks to their experience with the University administration and their good spirits.

I would be remiss if I did not send my thanks to the University Women's Club and the McLaughlin Dissertation Fellowship Fund at the University of Florida for their moral and financial support.

Also, a tremendous amount of thanks go to the other esteemed members of my thesis advisory committee: Dr.~James Fry, Dr.~Stephen Gottesman, Dr.~James Ipser, and Dr.~Bernard Whiting. It is an honor to have my name associated with these gentlemen, and I have benefited greatly from my interactions with them over the years.

Over the years of my endeavor, several businesses, products and groups of people I have never met helped to maintain my health and sanity. Special thanks to: The Boston Beer company; the fine folks who make Jose Cuervo tequila; Smith Kline Beecham (the makers of Tums antacids); Starbucks coffee; ESP custom-made guitars; Marshall Amplification; Roland Corporation; finally all the guys from Dream Theater and Metallica.

Last but not least, I would like to thank my cat Cleopatra. She was always willing to snuggle with me and keep me company during the long, lonely nights at home. She is a true friend, and I am excited at the prospects of where our lives will lead.


\tableofcontents

\mainmatter
\chapter{Introduction}

The conception of Einstein's Theory of General Relativity in 1915 was deemed a theorist's paradise, but an experimentalist's nightmare. In particular, the effects of gravitational radiation were believed to be too weak to ever have measurable effects on an Earth-based experiment, and were hence nothing more than an interesting prediction of the theory.

However, beginning with Weber's \cite{Weber} investigations using resonant bar detectors for the measurement of gravitational waves, the reality of actually measuring the effects of black holes and neutron stars on the surrounding universe became the primary goal of many investigators.

Today, with the construction of Earth-based gravitational wave detectors like LIGO \cite{Abramovici} it has become a matter of when, rather than if, gravitational waves will be detected. One of the remaining difficulties in detection of gravity waves is the construction of realistic waveform templates. Templates will allow investigators to sift through the avalanche of data to correctly and accurately identify gravitational wave signals from astrophysical sources.

Because of the need for waveform templates \cite{FlanaganHughes}, a great amount of effort has been spent on a fully relativistic treatment of the two-body problem. A solution to the two-body problem will yield detailed information of a binary's orbital parameters, which will enable investigators to construct realistic waveform templates.

This thesis documents an effort to describe the two-body problem in the framework of variational principles.

The first half of this thesis describes a variational approach to describing a binary system of neutron stars. This is an appealing method of approximating solutions to the Einstein equations and the equations describing the matter of the neutron stars---we find we only need to minimize a single function to arrive at an approximation, rather than solving the full set of nonlinear Einstein equations. We derive expressions for quantities we call the ``effective mass'' and the ``effective angular momentum'', which may be interpreted as the mass and angular momentum of the system under study, without a contribution from the gravitational radiation of the system. The variational principle dictates how we may derive the orbital angular frequency of the system, as well as the energy content of the gravitational waves---information vital to gravitational wave observatories.

The second half of the thesis is dedicated to a numerical method for describing binary black holes. The mathematical framework used to describe the black holes is known as the ``puncture'' method, and is attributed to Brandt and Br{\"u}gmann \cite{BB}. This method greatly simplifies the analysis of the geometry, allowing us to side-step many complications previously encountered when studying binary black holes. We develop a numerical algorithm employing adaptive multigrid techniques to solve a nonlinear elliptic equation describing the geometry of the holes, which in turn allows us to apply a variational principle for the energy of the system. Once again, the variational principle indicates how we may approximate the orbital angular momentum of the system, as well as additional geometrical and gravitational wave information. 

As mentioned above, both the analytical and numerical variational principles allow for accurate determination of orbital parameters, which in turn may be used to construct waveform templates for gravitational wave observatories. With accurate templates investigators can open a new window to the universe and ``listen'' to the symphony of black holes and binary neutron star systems.

\chapter[The Quasi-equilibrium Approach]{The Quasi-equilibrium Approach: A Variational Principle for Binary Neutron Stars}

In this chapter, we develop a variational principle for binary neutron stars. We begin with a Newtonian variational principle for neutron stars to give us a simple example of the formulation and power of variational principles in general. We then briefly discuss the importance of surface integrals in variational principles, and then begin developing our fully relativistic variational principle by entering the realm of the 3+1 formalism of the Einstein equations. This in turn leads us to a somewhat lengthy discussion of various aspects required in our variational principle: irrotational fluid flow in relativity, the quasi-equilibrium approach, symmetric trace-free tensors, linearized solutions to the Einstein equations, and finally the full formulation of our variational principle.

\section{Newtonian Analysis of Binary Neutron Stars}

We begin our Newtonian analysis by first demanding our neutron stars be composed of a perfect fluid: one in which there are no stresses or shears. We also require an equation of state which relates the fluid density $\rho$ to the pressure $p$ such that $p = p(\rho)$. We also assume the binary system has an orbital frequency of $\Omega$. We then define a quantity $S$, which is evaluated in the uniformly rotating frame:
\begin{eqnarray}
\label{newt1}
  S &=& \int d^{3}x \left[ \frac{1}{2} \rho \vec{u} \cdot \vec{u} + \frac{1}{8 \pi G} (\vec{\nabla} \Phi)^{2}
                    + \rho \Phi + \rho \int^{p}_{0} \frac{dp}{\rho} + p \right] \nonumber \\
		    & &~ + \lambda \left[ \int d^{3}x \rho - M \right],
\end{eqnarray}
where $\rho$ is the density of our perfect fluid neutron stars, $\vec{u}$ is the velocity of the fluid in the rotating frame, $\Phi$ is the Newtonian gravitational potential, $p$ is the pressure of the fluid, and
$\lambda$ is a Lagrange multiplier which ensures that the mass of each star is the integral of the fluid density.

Because we are evaluating $S$ in a non-inertial frame which is rotating at a uniform rate $\Omega$, we may relate $\vec{u}$ to the inertial fluid velocity $\vec{v}$ by
\begin{equation}
\label{newt2}
  \vec{u} = - \vec{\Omega} \times \vec{r} + \vec{v},
\end{equation}
where $\vec{r}$ is simply the radius of the orbit. It has been suggested \cite{Kochanek,Bildsten} and is widely held that neutron stars in binary orbits will not be tidally locked, or \it{corotating}\rm. It has been found that tidally locked neutron stars would have unrealisticly large viscosities \cite{Bildsten}. It is believed, therefore, the stars will have very little, if any, intrinsic spin. This scenario is called \it{irrotational fluid flow}\rm, which translates to saying the fluid velocity $\vec{v}$ has zero curl:
\begin{equation}
\label{newt3}
  \vec{\nabla} \times \vec{v} = 0.
\end{equation}
This implies we may introduce a vector potential $\psi$ via
\begin{equation}
\label{newt4}
  \vec{v} = \vec{\nabla} \psi.
\end{equation}
We may rewrite our equation for the fluid velocity in the rotating frame to be 
\begin{equation}
\label{newt5}
  \vec{u} = - \vec{\Omega} \times \vec{r} + \vec{\nabla} \psi.
\end{equation}

We now rewrite our function $S$, using the expression for the fluid velocity given by Eq.~(\ref{newt5}):
\begin{eqnarray}
\label{newt6}
  S & =& \int d^{3}x 
     \left[ \frac{1}{2} \rho \left(- \vec{\Omega} \times \vec{r} + \vec{\nabla} \psi \right) \cdot 
                        \left(- \vec{\Omega} \times \vec{r} + \vec{\nabla} \psi \right) \right. \nonumber \\ 
			& &~ \left. + \frac{1}{8 \pi G} \left(\vec{\nabla} \Phi \right)^{2}
                        + \rho \Phi + \rho \int^{p}_{0} \frac{dp}{\rho} + p \right] 
			+ \lambda \left[ \int d^{3}x \rho - M \right].
\end{eqnarray}

We now allow infinitesimal, arbitrary variations of the velocity potential $\psi$, the fluid density $\rho$, and the gravitational potential $\Phi$. For clarity, we analyze each variation separately, starting with the variation due to the velocity potential $\psi$:
\begin{equation}
\label{newt7}
  \delta S_{\psi} =  \int d^{3}x \left[ \rho \vec{\nabla} \psi \cdot \vec{\nabla} \delta \psi 
                                  - \rho \left(\vec{\Omega} \times \vec{r} \right) \cdot \vec{\nabla} \delta \psi \right].
\end{equation} 
After some rearranging and integrating by parts, $\delta S_{\psi}$ may be written as
\begin{eqnarray}
\label{newt8}
  \delta S_{\psi} & = & \int d^{3}x \left[ \left( \vec{\Omega} \times \vec{r} \right) \cdot \vec{\nabla} \rho
                                 - \vec{\nabla} \psi \cdot \vec{\nabla} \rho
				 - \rho \nabla^{2} \psi \right] \delta \psi  \nonumber \\
                   & &~+ \oint d^{2}x \left\{ \rho \vec{n} \cdot \left[\vec{\nabla} \psi 
		                 - \left( \vec{\Omega} \times \vec{r} \right) \right] \right\} \delta \psi ,
\end{eqnarray}
where the surface integral is evaluated on the surface of the star, and $\vec{n}$ is a unit vector normal to the surface of the star. If $S$ is to be an extremum the coefficients of Eq.~(\ref{newt8}) demand
\begin{equation}
\label{newt9}
  \nabla^{2} \psi = - \left[ \vec{\nabla} \psi 
                    - \left( \vec{\Omega} \times \vec{r} \right) \right] \cdot \frac{\vec{\nabla} \rho}{\rho},
\end{equation}
which is the equation for irrotational fluid flow \cite{Teuk1}. The variation of $\psi$ in $S$ also dictates Eq.~(\ref{newt9}) is subject to a boundary condition, specifically
\begin{equation}
\label{newt10}
  \left[ \vec{\nabla} \psi - \left( \vec{\Omega} \times \vec{r} \right) \right] \cdot \vec{n} = 0,
\end{equation}
which again is imposed on the surface of the star. Note our Eqs.~(\ref{newt9}) and (\ref{newt10}) are the same as Teukolsky's Eqs.~(13) and (14) \cite{Teuk1}.

We now turn our attention to the variation in $S$ due to variations in the fluid density $\rho$. We find the variation may be written as
\begin{equation}
\label{newt11}
  \delta S_{\rho} = \int d^{3}x \left[ \frac{1}{2} \left( - \vec{\Omega} \times \vec{r} + \vec{\nabla} \psi \right) \cdot 
                        \left( - \vec{\Omega} \times \vec{r} + \vec{\nabla} \psi \right) 
			+ \Phi + \int^{p}_{0} \frac{dp}{\rho} + \lambda \right] \delta \rho.
\end{equation}
If we demand $S$ be an extremum, then Eq.~(\ref{newt11}) implies
\begin{equation}
\label{newt12}
  \frac{1}{2} \left( - \vec{\Omega} \times \vec{r} + \vec{\nabla} \psi \right) \cdot 
    \left(- \vec{\Omega} \times \vec{r} + \vec{\nabla} \psi \right) 
     + \Phi + \int^{p}_{0} \frac{dp}{\rho} = - \lambda,
\end{equation}
which is simply the Bernoulli equation for fluid flow given by Teukolsky's Eq.~(12) \cite{Teuk1}.

Finally, we examine the variation of $S$ under changes in the gravitational potential $\Phi$. This variation is
\begin{equation}
\label{newt13}
  \delta S_{\Phi} = \int d^{3}x \left[ \frac{1}{4 \pi G} \vec{\nabla} \Phi \cdot \vec{\nabla} \delta \Phi
                                    + \rho \delta \Phi \right].
\end{equation} 
With some rearranging and integration by parts, we find this variation may be written as the sum of two terms:
\begin{equation}
\label{newt14}
  \delta S_{\Phi} = \int d^{3}x \left[ \rho - \frac{1}{4 \pi G} \nabla^{2} \Phi \right] \delta \Phi
                      + \oint d^{2}x \left[ \frac{1}{4 \pi G} \vec{n} \cdot \vec{\nabla} \Phi \right] \delta \Phi.
\end{equation}
The location of the surface integral is somewhat ambiguous, and depends on the conditions one wishes to impose. For instance, if the location of the boundary is on the surface of the star, $\vec{n}$ is interpreted as a vector normal to the surface of the star. One then specifies the value of the gravitational potential on the surface of the star. On the other hand, one could choose a different surface in which to specify the value of the potential $\Phi$ and the vector $\vec{n}$. For instance, a different choice would demand the potential fall off sufficiently rapidly as $r \rightarrow \infty$ so as to eliminate the surface integral altogether. 

For whatever method one chooses to handle the surface integral for the gravitational potential, if we demand $S$ be an extremum, then Eq.~(\ref{newt14}) demands
\begin{equation}
\label{newt15}
  \nabla^{2} \Phi = 4 \pi G \rho,
\end{equation}
which is simply the Poisson equation describing the gravitational field of the star.

We may therefore describe the total variation of $S$ as the sum of the individual variations:
\begin{equation}
\label{newt16}
  \delta S = \delta S_{\psi} + \delta S_{\rho} + \delta S_{\Phi},
\end{equation}
where the individual variations are given by Eq.~(\ref{newt8}), Eq.~(\ref{newt11}), and Eq.~(\ref{newt14}), respectively. $S$ is an extremum if and only if under arbitrary variations of $\psi$, $\rho$, and $\Phi$, the fluid velocity potential, fluid density, and gravitational potential are described by Eq.~(\ref{newt9}), Eq.~(\ref{newt12}), and Eq.~(\ref{newt15}), and are subject to the boundary conditions given by Eq.~(\ref{newt10}) and Eq.~(\ref{newt14}).

\section[The Importance of Surface Integrals]{The Importance of Surface Integrals}

Before we delve into a fully relativistic treatment of binary neutron stars, it is important to point out the role surface integrals (or boundary terms) play in variational principles. We demonstrate this with the aid of an over-simplified toy model. Imagine we are studying a one-dimensional thin wire of length $L$, and wish to formulate a variational principle which properly describes the temperature along the length of this wire. Let us define a quantity $I$ to be
\begin{equation}
\label{toy1}
  I = \int \left[ \frac{1}{2} \left(\frac{d T(x)}{dx} \right)^{2} \right] dx,
\end{equation}
where $T(x)$ is the position dependent temperature of the wire. If we allow for infinitesimal, arbitrary variations in this temperature, then the variation in $I$ becomes
\begin{equation}
\label{toy2}
  \delta I = - \int \left[ \frac{ d^{2} T(x) }{ dx^{2} } \right] \delta T(x) dx 
                  + \left. \delta T(x) \frac{d T(x)}{dx} \right\vert_{L}
		  - \left. \delta T(x) \frac{d T(x)}{dx} \right\vert_{0},
\end{equation}
where the boundary terms are the result of integration by parts. Note the coefficient of the integral is simply the one-dimensional Laplacian---the correct equation that describes the temperature in a source-free wire. The variation in $I$ given by Eq.~(\ref{toy2}) indicates $I$ gives the correct formulation of the variational principle as long as we fix the value of the temperature on the boundaries (i.e., $ \delta T(x) \vert_{Boundary} = 0 $ ). 

However, the quantity $I$ given by Eq.~(\ref{toy1}) does not give the correct formulation of the variational principle if we wish to specify the value of some other physical quantity on the boundary, say the derivative of the temperature. To formulate our new variational principle, we must modify $I$ by the inclusion of an additional term. We define the quantity $S$ as
\begin{eqnarray}
\label{toy3}
  S &\equiv& I - \int \frac{d}{dx} \left( T(x) \frac{d T(x)}{dx} \right) dx \nonumber \\
    &=& \int \left[\frac{1}{2} \left( \frac{d T(x)}{dx} \right)^{2} 
          - \frac{d}{dx}\left( T(x) \frac{d T(x)}{dx} \right) \right].
\end{eqnarray}

If we now allow for infinitesimal, arbitrary variations in the temperature, we find the total variation in $S$ to be of the form
\begin{equation}
\label{toy4}
  \delta S = - \int \left[ \frac{ d^{2} T(x) }{ dx^{2} } \right] \delta T(x) dx 
                  - \left. T(x) \delta \left( \frac{d T(x)}{dx} \right) \right\vert_{L}
		  + \left. T(x) \delta \left( \frac{d T(x)}{dx} \right) \right\vert_{0}.
\end{equation}

Note the differences and similarities between Eq.~(\ref{toy4}) and Eq.~(\ref{toy2}). Specifically, the coefficients in the integral are exactly the same---the one-dimensional Laplacian. However, the boundary terms are markedly different. The variation of $S$ given by Eq.~(\ref{toy4}) indicates $S$ gives the proper formulation of the variational principle if we hold the value of the derivative of the temperature fixed on the boundaries (i.e., $\delta \left( \frac{d T(x)}{dx} \right) \vert_{Boundary} = 0$).

There is a moral to this story: in general, the formulation of the variational principle depends on the quantities one wishes to specify on the boundaries of the system. This idea will play a major role in our treatment of binary neutron stars.

It is also important to note, in general, the boundary terms of a variational principle have a physical interpretation \cite{Regge}. They may correspond to a mass, angular momentum, or other physical quantity which helps to specify the system. This will also become apparent in the coming discussion.

\section{The 3+1 Formalism}

We now develop the necessary tools to formulate our variational principle for binary neutron stars, starting with a discussion of the 3+1 formalism. We decompose our four-dimensional metric $g_{ab}$ in the typical ADM fashion \cite{ADM} in which we foliate spacetime into spacelike hypersurfaces of constant time. The foliated metric then has the form
\begin{equation}
\label{3+1_1}
  g_{ab} dx^{a} dx^{b} = - N^{2} dt^{2} + \gamma_{ab} \left( dx^{a} + N^{a} dt \right) \left( dx^{b} + N^{b} dt \right),
\end{equation}
where $\gamma_{ab}$ is the three-metric on a spacelike hypersurface $\Sigma_{t}$, $N$ is the lapse function, and $N^{a}$ is the shift vector. The lapse function $N$ is a measure of the proper time elapsed as one translates from one time slice to the next, and is related to the four-metric via
\begin{equation}
\label{3+1_2}
  g^{00} = - \frac{1}{ N^{2} }.
\end{equation}

The shift vector $N^{a}$ is a measure of how coordinates shift from one hypersurface to the next, and is related to the four-metric via
\begin{equation}
\label{3+1_3}
  g^{0j} = \frac{ N^{j} }{ N^{2} }.
\end{equation}

For completeness, we express the spatial part of the contravariant four-metric as
\begin{equation}
\label{3+1_4}
  g^{ij} = \gamma^{ij} - N^{i} N^{j} / N^{2}.
\end{equation}

The time translation vector $t^{a}$ is related to the lapse and shift via
\begin{equation}
\label{3+1_5}
  t^{a} = N u^{a} + N^{a},
\end{equation}
where $u^{a}$ is a timelike unit normal ($u_{a} u^{a} = -1$) to the hypersurface $\Sigma_{t}$. Finally, we may also write the three-metric in a way which captures the flavor of a projection operator. Specifically,
\begin{equation}
\label{3+1_6}
  \gamma_{ab} = g_{ab} + u_{a} u_{b}.
\end{equation}

We must determine a covariant derivative operator, denoted as $D_{a}$, which is compatible with this metric. This is achieved by projecting all of the indices of the four-dimensional covariant derivative $\nabla_{a}$ to ensure a completely spatial quantity:
\begin{equation}
\label{3+1_7}
  D_{a} V^{b} = \gamma^{\;c}_{a} \gamma^{\;b}_{d} \nabla_{c} V^{d}
\end{equation}
for any spatial vector $V^{a}$ and
\begin{equation}
\label{3+1_8}
  D_{a} \phi = \gamma^{\;b}_{a} \nabla_{b} \phi
\end{equation}
for any scalar field $\phi$.

With the fundamental geometrical quantities thus defined, then we may construct the Ricci tensor $R_{ab}$ and the Ricci scalar $R$ from the three-metric $\gamma_{ab}$.

The foliation of the four-dimensional spacetime into three-dimensional hypersurfaces introduces a new geometrical quantity, the extrinsic curvature ${\cal K}_{ab}$, defined by
\begin{equation}
\label{3+1_9}
  {\cal K}_{ab} \equiv - \frac{1}{2} \cal{L}\it_{u} \gamma_{ab} = - \gamma^{\;c}_{a} \nabla_{c} u_{b},
\end{equation}
where $ \cal{L}\it_{u} $ is the Lie derivative with respect to the vector $u^{a}$. Another geometrical result of the foliation is that we may write an expression for how the three-metric changes as one moves off of a particular hypersurface and onto another:
\begin{equation}
\label{3+1_10}
  \cal{L}\it_{t} \gamma_{ab} = -\rm{2}\it N {\cal K}_{ab} + \rm{2}\it D_{(a} N_{b)} \equiv \cal{G}\it_{ab},
\end{equation}
which also defines ${\cal G}_{ab}$.

In the Hamiltonian formulation of General Relativity, the momentum conjugate to $\gamma_{ab}$ is $\pi^{ab} / 16 \pi$, where $\pi^{ab}$ is known as the field momentum \cite{ADM,MTW,York3+1}, and is related to the extrinsic curvature via
\begin{equation}
\label{3+1_11}
  \pi^{ab} \equiv - \gamma^{1/2} \left( {\cal K}^{ab} - \gamma^{ab} {\cal K}^{\;c}_{c} \right),
\end{equation}
where $\gamma$ is the determinant of the three-metric.

We may now write the Einstein equations in the 3+1 formalism. Two contractions of the four-dimensional Einstein equations with the unit normal ($u^{a}u^{b}G_{ab}$) yields the Hamiltonian constraint, and must be satisfied on every hypersurface. It may be written as
\begin{equation}
\label{3+1_12}
  R + \gamma^{-1} \left( \frac{1}{2} \pi^{\;a}_{a} \pi^{\;b}_{b} 
                    - \pi^{ab} \pi_{ab} \right) = 16 \pi \rho_{H} \equiv \cal{N}\rm,
\end{equation}
which also defines $\cal{N}\rm$. The energy density $\rho_{H}$ depends on the stress-energy tensor $T^{ab}$ via
\begin{equation}
\label{3+1_13}
  \rho_{H} \equiv u^{a} u^{b} T_{ab}.
\end{equation}

Likewise, a single contraction of the four-dimensional Einstein equations with the unit normal, and then a contraction with the three-metric ($u_{b}\gamma_{c}^{\;a}G^{cb}$) yields the momentum constraint, and must also be satisfied on every hypersurface. It is given as
\begin{equation}
\label{3+1_14}
  D_{b} \left( \pi^{ab} / \gamma^{1/2} \right) = -8 \pi j^{a} \equiv \cal{N}\it^{a},
\end{equation}
which also defines $\cal{N}\it^{a}$. The momentum density $j^{a}$ is related to the stress-energy tensor via
\begin{equation}
\label{3+1_15}
  j^{a} \equiv u_{b} \gamma^{\;a}_{c} T^{cb}.
\end{equation}

The remaining spatial part of the Einstein equations contains the dynamics, and is given by
\begin{equation}
\label{3+1_16}
  {\cal L}_{t} \pi^{ab} = {\cal P}^{ab} + 8 \pi N \gamma^{1/2} S^{ab}, 
\end{equation}
where
\begin{eqnarray}
\label{3+1_16a}
  {\cal P}^{ab} & \equiv & - N \gamma^{1/2} \left( R^{ab} - \frac{1}{2} \gamma^{ab} R \right) 
                         + \frac{1}{2} N \gamma^{-1/2} \gamma^{ab} \left( \pi^{cd} \pi_{cd} 
			                    - \frac{1}{2} \pi^{\;c}_{c} \pi^{\;d}_{d} \right) \nonumber \\ 
			 & & ~- 2 N \gamma^{-1/2} \left( \pi^{ac} \pi^{\;b}_{c} - \frac{1}{2} \pi^{\;c}_{c} \pi^{ab} \right)
			 + \gamma^{1/2} \left( D^{a} D^{b} N - \gamma^{ab} D^{c} D_{c} N \right) \nonumber \\
			 & & ~+ \gamma^{1/2} D_{c} \left( \gamma^{-1/2} \pi^{ab} N^{c} \right)
			 - \pi^{ac} D_{c} N^{b} - \pi^{bc} D_{c} N^{a}
\end{eqnarray}
and $S^{ab}$ is the spatial stress tensor, defined as
\begin{equation}
\label{3+1_17}
  S_{ab} \equiv \gamma^{\;c}_{a} \gamma^{\;d}_{b} T_{cd}.
\end{equation}

If one wishes to analyze solutions to Einstein's equations which are stationary or approximately stationary (i.e., not evolving with time), this corresponds to the constraint equations given by Eq.~(\ref{3+1_12}) and Eq.~(\ref{3+1_14}), as well as the dynamical equations ${\cal G}_{ab} = 0$ and ${\cal P}^{ab} + 8 \pi N \gamma^{1/2} S^{ab} = 0$. Solutions to this set of equations are known as quasi-stationary, or quasi-equilibrium solutions \cite{Det2}. Physically, they may approximately correspond to a binary neutron star system in a circular orbit, where the radiation reaction time scale is much larger than the orbital time scale. It is these very types of systems we will explore. Let us first investigate the consequences of modeling our neutron stars as perfect fluids, with the constraint of irrotational motion placed on these stars.

\section{Irrotational Fluid Flow in General Relativity}
\label{irrfluid}
There are three time scales involved in orbital dynamics in the framework of General Relativity: the orbital period $\tau_{\text{orbit}}$, the time scale of circularization of the orbit due to the radiation of gravitational waves $\tau_{\text{circle}}$, and the time scale in which the radius of the orbit decreases due to radiation reaction $\tau_{\text{rr}}$ \cite{MTW}. These time scales are related by $\tau_{\text{orbit}} \ll \tau_{\text{circle}} \ll \tau_{\text{rr}}$. Because of the relative sizes of the time scales involved, it is believed that a good approximation to the physical system is a sequence of binary systems in circular orbits in which the effects of gravitational radiation are small. These systems are said to be in quasi-stationary or quasi-equilibrium circular orbits. 
As was mentioned in the Newtonian treatment of our binary system, it is also believed that binary neutron stars will have very little intrinsic spin with respect to a local inertial frame; the fluid viscosities would be unrealistically large if this were not the case. Hence, it is believed binary stars exhibiting irrotational fluid flow serve as a well-justified approximation to reality \cite{Bildsten}.

In this section, we first derive the pertinent equations describing irrotational fluid flow following the methodology of Teukolsky \cite{Teuk1}, and then we rewrite these equations in the 3+1 formalism.

We begin by defining the relativistic enthalpy $h$ as
\begin{equation}
\label{irr1}
  h = \frac{\rho + p}{\rho_{0}},
\end{equation}
where $\rho$ is the total energy density of the fluid, $p$ is the pressure, and $\rho_{0} = m_{B} n$ is the rest mass density. The baryon rest mass is $m_{B}$ and the baryon number density is $n$. We also assume an equation of state relating the pressure and the density via $p = p(\rho)$.

The stress-energy tensor of an isentropic perfect fluid is
\begin{equation}
\label{irr2}
  T^{ab} = (\rho + p) U^{a} U^{b} + p g^{ab},
\end{equation}
where $U^{a}$ is the fluid four-velocity (not to be confused with the unit normal to the hypersurface $u^{a}$) and $g^{ab}$ is the four-metric of the spacetime. Conservation of stress-energy $(\nabla_{a} T^{ab} = 0)$ allows us to write the conservation of energy equation for the fluid by projecting the conservation equation along the four-velocity:
\begin{equation}
\label{irr3}
 - U_{a} \nabla_{b} T^{ab} = \nabla^{a} \left[ (\rho + p) U_{a} \right] - U^{a} \nabla_{a} p 
                           = U^{a} \nabla_{a} \rho + (\rho + p) \nabla^{a} U_{a} = 0.
\end{equation}

Likewise, we may determine Euler's equation for the fluid by projecting the conservation of stress-energy in a direction perpendicular to the fluid four-velocity:
\begin{equation}
\label{irr4}
  \left( g^{\;b}_{a} + U_{a} U^{b} \right) \nabla^{c} T_{bc} = (\rho + p) U^{c} \nabla_{c} U_{a} 
                                                + \left( g^{\;b}_{a} + U_{a} U^{b} \right) \nabla_{b} p = 0.
\end{equation}
In addition to the conservation of energy and Euler's equation, we may also write the equation for conservation of rest mass:
\begin{equation}
\label{irr5}
  \nabla_{a} \left( \rho_{0} U^{a} \right) = 0.
\end{equation}
We note Euler's equation, written in terms of the relativistic enthalpy, is given as
\begin{equation}
\label{irr6}
  U^{a} \nabla_{a} ( h U_{b} ) + \nabla_{b} h = 0.
\end{equation}

At this point, we introduce the concept of irrotational fluid flow in relativity. In terms of the relativistic enthalpy and the fluid four-velocity, the relativistic vorticity tensor is defined as
\begin{equation}
\label{irr7}
  \omega_{ab} \equiv P^{\;c}_{a} P^{\;d}_{b} \left[ \nabla_{c} ( h U_{d} ) - \nabla_{d} ( h U_{c} ) \right],
\end{equation}
where $P^{\;a}_{b} \equiv \delta^{\;a}_{b} + U^{a} U_{b}$ is a projection tensor. Using the form of the Euler equation given by Eq.~(\ref{irr6}), the relativistic vorticity tensor \cite{Teuk1} may be written as
\begin{equation}
\label{irr8}
  \omega_{ab} = \nabla_{a} ( h U_{b} ) - \nabla_{b} ( h U_{a} ).
\end{equation}

If a fluid is irrotational, then the vorticity tensor is equal to zero. The physical consequences of this indicate  the neutron stars have no intrinsic spin with respect to a local inertial frame. If we insist on a vanishing vorticity tensor, this implies we may introduce a velocity potential \cite{Teuk1,Shibata1}, denoted as $\psi$, in a fashion similar to that of the Newtonian problem mentioned above:
\begin{equation}
\label{irr9}
  h U_{a} = \nabla_{a} \psi.
\end{equation}
The introduction of the velocity potential allows us to rewrite the conservation of rest mass equation, Eq.~(\ref{irr5}), as
\begin{equation}
\label{irr10}
  \nabla^{a} \left( h^{-1} \rho_{0} \nabla_{a} \psi \right) = 0,
\end{equation}
and we note the normalization of the fluid four velocity, $U^{a} U_{a} = -1$, results in
\begin{equation}
\label{irr11}
  h^{2} = - \left( \nabla_{a} \psi \right) \nabla^{a} \psi.
\end{equation}

In general, if a fluid is exhibiting irrotational motion, then in addition to being described by some equation of state $p=p(\rho)$, the fluid also satisfies Eq.~(\ref{irr1}), Eq.~(\ref{irr10}), and Eq.~(\ref{irr11}).

In anticipation of describing the irrotational fluid in the Hamiltonian formalism of General Relativity, we now decompose the fluid equations in the 3+1 formalism under the assumption that the stars are in quasi-equilibrium.

In the quasi-equilibrium approximation, we require the existence of a Killing vector in a particular coordinate system of the form
\begin{equation}
\label{irr12}
  \vec{t} = \frac{\partial}{\partial t} + \Omega \frac{\partial}{\partial \phi}.
\end{equation}
This Killing vector is a symmetry generator for the matter fields, which implies
\begin{equation}
\label{irr13}
  \cal{L}\it_{\vec{t}} ( h U_{a} ) = {\rm 0} = \cal{L}\it_{\vec{t}} ( \nabla_{a} \psi ) 
                                    = \nabla_{a} \left( \cal{L}\it_{\vec{t}} \psi \right),
\end{equation}
where $\cal{L}\it$ once again denotes the Lie derivative. The above condition in turn implies 
\begin{equation}
\label{irr14}
  {\cal L}_{\vec{t}} \psi = t^{a} \nabla_{a} \psi 
                         = \frac{\partial \psi}{\partial t} + \Omega \frac{\partial \psi}{\partial \phi}
			 = - C,
\end{equation}
where $C$ is a positive constant related to the Bernoulli integral.

Before we write down the fluid equations in the 3+1 language, it is useful to note some relations which are used in the derivations. The derivative of the vector normal to a spacelike hypersurface is
\begin{equation}
\label{irr15}
  \nabla_{a} u_{b} = - {\cal K}_{ab} - u_{b} D_{a} \ln N.
\end{equation}
Recall $ {\cal K}_{ab}$ is the extrinsic curvature of the spacelike hypersurface, $ D_{a}$ is the derivative compatible with $\gamma_{ab}$, and $N$ is the lapse function. We also note the result of the derivative acting on a scalar field $\Phi$ is
\begin{equation}
\label{irr16}
  \nabla_{a} \Phi = D_{a} \Phi - u_{a} u^{b} \nabla_{b} \Phi.
\end{equation}
Likewise, we note the result for any spatial vector to be
\begin{equation}
\label{irr17}
  \nabla_{a} V_{b} = D_{a} V_{b} - u_{a} u^{c} \nabla_{c} V_{b} - {\cal K}_{ac} V^{c} u_{b}.
\end{equation}

Armed with the above useful expressions, we may begin to derive the fluid equations, specifically Eq.~(\ref{irr10}) and Eq.~(\ref{irr11}), in the 3+1 formalism. We begin by noting
\begin{equation}
\label{irr18a}
  t^{a} \nabla_{a} \psi = N u^{a} \nabla_{a} \psi + N^{a} D_{a} \psi = -C,
\end{equation}
which directly leads to
\begin{equation}
\label{irr18}
  u^{a} \nabla_{a} \psi = - \frac{1}{N} \left( C + N^{a} D_{a} \psi \right).
\end{equation}
We also note for any scalar $f$ which satisfies
\begin{equation}
\label{irr18bb}
  {\cal L}_{\vec{t}} f = 0
\end{equation}
then we may use a variant of Eq.~(\ref{irr18a}) to derive yet another useful expression. Specifically,
\begin{equation}
\label{irr18cc}
  t^{a} \nabla_{a} f = N u^{a} \nabla_{a} f + N^{a} D_{a} f = 0,
\end{equation}
which implies
\begin{equation}
\label{irr18dd}
  u^{a} \nabla_{a} f = - \frac{1}{N} N^{a} D_{a} f.
\end{equation}

In light of Eq.~(\ref{irr16}), this implies
\begin{equation}
\label{irr19}
  \nabla_a \psi = D_{a} \psi + u_{a} \left( C + N^{b} D_{b} \psi \right) / N.
\end{equation}
Therefore, we may write
\begin{equation}
\label{irr20}
  \nabla^{a} \nabla_{a} \psi = \nabla^{a} D_{a} \psi 
                             + \left( \nabla^{a} u_{a} \right) \left( C + N^{b} D_{b} \psi \right)/N 
			     + u^{a} \nabla_{a} \left[ \left( C + N^{b} D_{b} \psi \right)/N \right].
\end{equation}

Using Eq.~(\ref{irr17}) with the identification $V_{a} \rightarrow D_{a} \psi$,  the first term of Eq.~(\ref{irr20}) may be written as
\begin{equation}
\label{irr21}
  \nabla^{a} D_{a} \psi  = D^{a} D_{a} \psi + (D_{a} \psi) D^{a} \ln N.
\end{equation}

The second term in Eq.~(\ref{irr20}), $\nabla^{a} u_{a}$, is simply replaced by minus the trace of the extrinsic curvature, $ - {\cal K}^{\;a}_{a} $. Using Eq.~(\ref{irr18dd}), we may rewrite the third term of Eq.~(\ref{irr20}) as
\begin{equation}
\label{irr22}
  u^{a} \nabla_{a} \left[ \frac{1}{N} \left( C + N^{b} D_{b} \psi \right) \right]
                    = - \frac{1}{N} N^{a} D_{a} \left[ \frac{1}{N} \left( C + N^{b} D_{b} \psi \right) \right].
\end{equation}

Expanding Eq.~(\ref{irr10}) and using the simplifications indicated by Eq.~(\ref{irr20}), Eq.~(\ref{irr21}) and Eq.~(\ref{irr22}), we may express the equation for irrotational fluid flow in the 3+1 formalism as
\begin{equation}
\label{irr23}
  \Psi \equiv D_{a} \left[ \frac{\rho_{0}}{N h} \left( C + N^{b} D_{b} \psi \right) N^{a} 
                           - \frac{N \rho_{0}}{h} D^{a} \psi \right] = 0,
\end{equation}
which also defines $\Psi$. Note this equation is equivalent to Teukolsky's Eq.~(50).

We also note the 3+1 decomposition of the normalization of the fluid four velocity, previously given by Eq.~(\ref{irr11}), which defines $\cal{A}\rm$:
\begin{equation}
\label{irr24}
  {\cal A} \equiv h^{2} + \left( D^{a} \psi \right) D_{a} \psi 
                  - \frac{1}{N^{2}} \left( C + N^{b} D_{b} \psi \right)^{2} = 0.
\end{equation}

In addition to Eq.~(\ref{irr23}) and Eq.~(\ref{irr24}), the fluid four-velocity is subject to a boundary condition at the surface of the star. In particular, the four-velocity must be perpendicular to the normal vector of the star's surface. We define the surface to be the location in which the pressure $p$ vanishes, so the boundary condition may be written as
\begin{equation}
\label{irr25}
  \Psi_{\partial \text{Star}} \equiv U^{a} \nabla_{a} p \vert_{\partial \text{Star}}
                       = h^{-1} \left[ D^{a} \psi 
		         - \frac{1}{N^{2}} \left( C + N^{b} D_{b} \psi \right) N^{a} \right] D_{a} p \vert_{\partial \text{Star}}=0,
\end{equation}
which also defines $ \Psi_{\partial \text{Star}} $.

For completeness, the decomposition of the stress-energy tensor for an irrotational fluid is given by Eqs.~(\ref{3+1_13}), (\ref{3+1_15}) and (\ref{3+1_17}) \cite{Det3}. Specifically,
\begin{equation}
\label{irr26}
  \rho_{H} = \frac{\rho_{0}}{N^{2} h} \left( C + N^{a} D_{a} \psi \right)^{2} - \left( p - \frac{\rho_{0}}{2h} \cal{A} \right),
\end{equation}
\begin{equation}
\label{irr27}
  j^{a} = \frac{\rho_{0}}{N h} \left( C + N^{b} D_{b} \psi \right) D^{a} \psi,
\end{equation}
and
\begin{equation}
\label{irr28}
  S_{ab} = \frac{\rho_{0}}{h} \left( D_{a} \psi \right) D_{b} \psi + \gamma_{ab} \left( p - \frac{\rho_{0}}{2h} \cal{A} \right).
\end{equation}

Hence, for a given equation of state $p = p(\rho)$, a solution of the quasi-stationary Einstein equations with irrotational fluid flow is a set of $N$, $N^{a}$, $\gamma_{ab}$, $\pi^{ab}$, $\rho_{0}$, $\psi $ and $C$ which satisfy Eqs.~(\ref{irr23}), ~(\ref{irr24}), and  ~(\ref{irr25}) in addition to the quasi-stationary Einstein equations given by Eqs.~(\ref{3+1_10}),~(\ref{3+1_12}),~(\ref{3+1_14}) and ~(\ref{3+1_16}) with sources given by Eqs.~(\ref{irr26}), ~(\ref{irr27}), and  ~(\ref{irr28}).

\section{Foundations of the Variational Principle}

Before we proceed with the details of the variational principle, we first describe the physics and the implications of the quasi-equilibrium condition on our binary system. In a realistic binary star system, as the two stars orbit, the system will radiate gravitational waves. The back reaction of these waves will in turn ``push'' the orbiting stars into closer orbits. The system will continue to radiate, and the stars will slowly spiral inward until tidal forces distort the objects and cause material to be transferred from one star to another \cite{Uryu3,Shibata1,Shibata2}. If the stars are not completely tidally disrupted at this point, they will eventually merge to form a single object, which may collapse to form a black hole.

By demanding that our system obey the quasi-equilibrium Einstein equations, we are effectively disallowing any time evolution of our system. Physically, this would correspond to demanding the radiation reaction force is negligible, so the orbiting stars would forever remain at a constant radius orbit. Again, this is a realistic approximation because of the long time scales of radiation reaction forces. 

However, rather than ``turning off'' the radiative terms (and potentially losing some interesting physics), we choose to place our binary system in quasi-equilibrium by setting up a boundary which encompasses our system \cite{Det2}. We then carefully choose the amplitudes and phases of gravitational radiation propagating from this boundary \it{inward}\rm ~towards the binary system. The ingoing radiation feeds exactly the same amount of energy into the system as the energy lost because of the radiation emitted by the binary system. This physically unrealistic construct greatly simplifies a large portion of the analysis to come, and it also allows us to extract useful information about realistic binary systems. One may be tempted to set the location of this boundary at spatial infinity, but it has been shown \cite{GibbonsStewart} periodic geometries containing radiation are not asymptotically flat. This effectively limits our ability to describe the gravitational field at infinity via the linearized Einstein equations. Hence, the location of this artificial boundary is subject to two constraints: first, the energy content of the gravitational radiation within the region of space bounded by this surface must be much less than the energy content of the sources themselves; secondly, we demand in the vicinity of this boundary the gravitational field is described by the linearized Einstein equations. 

The following discussion is somewhat lengthy and mathematically intense. To help simplify matters, we delay any inclusion of the irrotational fluid until we have the full mathematical machinery of the variational principle at our disposal.

\subsection[The Initial Formulation of the Variational Principle]{The Initial Formulation of the Variational Principle}
\label{vpfoundation}
This section is devoted to deriving an expression we will eventually employ in our variational principle, as well as some mathematical conventions and other useful formulae. We begin by noting several useful conventions \cite{Det3} related to the variation of some of the fundamental geometrical quantities:
\begin{eqnarray}
\label{appa1}
  \delta \gamma_{ab} & \equiv & \delta ( \gamma_{ab} ), \\
  \delta \gamma^{ab}  & \equiv & \gamma^{ac} \gamma^{bd} \delta \gamma_{cd} = - \delta \left( \gamma^{ab} \right), \\
  \gamma & \equiv &  {\rm det} ( \gamma_{ab} ), \\
  \delta \gamma / \gamma  & = &  \gamma^{ab} \delta \gamma_{ab} = \delta \gamma^{\;a}_{a}.
\end{eqnarray}

The unit normal to a surface is defined by a scalar field $r = $ constant via
\begin{equation}
\label{appa2}
  n_{a} \equiv \rho D_{a} r,
\end{equation}
where in this context $\rho$ is not the fluid density, but is defined by
\begin{equation}
\label{appa3}
  \rho^{-2} \equiv \gamma^{ab} D_{a} r D_{b} r.
\end{equation}
It is important to note $r$ is not necessarily related to the flat space $r$ coordinate. The metric of a two-dimensional boundary, as well as the projection operator onto this boundary, is given by
\begin{equation}
\label{appa4}
  \sigma^{ab} \equiv \gamma^{ab} - n^{a} n^{b},
\end{equation}
and $\sigma$ is the determinant of $\sigma_{ab}$. We also note the useful relationship
\begin{equation}
\label{appa5}
  D_{a} n_{b} = \sigma_{ (a }^{\;\;c} \sigma_{ b) }^{\;\;d} D_{c} n_{d} - \rho^{-1} n_{a} \sigma_{b}^{\;d} D_{d} \rho.
\end{equation}

Some useful expressions for variations of the geometry, holding the location of the boundary fixed, are
\begin{equation}
\label{appa6}
  \delta ( n_{a} ) = \delta \rho D_{a} r = n_{a} \delta \rho / \rho,
\end{equation}
\begin{equation}
\label{appa7}
  n^{a} \delta( n_{a} ) = - \delta ( n^{a} ) n_{a} = \delta \rho / \rho,
\end{equation}
\begin{equation}
\label{appa8}
  \delta n^{a} \equiv \delta ( n^{a} ) = \sigma^{\;a}_{b} \delta n^{b} - n^{a} \delta \rho / \rho, 
\end{equation}
\begin{equation}
\label{appa9}
  \delta \left( \sigma^{ab} \right) n_{b} = 0,
\end{equation}
and finally
\begin{equation}
\label{appa10}
  \delta \left( \gamma^{ab} \right) = \delta \left( \sigma^{ab} \right) + n^{a} \sigma^{\;b}_{c} \delta n^{c} 
                                                  + n^{b} \sigma^{\;a}_{c} \delta n^{c} 
						  - 2 n^{a} n^{b} \delta \rho / \rho.
\end{equation}

We also note a variety of useful forms of Gauss' law:
\begin{eqnarray}
\label{appa11}
  \int \left( D_{b} A^{b} \right) \gamma^{1/2} d^{3} x & = & 
  \int \frac{\partial}{ \partial x^{b} } \left( \gamma^{1/2} A^{b} \right) d^{3} x
                                              =  \oint ( D_{b} r ) A^{b} \gamma^{1/2} d^{2} x \nonumber \\
					    & = & \oint \frac{ n_{b} A^{b} }{\rho} \gamma^{1/2} d^{2} x
					      = \oint n_{b} A^{b} \sigma^{1/2} d^{2} x,  
\end{eqnarray}
where we have employed Eq.~(\ref{appa2}) and used the fact that
\begin{equation}
\label{appa12}
  \gamma = \sigma \rho^{2}.
\end{equation}

We may now begin to lay the foundations of our variational principle. A variational principle for the Einstein equations may be constructed by integrating the scalar curvature of the four-metric \cite{MTW}. In the 3+1 formalism, this is expressed as
\begin{equation}
\label{appa13}
  16 \pi H_{0} \equiv - \int \left\{ N \left[ R + \gamma^{-1} \left( \frac{1}{2} \pi^{\;a}_{a}  \pi^{\;b}_{b} 
                              - \pi^{ab} \pi_{ab} \right) \right] 
                               + 2 N^{a} D_{b} \left( \gamma^{-1/2} \pi^{\;b}_{a} \right) \right\} \gamma^{1/2} d^{3} x,
\end{equation}
where the integrand of $16 \pi H_{0}$ is simply the 3+1 form of the Ricci scalar \cite{ADM}.

The scalar derived from the three-metric, $R = \gamma^{ab} R_{ab}$, has the following variation
\begin{equation}
\label{appa14}
  \delta R = - \delta \gamma_{ab} R^{ab} + D^{a} D^{b} \delta \gamma_{ab} - D^{a} D_{a} \delta \gamma_{c}^{\;c},
\end{equation}
which allows us to write the variations of the different parts of the integral given by Eq.~(\ref{appa13}). Specifically, if we allow variations of $N$, $N^{a}$, $\gamma_{ab}$, and $\pi^{ab}$, we find
\begin{eqnarray}
\label{appa15}
  \gamma^{-1/2} \delta \left( N R \gamma^{1/2} \right) & = & 
                        \delta N R - N \delta \gamma_{ab} \left( R^{ab} - \frac{1}{2} \gamma^{ab} R \right) \nonumber \\
                        & + & \delta \gamma_{ab} \left( D^{a} D^{b} N - \gamma^{ab} D^{c} D_{c} N \right) \nonumber \\
			                    & + & D^{a} \left( N D^{b} \delta \gamma_{ab} \right) 
					                  - D^{a} \left( N D_{a} \delta \gamma^{\;b} _{b} \right) \nonumber \\
					    & - & D^{a} \left( \delta \gamma_{ab} D^{b} N \right) 
					                  + D^{a} \left( \delta \gamma^{\;b}_{b} D_{a} N \right),
\end{eqnarray}
\begin{eqnarray}
\label{appa16}
  \gamma^{-1/2} \delta \left[ \frac{N}{\gamma} \left( \frac{1}{2} \pi^{\;a}_{a}  \pi^{\;b}_{b} 
                     - \pi^{ab} \pi_{ab} \right) \gamma^{1/2} \right]
     & = &  \left( \frac{\delta N}{\gamma} - \frac{ N \delta \gamma^{\;c}_{c} }{ 2 \gamma } \right) 
               \left( \frac{1}{2} \pi^{\;a}_{a}  \pi^{\;b}_{b} - \pi^{ab} \pi_{ab} \right) \nonumber \\
     & + & \frac{N}{\gamma} \left( \pi^{\;c}_{c} \pi^{ab} \delta \gamma_{ab} 
                                - 2 \pi^{ac} \pi^{\;b}_{c} \delta \gamma_{ab} \right) \nonumber \\
     & + & \frac{N}{\gamma} \left( \delta \pi^{ab} \gamma_{ab} \pi^{\;c}_{c} - 2 \delta \pi^{ab} \pi_{ab} \right),             
\end{eqnarray}
and finally
\begin{eqnarray}
\label{appa17}
  \gamma^{-1/2} \delta \left[ 2 N^{a} D_{b} \left( \pi^{\;b}_{a} / \gamma^{1/2} \right) \gamma^{1/2} \right]
     & = & 2 \delta N^{a} D_{b} \left( \pi^{\;b}_{a} / \gamma^{1/2} \right) 
     - 2 \delta \pi ^{ab} D_{a} N_{b} / \gamma^{1/2} \nonumber \\
     & - & \delta \gamma_{ab} \left[ 2 \pi^{bc} D_{c} N^{a} / \gamma^{1/2} 
     - D_{c} \left( N^{c} \pi^{ab} / \gamma^{1/2} \right) \right] \nonumber \\
     & - & D_{a} \left( N^{a} \pi^{bc} \delta \gamma_{bc} / \gamma^{1/2} \right) \nonumber \\
     & + & D_{a} \left( 2 N_{b} \delta \pi^{ab} / \gamma^{1/2} \right) \nonumber \\
     & + & D_{a} \left( 2 N^{b} \pi^{ac} \delta \gamma_{bc} / \gamma^{1/2} \right).
\end{eqnarray}
After substituting the above three variations into $16 \pi \delta H_{0}$, and applying Gauss' law, the total variation in $H_{0}$ becomes
\begin{eqnarray}
\label{appa18}
  16 \pi \delta H_{0} & = & - \int \left( \delta N {\cal N} \gamma^{1/2}+ 2 \delta N^{a} {\cal N}_{a} \gamma^{1/2}
                        + \delta \gamma_{ab} {\cal P}^{ab} - \delta \pi^{ab} {\cal G}_{ab}  \right) d^{3} x \nonumber \\           && +  \oint \left( n_{a} N^{a} \pi^{bc} \delta \gamma_{bc} / \gamma^{1/2}
	- 2 n_{a} N_{b} \delta \pi^{ab} / \gamma^{1/2} \right. \nonumber \\
	& &~~~~~~~~ \left. - 2 n_{a} N^{b} \pi^{ac} \delta \gamma_{bc} / \gamma^{1/2} \right) \sigma^{1/2} d^{2} x \nonumber \\
	&& + \oint \left( -N n^{a} D^{b} \delta \gamma_{ab} + N n^{a} D_{a} \delta \gamma^{\;b}_{b} \right. \nonumber \\
		& &~~~~~~~~ \left. + n^{a} \delta \gamma_{ab} D^{b} N 
				   - n^{a} \delta \gamma^{\;b}_{b} D_{a} N \right) \sigma^{1/2} d^{2} x.
\end{eqnarray}

It would be useful to re-express the boundary terms of $16 \pi \delta H_{0}$ in terms of only the variations of the geometry, and not derivatives of the variations \cite{Wald}. With this goal in mind, we proceed to manipulate the surface integrals of Eq.~(\ref{appa18}).

We begin by noting the surface integrals involving $\pi^{ab}$ may be rewritten as
\begin{eqnarray}
\label{appa19}
  16 \pi \delta H_{0}^{\pi \partial} & \equiv & \oint \left[ n_{a} N^{a} \pi^{bc} \delta \gamma_{bc} / \gamma^{1/2}
         + 2 n_{a} \delta N^{b} \pi^{\;a}_{b} / \gamma^{1/2} \right] \sigma^{1/2} d^{2} x \nonumber \\
         &&\; \;  -  2 \delta \left[ \oint \left( n_{a} N^{b} \pi^{\;a}_{b} / \gamma^{1/2} \right) \sigma^{1/2} d^{2} x \right].
\end{eqnarray}

Turning our attention to the surface terms involving $\delta \gamma_{ab}$, we note the first two terms of the last surface integral of Eq.~(\ref{appa18}) may be written as
\begin{eqnarray}
\label{appa20}
  \oint \left[ - N n^{a} \gamma^{bc} D_{b} \delta \gamma_{ac} 
       + N n^{a} \gamma^{bc} D_{a} \delta \gamma_{bc} \right] \sigma^{1/2} d^{2} x &&\nonumber \\
        =  \oint \left[ N n_{a} \sigma^{\;b}_{c} D_{b} \delta \left( \gamma^{ac} \right) \
       - N n^{a} \sigma_{bc} D_{a} \delta \right. \! \! \! \! \!&& \! \! \! \! \! \left. \left( \gamma^{bc} \right) \right] \sigma^{1/2} d^{2} x.
\end{eqnarray}
The first term on the right hand side of Eq.~(\ref{appa20}) yields
\begin{eqnarray}
\label{appa21}
  \oint  N n_{a} \sigma^{\;b}_{c} D_{b} \delta \left( \gamma^{ac} \right) \sigma^{1/2} d^{2} x & = & 
     \oint N \left[ - \sigma^{\;b}_{c} \delta \left( \sigma^{ac} \right) D_{b} n_{a} 
     + D^{'}_{c} \left( \sigma^{\;c}_{d} \delta n^{d} \right) \right. \nonumber \\
     && \left. \; \; \; \; \; - 2 \rho^{-1} \delta \rho \sigma^{bc} D_{b} n_{c} \right] \sigma^{1/2} d^{2} x,
\end{eqnarray}
where $D^{'}_{c}$ is the two dimensional derivative operator on the boundary. We may further simplify our expression by noting $N D^{'}_{c} ( \sigma^{\;c}_{d} \delta n^{d} ) = D^{'}_{c} ( N \sigma^{\;c}_{d} \delta n^{d} ) - \sigma^{\;c}_{d} \delta n^{d} D^{'}_{c} N$. The realization that the divergence term contributes to an integral of a two dimensional divergence over the two dimensional boundary, and is indeed zero by Stokes' theorem, allows us to ultimately express the first term of Eq.~(\ref{appa20}) as
\begin{eqnarray}
\label{appa22}
  \oint  N n_{a} \sigma^{\;b}_{c} D_{b} \delta ( \gamma^{ac} ) \sigma^{1/2} d^{2} x 
     & = & \oint [ - N \delta ( \sigma^{ab} ) D_{a} n_{b} - \sigma^{\;a}_{b} \delta n^{b} D_{a} N \nonumber \\
     & &~~~~~ - 2 N \rho^{-1} \delta \rho \sigma^{bc} D_{b} n_{c} ] \sigma^{1/2} d^{2} x.
\end{eqnarray}

Now, we write the second term of the right hand side of Eq.~(\ref{appa20}) as
\begin{eqnarray}
\label{appa23}
 \!\!\!\!\! - \oint N n^{a} \sigma_{bc} D_{a} \delta \left( \gamma^{bc} \right) \sigma^{1/2} d^{2} x 
   \!\!\!\!\!&=&\!\!\!\!\! - \oint N \left\{ n^{a} D_{a} \left[ \sigma_{bc} \delta ( \sigma^{bc} ) \right] \right. \nonumber \\
    && \,\,\,\, \left. + 2 n^{a} \sigma^{\;c}_{b} D_{a} \left( n_{c} \sigma^{\;b}_{d} \delta n^{d} \right) \right\}\sigma^{1/2} d^{2} x. \nonumber \\
\end{eqnarray}
Substituting $ 2 \delta ( \sigma^{ab} D_{a} n_{b} ) = - \gamma^{-1} \delta \gamma - D_{a} [ \sigma_{bc} \delta ( \sigma^{bc} ) n^{a} - 2 \sigma^{\;a}_{b} \delta n^{b} ] $ into the result of Eq.~(\ref{appa23}) yields
\begin{eqnarray}
\label{appa24}
  - \oint N n^{a} \sigma_{bc} D_{a} \delta ( \gamma^{bc} ) \sigma^{1/2} d^{2} x 
    & = & \oint N \left[ 2 \delta ( \sigma^{ab} D_{a} n_{b} ) + 2 \rho^{-1} \delta \rho D_{a} n^{a} \right. \nonumber \\
       & &~~~~ \left. - 2 \sigma^{\;a}_{c} D_{a} ( \sigma^{\;c}_{b} \delta n^{b}) \right] \sigma^{1/2} d^{2} x.
\end{eqnarray}
However, Stokes' theorem dictates
\begin{eqnarray}
\label{appa25}
  \oint N \left[ - 2 \sigma^{\;a}_{c} D_{a} ( \sigma^{\;c}_{b} \delta n^{b}) \right] \sigma^{1/2} d^{2} x
    & = & \oint \left[-2 D^{'}_{a} (N \sigma^{\;a}_{b} \delta n^{b} ) 
             + 2 \sigma^{\;a}_{b} \delta n^{b} D_{a} N \right] \sigma^{1/2} d^{2} x \nonumber \\
    & = & \oint \left[ 2 \sigma^{\;a}_{b} \delta n^{b} D_{a} N \right] \sigma^{1/2} d^{2} x. 
\end{eqnarray}
At long last, we may write
\begin{eqnarray}
\label{appa26}
  - \oint N n^{a} \sigma_{bc} D_{a} \delta \left( \gamma^{bc} \right) \sigma^{1/2} d^{2} x 
  & = & 2 \delta \left( \oint N D_{a} n^{a} \sigma^{1/2} d^{2} x \right) \nonumber \\
  & &\; \; + \oint \left[ (-N \sigma^{-1} \delta \sigma 
       - 2 \delta N + 2 N \rho^{-1} \delta \rho ) D_{a} n^{a} \right. \nonumber \\
  & & \; \; \; \; \; \; \left. + 2 \sigma^{\;a}_{b} \delta n^{b} D_{a} N \right] \sigma^{1/2} d^{2} x.
\end{eqnarray}

Finally, we may also note
\begin{eqnarray}
\label{appa27}
  \oint ( \sigma^{\;a}_{b} \delta n^{b} D_{a} N \!\!\!\!& + &\!\!\!\! n^{a} \delta \gamma_{ab} D^{b} N 
         - \gamma^{-1} \delta \gamma n^{a} D_{a} N ) \sigma^{1/2} d^{2} x \nonumber \\
	 & = & - \oint \sigma^{-1} \delta \sigma n^{a} D_{a} N \sigma^{1/2} d^{2} x.
\end{eqnarray}

Substituting Eqs.~(\ref{appa22}) and (\ref{appa26}) into Eq.~(\ref{appa20}), coupled with the usage of Eqs.~(\ref{appa12}) and (\ref{appa27}), we may reduce the $\delta \gamma$ surface terms of $16 \pi \delta H_{0}$ to
\begin{eqnarray}
\label{appa28}
 16 \pi \delta H^{\gamma \partial}_{0} & = & 2 \delta \left( \oint N D_{a} n^{a} \sigma^{1/2} d^{2} x \right) \nonumber \\
    & &\,\, - \oint \left[ N \delta ( \sigma^{ab} ) D_{a} n_{b} 
         + ( 2 \delta N + N \sigma^{-1} \delta \sigma ) D_{a} n^{a} \right. \nonumber \\
    & &\,\,\,\,\,\,\,\,\,\,\,\,\,\,\,\, \left. + \sigma^{-1} \delta \sigma n^{a} D_{a} N \right] \sigma^{1/2} d^{2} x. 
\end{eqnarray}

Using Eq.~(\ref{appa18}) with Eqs.~(\ref{appa19}) and (\ref{appa28}), we now define a new quantity $H_{1}$:
\begin{equation}
\label{vpfound2}
  16 \pi H_{1} = 16 \pi H_{0} + \oint 2 n^{a} N^{b} \gamma^{-1/2} \pi_{ab} \sigma^{1/2} d^{2} x 
                        - \oint 2 N D_{a} n^{a} \sigma^{1/2} d^{2} x.
\end{equation}
We are ultimately interested in how $H_{1}$ varies under arbitrary, infinitesimal variations of the geometric quantities $N$, $N^{a}$, $\gamma_{ab}$ and $\pi^{ab}$, while the location of the boundary $\Sigma_{t}$ remains fixed. The resulting variation is
\begin{eqnarray}
\label{vpfound4}
  16 \pi \delta H_{1} & = & - \int \left( \delta N {\cal N} \gamma^{1/2}
                                  + 2 \delta N^{a} {\cal N}_{a} \gamma^{1/2}
				  + \delta \gamma_{ab} {\cal P}^{ab}
				  - \delta \pi^{ab} {\cal G}_{ab} \right) d^{3}x \nonumber \\
			& &~ - \oint \left\{ \delta \sigma^{ab} [ N D_{a} n_{b} - \sigma_{ab} D_{c} (N n^{c}) ] 
			           + 2 \delta N D_{a} n^{a} \right\} d^{2}x \nonumber \\
			& &~ + \oint \left( n_{a} N^{a} \pi^{bc} \gamma^{-1/2} \delta \gamma_{bc} 
			           + 2 n^{a} \delta N^{b} \gamma^{-1/2} \pi_{ab} \right) \sigma^{1/2} d^{2}x.  
\end{eqnarray}

Equation ~(\ref{vpfound4}) indicates how $H_{1}$ may be used in a variational principle. If the boundary conditions of the system specify the values of $\sigma^{ab}, N, N^{a}$, and if the normal vector $n_{a}$ is perpendicular to the shift vector $N^{a}$ \cite{Hayward}, then all of the surface integrals of $\delta H_{1}$ vanish. It then follows that $H_{1}$ is an extremum under arbitrary, infinitesimal variations of $N$, $N^{a}$, $\gamma_{ab}$ and $\pi^{ab}$ if and only if the variation of these quantities is about a solution to the quasi-stationary vacuum Einstein equations.

In principle, it is perfectly legitimate to specify the values of $\sigma^{ab}, N, N^{a}$ and $n_{a}$ on the boundary, but in practice this boundary data does not have a direct physical interpretation in the sense of being directly correlated to quantities such as angular momentum or mass. Also, it may not be an ``easy'' task to determine the values of these geometrical quantities which are to be specified on the boundary. If we are to seek an eloquent, physically intuitive variational principle, we must exert some effort and re-formulate the variational principle for $H_{1}$. The quest for more physical insight leads us into a discussion of symmetric trace-free tensors, the very nature of the quasi-equilibrium approximation, and a generalized solution to the linearized Einstein equations credited to Thorne.

\subsection[Symmetric Trace-Free Tensors]{Symmetric Trace-Free Tensors}

We begin our efforts to re-formulate our variational principle for $H_{1}$ by taking inspiration from Thorne's generalized solution of the linearized Einstein equations \cite{Thorne1}. Thorne chooses a particular gauge and proceeds to decompose his solution in terms of symmetric trace-free tensors. Let us first discuss some of the properties of symmetric trace-free (or STF) tensors.

Tensors with a large number of indices, say $\ell$, are common. We introduce a convenient short-hand notation as
\begin{eqnarray}
\label{STF1}
  T_{L} \equiv T_{j_{1} j_{2} \ldots j_{\ell}},  \;\;\;\; 
  T_{L-2} \equiv T_{j_{1} j_{2} \ldots j_{\ell - 2}}. 
\end{eqnarray}
The indices represent tensor components in a Cartesian coordinate system, and the tensors are both symmetric on all pairs of indices and are trace-free. The Cartesian components of these STF tensors are functions of $t$ and $r$ only, are independent of $\theta$ and $\phi$, and are denoted with capital script letters. For example,
\begin{eqnarray}
\label{STF2}
  {\cal T}_{L} = {\cal T}_{j_{1} j_{2} \ldots j_{\ell}} = {\cal T}_{ ( j_{1} j_{2} \ldots j_{\ell}) }, \;\;\;\;
  {\cal T}_{i i L-2} = 0  
\end{eqnarray}
where
\begin{eqnarray}
\label{STF3}
  \partial \cal{T}\it_{L} / \partial \theta = {\rm 0}, \;\;\;\;
   \partial \cal{T}\it_{L} / \partial \phi = {\rm 0}.  
\end{eqnarray}

It is also useful to use the abbreviation
\begin{equation}
\label{STF4}
  n_{L} \equiv n_{j_{1}} n_{j_{2}} \ldots n_{j_{\ell}}
\end{equation}
to represent the outer product of $\ell$ unit radial vectors.

Another useful property of the STF tensors which are only dependent on $t$ and $r$ is
\begin{equation}
\label{STF5}
  \oint {\cal A}_{L} n^{L} {\cal B}_{L'} n^{L'} d \Omega 
                 = \frac{ 4 \pi \ell !}{ ( 2 \ell + 1)!! } {\cal A}_{L} {\cal B}_{L} \delta_{\ell \ell'}
\end{equation}
for arbitrary STF tensors $\cal{A}\it_{L}$ and $\cal{B}\it_{L}$, where the repeated multi-index $L$ implies summation and the integral is performed over a unit sphere with area element $d \Omega$.

A convenient representation of a STF tensor on a two-sphere with $\ell$ indices is via a basis set of $2 \ell + 1$ tensors $\cal{Y}\it^{\ell m}_{L}$, with $ - \ell \leq m \leq \ell$, as defined by Thorne for a Cartesian coordinate system.  It is interesting to note
\begin{equation}
\label{STF6}
  \cal{Y}\it^{\ell m}_{L} n^{L} = Y ^{\ell m},
\end{equation}
where $Y^{\ell m}$ is the well-known spherical harmonic function. It also follows
\begin{equation}
\label{STF7}
  \cal{Y}\it^{\ell m}_{L} = (\rm{-1}\it)^{m} (\cal{Y}\it^{\ell ~-m}_{L})^{\ast},
\end{equation}
and that the $\cal{Y}\it^{\ell m}_{L}$ are orthogonal,
\begin{equation}
\label{STF8}
  {\cal Y}^{\ell m_{1}}_{L} {\cal Y}^{\ell m_{2} \ast}_{L} 
                 = \frac{( 2 \ell + 1)!!}{ 4 \pi \ell !} \delta^{ m_{1} m_{2} }.
\end{equation}

We also note the relationship between the $\cal{Y}\it^{\ell m}_{L}$ for two different Cartesian coordinate systems, where the primed coordinates are rotated by an angle $\Phi$ about the z-axis with respect to the unprimed coordinates:
\begin{equation}
\label{STF9}
  \cal{Y}\it^{\ell m'}_{L} = e^{ - i m \Phi } \cal{Y}\it^{\ell m}_{L}.
\end{equation}

With a basis set of STF tensors given by $\cal{Y}\it^{\ell m}_{L}$ we may decompose $\cal{T}\it_{L}(t,r)$ as
\begin{equation}
\label{STF10}
 \cal{T}\it_{L}(t,r) = \sum^{\ell}_{m = - \ell} T_{\ell m}(t,r) \cal{Y}\it^{\ell m}_{L},
\end{equation}
where the coefficients $T_{\ell m}(t,r)$ are determined from the orthogonality condition described by Eq.~(\ref{STF8}).

If our STF tensor is a combination of ingoing and outgoing radiation, then the radiative terms may be separated as
\begin{equation}
\label{STF11}
  {\cal T}_{L}(t,r) = {\cal T}^{\,\rm{in}\it}_{L}(t+r) + {\cal T}^{\,\rm{out}\it}_{L}(t-r),
\end{equation}
and we may naturally decompose the individual radiative terms as
\begin{equation}
\label{STF12}
 \cal{T}\it^{\,\rm{in/out}\it}_{L} (t \pm r) = \sum^{\ell}_{m = - \ell} T^{\,\rm{in/out}\it}_{\ell m}(t \pm r) \cal{Y}\it^{\ell m}_{L}.
\end{equation}

If $T_{L}(t,r)$ is quasi-stationary in a rotating frame of reference, where $\phi' = \phi - \Omega t$, then
\begin{equation}
\label{STF13}
  T^{\,\rm{in/out}\it}_{\ell m} (t \pm r) = \frac{1}{2} T_{\ell m} e^{ [ \pm i \vartheta_{\ell m} + i \omega_{m} (t \pm r) ] },
\end{equation}
for some constant complex amplitudes $T_{\ell m}$ and phases $\vartheta_{\ell m}$, where $\omega_{m} = - m \Omega$. If the waves have equal amounts of ingoing and outgoing radiation for every value of $\ell$ and $m$, then the phases $\vartheta_{\ell m}$ are real. If the STF tensor $\cal{T}\it_{L}$ is itself real, this implies
\begin{equation}
\label{STF14}
  \vartheta_{\ell , m} = - \vartheta_{\ell, -m}
\end{equation}
and
\begin{equation}
\label{STF15}
  T_{\ell, m} = (-1)^{m} T^{\ast}_{\ell, -m}.
\end{equation}

The usefulness of the STF decomposition will become more apparent as we proceed through our analysis. Until then, it is important to point out that if $\cal{T}\it_{L}(t,r)$ is a real valued tensor field which is quasi-stationary when viewed in a rotating frame, and if it has equal ingoing and outgoing radiative terms, then the field is completely described in terms of the amplitudes $T_{\ell m}$ and phases $\vartheta_{\ell m}$, subject to Eq.~(\ref{STF14}) and Eq.~(\ref{STF15}). For this scenario, $\cal{T}\it_{L}(t,r)$ is described by
\begin{equation}
\label{STF16}
  \cal{T}\it_{L}(t,r) = \sum^{\ell}_{m = -\ell} T_{\ell m} \cal{Y}\it^{\ell m}_{L} 
                                              \cos(\omega_{m} r + \vartheta_{\ell m}) e^{i \omega_{m} t}.
\end{equation}
The above decomposition for ${\cal T}_{L}(t,r)$ holds for all values of $m$.

\subsection[Thorne's Generalized Solution]{Thorne's Generalized Solution to the Linearized \\Einstein Equations}
\label{thornesoln}
Recall the decomposition of the STF tensor fields, described by Eq.~(\ref{STF16}), is made possible by the introduction of a region in which ingoing radiation is carefully matched with the amplitude and phases of the outgoing radiation. Also recall we stipulated the location of this region is subject to the condition that the energy content of the gravitational waves within this volume of space is much less than that of the sources alone. In addition to this, we also require the gravitational field to be described by the linearized Einstein equations. 

Thorne \cite{Thorne1} gives a general solution to the metric perturbation $g^{1}_{ab}$ in the DeDonder gauge. He expresses his solution in terms of the retarded mass and current moments, $\cal{I}\it_{L}(t-r)$ and $\cal{S}\it_{L}(t-r)$. For our system in which the metric is quasi-stationary when viewed from a rotating frame of reference and the ingoing and outgoing gravitational radiation have equal amplitudes and phases, we may express the mass and current moments in terms of the amplitudes, $I_{\ell m}$ and $S_{\ell m}$, and the respective phases $\vartheta^{I}_{\ell m}$ and $\vartheta^{S}_{\ell m}$, subject to the conditions laid out by Eq.~(\ref{STF14}) and Eq.~(\ref{STF15}). For this solution, the mass monopole is denoted as $I_{0,0}$. In this mass-centered coordinate system $I_{1,m} = 0$, $S_{1,0}$ is the angular momentum and $S_{1,\pm 1} = 0$. In terms of the amplitudes and the phases, Thorne's solution is
\begin{equation}
\label{thorne1}
  g^{1}_{00} = 2 \sum_{\ell m} \frac{ (-1)^{\ell} }{ \ell ! } I_{\ell m} {\cal Y}^{\ell m}_{L}
                  \nabla_{L} \left[ r^{-1} \cos \left(\omega_{m} r + \vartheta^{I}_{\ell m} \right) \right] e^{i \omega_{m} t},
\end{equation}
\begin{eqnarray}
\label{thorne2}
  g^{1}_{0j} & = & 4 \sum_{\ell m} \frac{ (-1)^{\ell} }{\ell !} i \omega_{m} I_{\ell m} {\cal Y}^{\ell m}_{j L-\rm{1}\it}
      \nabla_{L-\rm{1}\it} \left[ r^{-\rm{1}\it} \cos \left(\omega_{m} r + \vartheta^{I}_{\ell m} \right) \right] 
             e^{i \omega_{m} t} \nonumber \\
      &&-4 \sum_{\ell m} \frac{(-1)^{\ell} \ell}{ (\ell + 1)! } S_{\ell m } \epsilon_{jpq} 
	{\cal Y}^{\ell m}_{p L-\rm{1}\it} 
	\nabla_{q L-\rm{1}\it} \left[ r^{-\rm{1}\it} \cos \left(\omega_{m} r + \vartheta^{S}_{\ell m}\right) \right] 
	e^{i \omega_{m} t}, \nonumber \\
\end{eqnarray}
and
\begin{eqnarray}
\label{thorne3}
  g^{1}_{ij} & = & \sum_{\ell m} \frac{ (-1)^{\ell} }{ \ell ! } 
         I_{\ell m} \left\{ 2 f_{ij} {\cal Y}^{\ell m}_{L} \nabla_{L} 
         - \rm{4}\it \omega^{\rm{2}\it}_{m} {\cal Y}^{\ell m}_{ij L-\rm{2}\it} \nabla_{L-\rm{2}\it} \right\}
         \left[ r^{-\rm{1}\it} \cos \left(\omega_{m} r + \vartheta^{I}_{\ell m}\right) \right] e^{i \omega_{m} t} \nonumber \\
         & & + 8 \sum_{\ell m} \frac{(-1)^{\ell + 1} \ell}{ (\ell + 1)! } 
	     i \omega_{m} S_{\ell m} \epsilon_{pq(i} {\cal Y}^{\ell m}_{j)p L-\rm{2}\it}
	     \nabla_{q L-\rm{2}\it} \left[ r^{-\rm{1}\it} \cos \left(\omega_{m} r + \vartheta^{S}_{\ell m} \right)                    \right] e^{i \omega_{m} t}. \nonumber \\ 
\end{eqnarray}
In the above equations, $f_{ij}$ denotes the flat space three-metric.

We choose a slightly different gauge in which all of the time dependent terms, those with $m \neq 0$, are removed from $g^{1}_{00}$ and $g^{1}_{0j}$. A particular choice of gauge field \cite{Det2} $\xi_{a}$ modifies the metric according to $g^{1\, \rm{new}\it}_{ab} = g^{1\, \rm{old}\it}_{ab} - 2 \nabla_{(a} \xi_{b)}$, where
\begin{eqnarray}
\label{thorne4}
  \partial \xi_{t} / \partial t & = & \frac{1}{2} [m \neq 0 ~\rm{part ~of}\it~ g^{\rm{1}\it}_{\rm 00}]  \nonumber \\
                & = & \sum_{\ell, m \neq 0} \frac{ (-\rm{1}\it)^{\ell} }{\ell !}
		I_{\ell m} {\cal Y}^{\ell m}_{L}
		\nabla_{L} \left[ r^{-1} \cos \left(\omega_{m} r + \vartheta^{I}_{\ell m} \right) \right] e^{i \omega_{m} t}
\end{eqnarray}
and
\begin{equation}
\label{thorne5}
  \partial \xi_{j} / \partial t = - \nabla_{j} \xi_{t} - [m \neq 0 ~\rm{part ~of}\it~ g^{\rm{1}\it}_{{\rm 0}j}].
\end{equation}

With this particular choice of gauge, we may write the lapse function $N$ as
\begin{equation}
\label{thorne6}
  N = (-g^{00})^{-1/2} = 1 - I/2,
\end{equation}
where
\begin{equation}
\label{thorne7}
  I \equiv \frac{2 I_{0,0} }{r} 
            + \sum^{\infty}_{\ell = 2} \frac{2(2 \ell - 1)!!}{\ell ! r^{\ell + 1}} I_{\ell 0} Y^{\ell 0}.
\end{equation}

We also define the shift vector $S^{j} \equiv N^{2} g^{0j}$, and our gauge choice reduces it to
\begin{equation}
\label{thorne8}
  S^{j} = \frac{-2 f^{ji} \epsilon_{ipq} S_{1,0} \cal{Y}\it^{\rm{1,0}\it}_{p} n_{q} }{ r^{2} }
            - f^{ji} \sum^{\infty}_{\ell = 2} \frac{4 \ell ( 2 \ell - 1)!! }{ (\ell + 1)! r^{\ell + 1} }
	         \epsilon_{ipq} S_{\ell 0} \cal{Y}\it^{\ell 0}_{p L-\rm{1}\it} n_{q L-\rm{1}\it}. 
\end{equation}

The three-metric $\gamma_{ij}$ is given by
\begin{equation}
\label{thorne9}
  \gamma_{ij} = (1 + I) f_{ij} + h_{ij},
\end{equation}
where all of the information pertaining to the radiation is contained within $h_{ij}$. To facilitate the description of $h_{ij}$, it is useful to introduce a basis set of solutions $H^{\ell m Q}_{ij}$ which obey the tensor wave equation in flat space
\begin{equation}
\label{thorne10}
  \nabla_{k} \nabla^{k} H^{\ell m Q}_{ij} + \omega^{2}_{m} H^{\ell m Q}_{ij} =0,
\end{equation}
where $Q$ runs over $I$ and $S$. This basis set of solutions are also transverse
\begin{equation}
\label{thorne11}
  \nabla^{i} H^{\ell m Q}_{ij} = 0,
\end{equation}
and trace-free
\begin{equation}
\label{thorne12}
  f^{ij} H^{\ell m Q}_{ij} = 0.
\end{equation}

Written out explicitly, this basis set of solutions is
\begin{eqnarray}
\label{thorne13}
  H^{\ell m I}_{ij} & = & \frac{ 2 (-1)^{\ell} }{\ell !} 
          \left[ f_{ij} {\cal Y}^{\ell m}_{L} \nabla_{L} 
	  - 2 \omega^{2}_{m} {\cal Y}^{\ell m}_{ij L-2\it} \nabla_{L-2} \right. \nonumber \\
	  & & \left. - \omega^{-2}_{m} {\cal Y}^{\ell m}_{L} \nabla_{L ij}
			  - 4 {\cal Y}^{\ell m}_{L-1 (i} \nabla_{j) L-1} \right] 
			    \left[ r^{-1} \cos \left(\omega_{m} r + \vartheta^{I}_{\ell m} \right) \right] \nonumber \\
\end{eqnarray}
and
\begin{eqnarray}
\label{thorne15}
  H^{\ell m S}_{ij} & = & \frac{ 8 (-1)^{\ell + 1} \ell }{ (\ell + 1)! } 
       \left[ i \omega_{m} \epsilon_{pq(i} \cal{Y}\it^{\ell m}_{j)p L-\rm{2}\it} \nabla_{q L-\rm{2}\it} \right. \nonumber \\
	& & \left. - (i \omega_{m})^{-1} \cal{Y}\it^{\ell m}_{p L-\rm{1}\it} \epsilon_{pq(i} \nabla_{j)q L-\rm{1}\it} \right] 
                           \left[ r^{-\rm{1}\it} \cos \left(\omega_{m} r + \vartheta^{S}_{\ell m} \right) \right]. \nonumber \\
\end{eqnarray}

In the local wave zone, where the radiation can be regarded as a small perturbation about a flat background metric, the leading $1/r$ behavior of the basis solutions is
\begin{equation}
\label{thorne16}
  H^{\ell m I}_{ij} = \frac{ 4(\omega_{m})^{\ell} }{ r \ell ! } 
                         \left( \sigma^{p}_{(i} \sigma^{q}_{j)} - \frac{1}{2} \sigma^{pq} \sigma_{ij} \right) 
			 \cal{Y}\it^{\ell m}_{pq L-\rm{2}\it} n_{L-\rm{2}\it} 
			 \cos \left(\omega_{m} r + \vartheta^{I}_{\ell m} - \ell \pi / \rm{2}\it \right)
\end{equation}
and
\begin{equation}
\label{thorne17}
  H^{\ell m S}_{ij} = \frac{ 8 \ell (\omega_{m})^{\ell} }{ r (\ell + 1)! } 
                         \sigma^{p}_{(i} \sigma^{q}_{j)} \epsilon_{knp} \cal{Y}\it^{\ell m}_{qk L-\rm{2}\it} n_{n L-\rm{2}\it}
			 \cos \left(\omega_{m} r + \vartheta^{S}_{\ell m} - \ell \pi / \rm{2}\it \right). 
\end{equation}

We complete the decomposition of the radiative part of the metric $h_{ij}$ by relating it to the basis solutions via
\begin{equation}
\label{thorne18}
  h_{ij} = \sum_{Q; \ell; m \neq 0} h^{\ell m Q}_{ij} e^{ i \omega_{m} t },
\end{equation}
where
\begin{equation}
\label{thorne19}
  h^{\ell m Q}_{ij} = Q_{\ell m} H^{\ell m Q}_{ij}.
\end{equation}

With our choice of gauge, a number of useful identities hold at linearized order:
\begin{equation}
\label{thorne20}
  \nabla_{i} h^{i}_{j} = 0,
\end{equation}
\begin{equation}
\label{thorne21}
  f^{ij} h_{ij} = 0,
\end{equation}  
\begin{equation}  
\label{thorne22}
  \nabla_{k} \nabla^{k} I = 0,
\end{equation}   
\begin{equation}    
\label{thorne23}
  n_{i} S^{i} = 0,
\end{equation}   
\begin{equation}   
\label{thorne24}
  \nabla_{i} S^{i} = 0,
\end{equation}
\begin{equation}   
\label{thorne25}
  \nabla_{k} \nabla^{k} S^{i} = 0,
\end{equation}
and
\begin{equation}   
\label{thorne26}
  \nabla_{k} \nabla^{k} h_{ij} - \frac{ \partial^{2} h_{ij} }{\partial t^{2}} = 0.
\end{equation}

With this particular choice of gauge, we also note the linearized geometry's extrinsic curvature ${\cal K}_{ij}$ obeys the following equations
\begin{equation}
\label{thorne27}
  {\cal K}_{ij} = -\frac{1}{2} \frac{\partial h_{ij}}{\partial t} + \nabla_{(i} S_{j)},
\end{equation}
\begin{equation}
\label{thorne28}
  f^{ij} {\cal K}_{ij} = 0,
\end{equation}
and finally
\begin{equation}
\label{thorne29}
  \pi^{ij} = - \gamma^{1/2} {\cal K}^{ij}.
\end{equation}

We use several of the results from this section to aid in the evaluation of the surface integrals of our variational principle, given by Eq.~(\ref{vpfound4}). The following section will be dedicated to the explicit evaluation of these surface terms.

\subsection{Evaluation of the Surface Terms}
\label{surfterms}
Let us refresh our memory as to the purpose of studying Thorne's generalized solution to the linearized Einstein equations, and our introduction of the concept of the weak field zone boundary. In \S (\ref{vpfoundation}), we defined a function $H_{1}$ to be
\begin{eqnarray}
\label{surf1}
  16 \pi H_{1} & \equiv & - \int \left(N \cal{N}\it 
  + \rm{2}\it N^{a} \cal{N}\it_{a} \right) \gamma^{\rm{1/2}\it} d^{\rm{3}\it}x \nonumber \\
                      & &~~ + \oint 2 n^{a} N^{b} \gamma^{-1/2} \pi_{ab} \sigma^{1/2} d^{2}x   \nonumber \\
		      & &~~ - \oint 2 N D_{a} n^{a} \sigma^{1/2} d^{2}x,                        
\end{eqnarray} 
and the total variation of $H_{1}$ resulted in
\begin{eqnarray}
\label{surf2}
  16 \pi \delta H_{1}& = & - \int \left( \delta N \cal{N}\it \gamma^{\rm{1/2}\it}
                                  + \rm{2}\it \delta N^{a} \cal{N}\it_{a} \gamma^{\rm{1/2}\it}
				  + \delta \gamma_{ab} \cal{P}\it^{ab}
				  - \delta \pi^{ab} \cal{G}\it_{ab} \right) d^{\rm{3}\it}x \nonumber \\
			& &~ - \oint \left\{ \delta \sigma^{ab} \left[ N D_{a} n_{b} - \sigma_{ab} D_{c} (N n^{c}) \right] 
			           + 2 \delta N D_{a} n^{a} \right\} d^{2}x \nonumber \nonumber \\
			& &~ + \oint \left( n_{a} N^{a} \pi^{bc} \gamma^{-1/2} \delta \gamma_{bc}
			           + 2 n^{a} \delta N^{b} \gamma^{-1/2} \pi_{ab} \right) \sigma^{1/2} d^{2}x.  
\end{eqnarray}

Recall a question arose as to the feasibility of holding geometrical quantities, such as the lapse $N$ or the two-metric $\sigma_{ab}$, fixed on the boundary $\Sigma_{t}$. As stated above, it may be more desirable to specify the value of physical observables on the boundary. We now use the linearized solution described in \S (\ref{thornesoln}) to rewrite the surface integrals in Eq.~(\ref{surf2}), in hopes that a more physically intuitive variational principle will result.

We first mention some notational conventions. At the location of the boundary, which is in the weak field zone, the geometry is nearly flat and is described by the linearized Einstein equations. In this region, indices are raised and lowered by the flat space metric $f_{ij}$, and summation is implied for repeated indices. It is important to note some repeated indices may appear as both being raised or both lowered, but the summation convention is still implied. The flat space derivative operator is denoted by $\nabla^{0}_{i}$, where the sub- or super-script $0$ denotes a quantity associated with the flat space metric. We further assume our weak field boundary is spherically symmetric, and is located at a constant Cartesian radius of $r$. The outward pointing unit normal to this surface is denoted as $n^{0}_{i} \equiv \nabla^{0}_{i} r$, which is not to be confused with the vector $n_{a}$ which is normalized with the three-metric $\gamma_{ab}$. Likewise, the two-metric associated with the spherical boundary embedded in flat space is $\sigma^{0}_{ij} = r \nabla^{0}_{i} n^{0}_{j}$, and has determinant $\sigma_{0}$. Often times, quantities are contracted with the flat space vector $n^{0}_{i}$, in which case a subscript $r$ is used ( for example, $h_{ir} = h_{ij} n^{j}_{0}$ ). Also, the radial partial derivative is denoted as $\partial_{r} = n^{i}_{0} \nabla^{0}_{i}$. We also note the shift vector in an inertial reference frame is $S^{i}$, defined by Eq.~(\ref{thorne8}). In a frame rotating at a rate of $\Omega$, the rotating shift vector may be written as 
\begin{equation}
\label{surf3}
  N^{i} = \Omega \Phi^{i} + S^{i},
\end{equation}
where $\Phi^{i} \partial / \partial x^{i} = \partial / \partial \phi$. 

We may now turn our attention to the surface integrals of the variation of $H_{1}$, given in Eq.~(\ref{surf2}). The first surface integral of Eq.~(\ref{surf2}) may be rewritten as
\begin{eqnarray}
\label{surf4}
  & & - \oint_{r_{0}} \left\{ \delta \sigma^{ab} \left[ N D_{a} n_{b} - \sigma_{ab} D_{c} (N n^{c}) \right] 
	 + 2 \delta N D_{a} n^{a} \right\} d^{2}x \nonumber \\
	 & &~~ = - \delta \left( \oint_{r_{0}} N^2 \nabla^{0}_{i} n^{i}_{0} \sigma^{1/2} d^{2}x \right) \nonumber \\
	 & &~~~~ - \oint_{r_{0}} \left[ \delta \left( \sigma^{1/2} \sigma^{ab} \right) N \left( D_{a} n_{b} 
	 - \frac{1}{2} \sigma_{ab} D_{c} n^{c} \right) \right. \nonumber \\
	 & &~~~~~~~ \left. + \delta \left( \sigma^{1/2} N^{2} \right) \left( N^{-1} D_{a} n^{a} 
	 - \nabla^{0}_{i} n^{i}_{0} \right) + 2 \delta \left( \sigma^{1/2} \right) n^{a} D_{a} N \right] d^{2}x.\nonumber \\
\end{eqnarray}

The total variation of the surface integral on the right hand side will eventually be absorbed into a new definition for $H_{1}$. The remaining terms are of second order in the wave amplitudes $I_{\ell m}$ and $S_{\ell m}$, and will require some effort to rewrite.

First, we rewrite the term involving $n^{a} \delta N^{b} \pi_{ab}$ in Eq.~(\ref{surf2}). Specifically,
\begin{eqnarray}
\label{surf5}
 \lefteqn{ \oint_{r_{0}} 2 n^{a} \delta N^{b} \gamma^{-1/2} \pi_{ab} \sigma^{1/2} d^{2}x } & & \nonumber \\
     &&~~ = \delta \left( 2 \Omega \oint_{r_{0}} n^{a} \Phi^{b} \gamma^{-1/2} \pi_{ab} \sigma^{1/2} d^{2}x \right) \nonumber \\
       &&~~~~ + 16 \pi \Omega \delta J 
            + \oint_{r_{0}} 2 n^{a} \delta S^{b} \gamma^{-1/2} \pi_{ab} \sigma^{1/2} d^{2}x, 
\end{eqnarray}
where
\begin{equation}
\label{surf6}
  16 \pi J \equiv - \oint_{r_{0}} 2 n^{a} \Phi^{b} \gamma^{-1/2} \pi_{ab} \sigma^{1/2} d^{2}x.
\end{equation}

In the vicinity of the weak field zone boundary, $J$ is approximately the angular momentum $S_{10}$ of the linearized geometry up to first order in the wave amplitudes. Once again, the total variation of the surface integral on the right hand side of Eq.~(\ref{surf5}) will be absorbed into a new definition of $H_{1}$, and the remaining surface terms are of second order in the wave amplitudes.

We now define a new function $H_{2}$, which is similar to $H_{1}$, but includes the total variations of Eq.~(\ref{surf4}) and Eq.~(\ref{surf5}):
\begin{eqnarray}
\label{surf7}
  16 \pi H_{2} & \equiv & - \int \left(N \cal{N}\it 
            + \rm{2}\it N^{a} \cal{N}\it_{a} \right) \gamma^{\rm{1/2}\it} d^{\rm{3}\it}x \nonumber \\
            & &~ - \oint_{r_{0}} N \left( 2 D_{a} n^{a} - N \nabla^{0}_{i} n^{i}_{0} \right) \sigma^{1/2} d^{2}x  \nonumber \\
	    & &~ + \oint_{r_{0}} \nabla^{0}_{i} n^{i}_{0} \sigma^{1/2}_{0}  d^{2}x  \nonumber \\
	    & &~ + \oint_{r_{0}} 2 n^{a} S^{b} \gamma^{-1/2} \pi_{ab} \sigma^{1/2} d^{2}x.
\end{eqnarray}
The second surface integral evaluates to $8 \pi r_{0}$ and does not change under a variation. It is included so the surface integrals of $H_{2}$ evaluate to the mass monopole $M = I_{00}$ at first order in the wave amplitudes.

Arbitrary, infinitesimal variations in $N$, $N^{a}$, $\gamma_{ab}$, and $\pi^{ab}$ result in a variation in $H_{2}$, given by
\begin{eqnarray}
\label{surf8}
  16 \pi \delta H_{2} & = & - \int \left[\delta N \cal{N}\it \gamma^{\rm{1/2}\it}
                                  + \rm{2}\it \delta N^{a} \cal{N}\it_{a} \gamma^{\rm{1/2}\it}
				  + \delta \gamma_{ab} \cal{P}\it^{ab}
				  - \delta \pi^{ab} \cal{G}\it_{ab} \right] d^{\rm{3}\it}x  \nonumber \\
        & & - \oint_{r_{0}} \left[ \delta \left( \sigma^{1/2} \sigma^{ab} \right) N \left( D_{a} n_{b} 
	                                - \frac{1}{2} \sigma_{ab} D_{c} n^{c}  \right)  \right. \nonumber \\
	& &~~~~ + \delta \left( \sigma^{1/2} N^2 \right) \left( N^{-1} D_{a} n^{a} 
	                              - \nabla^{0}_{i} n^{i}_{0} \right)  \nonumber \\
	& &~~~~ \left. + 2 \delta \left( \sigma^{1/2} \right) n^{a} D_{a} N \right] d^{2}x \nonumber \\
	& & + \oint_{r_{0}} \left[ n_{a} N^{a} \gamma^{-1/2} \pi^{bc} \delta \gamma_{bc} 
	   + 2 n^{a} \delta S^{b} \gamma^{-1/2} \pi_{ab} \right] \sigma^{1/2} d^{2}x  \nonumber \\
	& & + 16 \pi \Omega \delta J.
\end{eqnarray}
Note that each of the terms in the surface integrals are of second order in the wave amplitudes $I_{\ell m}$ and $S_{\ell m}$.

At this point, we now attempt to rewrite the surface integrals in terms of the parameters of the linearized geometry, namely $I_{\ell m}$, $S_{\ell m}$, $\vartheta^{I}_{\ell m}$, $\vartheta^{S}_{\ell m}$, $\Omega$, and the variations of these parameters.

Let us first note the following relationship, valid up to first order in the linearized parameters:
\begin{equation}
\label{surf9}
  \gamma^{-1/2} \pi_{ab} = \frac{1}{2} \frac{\partial h_{ab}}{\partial t} - \nabla_{(a} S_{b)}.
\end{equation} 
We now substitute Eq.~(\ref{surf9}) into the last surface term of Eq.~(\ref{surf8}) to yield
\begin{equation}
\label{surf10}
  \oint_{r_{0}} 2 n^{a} \delta S^{b} \gamma^{-1/2} \pi_{ab} \sigma^{1/2} d^{2}x
         = - \oint_{r_{0}} n^{j} \delta S^{i} \left( \nabla_{i} S_{j} + \nabla_{j} S_{i} \right) \sigma^{1/2} d^{2}x.
\end{equation}

Note the $\partial h_{ij} / \partial t$ term vanishes on integration because $S^{i}$ is axisymmetric ($m=0$), but $h_{ij}$ is necessarily not axisymmetric ($m\neq0$). Using the fact $n_{i} S^{i} = 0$ in the linearized geometry, and performing an integration by parts, we find Eq.~(\ref{surf10}) may be written as
\begin{equation}
\label{surf11}
  \oint_{r_{0}} \left( \delta S^{i} S_{i} / r - \delta S^{i} n^{j} \nabla_{j} S_{i} \right) \sigma^{1/2} d^{2}x.
\end{equation}

Using the linearized form of the shift vector $S^{i}$ given by Eq.~(\ref{thorne8}), and the orthogonality condition of the $\cal{Y}\it^{\ell m}_{L}$ given by Eq.~(\ref{STF8}), we find the second term on the right hand side of the above equation is symmetric in $\delta S^{i}$, so we may ultimately write
\begin{equation}
\label{surf12}
  \oint_{r_{0}} 2 n^{a} \delta S^{b} \gamma^{-1/2} \pi_{ab} \sigma^{1/2} d^{2}x
       = \frac{1}{2} \delta \left[ \oint_{r_{0}} \left( S^{i} S_{i} /r 
                 - S^{i} \partial_{r} S_{i} \right) \sigma^{1/2} d^{2}x \right].
\end{equation}
We employ the result of Eq.~(\ref{surf12}) after we rewrite the remaining surface terms of Eq.~(\ref{surf8}).

Next, we note the following useful identities, all accurate up to first order in the radiation amplitudes $I_{\ell m}$ and $S_{\ell m}$:
\begin{equation}
\label{surf13}
  N^{2} \sigma^{1/2} = \sigma^{1/2}_{0} \left( 1 + \frac{1}{2} \sigma^{ij} h_{ij} \right),
\end{equation}
\begin{equation}
\label{surf14}
  N^{-1} D_{a} n^{a} - \nabla^{0}_{i} n^{i}_{0} = \left[ \frac{ 2 h_{rr} }{r} 
                    + n^{i} \nabla_{i} \left(I + \frac{1}{2} h_{rr} \right) \right],
\end{equation}
\begin{equation}
\label{surf15}
  \sigma^{1/2} = \sigma^{1/2}_{0} \left( 1 + I + \frac{1}{2} \sigma^{ij} h_{ij} \right),
\end{equation}
and
\begin{equation}
\label{surf16}
  n^{a} D_{a} N = - \frac{1}{2} n^{i} \nabla_{i} I.
\end{equation}

Armed with the above identities, we may analyze yet another portion of the surface integrals of Eq.~(\ref{surf8}). Specifically,
\begin{eqnarray}
\label{surf17}
 & & - \oint_{r_{0}} \left[\delta \left( \sigma^{1/2} N^{2} \right) \left( N^{-1} D_{a} n^{a} - \nabla^{0}_{i} n^{i}_{0} \right)  
	+ 2 \delta \left( \sigma^{1/2} \right) n^{a} D_{a} N \right] d^{2}x \nonumber \\
	& &~~~~ = \oint_{r_{0}} \left[ \delta h_{rr} \left( \frac{h_{rr}}{r} + \frac{1}{4} \partial_{r} h_{rr} \right) 
	  + \delta I \partial_{r} I \right] \sigma^{1/2} d^{2}x.	
\end{eqnarray}

Noting the expression for $I$ given by Eq.~(\ref{thorne7}) and the orthogonality condition of the $\cal{Y}\it^{\ell m}_{L}$ given by Eq.~(\ref{STF8}), we find the last term in Eq.~(\ref{surf17}) is symmetric in $I$ and $\delta I$, and may be written as a total variation. The remaining terms of the above equation will eventually be combined with other terms yet to be evaluated from the surface integrals of Eq.~(\ref{surf8}).

We still have one more surface integral in Eq.~(\ref{surf8}) to rewrite. We first note the following useful identities, valid up to first order in the linearized parameters:
\begin{equation}
\label{surf18}
  \sigma^{1/2} \sigma^{ab} = \sigma^{1/2}_{0} \left[ \sigma^{ab}_{0} 
                               - \left( \sigma^{\;a}_{k} \sigma^{\;b}_{l} 
			       - \frac{1}{2} \sigma^{ab} \sigma_{kl} \right) h^{kl} \right],
\end{equation}
and 
\begin{eqnarray}
\label{surf19}
  N ( \sigma^{\;c}_{a} \sigma^{\;d}_{b}  \!\!&-&\!\!  \frac{1}{2} \sigma_{ab} \sigma^{cd} ) D_{c} n_{d} \nonumber \\ 
	 && \!\!\!\!\!\!\!\!\!\!\!\!\!\!\!\!\!\!\!\!\!\!\!\! = \left( \sigma^{\;k}_{a} \sigma^{\;l}_{b} - \frac{1}{2} \sigma_{ab} \sigma^{kl} \right) 
	     \left[ -\frac{1}{r} h_{kl} + \frac{1}{2} n^{p} \nabla_{p} h_{kl} - n^{p} \nabla_{(k} h_{l)p} \right].
\end{eqnarray}

With these two identities in hand, along with the fact that $h_{ij}$ is transverse and traceless, we find we may write
\begin{eqnarray}
\label{surf20}
 & & - \oint_{r_{0}} \delta \left( \sigma^{1/2} \sigma^{ab} \right) N \left( D_{a} n_{b} 
   - \frac{1}{2} \sigma_{ab} D_{c} n^{c} \right) d^{2}x \nonumber \\
    & &~~~ = \oint_{r_{0}} \left[ \frac{3}{2 r} \delta h_{rr} h_{rr} - \frac{1}{r} \delta h^{ij} h_{ij}  
        + \frac{1}{r} \delta h^{ri} h_{ri} - \frac{1}{4} \delta h_{rr} \partial_{r} h_{rr}
	+ \frac{1}{2} \delta h^{ij} \partial_{r} h_{ij} \right. \nonumber \\
	& &~~~~~~~~~~~~~~ \left. - n_{j} \nabla_{i} \left( \delta h^{ij} \sigma_{kl} h^{lj} \right) \right] \sigma^{1/2} d^{2}x.
\end{eqnarray}

The first five terms of the above equation will be combined with other terms we have previously evaluated. However, we still need to focus on the last term of Eq.~(\ref{surf20}). If we bring the $n_{j}$ inside of the derivative and project the $i$ indices parallel and perpendicular to $n_{i}$, we find we may write
\begin{eqnarray}
\label{surf21}
 - \!\!\!\!\!\!&&\!\!\!\!\!\!\! \oint_{r_{0}} n_{j} \nabla_{i} \left( \delta h^{ij} \sigma_{kl} h^{lj} \right) \sigma^{1/2} d^{2}x  \nonumber \\
     & = & \oint_{r_{0}}\left[ - \sigma^{ij} \nabla_{i} \left( \sigma^{\;k}_{j} \delta h_{kl} \sigma^{lp} h_{pr} \right)- \sigma^{ij} \nabla_{i} \left( n_{j} \delta h_{rl} \sigma^{lp} h_{pr} \right) \right. \nonumber \\
	& &~~~~~\left. - n^{i} n^{j} \nabla_{i} \left( \delta h_{jk} \sigma^{kl} h^{lr} \right) 
		       + r^{-1} \sigma_{ij} \delta h^{ik} \sigma_{kl} h^{lj} \right] \sigma^{1/2} d^{2}x.
\end{eqnarray}

Invoking Stokes' theorem, we note the first term on the right hand side is a two-dimensional divergence of a two-dimensional vector on a two-dimensional boundary, and hence vanishes. We may then reduce Eq.~(\ref{surf21}) to
\begin{equation}
\label{surf22}
  \oint_{r_{0}} \left[ - 2 r^{-1} \delta h_{rl} \sigma^{lp} h_{pr} 
                  - n^{i} \nabla_{i} \left( \delta h_{rk} \sigma^{kl} h_{lr} \right) 
		  + r^{-1} \sigma_{ij} \delta h^{ik} \sigma_{kl} h^{lj}  \right] \sigma^{1/2} d^{2}x,
\end{equation}
which is symmetric in $h_{ij}$ and $\delta h_{ij}$.

Finally, we may combine many of our results. Specifically, if we combine the results of Eq.~(\ref{surf17}), the first five terms of the last integral in Eq.~(\ref{surf20}), and the result given by Eq.~(\ref{surf22}), we find the first surface integral of Eq.~(\ref{surf8}) is equivalent to
\begin{eqnarray}
\label{surf23}
  \oint_{r_{0}} \left[ \frac{1}{2} \delta h^{ij} \partial_{r} h_{ij} \right.
                  \!\!\!&-&\!\!\! \left. \partial_{r} ( \sigma^{ij} \delta h_{ir} h_{jr} )
		  - \frac{3}{r} \sigma^{ij} \delta h_{ir} h_{jr}
		  + \frac{5}{2r} \delta h_{rr} h_{rr} + \delta I \partial_{r} I \right] \sigma^{1/2} d^{2}x \nonumber \\
    & = & \!\! \frac{1}{2} \delta \left\{ \oint_{r_{0}} \left[ \frac{1}{2} h^{ij} \partial_{r} h_{ij} 
                         - \partial_{r} ( \sigma^{ij} h_{ir} h_{jr} ) \right. \right. \nonumber \\
    &&~~~~~~~~~~~~~~~~ \left. \left. - \frac{3}{r} \sigma^{ij} h_{ir} h_{jr} 
                         + \frac{5}{2r} h_{rr} h_{rr} + I \partial_{r} I \right] \sigma^{1/2} d^{2}x \right\} \nonumber \\
    & &~ + \frac{1}{4} \oint_{r_{0}} \left[ \delta h^{ij} \partial_{r} h_{ij} 
                                     - h^{ij} \partial_{r} \delta h_{ij} \right] \sigma^{1/2} d^{2}x.
\end{eqnarray}
The total variation will be used when we write down the final form of the variational principle, but we still need to rewrite the second surface integral of Eq.~(\ref{surf23}).

We note $h_{ij}$ in the final surface integral of Eq.~(\ref{surf23}) is described by Eq.~(\ref{thorne18}) and Eq.~(\ref{thorne19}). When expressed in terms of $\delta Q_{\ell m}$, $\delta \vartheta^{Q}_{\ell m}$, and $\delta \Omega$, the variation in $h_{ij}$ may be written as
\begin{equation}
\label{surf24}
  \delta h^{\ell m Q}_{ij} = \delta Q_{\ell m} H^{\ell m Q}_{ij} 
         + Q_{\ell m} \frac{ \delta H^{\ell m Q}_{ij} }{ \delta \vartheta^{Q}_{\ell m} } \delta \vartheta^{Q}_{\ell m}
	 + Q_{\ell m} \frac{ \delta H^{\ell m Q}_{ij}}{ \delta \Omega } \delta \Omega.
\end{equation}

We now may substitute Eq.~(\ref{thorne18}), Eq.~(\ref{thorne19}) and  Eq.~(\ref{surf24}) into the final surface integral of Eq.~(\ref{surf23}). Integrating over the boundary using Eq.~(\ref{STF5}) along with the orthogonality of the $\cal{Y}\it^{\ell m}_{L}$ allows the surface integral to be written as a sum of terms each involving a single choice of $\ell$, $m$, and $Q$, namely
\begin{eqnarray}
\label{surf25}
  \oint_{r_{0}} \!\!\!\!\!\!\!&&\!\!\!\!\!\!\! \left[ \delta h^{ij} \partial_{r} h_{ij} 
                - h^{ij} \partial_{r} \delta h_{ij} \right] \sigma^{1/2} d^{2}x \nonumber \\
        & = & \sum_{Q;\ell;m \neq 0} \oint_{r_{0}} 
	         \left( \delta h^{\ell m Q}_{ij} \partial_{r} h^{\ast ij}_{\ell m Q} 
                      - h^{\ast ij}_{\ell m Q} \partial_{r} \delta h^{\ell m Q}_{ij} \right) \sigma^{1/2} d^{2}x.
\end{eqnarray}

Paying particular attention to the first term on the right hand side, specifically the portion that involves $\delta \Omega$ after substitution of Eq.~(\ref{surf24}), we note the product $r H^{\ell m Q}_{ij}$ may be written as $\Omega^{\ell}$ times a sum of terms, each of which is a product of a function of $r \Omega$ and a function of $\theta$ and $\phi$. In particular, we may write
\begin{equation}
\label{surf26}
  \delta \Omega \frac{ \delta H^{\ell m Q}_{ij} }{ \delta \Omega } 
     = \delta \Omega \frac{ \delta \left( r H^{\ell m Q}_{ij} \right) }{ \delta \left( r \Omega \right) } + \frac{ \ell \delta \Omega }{ \Omega } H^{\ell m Q}_{ij}.
\end{equation}

It can be shown, with some effort, the terms proportional to $\delta Q_{\ell m}$ in Eq.~(\ref{surf24}) and proportional to $\ell \delta \Omega$ in Eq.~(\ref{surf26}) do not contribute to the surface integral we are attempting to evaluate. For example, the term involving $\delta Q_{\ell m}$ vanishes because of the fact its coefficient is proportional to the Wronskian of two dependent solutions of the same linear equation. This leaves us with only the $\delta \vartheta^{I}_{\ell m}$ and $\delta \vartheta^{S}_{\ell m}$ contributions to the final surface integral of Eq.~(\ref{surf23}).

Once again using the properties given by Eq.~(\ref{STF5}) and Eq.~(\ref{STF8}), and evaluating in the wave zone, we find the $\delta \vartheta^{I}_{\ell m}$ contribution is
\begin{eqnarray}
\label{surf27}
  \oint_{r_{0}} \left( \frac{ \delta H^{\ell m I}_{ij} }{\delta \vartheta^{I}_{\ell m}} \partial_{r} H^{\ast ij}_{\ell m I} 
                  - H^{\ast ij}_{\ell m I} \partial_{r} \frac{\delta H^{\ell m I}_{ij}}{\delta \vartheta^{I}_{\ell m}} \right)                      \sigma^{1/2} d^{2}x \nonumber \\
  = \frac{ 8 \vert \omega_{m} \vert^{2 \ell + 1} (\ell + 1)(\ell + 2) }{ (\ell !)^{2} (\ell - 1) \ell } \vert I_{\ell m} \vert^{2}.
\end{eqnarray}

Likewise, using the same identities and evaluating the integral in the wave zone, we find the $\delta \vartheta^{S}_{\ell m}$ contribution to be
\begin{eqnarray}
\label{surf28}
  \oint_{r_{0}} n^{k} \left( \frac{ \delta H^{\ell m S}_{ij} }{\delta \vartheta^{S}_{\ell m}} \nabla_{k} H^{\ast ij}_{\ell m S} 
                  - H^{\ast ij}_{\ell m S} \nabla_{k} \frac{\delta H^{\ell m S}_{ij}}{\delta \vartheta^{S}_{\ell m}} \right)                      \sigma^{1/2} d^{2}x \nonumber \\
    = \frac{ 32 \vert \omega_{m} \vert^{2\ell + 1} \ell (\ell + 2) }{ (\ell !)^{2} (\ell-1) (\ell+1) } \vert S_{\ell m} \vert^{2}.
\end{eqnarray}

Finally we see the fruits of our labor by combining many of the results of this section to re-state the first surface integral of Eq.~(\ref{surf8}), namely
\begin{eqnarray}
\label{surf29}
  - \oint_{r_{0}} \left[ \delta \left( \sigma^{1/2} \right. \right. \!\!\!\!\!&&\!\!\!\!\!\! \left. \sigma^{ab} \right) N \left( D_{a} n_{b} 
          - \frac{1}{2} \sigma_{ab} D_{c} n^{c}  \right)  
         + \delta \left( \sigma^{1/2} N^{2} \right) \left( N^{-1} D_{a} n^{a} 
	 - \nabla^{0}_{i} n^{i}_{0} \right) \nonumber \\ 
       && \left. + 2 \delta \left( \sigma^{1/2} \right) n^{a} D_{a} N \right] d^{2}x  \nonumber \\ 
       && = \frac{1}{2} \delta \left\{ \oint_{r_{0}} [ - \partial_{r} ( \sigma^{ij} h_{ir} h_{jr} ) 
                              - \frac{3}{r} \sigma^{ij} h_{ir} h_{jr} 
			      + \frac{5}{2r} h_{rr} h_{rr} + I \partial_{r} I \right. \nonumber \\
	&&~~~~~~~~~~~~~~~~ \left. + \frac{r}{2} (\partial_{r} h^{ij}) \partial_{r} h_{ij}
			      - \frac{r}{2} h^{ij} \partial_{r} \partial_{r} h_{ij} ] \sigma^{1/2} d^{2}x \right\} \nonumber \\
      & &~~~~ - \frac{\Omega}{4} \delta \left\{ \oint_{r_{0}} \Omega^{-1} [ r ( \partial_{r} h^{ij} ) \partial_{r} h_{ij}
                              - r h_{ij} \partial_{r} ( r \partial_{r} h^{ij} ) ] \sigma^{1/2} d^{2}x \right\} \nonumber \\
      & &~~~~ - 8 \sum_{\ell;m > 0} ( E^{I}_{\ell m} \delta \vartheta^{I}_{\ell m} + E^{S}_{\ell m} \delta \vartheta^{S}_{\ell m} ),
\end{eqnarray}
where
\begin{equation}
\label{surf30}
  E^{I}_{\ell m} = 
   \frac{ \vert \omega_{m} \vert^{2 \ell + 1} (\ell + 1)(\ell + 2) }{ 2 (\ell !)^{2} (\ell - 1) \ell } 
   \vert I_{\ell m} \vert^{2}
\end{equation}
and
\begin{equation}
\label{surf31}
  E^{S}_{\ell m} = 
    \frac{ 2 \vert \omega_{m} \vert^{2 \ell + 1} \ell (\ell + 1) (\ell + 2) }{ [(\ell + 1)!]^{2} (\ell-1) } 
    \vert S_{\ell m} \vert^{2}.
\end{equation}
The coefficients $E^{Q}_{\ell m}$ are so named because they are the amount of energy in a gravitational wave for a $(\ell, m)$ and $(\ell, -m)$ multipole in a spherical shell in the wave zone which is one wavelength thick \cite{Det2}.

We are almost to the point of re-writing our variational principle for $H_{2}$, given by Eq.~(\ref{surf7}). However, before we do, it is convenient to define some new quantities. In particular, we absorb the last surface integral of Eq.~(\ref{surf29}) into the definition of $J$ to define the \it{effective angular momentum}\rm, given by $J^{\dag}$:
\begin{eqnarray}
\label{surf32}
  16 \pi J^{\dag} \!\!\!\!\!&&\!\!\!\!\! \equiv - \oint_{r_{0}} 2 n^{a} \Phi^{b} \gamma^{-1/2} \pi_{ab} \sigma^{1/2} d^{2}x \nonumber \\
                        &&~ - \frac{1}{4} \oint_{r_{0}} \Omega^{-1} \left[ r ( \partial_{r} h^{ij} ) \partial_{r} h_{ij}
                              - h^{ij} \partial_{r} ( r \partial_{r} h_{ij} ) \right] \sigma^{1/2} d^{2}x.
\end{eqnarray}
We note $J^{\dag}$ is defined through second order in the wave amplitudes $I_{\ell m}$ and $S_{\ell m}$, and is independent of the location of the boundary up to this order. In this sense, $J^{\dag}$ is the angular momentum of the source, without a contribution from the gravitational radiation. The necessity of this definition for $J^{\dag}$ will soon be apparent when we formulate the variational principle for neutron stars.

Rather than redefine $H_{2}$, we introduce the \it{effective mass}\rm, denoted by $M^{\dag}$. The effective mass consists of surface integrals of $H_{2}$, modified by the surface integrals whose total variations appear in Eq.~(\ref{surf29}) and Eq.~(\ref{surf12}):
\begin{eqnarray}
\label{surf33}
  16 \pi M^{\dag} \!\!\!\!\!&&\!\!\!\!\! \equiv - \oint_{r_{0}} N \left( 2 D_{a} n^{a} - N \nabla^{0}_{i} n^{i}_{0} \right) \sigma^{1/2}                  d^{2}x + \oint_{r_{0}} \nabla^{0}_{i} n^{i}_{0} \sigma^{1/2}_{0}  d^{2}x  \nonumber \\
		 &&~~ - \frac{1}{2} \oint_{r_{0}} \left[ - \partial_{r} \left( \sigma^{ij} h_{ir} h_{jr} \right) 
                              - \frac{3}{r} \sigma^{ij} h_{ir} h_{jr} 
			      + \frac{5}{2r} h_{rr} h_{rr} + I \partial_{r} I \right. \nonumber \\
		 &&~~~~~~~~~~~~~~~~~~ \left. + \frac{r}{2} \left(\partial_{r} h^{ij}\right) \partial_{r} h_{ij}
			      - \frac{r}{2} h^{ij} \partial_{r} \partial_{r} h_{ij} \right] \sigma^{1/2} d^{2}x  \nonumber \\
		 &&~~ + \frac{1}{2} \oint_{r_{0}} \left( S^{i} S_{i} /r - S^{i} \partial_{r} S_{i} \right) \sigma^{1/2} d^{2}x. 
\end{eqnarray}
We note $M^{\dag}$ is defined through second order in the wave amplitudes $I_{\ell m}$ and $S_{\ell m}$, and does not depend on the location of the boundary up to this order. In this sense, $M^{\dag}$ is the mass of the source, without a contribution from the gravitational radiation. We may now complete our description of a variational principle for irrotational binary neutron stars, in which $M^{\dag}$ plays a central role.

\section[Formulation of the Variational Principle]{Formulation of the Variational Principle for \\Irrotational Binary Neutron Stars}

We may now combine all of our results from the previous sections and present the variational principle for irrotational binary neutron stars. We first formulate a variational principle which satisfies only the fluid equations for irrotational motion described in \S (\ref{irrfluid}), will incorporate the variational principle for the matter into the variational principle for Einstein's equations described in \S (\ref{surfterms}), and will then finish with discussions of the properties of the variational principle, as well as methods of implementation.

We begin by noting a variational principle for the fluid \cite{Det3} may be described by 
\begin{equation}
\label{irrVP1}
  H_{\text{mat}} = \int_{\text{mat}} \left( N \rho_{H} - N^{a} j_{a} \right) \gamma^{1/2} d^{3}x,
\end{equation}
where $\rho_{H}$ and $j_{a}$ are given by Eq.~(\ref{irr26}) and Eq.~(\ref{irr27}), respectively. In terms of the fluid variables, the integrand may be written as
\begin{equation}
\label{irrVP2}
  N \rho_{H} - N^{a} j_{a} = \frac{ \rho_{0} }{2 N h} \left[ C^{2} - \left(N^{a} D_{a} \psi \right)^{2} \right] 
                           + \frac{ N \rho_{0} }{2 h} \left(D_{a} \psi \right) D^{a} \psi 
			   + \frac{N}{2} \left(\rho - p \right).
\end{equation}

For arbitrary variations of $N$, $N^{a}$, $\gamma_{ab}$, $\psi$, $\rho_{0}$, and $C$ the change in $H_{\text{mat}}$ is
\begin{eqnarray}
\label{irrVP3}
  \delta H_{\text{mat}} = C \delta M_{B} + \int_{\text{mat}} \left[ \delta N \rho_{H} - \delta N^{a} j_{a}
                  - \frac{1}{2} \delta \gamma_{ab} S^{ab} \right. \nonumber \\ 
		  \left. - \delta \psi \frac{N \rho_{0} }{h} \Psi
		  + \delta \left(\rho_{0} / h \right) \frac{N}{2} \cal{A}\rm \right] \gamma^{1/2} d^{3}x \nonumber \\
		  + \oint_{\partial \text{mat}} \frac{N \rho_{0} }{h} \delta \psi \Psi_{\partial \text{mat}} \sigma^{1/2}_{\partial \text{mat}} d^{2}x	  
\end{eqnarray}
where $S^{ab}$, $\Psi$, $\cal{A}\it$, and $\Psi_{\partial \text{mat}}$ are described by Eq.~(\ref{irr28}), Eq.~(\ref{irr23}), Eq.~(\ref{irr24}) and Eq.~(\ref{irr25}), respectively. The rest mass of the star is denoted as $M_{B}$, and is defined as
\begin{equation}
\label{irrVP4}
  M_{B} \equiv - \int_{\text{mat}} \rho_{0} n_{a} U^{a} \gamma^{1/2} d^{3}x
               = \int_{\text{mat}} \frac{\rho_{0}}{N h} \left( C + N^{a} D_{a} \psi \right) \gamma^{1/2} d^{3}x. 
\end{equation}

We note if the fluid equations are satisfied, then $H_{mat}$ evaluates to
\begin{equation}
\label{irrVP5}
  H_{\text{mat}} = C M_{B} - \int_{\text{mat}} N p \gamma^{1/2} d^{3} x.
\end{equation}

The variational principle for the fluid equations, as well as the Einstein equations, is based on the following definition of $M^{\dag}$, given as
\begin{eqnarray}
\label{irrVP6}
  16 \pi M^{\dag} & \equiv&  - \int [ N ( \cal{N}\it - \rm{16}\it \pi \rho_{H} ) 
                                 + \rm{2}\it N^{a} ( \cal{N}\it_{a} + \rm{8}\it \pi j_{a} ) ] \gamma^{\rm{1/2}\it}                                                d^{\rm{3}\it}x \nonumber \\
		      & &~ - \oint_{r_{0}} N ( 2 D_{a} n^{a} - N \nabla^{0}_{i} n^{i}_{0} ) \sigma^{1/2} d^{2}x 
		        + \oint_{r_{0}} \nabla^{0}_{i} n^{i}_{0} \sigma^{1/2}_{0}  d^{2}x  \nonumber \\
		      & &~  - \frac{1}{2} \oint_{r_{0}} [ - \partial_{r} ( \sigma^{ij} h_{ir} h_{jr} ) 
                              - \frac{3}{r} \sigma^{ij} h_{ir} h_{jr} 
			      + \frac{5}{2r} h_{rr} h_{rr} + I \partial_{r} I \nonumber \\
		      & &~~~~~~~~~~~~ + \frac{r}{2} (\partial_{r} h^{ij}) \partial_{r} h_{ij}
			      - \frac{r}{2} h^{ij} \partial_{r} \partial_{r} h_{ij} ] \sigma^{1/2} d^{2}x  \nonumber \\
		      & &~ + \frac{1}{2} \oint_{r_{0}} ( S^{i} S_{i} /r - S^{i} \partial_{r} S_{i} )\sigma^{1/2} d^{2}x.	 
\end{eqnarray}
Under arbitrary variations of the geometry as well as the fluid variables, we find
\begin{eqnarray}
\label{irrVP7}
16 \pi \delta M^{\dag} \!\!\!\!\!& = &\!\!\!\!\! - \int \left[ \delta N \left( {\cal N} 
               - 16 \pi \rho_{H} \right) \gamma^{1/2} 
	       + 2 \delta N^{a} \left( {\cal N}_{a} + 8 \pi j_{a} \right) 
	       \gamma^{1/2} \right. \nonumber \\
	& &~~  + \delta \gamma_{ab} \left({\cal P}^{ab} + 8 \pi N \gamma^{1/2} S^{ab}\right)- \delta \pi^{ab} {\cal G}_{ab} 
	       - 16 \pi \delta \psi \frac{N \rho_{0} }{h} \Psi \gamma^{1/2} \nonumber \\
	& &~~~~~~~~ \left. + 8 \pi \delta \left(\rho_{0} / h \right) N 
	       {\cal A} \gamma^{1/2} \right] d^{3}x \nonumber \\
	& &~~ + \oint_{\partial \text{mat}} \frac{N \rho_{0} }{h} \delta \psi \Psi_{\partial \text{mat}} 
	        \sigma^{1/2}_{\partial \text{mat}} d^{2}x \nonumber \\
	& &~~ - 8 \sum_{\ell; m > 0} (E^{I}_{\ell m} \delta \vartheta^{I}_{\ell m} 
	       + E^{S}_{\ell m} \delta \vartheta^{S}_{\ell m})\nonumber \\ 
	& &~~ + 16 \pi \Omega \delta J^{\dag} + 16 \pi C \delta M_{B}.     
\end{eqnarray}

We can see from Eq.~(\ref{irrVP7}) how we may apply a variational principle: for fixed values of the effective angular momentum $J^{\dag}$, the baryonic rest mass $M_{B}$, and the radiation phases $\vartheta_{\ell m}$, the effective mass $M^{\dag}$ is an extremum under infinitesimal variations of the matter and geometry precisely when the matter and geometry are solutions to the quasi-stationary Einstein equations and the equations for irrotational fluid flow.

The variational principle also yields an accurate estimate of the value of $M^{\dag}$. Assume a geometry and matter distribution which is an approximate solution of the quasi-stationary Einstein equations, and differs from an exact solution by $O(\delta)$. If so, then the difference between $M^{\dag}$ for the exact solution and $M^{\dag}$ for the approximate solution is $O(\delta^{2})$, as long as the parameters $J^{\dag}$, $M_{B}$, and the phases $\vartheta_{\ell m}$ are the same for the two geometries.

The variational principle also allows one to calculate values for the parameters $\Omega$, $C$, and the radiation coefficients $E_{\ell m}$ in the following fashion. Assume a geometry and matter distribution differs from an exact solution by $O(\delta)$, with parameters $J^{\dag}$, $M_{B}$, and phases $\vartheta_{\ell m}$. This configuration yields some value of the effective mass $M^{\dag}$. Now, alter the geometry by changing the value of the effective angular momentum by some amount $\Delta J^{\dag}$, holding all other quantities fixed. This change in the effective angular momentum will correspond to a change in the effective mass with error of $O(\delta^2)$ via
\begin{equation}
\label{irrVP8}
  \Delta M^{\dag} = \Omega \Delta J^{\dag} + O \left(\delta^{2} \right).
\end{equation}
{}From Eq.~(\ref{irrVP8}), one can easily see estimates of $\Omega$, accurate to order $O(\delta^2)$, may be determined via
\begin{equation}
\label{irrVP9}
  \Omega = \Delta M^{\dag} / \Delta J^{\dag} + O \left(\delta^{2} \right).
\end{equation}
In a similar fashion, estimates of $C$ and of the energies $E_{\ell m}$ may also be determined, accurate to  $O(\delta^{2})$. This procedure leads us to a technical definition of the constant $C$: it is the derivative of the effective mass with respect to the rest mass of the neutron star holding all other physical parameters fixed. A more intuitive explanation of the role of $C$ may be seen from Eq.~(\ref{irrVP7}). Imagine injecting a single baryon of mass $m_{B}$ into our system, holding the angular momentum fixed---Eq.~(\ref{irrVP7}) indicates $C m_{B}$ is the mass increase per baryon of the system for fixed angular momentum. This is sometimes referred to as the relativistic chemical potential or the injection energy \cite{MTW}.

\section{Discussion}

In closing, we point out some of the highlights of our analysis of binary neutron stars undergoing irrotational motion. The expressions for the effective mass $M^{\dag}$, given by Eq.~(\ref{irrVP6}), and the effective angular momentum $J^{\dag}$, given by Eq.~(\ref{surf32}), are of great interest. As far as quantities such as mass or angular momentum can be defined in the framework of General Relativity, they can be viewed as the mass and angular momentum due to the sources alone, without a contribution from the gravitational radiation present in the system. We assert these quantities are not a measure of the energy or the angular momentum of the radiation because $M^{\dag}$ and $J^{\dag}$ are independent of the location of the boundary in which they are evaluated.

The expression for $M^{\dag}$ not only yields a measure of the mass of the system, but also generates a variational principle for our neutron stars. We found, as shown in Eq.~(\ref{irrVP7}), an extremum in $M^{\dag}$ corresponds to a solution of the quasi-stationary Einstein equations and the fluid equations for irrotational motion for fixed values of the effective angular momentum $J^{\dag}$, the baryonic mass of the stars $M_{B}$, and the phases of the gravitational radiation $\vartheta_{\ell m}$. Hence, $M^{\dag}$ provides a means of generating realistic solutions which satisfy the quasi-stationary Einstein equations and the irrotational fluid equations.

We also found the variational principle allows a means for determining accurate estimates of the orbital angular frequency $\Omega$, the gravitational wave energies $E_{\ell m}$, as well as the relativistic chemical potential parameter $C$, which are of interest to those investigating gravitational wave signals. Knowledge of $\Omega$ can be used to yield estimates of the time evolution of the angular frequency, perhaps to a time very late in the evolution of the system. Knowledge of the relativistic chemical potential may be used to extract useful information about the structure and state of matter of neutron stars.

The variational principle for $M^{\dag}$ is justification for the numerical techniques employed by some investigators \cite{Bonna3,Bonna4,Baumgarte7,Shibata2,Uryu3,Wilson4}. In past numerical studies, it was not clear what mathematical framework was used when investigators formulated their variational principles. Now, with this work, investigators are given a clear road map to approach numerical studies---specifically, which quantities should be held fixed as they generate their sequence of solutions to the quasi-stationary Einstein equations and the fluid equations for irrotational motion. For systems in which gravitational radiation is completely ``shut off'', it is clear the effective angular momentum and the baryonic mass of the system should be held fixed in order to determine a realistic solution to the equations of interest. However, if investigators choose to include the effects of radiation in their analysis, they must also ensure the phases of the radiation are fixed as they generate their solution.

The variational principle for $M^{\dag}$ acts as a proving ground for those numerical studies in which the conformally flat conjecture is used \cite{Bonna3,Bonna4,Shibata2,Uryu3,Wilson4}. The conformally flat conjecture assumes the geometry is conformally flat and that the effects of gravitational radiation may be neglected. In principle, one can employ the variational principle for $M^{\dag}$, using the same parameters as the conformally flat geometries used by other investigators. This will allow the angular frequencies and energies of the radiation to be compared to those derived from the conformally flat studies. If a large variation in the comparison of these quantities is observed, then it would indicate the conformally flat metric, and their assertions that gravitational radiation effects can be neglected, are erroneous. However, if little variation is observed in these quantities, it would indicate the conformally flat studies are well-justified in their assumptions.

\chapter[The Initial Value Problem]{The Initial Value Problem: \\A Variational Principle for Binary Black Holes}
\label{InitValue}
One of the remaining unresolved issues in numerical relativity is known as the initial value problem. Specifically, how does one determine physically realistic initial data for binary black hole systems? It is this initial data which in turn is evolved in time via the Einstein equations, so if one wishes to generate reliable predictions about the location and angular frequency of the innermost stable circular orbit, then it is vital that the initial data be physically realistic.

The vacuum constraint equations are the cornerstone of the initial value problem. Recall from Eq.~(\ref{3+1_12}) the Hamiltonian constraint is
\begin{equation}
\label{vp2-1a}
  R + \cal{K}\it^{\;a}_{a} \cal{K}\it^{\;b}_{b} - \cal{K}\it_{ab} \cal{K}\it^{ab} = {\rm 0},
\end{equation}
where $\cal{K}\it_{ab}$ is the extrinsic curvature associated with the three-metric $\gamma_{ab}$, and $R$ is the associated Ricci scalar. Also recall from Eq.~(\ref{3+1_14}) the momentum constraint is
\begin{equation}
\label{vp2-1b}
  D^{a} \left( \cal{K}\it_{ab} - \cal{K}\it^{\;c}_{c} \gamma_{ab} \right) = 0,
\end{equation}
where $D^{a}$ is the covariant derivative compatible with the three-metric $\gamma_{ab}$. These constraint equations must be satisfied on every spacelike hypersurface, but do not determine the dynamics of the geometry. 

A solution to the initial value problem consists of a three-metric $\gamma_{ab}$ and an extrinsic curvature $\cal{K}\it_{ab}$ which satisfy Eqs.~(\ref{vp2-1a}) and (\ref{vp2-1b}). However, one notes the equations are coupled, nonlinear equations---this greatly complicates any efforts to generate either analytic or numerical solutions to the initial value equations. To simplify the analysis, we assume the three-metric $\gamma_{ab}$ is conformally flat. This introduces the conformal factor $\psi$ via the relationship
\begin{equation}
\label{vp2-3}
 \gamma_{ab} = \psi^{4} g_{ab}, 
\end{equation}
where $g_{ab}$ is the flat three-metric. Under the conformal flatness assumption, the extrinsic curvature transforms as well according to
\begin{equation}
\label{vp2-4}
 \cal{K}\it_{ab} = \psi^{\rm -2} K_{ab}. 
\end{equation}
We call $K_{ab}$ the conformal extrinsic curvature.

In addition to the conformally flat assumption, we choose the maximal slicing gauge \cite{Bowen}, which corresponds to the conformal extrinsic curvature being trace-free:
\begin{equation}
\label{vp2-1c}
  K^{\;a}_{a} = 0.
\end{equation}

By assuming a conformally flat three-metric, as well as choosing the maximal slicing gauge, we effectively decouple the constraint equations, making them more amenable to study. In particular, the Hamiltonian constraint equation becomes
\begin{equation}
\label{vp2-5}
  \nabla^{2} \psi + \frac{1}{8} K_{ab} K^{ab} \psi^{-7} = 0,
\end{equation}
where the $\nabla^{2}$ operator is the flat space Laplacian. Likewise, the momentum constraint simplifies to 
\begin{equation}
\label{vp2-6}
  \nabla_{b} K^{ab}=0. 
\end{equation}
Hence, determination of $K_{ab}$ and $\psi$ via Eq.~(\ref{vp2-5}) and Eq.~(\ref{vp2-6}) completely specify the geometry on an initial space-like hypersurface. In practice, one would take this initial data and evolve it via the dynamical Einstein equations.

However, a question arises: how does one determine \it{physically realistic}\rm~ initial data---data which is truly representative of binary black holes in circular orbits? The answer to this question has a long and interesting history---one which we briefly recount.

The first attempt at realistic initial data is the result of work by Misner and Wheeler \cite{Wheeler1,Misner1}. Misner studied the situation in which two masses were initially at rest, and found that the topology was analogous to that of a Wheeler wormhole---a single ``sheet'' of spacetime in which two separate locations on the sheet are connected by a ``throat''. It was later discovered that the Misner data set, when expressed in isotropic coordinates, could be viewed as a topology consisting of two sheets \cite{Misner1}. Each sheet corresponds to a separate asymptotically flat region, where the two sheets are connected by two ``throats''. In 1963 Brill and Lindquist \cite{Brill1} generated an initial data set which corresponded to a three-sheeted geometry, each sheet again corresponding to a separate asymptotically flat region. 

During the intervening years, many have investigated the properties of the two- and three-sheeted geometries. For instance, Bowen and York \cite{Bowen} studied black holes using the two-sheeted geometry. However, the two-sheeted analysis introduces several complications, most notably the need for boundary conditions not only at spatial infinity of the upper sheet, but also at the locations of the throats of the holes. These conditions at the throats are typically based on how the conformal factor $\psi$ and the extrinsic curvature $K_{ab}$ behave under a coordinate inversion through the throats. Despite the added complications associated with the Misner data set, it has been widely used by a number of investigators to study black hole collisions \cite{Cook1,Cook4,Cook3,Bowen,Bonna1,Friedman1}. 

In 1997 Brandt and Br\"ugmann \cite{BB} proposed a method which would reduce the number of complications introduced by the Misner data set. They choose a topology in accordance with the Brill data, and proposed a method to compactify the lower sheets asymptotically flat regions. This compactification effectively simplifies the domain of integration in which the constraint equations must be solved. Simply stated, the compactification of an asymptotically flat region to a single point in $R^{3}$ eliminates the need to impose boundary conditions at the throat and eliminates the need to excise domains of integration.

Solutions to the momentum constraint, Eq.~(\ref{vp2-6}), are well-known for single black holes with arbitrary linear and spin angular momentum \cite{Bowen}. The trace-free extrinsic curvature which satisfies the momentum constraint is of the form
\begin{eqnarray}
\label{vp2-7}
  K^{ab}_{PS(i)} & = & \frac{3}{2r^2} \left[P^a n^b + P^b n^a - \left(g^{ab} - n^a n^b \right) P^c n_c \right] \nonumber \\
                   & &~  + \frac{3}{r^3} \left[ \epsilon^{acd} S_c n_d n^b + \epsilon^{bcd} S_c n_d n^a \right],
\end{eqnarray}
where $P^a$ and $S^a$ are the linear momentum and spin angular momentum of the black hole measured on the upper sheet of the geometry. Because the conformal metric is flat, we may use normal Cartesian coordinates, where $n^a$ is a radial normal vector of the form $n^a = x^a / r$, where $r^2 = x^2 + y^2 + z^2$. For a geometry which consists of $N$ black holes, the extrinsic curvature is given by
\begin{equation}
\label{vp2-8}
  K^{ab} = \sum^{N}_{i=1} K^{ab}_{PS(i)},
\end{equation}
with each term in the sum having its own coordinate origin. Note for $N$ black holes on the upper sheet, there are a total of $N+1$ sheets in the geometry, each with it's own separate asymptotically flat region.

Brandt and Br\"ugmann then proceed to solve the Hamiltonian constraint, given by Eq.~(\ref{vp2-5}), rewriting the conformal factor $\psi$ as
\begin{equation}
\label{vp2-9}
  \psi = \frac{1}{\alpha} + U,
\end{equation}
where $1/\alpha$ is given via
\begin{equation}
\label{vp2-10}
  \frac {1}{\alpha} = \sum^N _{i=1} \frac{M_{(i)}}{2 \vert \vec{r} - \vec{r}_{(i)} \vert }, 
\end{equation}
where $M_{i}$ is the mass of the $i^{th}$ hole, and $\vec{r}_{(i)}$ is the coordinate location of the corresponding hole.

With these identifications, the Hamiltonian constraint becomes a nonlinear elliptic equation for $U$:
\begin{equation}
\label{vp2-11}
  \nabla^2 U + \beta \left( 1 + \alpha U \right)^{-7} = 0,
\end{equation}
where $ \beta $ is related to the conformal extrinsic curvature $ K^{ab} $ via
\begin{equation}
\label{vp2-12}
  \beta = \frac{1}{8} \alpha^7 K^{ab} K_{ab}.
\end{equation}

Recall we demand asymptotic flatness on the upper sheet, which determines the boundary condition on $U$ at spatial infinity. This Robin boundary condition may be stated as
\begin{equation}
\label{vp2-13}
  \frac{\partial U}{\partial r} = \frac{1 - U}{r},
\end{equation}
which may be interpreted as demanding that $U$ consists primarily of a monopole term as one recedes far from the black holes.

Hence, Eq.~(\ref{vp2-7}) which satisfies the momentum constraint, coupled to Eq.~(\ref{vp2-11}) and Eq.~(\ref{vp2-13}) complete the mathematical description of the initial value problem. It is important to reiterate Eq.~(\ref{vp2-11}) may be ``easily'' solved on $R^3$ without the need of any boundary conditions at the throats of the punctures. Discussion of the method of solution and numerical results of Eq.~(\ref{vp2-11}) will be covered in the following chapters.

As motivation, let us describe the work of Baumgarte \cite{Baumgarte2} and his investigations into circular orbits of binary black hole systems. Baumgarte also chooses the three-sheeted topology to describe his binary system, and also adopts the ``puncture'' method of Brandt and Br\"ugmann. Baumgarte proceeds to formulate a minimization principle in the following fashion, which is in line with the method of Cook \cite{Cook4}.

For a system of equal mass black holes, and for a particular choice of angular momentum $J$ for the system, Baumgarte demands the momenta $P$ of the punctures are equal in magnitude and opposite in direction, and that the total separation distance is given by the quantity $D$. For circular orbits, he also demands $\vec{P} \cdot \vec{D} = 0$. The momenta $P$ are determined by the choice of the angular momentum $J$ and the separation distance $D$ via $J = P D$.

Baumgarte then identifies the irreducible mass of the individual punctures, approximately given by Christodolou's formula
\begin{equation}
\label{vp2-14}
  M \approx M_{\text{irr}} = \left( \frac{A}{16 \pi} \right)^{1/2},
\end{equation}
where $A$ is the area of the black hole's apparent horizon \cite{Christo}. He then goes on to define an effective potential, given by the binding energy of the system:
\begin{equation}
\label{vp2-15}
  E_{\text{b}} = E_{\text{ADM}} - 2 M,
\end{equation}
where $E_{\text{ADM}}$ is the ADM mass of the system (see \S (\ref{ADM Mass}) for a discussion of the ADM mass). Baumgarte then varies the separation distance between the black holes, and looks for minima in the binding energy for fixed values of angular momentum $J$ and for fixed values of the apparent horizon areas $A$. The minimum in the binding energy for some value of the coordinate separation distance $D$ on the initial slice is the necessary condition for circular orbits, given by 
\begin{equation}
\label{vp2-16}
 \left. \frac{ \partial{ E_{\text{b}} } } {\partial {D}} \right\vert_{ A,J }= 0.
\end{equation}
Likewise, the orbit's angular frequency $\Omega$ may be determined from
\begin{equation}
\label{vp2-17}
 \Omega = \left. \frac{ \partial{ E_{\text{b}} } } {\partial {J}} \right\vert_{ A }.
\end{equation}

A Newtonian example of the above two equations may help to solidify this concept. Assume we are studying a system of two equal mass point particles, each of mass $M$. The system has a total angular momentum $J$ and the point particles are separated by a distance $D$. The energy of the system may be written as
\begin{equation}
\label{vp2-18}
 E_{\text{b}} = \frac{J^2}{D^2 M} - \frac{M^2}{D}.
\end{equation}

Application of the condition for circular orbits, Eq.~(\ref{vp2-16}), to the energy given in Eq.~(\ref{vp2-18}) yields a relationship between the angular momentum $J$, the mass $M$, and the separation distance $D$:
\begin{equation}
\label{vp2-19}
  D = \frac{2 J^2}{M^3}.
\end{equation}

Hence, for specified values of $J$ and $M$, one may determine the location of the circular Newtonian orbit. We may also generate an expression for the orbital angular frequency $\Omega$, as determined by Eq.~(\ref{vp2-17}), which yields
\begin{equation}
\label{vp2-20}
  \Omega = \frac{2 J}{D^2 M}.
\end{equation}

We may combine the results of Eq.~(\ref{vp2-19}) and Eq.~(\ref{vp2-20}) to yield the familiar form of Kepler's third law of planetary motion:
\begin{equation}
\label{vp2-21}
  \Omega^{2} = \frac{2 M}{D^3}.
\end{equation}

In this fashion, Baumgarte proceeds to generate a sequence of circular orbits, terminating at the innermost stable circular orbit. One of his main conclusions is the agreement between his results (which are based on Brill data) and those of Cook (which are based on Misner data). He concludes the underlying topology is not a strong determining factor of the resulting physics.

Despite all of the simplifications introduced by the three sheeted topology and the puncture method of Brandt and Br\"ugmann, Baumgarte's analysis is complicated by holding the apparent horizon areas fixed as he generates his sequence. The algorithm Baumgarte et al. employ is described in \cite{Baumgarte1}, in which the location of the apparent horizon is expressed in terms of symmetric trace-free tensors. The coefficients of these tensors are varied until the equation describing the expansion of null geodesics drops below some specified tolerance level. The horizon-finding code appears to reproduce the horizons associated with the Schwarzschild geometry to arbitrarily high degrees of accuracy, and the authors indicate varying degrees of success with situations in which the horizons are only slightly distorted.

In the heuristic extremum principle used by Baumgarte there appears no specific reason for holding the areas of the apparent horizons fixed while searching for the extremum in the binding energy. This method could be legitimate for the two-sheeted Misner data of Cook \cite{Friedman1}, but our analysis indicates the apparent horizon area need not be held fixed for the three-sheeted Brill data.

As was demonstrated by Blackburn and Detweiler \cite{Det1}, a variational principle for a binary black hole system with the Brill topology has the form
\begin{equation}
\label{stevevp1}
  \delta E_{\text{ADM}} = - 2 N \delta \overline{m} + \Omega \delta J,
\end{equation}
assuming the quasi-stationary Einstein equations are satisfied, and the system does not contain any radiation. In the above equation, $E_{\text{ADM}}$ is the ADM mass of the system, as measured on the upper sheet, $N$ is the ratio of the lapse functions on the lower sheets to that of the upper sheet, $\overline{m}$ is the ``bare'' mass of the hole as measured on a lower sheet, $\Omega$ is the angular frequency of the binary system, and $J$ is the angular momentum of the binary system. The appearance of the ratio of the lapse $N$ may be somewhat surprising, but the quasi-equilibrium approximation (and the existence of the approximate Killing vector $\partial/\partial t + \Omega \partial/\partial \phi$) eliminate the arbitrary nature of the lapse. The variational principle of Blackburn and Detweiler demonstrates one should fix the values of the bare mass and the angular momentum of the black holes if they wish to determine an extremum in the ADM mass. More importantly, the extremum in the ADM mass indicates the geometry with angular momentum $J$ and bare mass $\overline{m}$ are solutions to the quasi-equilibrium Einstein equations. This insight is the foundation of our application of a variational principle for the ADM mass, employing the puncture method of Brandt and Br\"ugmann.

One question that arises is the following: given a particular configuration on the upper sheet of our geometry, how does one determine the bare masses on the lower sheets, and vice-versa? We first recall the metric may be written as 
\begin{equation}
\label{vp2-22}
  ds^{2} = \psi^{4} ( dr^{2} + r^{2} d\Omega^{2} ),
\end{equation}
where once again $\psi$ is the conformal factor, and $d\Omega$ represents an element of the solid angle. Also recall there exists an isometry given by $ \overline{r} = M^{2}/4r $, which leaves invariant the coordinate sphere at $r = M/2 $ \cite{Bowen}, and maps the entire exterior of the asymptotically flat region into the interior of that invariant sphere. Note that if we were to express the Schwarzschild metric in these isotropic coordinates, the conformal factor would take the form \cite{MTW}
\begin{equation}
\label{vp2-23}
  \psi = 1 + \frac{M}{2r},
\end{equation}
where $M$ is the mass associated with the Schwarzschild solution. Hence, we should expect a conformal factor in the coordinate inverted space to have a similar form, with a different mass parameter. Specifically, we perform the coordinate inversion and attempt to write the coordinate inverted conformal factor as
\begin{equation}
\label{vp2-24}
  \overline{\psi} = 1 + \frac{ \overline{m} }{ 2 \overline{r} },
\end{equation}
where $\overline{m}$ is the bare mass associated with that particular asymptotically flat region. We begin our coordinate inversion by substituting the expression given above for $\overline{r}$ into the metric given by Eq.~(\ref{vp2-22}). We note we may write the transformed metric as
\begin{equation}
\label{vp2-25}
  ds^{2} = \overline{\psi}^{\;4} ( d\overline{r}^{\;2} + \overline{r}^{\;2} d\Omega^{2} ),
\end{equation}
where 
\begin{equation}
\label{vp2-26}
  \overline{\psi} = \psi \frac{M}{2 \overline{r}}. 
\end{equation}

Substituting the definition of $\psi$ for equal mass punctures, given by Eq.~(\ref{vp2-9}), into Eq.~(\ref{vp2-25}), we find $\overline{\psi}$ takes the form
\begin{equation}
\label{vp2-27}
  \overline{\psi} = 1 + \frac{1}{2 \overline{r}} \left( M_{i} U_{i} + \frac{ M_{i} M_{j} }{2 D} \right),
\end{equation}
where $M_i$ is the ``Newtonian'' mass associated with the $i^{th}$ puncture as measured on the upper sheet, $U_{i}$ is the value of $U$ at the $i^{th}$ puncture, and $D$ is the separation distance between the punctures on the upper sheet. Comparison of Eq.~(\ref{vp2-27}) to Eq.~(\ref{vp2-24}) allows us to identify the bare mass associated with each puncture in each puncture's asymptotically flat region, denoted as $\overline{m}_{i}$:
\begin{equation}
\label{vp2-28}
 \overline{m}_{i} = M_{i} \left( U_{i} + \frac{ M_{j} }{2 D} \right),
\end{equation}
where $ i \neq j $. Blackburn and Detweiler's variational principle given by Eq.~(\ref{stevevp1}) tells us the bare mass $\overline{m}_{i}$ should be held fixed as one generates a sequence for circular orbits.

The application of the variational principle follows: Specify the value of the angular momentum $J$ and the ``bare'' masses $ \overline{m}_1 $ and $ \overline{m}_2 $. These quantities are to be held fixed during the analysis. Then, initialize the puncture separation distance $D$, which in turn determines the hole linear momentum $P$ via $J = P D$. Recall the linear momenta and hole separation specifies the conformal extrinsic curvature, given by Eq.~(\ref{vp2-7}). For some initial guess of the ``Newtonian'' masses $M_1$ and $M_2$, solve the Hamiltonian constraint equation given by Eq.~(\ref{vp2-11}) subject to the Robin boundary condition given by Eq.~(\ref{vp2-12}). Once $U$ is known, use it to determine new values for the Newtonian masses, as determined by Eq.~(\ref{vp2-28}).  Iterate the above procedure by re-solving the Hamiltonian constraint equation for the new Newtonian masses, and re-calculate the Newtonian masses. When the Newtonian masses have converged to within some predetermined tolerance, then we may assume we have properly solved the Hamiltonian constraint subject to fixed values of the angular momentum $J$ and the bare masses given by $\overline{m}$. At this point, calculate the ADM mass for the system, as described in \S (\ref{ADM Mass}). Then, decrement the hole separation distance $D$, holding the angular momentum $J$ and the bare masses $\overline{m}$ fixed. Repeat this procedure for a sequence of hole separations, calculating the ADM mass at each separation distance after reasonable convergence in the Newtonian masses has been obtained.

Once the ADM mass for the system has been calculated for a range of separation distances $D$, all the while holding the angular momentum $J$ and the bare masses $\overline{m}$ fixed, we may determine the location of circular orbits via
\begin{equation}
\label{vp2-29}
  \left. \frac{ \partial{ E_{\text{ADM}} } } {\partial {D}} \right\vert_{ \overline{m} ,J }= 0.
\end{equation}
Note the similarities between Eq.~(\ref{vp2-29}) and Eq.~(\ref{vp2-16}). Our variational principle also allows us to calculate the orbital angular frequency $\Omega$, given by
\begin{equation}
\label{vp2-30}
 \Omega = \left. \frac{ \partial{ E_{\text{ADM}} } } {\partial {J}} \right\vert_{ \overline{m} }.
\end{equation}
Again, note the similarities between Eq.~(\ref{vp2-30}) and Eq.~(\ref{vp2-17}).

The following chapter discusses the numerical methods employed to implement the variational principle, in which we solve the Hamiltonian constraint, given by Eq.~(\ref{vp2-11}), subject to the Robin boundary condition Eq.~(\ref{vp2-12}). We ultimately generate a sequence of constant angular momentum curves, determine the minima in the curves which correspond to circular orbits, and evaluate the orbital angular frequency in addition to other physically interesting quantities.

\chapter{Numerical Methods}
\label{Numerical Methods}
This chapter is dedicated to the development of the numerical methods employed to solve the Hamiltonian constraint equation for the ``puncture'' method in the previous chapter. We will begin by introducing the main ideas of linear multigrid techniques, building up to nonlinear multigrid techniques, and finally nonlinear adaptive multigrid techniques. We will close the chapter with brief discussions on methods of determining operators for numerical methods, as well as some tests of the computer code which we developed.

Multigrid methods are relatively new to the scene of numerical techniques, but their power and speed are widely recognized \cite{multigrid1,multigrid2,multigrid3,numrecipes}. The development of linear multigrid techniques will be presented via a toy problem, specifically determining an efficient and accurate way in which to solve Poisson-like equations with particular boundary conditions. Once the linear method is developed, we use it as a launching pad into the realms of nonlinear multigrid and nonlinear adaptive multigrid.

\section{Linear Multigrid}
\label{Linear Multigrid}
First, we indicate the notation used throughout the entire discussion. The problem at hand is the determination of the solution to the partial differential equation:
\begin{equation}
\label{mgeqn1}
  A^h u^h = f^h,
\end{equation}
\noindent where $A^h$ is some linear difference operator, $u^h$ is the exact solution to the equation, and $f^h$ is a source term. The superscript $h$ denotes that the operators and functions exist on a grid of spacing $h$.

It is possible, in the numerical sense, to solve the above equation exactly. This may be done via a direct solver, which effectively diagonalizes a matrix representing a system of equations for the numerical grid. However, these methods can be slow in arriving at a solution, especially if a large number of grid points are involved \cite{numrecipes}. 

In an effort to determine a more speedy method, we introduce an important quantity. A measure of how well our approximate solution is doing to solve the equation is found in the residual $r^h$ \cite{multigrid1}, defined by
\begin{equation}
\label{mgeqn2}
  r^h \equiv f^h - A^h v^h,
\end{equation}
\noindent where $v^h$ is the approximate solution. Obviously, if the residual is equal to zero, then we are assured $v^h$ is an exact solution to Eq.~(\ref{mgeqn1}).

In theory, one selects a numerical method (presumably iterative) in which to solve Eq.~(\ref{mgeqn1}). After several iterations, calculate the residual $r^h$. If the residual is ``small'' (where ``small'' means less than some tolerance level), then you have a reasonable approximate solution to Eq.~(\ref{mgeqn1}). If the residual is not ``small'', then simply iterate your numerical method some more and re-calculate the residual. This process is continued until one is satisfied with the ``smallness'' of the residual.

One such practical application of this type of numerical method is known as
red-black Gauss-Seidel relaxation, or simply red-black relaxation. In this
scheme, one imagines a numerical grid to consist of a checkerboard of
alternating colors (hence the name red-black) \cite{numrecipes}. The
significance of the alternating colors is when a particular grid point, say a
``red'' grid point, is being updated by the relaxation scheme, its value only
depends on the black grid points in the vicinity of the updated red grid point. The advantage of this is the red-black scheme allows one to update all of the red points first, and then all of the black points.

\begin{figure}[]
\begin{center}
\scalebox{0.5}[0.5]{
  \includegraphics{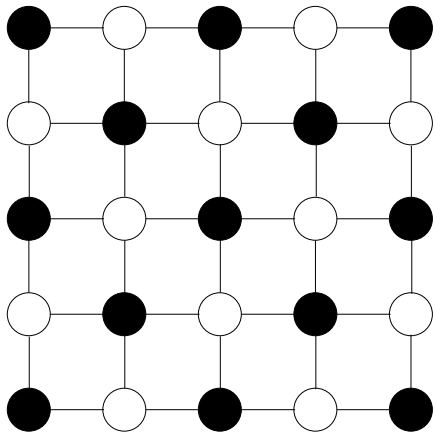}
}
\caption{A cartoon representation of a two-dimensional red-black numerical grid.}
\label{redblack.eps}
\end{center}
\end{figure}

\begin{figure}[]
\begin{center}
\scalebox{0.8}[0.8]{
  \includegraphics[]{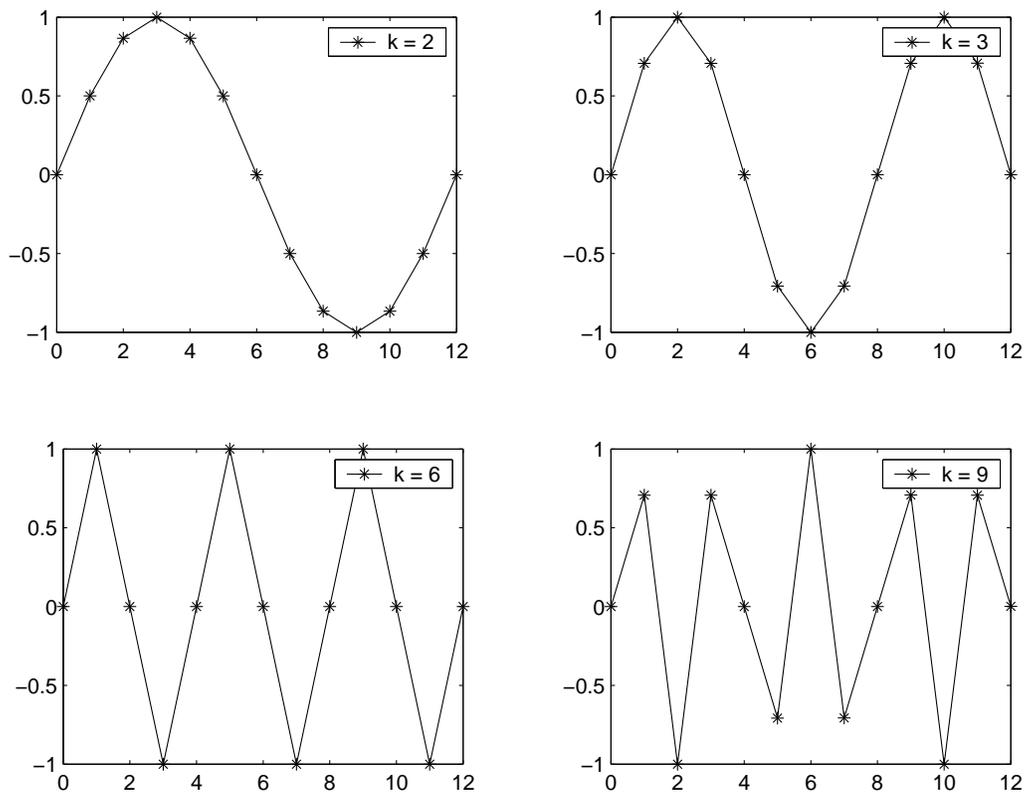}
}
\caption{Various Fourier modes on a one-dimensional grid consisting of thirteen grid points.}
\label{fmodes.eps}
\end{center}
\end{figure}

Despite the fact that red-black relaxation yields a speedy convergence rate in comparison to other relaxation schemes, it does have a major drawback which is related to the various Fourier modes which may exist in an approximate solution to Eq.~(\ref{mgeqn1}). As indicated in Fig.~\ref{fmodes.eps}, which depicts various Fourier modes of different mode numbers, for low values of the wavenumber $k$ the functions are relatively smooth across the grid. However, as soon as the wave number approaches a value comparable to half the total number of grid points, the function begins to look highly oscillatory. 

\begin{figure}[]
\begin{center}
\scalebox{0.8}[0.8]{
  \includegraphics[]{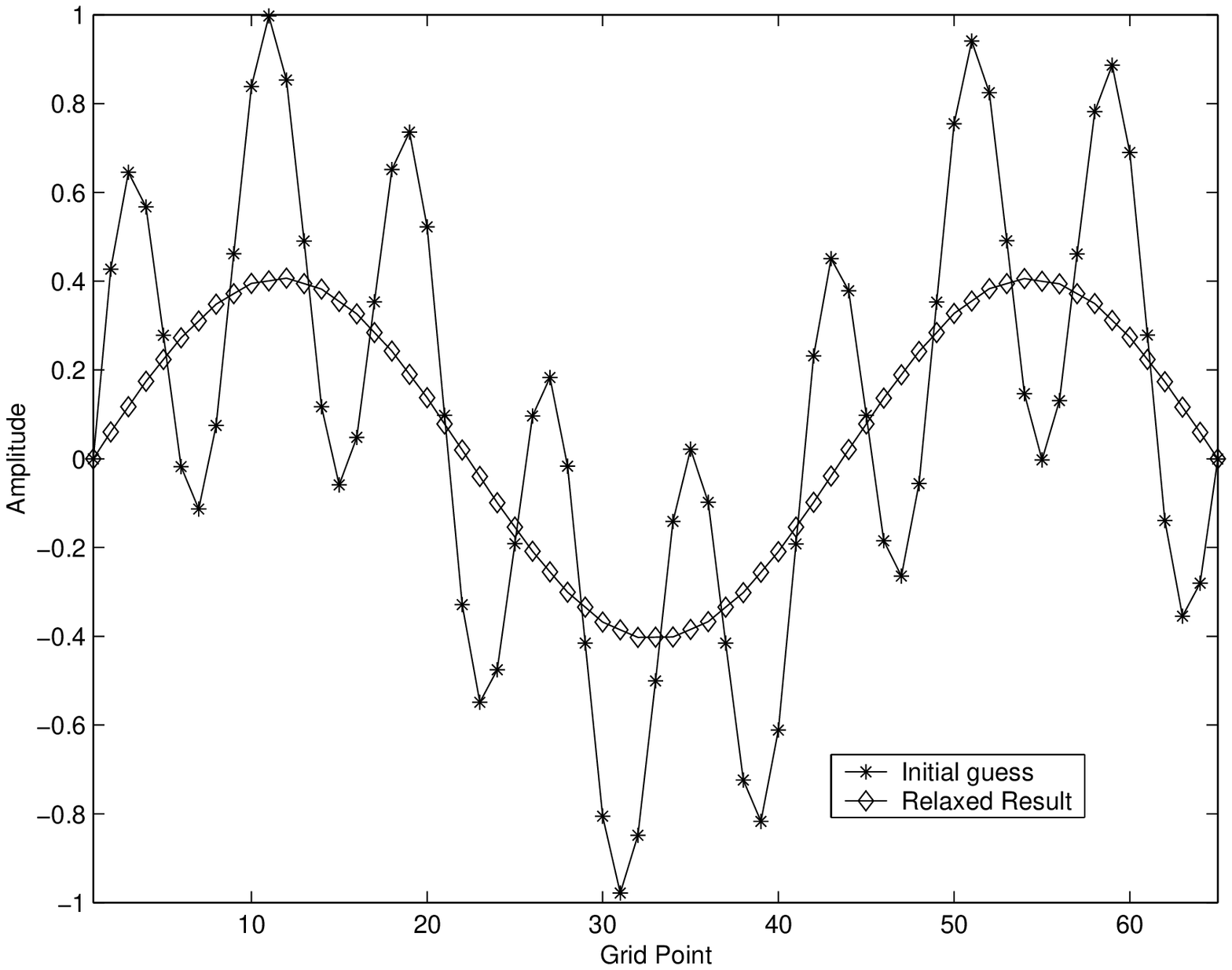}
}
\caption{Initial solution and relaxed solution for the one-dimensional Laplace equation.}
\label{onedrelax.eps}
\end{center}
\end{figure}

Figure~\ref{onedrelax.eps} shows the results of a simple one dimensional red-black relaxation scheme for Laplace's equation. The initial solution input into the relaxation scheme consists of a superposition of functions with wave numbers $k = 3$ and $k = 16$, respectively. Clearly, after twenty iterations of the relaxation scheme, the $k = 16$ mode has been completely suppressed, leaving only the $k = 3$ mode. This indicates red-black relaxation is exceedingly efficient at relaxing away high frequency modes, but apparently has difficulty suppressing low frequency modes.

\begin{figure}[]
\begin{center}
\scalebox{0.8}[0.8]{
  \includegraphics[]{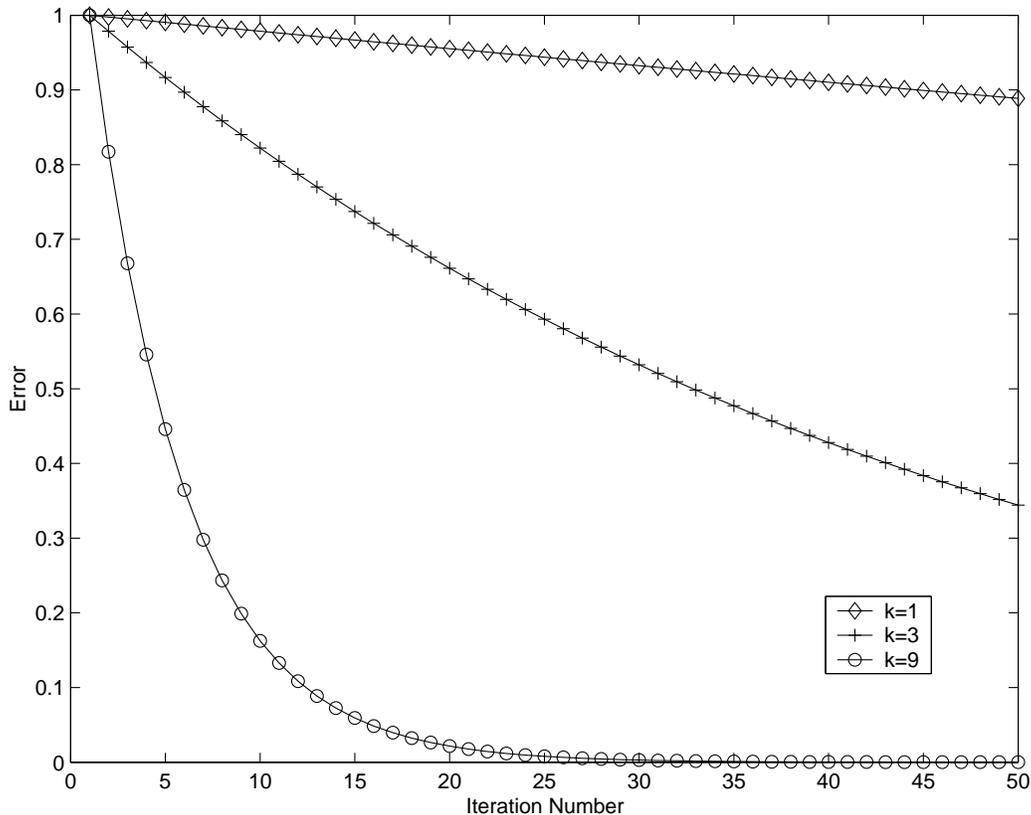}
}
\caption{Error of various Fourier modes as a function of iteration number.}
\label{ierror.eps}
\end{center}
\end{figure}

An even more dramatic demonstration of this effect is indicated in Fig.~\ref{ierror.eps}, in which three separate initial guesses to Laplace's equation were input into the red-black relaxation code, with wave numbers of $k = 1$, $k = 3$, and $k = 9$ respectively. The horizontal axis is the number of iterations, and the vertical axis is an rms measure of the residual (called the Error on this figure). As indicated in the figure, the function with the highest frequency (the $k = 9$ mode) relaxes to a solution after approximately 30 iterations. However, the convergence rate for the  $k = 3$ mode appears to be much slower, and the $k = 1$ mode's rate of convergence is horrendously slow. In other tests, even 1000 iterations had little effect on the convergence of the $k = 1$ mode. It is important to note the lack of suppression of low frequency modes is not inherent to the red-black scheme, but is manifest in all relaxation schemes \cite{numrecipes,multigrid1}.\\
\indent This presents a problem: if all relaxation schemes fail to suppress low frequency noise in a reasonable number of iterations, then how can one hope to solve Eq.~(\ref{mgeqn1}) quickly and efficiently? The answer lies in the following realization: low frequency modes on a grid of spacing $h$ look like high frequency modes on a grid of spacing 2$h$ \cite{multigrid1,multigrid2}. Stated another way, \it{low frequency modes on a fine grid look like high frequency modes on a coarse grid}\rm. If one could somehow transfer information about the approximate solution $v^h$ to a coarser grid, and then apply the red-black relaxation scheme on that coarser grid, the low frequency components of $v^h$ on the fine grid will be converted into high frequency components on the coarse grid, and will in turn be suppressed by the relaxation scheme. However, we must first discuss a method to transfer, or restrict, information from a fine grid to a coarse grid.

As an illustrative example, assume we wish to restrict the value of a function $v^h$ at a particular grid point \it{i}\rm, denoted as $v^h_i$, to a coarser grid level. A simple one-dimensional example of a restriction scheme is known as full-weighting restriction \cite{multigrid1}. This is a method in which the values from grid point $v^h_i$ as well as the values at grid points $v^h_{i-1}$ and $v^h_{i+1}$ help to determine the restricted value of the function on the coarser level, denoted by $v^{2h}$. Short hand notation for the restriction operator is $I^{2h}_h$, and can be seen as operating on the fine grid function $v^h$ in the following way:
\begin{equation}
\label{mgeqn3}
  v^{2h} = I^{2h}_h v^h.
\end{equation}

In general, it is somewhat arbitrary how one chooses the weighting associated with the value of the function at grid points $i,i \pm 1$ , etc..., but any choice is subject to the constraint that if the value of $v^h$ is equal to one everywhere, then the value of $v^{2h}$ should also be equal to one everywhere. For example, a legitimate choice \cite{multigrid1} of a one-dimensional restriction scheme is
\begin{equation}
\label{mgeqn4}
  v^{2h}_i = \frac{1}{2} v^h_i + \frac{1}{4} \left[ v^h_{i+1} + v^h_{i-1} \right],
\end{equation}
where an illegitimate choice of a one-dimensional restriction scheme would be
\begin{eqnarray}
\nonumber 
  v^{2h}_i = \frac{1}{2} v^h_i + \frac{1}{2} \left[ v^h_{i+1} + v^h_{i-1} \right].
\end{eqnarray}

\begin{figure}[]
\begin{center}
\scalebox{0.8}[0.8]{
  \includegraphics[]{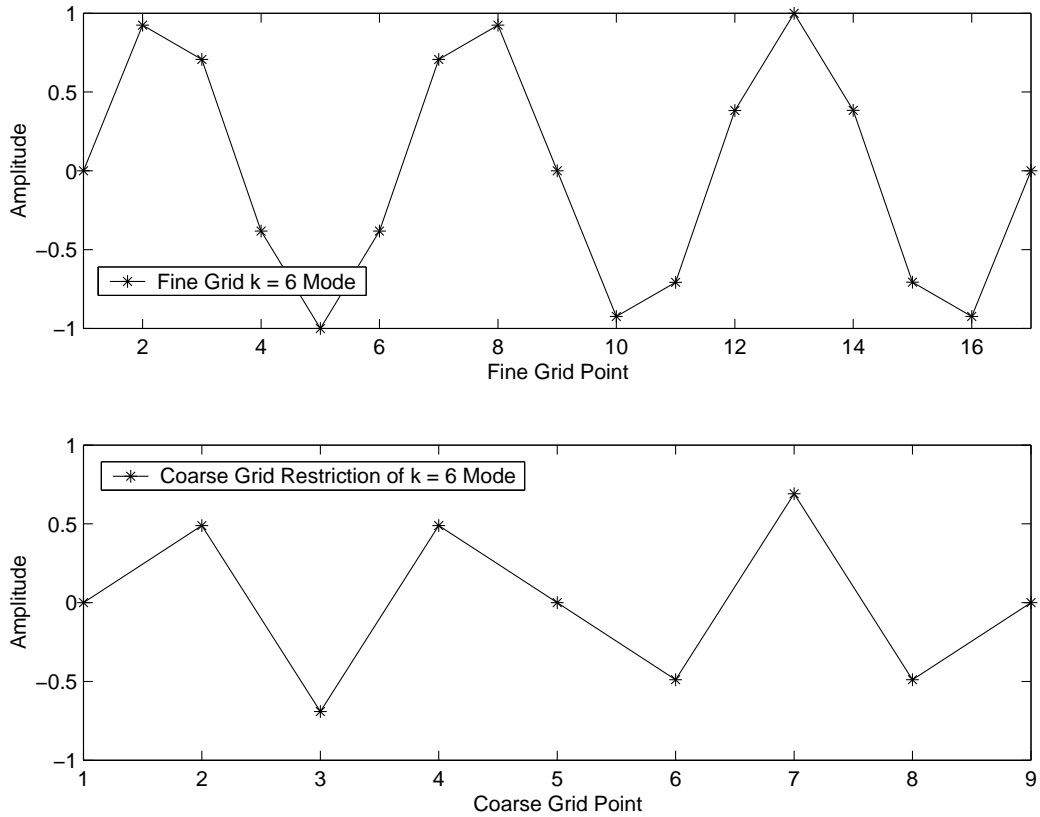}
}
\caption{Restriction of the k = 6 Fourier mode from fine to coarse grid levels.}
\label{restrict.eps}
\end{center}
\end{figure}

An implementation of the above (correct) full-weighting restriction is shown in Fig.~\ref{restrict.eps} for a $k = 6$ Fourier mode. One will note this mode is already of a relatively high frequency on the fine grid consisting of 17 grid points. However, when restricted to the coarse grid consisting of 9 grid points, the restricted function consists of even higher frequency modes. This is to our benefit, because we know that our red-black relaxation scheme will have no difficulty whatsoever in suppressing these high frequency modes on the coarse grid.

It is important to briefly note the following question: If one knows the form of the fine grid operator $A^h$, then is it proper to use the same operator on a coarse grid for subsequent relaxation? The answer, it would appear, is a resounding ``it depends''. In our investigations of systems involving linear operators, there were no ill effects when using the same operator on all grid levels, fine or coarse. We will find when dealing with nonlinear adaptive multigrid our coarse grid operator will be determined by what is called the \it{Gal\"erkin condition }\rm \cite{multigrid1,multigrid2}. However, we will defer any further discussion of the Gal\"erkin condition at this point, as it has little bearing on our development of multigrid methods for linear operators.

Let us review where we currently stand. We know relaxation on a fine grid will suppress the high frequency modes of a function, but does an unsatisfactory job of suppressing the low frequency modes on that level. We now know how to restrict information on that fine level to a coarse level. This is useful because we know low frequency modes on a fine level look like high frequency modes on a coarse level. This will allow us to suppress that low frequency mode by applying our relaxation scheme once again to the restricted function on the coarse level. There is one last piece of the puzzle remaining: transferring information from the coarse level back to the fine level.

Once again, let us use an illustrative example as a demonstration. Assume we wish to transfer, or interpolate, the value of a function $v^{2h}$ at a particular grid point \it{i}\rm, denoted as $v^{2h}_i$, back to the fine grid level. Short hand notation for the interpolation operator is $I^h_{2h}$, and can be seen as operating on the coarse grid function $v^{2h}$ in the following way:
\begin{equation}
\label{mgeqn5}
  v^h = I^h_{2h} v^{2h}.
\end{equation}

Unlike the restriction operator $I^{2h}_h$, we are constrained as to what form the interpolation operator takes. This condition may be expressed as
\begin{equation}
\label{mgeqn6}
  I^h_{2h} = c (I^{2h}_h)^T,
\end{equation}
where $c$ is some constant \cite{multigrid2}. In words, the interpolation operator is the transpose of the restriction operator, multiplied by some real-valued constant. A simple one dimensional example of an interpolation operator, which is consistent with Eq.~(\ref{mgeqn4}), is known as linear interpolation. Simply stated, fine grid points which correspond to coarse grid points take on the value of the coarse grid point. Intermediate fine grid points are simply an average of the neighboring fine grid points. Stated algorithmically,
\begin{equation}
\label{mgeqn7}
  v^h_{2i-1} = v^{2h}_i,
\end{equation}
where $i$ denotes the coarse grid number, and
\begin{equation}
\label{mgeqn8}
  v^h_i = \frac{1}{2} \left[ v^{2h}_{i+1} + v^{2h}_{i-1} \right],
\end{equation}
which is applied to the intermediate fine grid points.

\begin{figure}[]
\begin{center}
\scalebox{0.8}[0.8]{
  \includegraphics[]{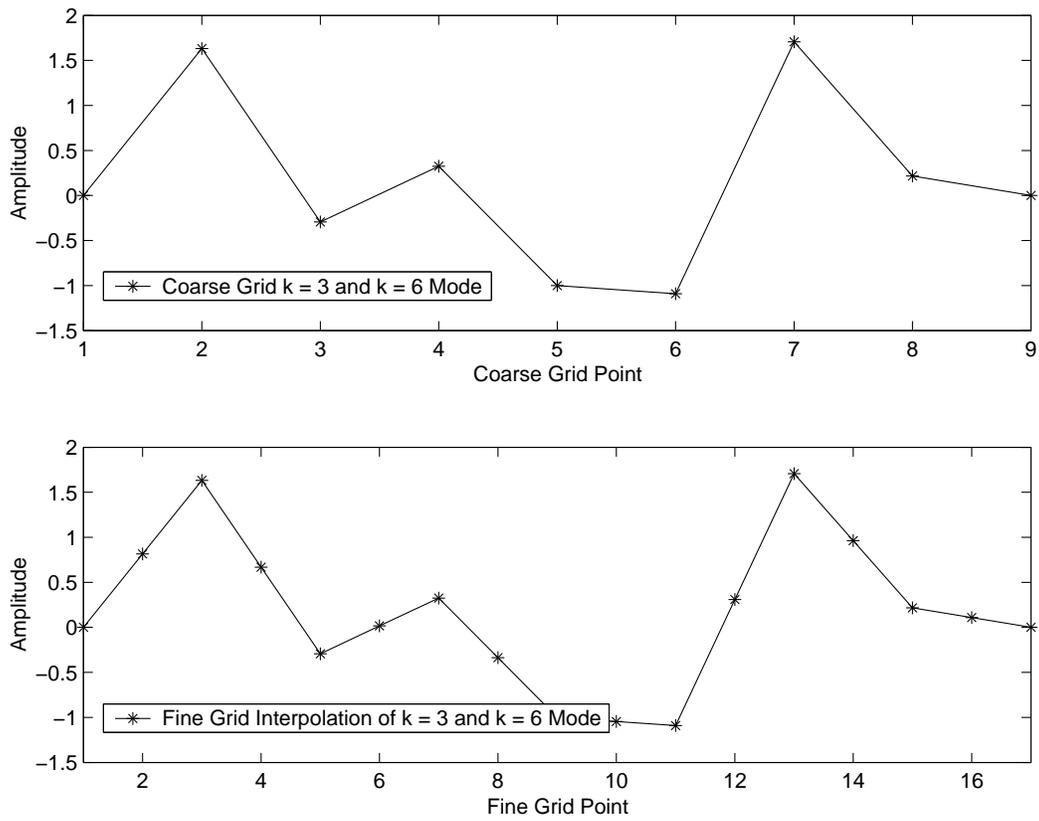}
}
\caption{Interpolation of the k = 3 and k = 6 Fourier modes from coarser to finer grid levels.}
\label{interpolate.eps}
\end{center}
\end{figure}

Figure~\ref{interpolate.eps} is a specific example of the one dimensional linear
interpolation scheme, as applied to a coarse grid function consisting of a
superposition of $k = 3$ and $k = 6$ Fourier modes. The figure shows the
information which is represented on the coarse grid is transferred to the fine
grid with little or no change---i.e., the coarse grid function is accurately represented by the interpolated function. This is important because we have made great gains in being able to suppress low frequency modes through the restriction and eventual relaxation on coarse grid levels. It would be an unfortunate circumstance if we were to lose a large portion of the relaxed coarse grid function on application of the interpolation operator.\\
\indent We now have all the foundations of multigrid laid in place. Now we simply need to determine a prescription for applying these tools to form the multigrid method for linear operators. First, we recall the equation describing the system we would like to solve, Eq.~(\ref{mgeqn1}), and also the equation defining the residual of the system, Eq.~(\ref{mgeqn2}):
\begin{eqnarray}
  A^h u^h = f^h, \\
  r^h = f^h - A^h v^h.
\end{eqnarray}
Recall $A^h$ is a linear differential operator, $u^h$ is the exact solution to the equation, $f^h$ is the source for the differential equation, $r^h$ is the residual, and $v^h$ is an approximate solution to the differential equation. We may substitute Eq.~(\ref{mgeqn1}) for $f^h$ into Eq.~(\ref{mgeqn2}) to yield
\begin{equation}
\label{mgeqn9}
  r^h = A^h u^h - A^h v^h,
\end{equation}
and because $A^h$ is a linear operator, we may rearrange Eq.~(\ref{mgeqn9}) into the following form:
\begin{equation}
\label{mgeqn10}
  A^h \left( u^h - v^h \right) = r^h.
\end{equation}

Defining the error \cite{multigrid1} of our solution $e^h$ as 
\begin{equation}
\label{mgeqn11}
 e^h = u^h - v^h,
\end{equation} 
we may substitute this definition into Eq.~(\ref{mgeqn10}) to yield what is known as the \it{residual equation }\rm \cite{multigrid1} given by
\begin{equation}
\label{mgeqn12}
  A^h e^h = r^h.
\end{equation}

The residual equation indicates that the error $e^h$ satisfies the same equation as the unknown solution $u^h$ when the source $f^h$ is replaced by the residual $r^h$. This is a very powerful statement and indicates a rudimentary method to improve our approximate solution $v^h$ to Eq.~(\ref{mgeqn1}): Assuming some approximate solution $v^h$ has been found by some means, calculate the residual $r^h$. Now, solve Eq.~(\ref{mgeqn12}) for the error, and update the approximate solution $v^h$ by using the definition of the error:
\begin{equation}
\label{mgeqn13}
 u^h = v^h + e^h.
\end{equation}

With this rudimentary idea in mind, we now develop a multigrid algorithm for linear operators known as a ``V cycle'' \cite{multigrid1}.

Assume you wish to solve a linear partial differential equation of the form of Eq.~(\ref{mgeqn1}) on a grid of dimension $d$ which consists of $n$ grid points on a side. Assume the form of the operator $A^h$ is known, and the source $f^h$ is known as well. The algorithm for a multigrid V cycle is as follows:
\begin{itemize}
  \item Given an initial guess $v^h$ on the finest grid, perform several iterations of your favorite relaxation scheme to determine an updated approximate solution $v^h$. 
  \item Calculate the residual due to this approximate solution using Eq.~(\ref{mgeqn2}).
  \item Restrict the residual from level $h$ to level $2h$ via $r^{2h} = I^{2h}_h r^h$.
  \item Relax Eq.~(\ref{mgeqn12}) on level $2h$, determining the error $e^{2h}$ for that level.
  \item Calculate the new residual on level $2h$ due to the error $e^{2h}$, and restrict this new residual to the next coarser level, level $4h$.
  \item Relax Eq.~(\ref{mgeqn12}) on level $4h$, determining the error $e^{4h}$ for that level.\\
  $ \vdots $\\
  This process continues from finer to coarser level...\\
  $ \vdots $
  \item On the coarsest level, denoted as $Nh$, either solve Eq.~(\ref{mgeqn12}) for $e^{Nh}$ exactly or relax until the resulting residual is sufficiently small.
  \item Interpolate the coarse grid error $e^{Nh}$ to the next finer level and update the fine level correction via $e^{(N-1)h} = e^{(N-1)h} + I^{(N-1)h}_{Nh} e^{Nh}$. Relax a few times to suppress any spurious high frequency modes.\\
  $ \vdots $\\
  This process continues from coarser to finer level...\\
  $ \vdots $
  \item Interpolate the coarse grid error $e^{2h}$ to the finest level and update the initial approximate solution $v^h$ via $v^h = v^h + I^h_{2h} e^{2h}$.
  \item Calculate the residual on the finest level due to this updated solution. If the residual is below an acceptable level, then $v^h$ is the numerical solution to Eq.~(\ref{mgeqn1}). Otherwise, the newly updated $v^h$ becomes the starting point for a new V cycle, in which one would simply follow the same procedure starting at the top of this list.
\end{itemize}

\begin{figure}[]
\begin{center}
\scalebox{0.8}[0.8]{
  \includegraphics[]{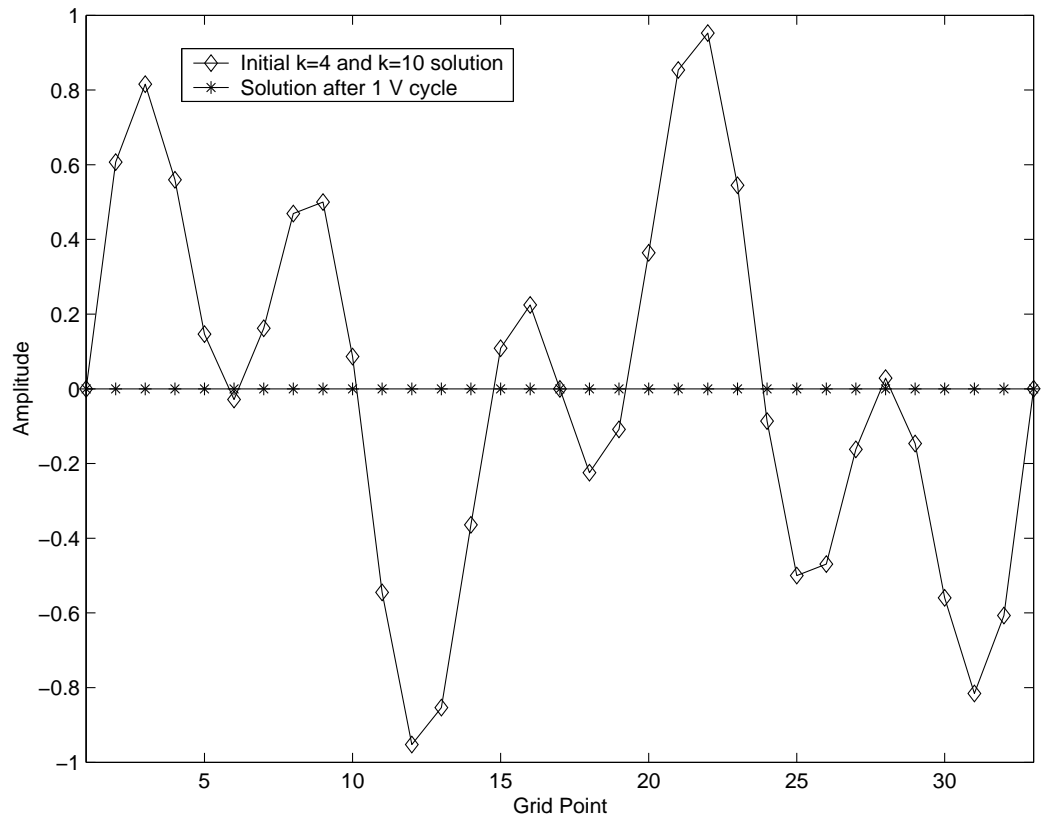}
}
\caption{The initial guess and subsequent solution to the one-dimensional Laplace equation after one multigrid V cycle.}
\label{mg1.eps}
\end{center}
\end{figure}

\begin{figure}[]
\begin{center}
\scalebox{0.8}[0.8]{
  \includegraphics[]{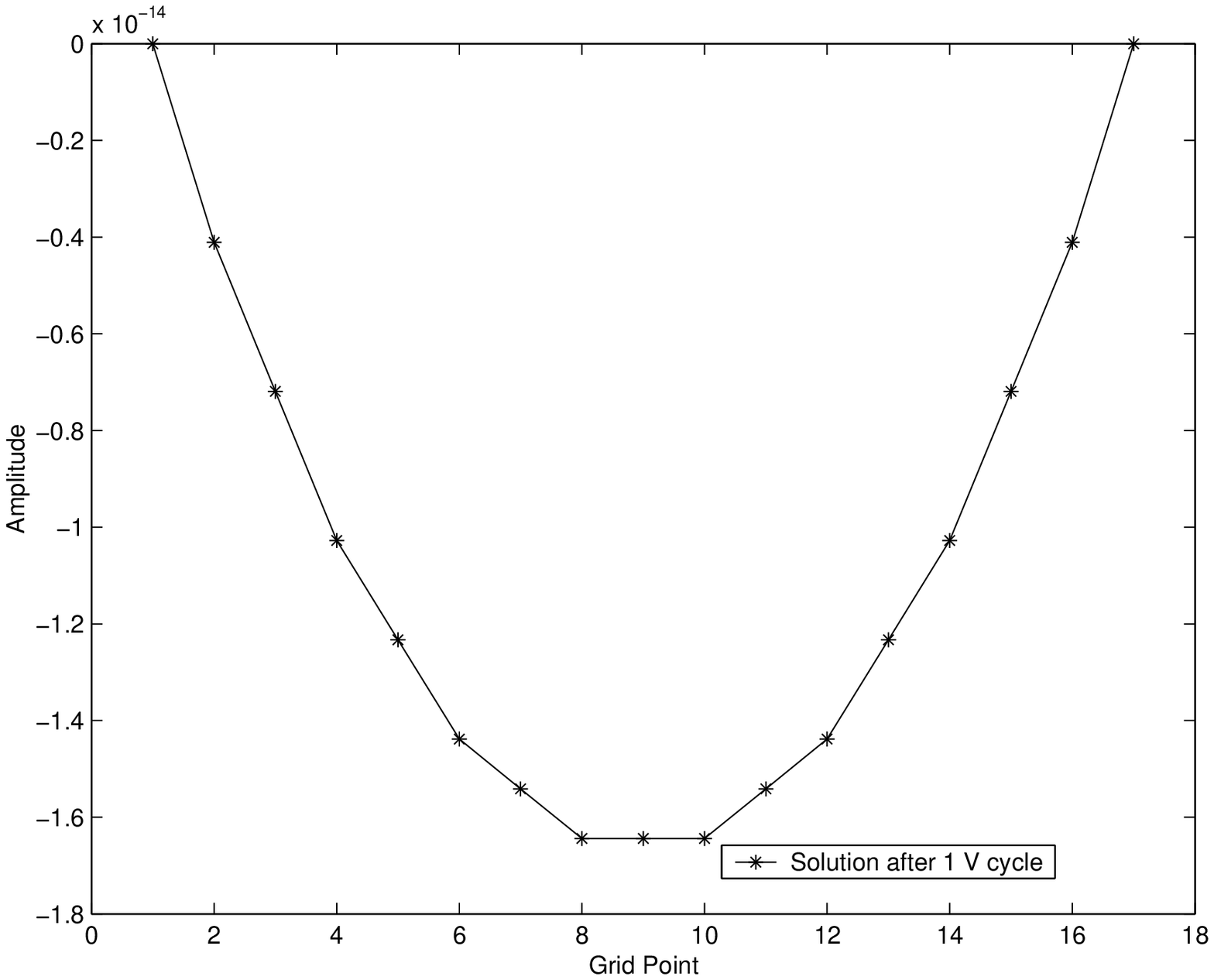}
}
\caption{Closer inspection of the solution to the one-dimensional Laplace equation after one multigrid V cycle.}
\label{mg2.eps}
\end{center}
\end{figure}

As a test of this basic V cycle process, a program was written in C++ to solve a generalized Poisson equation in one dimension. We chose to test the code by attempting to solve Laplace's equation on a one dimensional grid consisting of 17 grid points, using an initial solution consisting of Fourier modes with $k = 1$ and $k = 10$. The results of running the code through a single V cycle are shown in Fig.~\ref{mg1.eps}. It would appear that even after a single V cycle, the initial solution settles down to the desired solution ($u^h = 0$). On further inspection of the updated solution, as shown in Fig.~\ref{mg2.eps}, we can see that the error in the updated solution is on the order of $10^{-14}$. As one would expect, this error quickly approaches machine accuracy as one iterates through more V cycles. It is also important to note that for the one dimensional test code, a single V cycle is completed in a mere fraction of a second---what takes multigrid a mere blink of an eye takes traditional relaxation techniques minutes, if not hours, to accomplish.

The generalized algorithm for the linear multigrid V cycle was extended to two and three dimensions, with great success. Tests were run on the two and three dimensional codes for Laplace and Poisson equations; most notable were test codes written to solve Poisson's equation for a uniform density sphere and a uniform density thin shell. We found for a three-dimensional grid with $n$ = 33 grid points on a side, eight complete V cycles was more than sufficient to reduce the residual to the level of machine accuracy, and the resulting numerical solutions conformed to analytic solutions to well within 1 per cent.

With the foundations of linear multigrid in hand, we are now able to extend our treatment to the regime of nonlinear partial differential equations. We will find much of the ideas and analysis are parallel to linear multigrid, with a few subtle differences.

\section{Nonlinear Multigrid}
\label{Nonlinear Multigrid}
One notational difference we make is in the differential operator itself. Previously, for a linear operator acting on a function, we would denote it simply by $A^h u^h$. However, now that we are dealing with nonlinear differential operators, we must be more careful and will represent a nonlinear operator acting on a function simply by the inclusion of parenthesis: $A^h(u^h)$. With this notational change in mind, we set about analyzing the following situation: assume we wish to solve the nonlinear partial differential equation
\begin{equation}
\label{nlmg1}
  A^h (u^h) = f^h,
\end{equation}
where $A^h()$ now denotes a nonlinear differential operator, $u^h$ denotes the solution, and $f^h$ is the source. Recalling our original form for the residual,
\begin{equation}
\nonumber
  r^h = f^h - A^h (v^h),
\end{equation}
where once again $v^h$ denotes an approximate solution to our original equation, we may combine the above two equations to yield
\begin{equation}
\label{nlmg2}
  A^h (u^h) - A^h (v^h) = r^h.
\end{equation} 

Note that if we were still dealing with linear operators, we could invoke the definition of the error, $e^h = u^h - v^h$, and Eq.~(\ref{nlmg2}) would simply reduce to the linear version of the residual equation, Eq.~(\ref{mgeqn12}). However, due to the nonlinear nature of the operator $A^h ()$, we can no longer make that simplifying step. Hence, we must now use Eq.~(\ref{nlmg2}) as our new residual equation for nonlinear operators \cite{multigrid2}. Obviously, in the limit that the operator $A^h ()$ becomes linear, then the nonlinear residual equation reduces to the linear residual equation.

Now, the question of implementation arises. In theory, the algorithm for nonlinear multigrid is very similar to that for linear multigrid. In practice, as we see when we discuss the actual computer code used to solve the Hamiltonian constraint, there are some subtle points which must be addressed. We will delay discussion of these subtle points until we list the actual computer code in the Appendix. For now, we lay out a generalized algorithm for a nonlinear multigrid V cycle \cite{multigrid2}.

Assume you wish to solve a nonlinear partial differential equation of the form of Eq.~(\ref{nlmg1}) on a grid of dimension $d$ which consists of $n$ grid points on a side. Assume the form of the operator $A^h()$ is known, and the source $f^h$ is known as well. The algorithm for a nonlinear multigrid V cycle is as follows.
\begin{itemize}
  \item Given an initial guess $v^h$ on the finest grid, perform several iterations of your favorite relaxation scheme, using the nonlinear residual equation as the basis of that relaxation scheme. This results in an updated approximate solution denoted as $u^h$. 
  \item Calculate the residual due to the approximate solution using the nonlinear residual equation, Eq.~(\ref{nlmg2}).
  \item Restrict the residual from level $h$ to level $2h$ via $r^{2h} = I^{2h}_h r^h$, and restrict the updated approximate solution from level $h$ to level 2$h$ via $v^{2h} = I^{2h}_h u^h$. Note the updated fine grid approximation becomes the initial coarse grid approximation.
  \item Using the nonlinear residual equation as the basis for your relaxation scheme, relax Eq.~(\ref{nlmg2}) on level $2h$, resulting in an improved approximation $u^{2h}$.
  \item Calculate the new residual on level $2h$ using the nonlinear residual equation, and restrict this new residual to the next coarser level, level $4h$. Also restrict the updated approximate solution $u^{2h}$ to the coarser level initial approximate solution via $v^{4h} = I^{4h}_{2h} u^{2h}$.\\
  $ \vdots $\\
  This process continues from finer to coarser level...\\
  $ \vdots $
  \item On the coarsest level, denoted as $Nh$, relax Eq.~(\ref{nlmg2}) resulting in $u^{Nh}$. Calculate the error on level $Nh$ via the traditional error formula $e^{Nh} = u^{Nh} - v^{Nh}$.
  \item Interpolate the coarse grid error $e^{Nh}$ to the next finer level and update the fine level approximate solution via $u^{(N-1)h} = u^{(N-1)h} + I^{(N-1)h}_{Nh} e^{Nh}$. Relax on the updated solution a few times to suppress any spurious high frequency modes.\\
  $ \vdots $\\
  This process continues from coarser to finer level...\\
  $ \vdots $
  \item Interpolate the coarse grid error $e^{2h}$ to the finest level and update the approximate solution $u^h$ via $u^h = u^h + I^h_{2h} e^{2h}$.
  \item Calculate the residual on the finest level due to this updated solution. If the residual is below an acceptable level, then $u^h$ is the numerical solution to Eq.~(\ref{nlmg1}). Otherwise, the newly updated $u^h$ becomes the starting point for a new V cycle, in which one would simply follow the same procedure starting at the top of this list.
\end{itemize}

With the above algorithm laid out, we are now able, in theory, to tackle a litany of nonlinear partial differential equations. However, the nonlinear equation of the Hamiltonian constraint given by Eq.~(\ref{vp2-11}) requires a little more care. A natural question arises: How does one impose realistic boundary conditions on a field when solving a problem on a finite grid? This question leads us into the realm of nonlinear adaptive multigrid methods.

\section{Nonlinear Adaptive Multigrid}
\label{Nonlinear Adaptive Multigrid}
An unfortunate limitation of numerical modeling is based on attempting to gain as much resolution as possible (i.e., a high density of grid points) while also imposing realistic boundary conditions. For some applications, this does not present a problem. For example, solving Laplace's equation for a temperature distribution on a two-dimensional plate can be done relatively easily. Because the plate is of finite size, and because either the temperature distribution or the derivative must be specified on the boundary of the plate, one can develop code with relatively high grid point densities without regard for computational resources. 

On the other hand, imagine attempting to solve Poisson's equation for the Newtonian gravitational potential for a constant density star. Typically, the region of interest is close to the star, so one would like to have a relatively high density of grid points in that vicinity. However, the typical boundary condition imposed on the potential is that it falls off as $ 1/r $ when one is very far from the star. Ideally, we would like to impose this boundary condition an infinite distance from the star, but this is of course impossible in the world of numerical modeling. If we impose this boundary condition too close to the surface of the star, then the resulting solution will not model a physically realistic system.

One method of solving the problem of fine resolution with realistic boundary conditions is via a method called \it{adaptive multigrid }\rm \cite{multigrid2}. Many of the ideas previously developed for multigrid are still applicable here, but we add the feature of variable grid densities to increase or decrease the resolution around areas of interest. We do this in a way which allows us to have a high resolution around the area of interest, yet which keeps the total grid number at a level which will not exceed machine memory limitations.

\begin{figure}[]
\begin{center}
\scalebox{0.4}[0.4]{
  \includegraphics[]{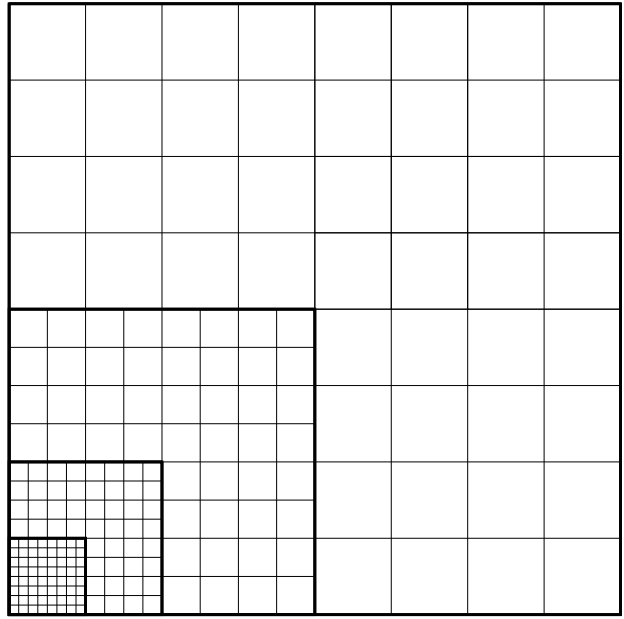}
}
\caption{A simple two-dimensional example of an adaptive grid.}
\label{adaptivegrid.eps}
\end{center}
\end{figure}

Figure~\ref{adaptivegrid.eps} is a simple illustration of the adaptive multigrid concept. It can roughly be thought of as a series of embedded adaptive levels, each with the same number of total grid points, but each consecutive adaptive level is twice as large as the previous adaptive level. As we can see in the figure, this allows for a relatively high grid density on the smallest adaptive level. It is important to note boundary conditions are not imposed on the smallest adaptive level, nor on any of the intermediate adaptive levels. Only on the largest adaptive level are the boundary conditions for the system imposed. Loosely speaking, information from the smallest level is transferred up to the next largest level. This process is continued until information from the last intermediate adaptive level is transferred up to the largest adaptive level. Once at this largest level, then regular multigrid methods can be employed, which solve the problem on the largest grid by imposing the appropriate boundary conditions on that level. Once the multigrid  process is completed, then the new information on the largest level is transferred to the smaller intermediate adaptive level. This process continues until information from the second smallest level is transferred to the smallest adaptive level. This whole process consists of one iteration step---in general the process may be continued for an arbitrary number of iterations. 

In practice, the procedure is quite a bit more delicate than the above paragraph would indicate. One must be very careful as to how and what information is transferred to particular grid points from one adaptive level to the next, so in the algorithm which follows, particular attention will be paid to each step in hopes of minimizing the ambiguities and confusion.

Assume you wish to solve a nonlinear partial differential equation of the form of Eq.~(\ref{nlmg1}) on a grid of dimension $d$ which consists of $n$ grid points on a side. We demand $n$ is of the form $n = 2^{MG} + 3$, where $MG$ is an integer representing the number of multigrid levels. Assume the form of the operator $A^h()$ is known, and the source $f^h$ is known as well. A natural length which plays an important role in the process is the physical length of the smallest adaptive level, which we denote as $LEN$. Another important length in the problem is denoted as $SCALE$, which is used to ensure that the grid spacings of a larger adaptive level are exactly twice that of the smaller adaptive level.

Once a small grid resolution has been determined via the choice of the parameter $MG$, one is free to choose the number of adaptive levels, denoted by the parameter $ADAPT$. The choices of this integer parameter may depend on computer resources, desired solution accuracy, or computational time constraints. Obviously, the larger the value of $ADAPT$, the more memory required, and the more time involved to execute the program.

It is of vital importance to be mindful of the physical sizes of the adaptive grids. For instance, the relationship between the natural scale of the problem, $SCALE$, and the physical length of the smallest adaptive level, $LEN$, is given by
\begin{equation}
\label{amg1}
  LEN = \frac{2^{MG}+2}{2^{MG}} SCALE.
\end{equation} 
Again, the relationship in Eq.~(\ref{amg1}) ensures the consistency in grid spacings from one adaptive level to the next. Once the smallest grid size is specified, then each larger adaptive level is simply twice as big as the previous adaptive level. For instance, if we denote the physical size of a particular adaptive level as $ d_{i} $, then $ d_0 = LEN $, $ d_1 = 2*LEN $, $ d_2 = 2*d_1 = 4*LEN $, etc.... This trend continues until we reach the largest adaptive level. Ensuring the grid spacings between the second largest and the largest levels are consistent, as well as ensuring we can perform straight nonlinear multigrid on this largest level, the physical size of the largest level is denoted as \it{maxsize}\rm,  and is given by
\begin{equation}
\label{amg2}
  maxsize = 2^{ADAPT} SCALE.
\end{equation} 

It is again important to point out that the number of grid points on all adaptive levels \it{except}\rm~ the largest adaptive level is the same, and is given by $ n = 2^{MG} + 3 $. The number of grid points on the largest adaptive level is given by $ n = 2^{MG} + 1 $. The reason for the difference in grid points numbers will become apparent when we discuss the full nonlinear adaptive multigrid algorithm, which we now present.
\begin{itemize}
  \item Begin by relaxing on the smallest adaptive level. It is important to only relax on the interior $ n - 3 $ grid points, without imposing any boundary conditions in the relaxation scheme. Denote the relaxed solution as $ u^{h} $, and the initial solution as $ v^{h} $.
  
  \item Calculate the residual due to the relaxed solution $ u^{h} $ from the standard nonlinear residual equation, given by Eq.~(\ref{nlmg2}), everywhere on the grid.
  
  \item Restrict the relaxed solution $ u^{h} $, which exists on the interior $ n - 3 $ grid points of the smallest adaptive level, onto the interior $ n/2 - 1 $ grid points of the next larger adaptive level, and denote it as $ u^{2h} $. One may think of this restricted ``interior'' solution as information which will eventually allow us to determine a correction to the smallest adaptive level. However, any solution information which exists outside of this ``interior'' grid region should be thought of as the best possible solution for that exterior region, and has nothing to do with a correction to the smaller level.
  
  \item The source for level $ 2h $ must now be calculated. This is achieved by first calculating the residual on level $ 2h $ due to the current approximate solution $ u^{2h} $, and then by restricting the residual $ r^h $ into the interior $ n/2 - 1 $ grid points of level $ 2h $. This ties into the concept that the interior region is related to the correction to be added to the smaller adaptive level, where as the exterior region deals with the best possible information on that particular level.
  
  \item Relax on the interior $ n - 3 $ grid points of level $ 2h $, without imposing any boundary conditions on the system.
  
  \item Keeping in mind the interior regions of larger adaptive levels consist of information related to a correction to be added back onto smaller adaptive solutions, and the exterior regions are the best possible information relating to the solution on that level, repeat the first five steps until the small grid solutions and residuals are eventually restricted to the interior $ n/2 $ grid points of the largest adaptive level. Note on none of the smaller adaptive levels are boundary conditions imposed on the solution. It is only on the largest adaptive level (and the associated multigrid levels) are boundary conditions imposed.
  
  \item Perform nonlinear multigrid on the largest adaptive level, using the algorithm set forth in \S(\ref{Nonlinear Multigrid}). Again, it is at this point of the algorithm, and only at this point, the boundary conditions are imposed on the system. On completion of a number of multigrid V cycles on the largest adaptive level, we are left with a solution which we denote as $ u^{ADAPT} $. Care must now be taken to ensure the interior region solution and the exterior region solution are treated in the proper fashion; specifically, the interior $ n/2 $ solution is related to the smaller grid correction, and the exterior $ n/2 $ solution is the best possible solution for that grid level.
  
  \item Calculate the error $ e^{ADAPT} = u^{ADAPT} - v^{ADAPT} $ on the interior $ n/2 - 1 $ grid points on the largest adaptive level. Interpolate this error onto the interior $ n - 3 $ grid points of the smaller adaptive level. This interpolated error, denoted as $ e^{ADAPT-1} $ should be added to the interior solution of the smaller adaptive level via $ u^{ADAPT-1} = u^{ADAPT-1} + e^{ADAPT-1} $. The exterior solution of $ u^{ADAPT} $, which represents the best solution on that level, should replace the exterior grid points of the next smaller adaptive level. This ensures the accurate solution information is transferred from one adaptive level to the next smaller level, and it also acts as an "anchor" for the relaxation scheme, fixing the field on the boundary of smaller adaptive levels to a particular value determined by the boundary conditions on the largest adaptive level.
  
  \item Continue interpolating the larger grid errors and the larger grid solutions to smaller and smaller grids. A few relaxation sweeps on each level may be employed, but again relaxation only occurs on the interior $ n - 3 $ grid points, and boundary conditions are not imposed on the smaller levels. Recall the information at grid point $ n - 2 $ coincides with the best solution at that particular physical distance---information which comes directly from the previous larger adaptive level. It is the solution at this point which acts, in effect, as the ``boundary condition'' for the relaxation of the interior $ n - 3 $ grid points, and assures that the solution on smaller adaptive levels is consistent with the boundary conditions imposed at the largest adaptive level.
  
  \item On reaching the smallest adaptive level the residual error, a measure of how well the approximate solution is solving the finite difference equation, and the truncation error, a measure of how much the approximate solution is changing from one iteration to the next, may be calculated. If the errors are below certain tolerances, then the equation of interest has been solved. If the errors are not below the tolerances, then simply return to the first step and reiterate the algorithm.   
\end{itemize}

This adaptive nonlinear multigrid algorithm is employed to solve the Hamiltonian constraint, which determines the conformal factor for a binary black hole system. We will briefly discuss some numerical tests of our code at the end of this chapter, but we defer discussion of numerical results of our variational principle until the next chapter. Before then, there are several other numerical issues which must be addressed.

\section{Other Numerical Issues}

\subsection[Reformulation of the Nonlinear PDE]{Reformulation of the Nonlinear Partial Differential \\Equation}
\label{nlpde}
Recall the motivation for our numerical endeavor. The Hamiltonian constraint of the Einstein equations is a nonlinear partial differential equation describing the conformal factor for a binary black hole system. Recall the conformal factor $ \psi $ is given by
\begin{equation}
\label{oni1}
  \psi = \frac{1}{\alpha} + U,
\end{equation}
where $1/ \alpha $ is given by
\begin{equation}
\label{oni2}
  \frac{1}{\alpha} = \sum^N _{i=1} \frac{M_{(i)}}{2 \vert \vec{r} - \vec{r}_{(i)} \vert }. 
\end{equation}
The ``Newtonian'' mass of the puncture is denoted as $ M_{(i)} $, and the puncture location is determined by the vector $ \vec{r}_{(i)} $. With this form for the conformal factor, the Hamiltonian constraint becomes an equation for $ U $, given by
\begin{equation}
\label{oni3}
  \nabla^2 U + \beta ( 1 + \alpha U )^{-7} = 0,
\end{equation}
where $ \beta $ is related to the conformal extrinsic curvature $ K^{ab} $ of the geometry via
\begin{equation}
\label{oni4}
  \beta = \frac{1}{8} \alpha^7 K^{ab} K_{ab}.
\end{equation}

Recall the extrinsic curvature must satisfy the momentum constraint equations of the Einstein equations. Due to the linearity of the momentum constraint, the extrinsic curvature is simply a sum of the conformal extrinsic curvatures of each puncture individually:
\begin{equation}
\label{oni5}
  K^{ab} = \sum^{N} _{i=1} K^{ab} _{PS(i)},
\end{equation}
where $ K^{ab} _{PS(i)} $ depends on the linear momentum $ P^{a} $ and spin angular momentum $ S_{b}$ of the punctures in the following way:
\begin{eqnarray}
\label{oni6}
  K^{ab} _{PS(i)} & = & \frac{3}{2r^2} [P^a n^b + P^b n^a - (g^{ab} - n^a n^b) P^c n_c ] \nonumber \\
                   & &~  + \frac{3}{r^3} [ \epsilon^{acd} S_c n_d n^b + \epsilon^{bcd} S_c n_d n^a ],
\end{eqnarray}
where $ r^2 = x^2 + y^2 + z^2 $ in normal Cartesian coordinates, $ n^a = x^a / r $, and $ g^{ab} $ is the normal flat space metric in Cartesian coordinates.

Now that we have refreshed our memory about the equation we wish to solve, we realize that if we are to have any success in solving this nonlinear equation numerically, we must linearize the equation in a way which will allow our numerical scheme to converge to the solution of Eq.~(\ref{oni3}). First, we must rewrite the equation in terms of a new variable for reasons related to the boundary condition we wish to impose. Further discussion on the nature of the boundary condition, known as the Robin condition, is deferred until \S(\ref{subsecrobin}). The first step is defining the following:
\begin{equation}
\label{oni7}
  U \equiv 1 + u .
\end{equation}

\noindent With this change of variables, Eq.~(\ref{oni3}) becomes
\begin{equation}
\label{oni8}
  \nabla^2 u + \beta ( 1 + \alpha + \alpha u )^{-7} = 0.
\end{equation}

Now, the task at hand is to linearize Eq.~(\ref{oni8}) in a way which preserves its nonlinear flavor. One way to achieve this is to imagine the quantity $ u $ consists of some value $ u_0 $ plus a perturbation $ \delta u $, namely $ u = u_0 + \delta u $. Plugging the full perturbed form of $ u $ into Eq.~(\ref{oni8}) yields
\begin{equation}
\label{oni9}
  \nabla^2 ( u_0 + \delta u ) + \beta ( 1 + \alpha + \alpha u_0 + \alpha \delta u)^{-7} = 0.
\end{equation}
In anticipation of linearization, we begin to isolate the terms involving $ \delta u $ in the following way:
\begin{equation}
\label{oni10}
  \nabla^2 ( u_0 + \delta u ) 
     + \beta ( 1 + \alpha + \alpha u_0 )^{-7} 
          \left( 1 + \frac{\alpha \delta u}{1 + \alpha + \alpha u_0} \right)^{-7} = 0.
\end{equation}

Now, by assuming $ \delta u \ll 1 $, we Taylor expand the equation to yield the following:
\begin{equation}
\label{oni11}
  \nabla^2 ( u_0 + \delta u ) 
     + \beta ( 1 + \alpha + \alpha u_0 )^{-7} 
          \left( 1 - \frac{7 \alpha \delta u}{1 + \alpha + \alpha u_0} \right) = 0.
\end{equation}

Now, we add and subtract terms in an attempt to regroup terms of the form $ u_0 + \delta u $, i.e., make the substitution $ \delta u \longrightarrow \delta u + u_0 - u_0 $. In doing so, and keeping only the $ u_0 + \delta u $ terms on the left hand side of the equation, we have
\begin{eqnarray}
\label{oni12}
  \nabla^2 ( u_0 + \delta u ) \!\!\!& - &\!\!\! 7 \alpha \beta ( 1 + \alpha + \alpha u_0 )^{-8} ( u_0 + \delta u ) \nonumber \\
     & & = - \beta ( 1 + \alpha + \alpha u_0)^{-7} 
              \left[ 1 + \frac{7 \alpha u_0}{1 + \alpha + \alpha u_0 } \right].
\end{eqnarray}
Once again using the identification of $ u = u_0 + \delta u $, we are left with our final equation describing  $ u $:
\begin{eqnarray}
\label{oni13}
  \nabla^2 u \!\!\!& - &\!\!\! 7 \alpha \beta ( 1 + \alpha + \alpha u_0 )^{-8} u \nonumber \\
     & & = - \beta ( 1 + \alpha + \alpha u_0)^{-7} 
              \left[ 1 + \frac{7 \alpha u_0}{1 + \alpha + \alpha u_0 } \right].
\end{eqnarray}

One way to interpret the meaning of this equation is to view $ u_0 $ as an ``old'' value of the field $ u $, so we can see in Eq.~(\ref{oni13}) how the field at a point depends on itself in a nonlinear fashion.

In anticipation of finite differencing the partial differential equation, we take inspiration from the nonlinear residual equation $ A(u) - A(v) = r $ to yield the final version of the equation in which we will apply our nonlinear adaptive multigrid scheme:
\begin{eqnarray}
\label{oni14}
  \nabla^2 u \!\!\!& - &\!\!\! 7 \alpha \beta ( 1 + \alpha + \alpha u_0 )^{-8} u \nonumber \\
      & & = s + \nabla^2 v + \beta ( 1 + \alpha + \alpha v )^{-7} \nonumber \\
      & &~ - \beta ( 1 + \alpha + \alpha u_0)^{-7} \left[ 1 + \frac{7 \alpha u_0}{1 + \alpha 
                        + \alpha u_0 } \right]  .
\end{eqnarray}
In the above equation, $s$ denotes the source on a particular level and $v$ denotes the previous best solution on a particular level. Equation~(\ref{oni14}) forms the basis of our relaxation scheme, as well as the method in which we calculate residuals. In practice, Eqn.~(\ref{oni14}) will become a finite-differenced equation for $u$. A subtle detail of Eqn.~(\ref{oni14}) is the interpretation of $s$. For instance, on the smallest adaptive level we are interested in a solution to Eqn.~(\ref{oni13}); this requires the source $s$ and the field $v$ to be initialized to zero on the smallest adaptive level. However, on subsequent adaptive levels, the source $s$ is roughly the restricted residual from a previous smaller adaptive level, and the field $v$ is the previous best solution which exists on that level. Details of the algorithm may be found in the Appendix.

\subsection{The Robin Boundary condition}
\label{subsecrobin}
As has been mentioned numerous times above, imposing realistic boundary conditions on the field of interest is of vital importance if one wishes to accurately model physically realistic systems.

Asymptotic flatness dictates the field $ U $ must behave as $U - 1 = O(r^{-1})$ for distances far from the punctures \cite{BB}. This translates to an approximate boundary condition, known as the Robin condition, which may be constructed by demanding when far from the source, the field $ U $ should be dominated by a monopole term \cite{Cook3}. This allows us to place a constraint on the radial derivative of $ U $ in the following way:
\begin{equation}
\label{oni15}
  \frac{\partial U}{\partial r} = \frac{1 - U}{r}.
\end{equation}

Recall in \S (\ref{nlpde}) we started with an equation describing the field $ U $, but then made a variable substitution to re-express the equation in terms of the field $ u $, which is related to $ U $ via $ U = 1 + u $. The reasoning behind this change of variables deals with the linearity of the Robin condition. To test the linearity of the Robin condition, we use two test functions, denoted as $ U_1 $ and $ U_2 $. Denoting the Robin condition operation as $ RC() $, we wish to see if $ RC(U_1) + RC(U_2) = RC(U_1 + U_2) $. Plugging $ U_1 $ and $ U_2 $ into Eq.~(\ref{oni15}), we find
\begin{equation}
\label{oni15-a}
  \frac{\partial U_1}{\partial r}  + \frac{\partial U_2}{\partial r} 
            = \frac{1 - U_1}{r} + \frac{1 - U_2}{r}
\end{equation}
which becomes
\begin{equation}
\label{oni16}	    
   \frac{\partial U_1}{\partial r}  + \frac{\partial U_2}{\partial r}
   = \frac{2 - U_1 - U_2}{r},
\end{equation}
which most assuredly is not the same as 
\begin{equation}
\label{oni17}
  \frac{\partial ( U_1 + U_2 ) }{\partial r} = \frac{1 - U_1 - U_2}{r}.
\end{equation}

Because of this nonlinear nature of the Robin condition, we must proceed with caution. Let us identify $ U \equiv 1 + u $ and  rewrite the Robin condition on $ u $ to read
\begin{equation}
\label{oni18}
  \frac{\partial u }{\partial r} = -\frac{u}{r}.
\end{equation}
This indicates the field $ u $ is dominated by a monopole term, and goes to zero far from the punctures, which preserves the asymptotic behavior of the field $ U $. Let us re-test the linearity of our reformulated Robin condition, Eq.~(\ref{oni18}), using the test fields $ u_1 $ and $ u_2 $. Writing down the term corresponding to $ RC(u_1 + u_2) $, we find
\begin{equation}
\label{oni19}
  \frac{\partial (u_1 + u_2) }{\partial r} = -\frac{(u_1 + u_2) }{r}.
\end{equation}
Now, comparing to $ RC(u_1) + RC(u_2) $ we find
\begin{equation}
\nonumber
  \frac{\partial u_1 }{\partial r} + \frac{\partial u_2 }{\partial r} 
        = -\frac{u_1}{r} - \frac{u_2}{r},
\end{equation}
which in turn becomes
\begin{equation}
\label{oni20}	
   \frac{\partial u_1 }{\partial r} + \frac{\partial u_2 }{\partial r}
   = -\frac{(u_1 + u_2)}{r}.
\end{equation}

Equation~(\ref{oni20}) is the same as Eq.~(\ref{oni19}), which indicates that the Robin condition specified by Eq.~(\ref{oni18}) is indeed a linear operator. Hence, the equation we wish to solve is Eq.~(\ref{oni14}) for the field $ u $ subject to the Robin boundary condition, given by Eq.~(\ref{oni18}).

As a final note on the Robin boundary condition, we must keep in mind we will employ our numerical scheme on a cubic lattice, so we must express the radial derivative using the following identification:
\begin{equation}
\label{oni21}	
  \frac{\partial}{\partial r} = n^a \nabla_a 
      = \frac{x}{r} \frac{\partial}{\partial x} + \frac{y}{r} \frac{\partial}{\partial y} 
                                      + \frac{z}{r} \frac{\partial}{\partial z}.
\end{equation}
The identity stated in Eq.~(\ref{oni21}), along with some minor simplification, yields the final form of the Robin condition employed in the numerical algorithm:

\begin{equation}
\label{oni22}	
  x \frac{\partial u }{\partial x} + y \frac{\partial u }{\partial y} 
         + z \frac{\partial u }{\partial z} = - u.
\end{equation}

\subsection{The Gal\"erkin Operator and Finite Differencing}
\label{galerkin}
Up to this point, we have avoided discussing the issue of determining the form of the coarse grid operator $A^{2h}$ with knowledge of the fine grid operator $A^h$. In general, there is no right or wrong way in which to choose the coarse grid operator, but its determination really depends on the type of problem one wishes to solve. 

Before we delve into the discussion of coarse grid operators and methods of finite differencing, let us first introduce some handy notation. It is often the case one can uniquely determine the form of an operator simply by its \it{stencil }\rm \cite{multigrid1}. The stencil of a particular operator is simply the numerical coefficients associated with an operator acting on a particular grid point. The notation we use for stencils will be the following (nonstandard) convention: 
\begin{equation}
\label{gal1}
  [0,1,2,3] = (A,B,C,D),
\end{equation}
where the numbers 0, 1, 2, and 3 in the square brackets denote the central grid point at which the operator is acting, grid points which are once removed from the central grid point, grid points twice removed from the central grid point, and grid points three times removed from the central grid point, respectively. The numerical coefficients $A$, $B$, $C$, and $D$ in the parenthesis correspond to the above mentioned grid points. As an example, assume we want to write the stencil for a simple identity operator. In this case, the stencil would be of the form:
\begin{equation}
\label{gal2}
  [0,1,2,3] = (1,0,0,0),
\end{equation}
simply because this identity operator only depends on the value at the central grid point. Another example is with the ``traditional'' $\nabla^2$ operator in three dimensions, whose stencil is given by:
\begin{equation}
\label{gal3}
  [0,1,2,3]_a = (-6,1,0,0).
\end{equation}
With this handy notation, we may now discuss the matter of determining coarse grid operators. 

As mentioned above, the method in which one determines the coarse grid operator $A^{2h}$ really depends on the problem at hand. In some instances, this issue may not even present itself as a problem; for our initial investigations into multigrid methods in which we were attempting to solve Poisson's equation in three dimensions, we gave little thought to the consequences of our choice of coarse grid operator. In effect, we used the stencil given by Eq.~(\ref{gal3}) as our fine and coarse grid operator. As was mentioned in previous sections, we had great success in solving Poisson's equations with this operator, but situations may arise (and they do!) in which this particular choice of coarse grid operator does not lead to convergence. A natural question arises: How does one determine the ``correct'' coarse grid operator?

The answer lies in what is known as the \it{Gal\"erkin condition}\rm \cite{multigrid2}, which was briefly mentioned in previous sections. Recalling our notation for restriction ($I^{2h}_h$) and interpolation ($I^h_{2h}$) operators, the Gal\"erkin condition may be stated mathematically as
\begin{equation}
\label{gal4}
  A^{2h} = I^{2h}_h A^h I^h_{2h},
\end{equation}
where once again $A^h$ is the fine grid operator, and $A^{2h}$ is the resulting coarse grid operator. The motivation for Eq.~(\ref{gal4}) may be viewed as follows: Imagine we have a coarse grid function $u^{2h}$ which is a solution to the problem of interest. If this function is indeed a solution on the coarse grid, then we would hope that it would also be a solution to the fine grid problem. Another way of thinking of this is the following: If we interpolate a coarse grid solution to the fine grid and then calculate the fine grid residual, we expect that fine grid residual to be zero. When we restrict the fine grid residual back to the coarse grid, we expect the residual to remain unchanged from its fine grid value (i.e., the residual should remain zero). If this condition is satisfied, then the coarse grid operator $A^{2h}$ is a Gal\"erkin operator.

One could calculate the Gal\"erkin operator ``on the fly''---that is, write computer code to generate the correct Gal\"erkin stencil for every coarse grid encountered during the multigrid or adaptive multigrid V cycles. However, we believe it is more beneficial to determine a Gal\"erkin operator which has exactly the same form on all levels, from the finest to the coarsest. This allows for the determination of a single operator, valid on all levels, which facilitates writing code, testing, and debugging.

Because the class of problems we are interested in can be deemed ``Poisson-like'', we endeavor to determine a stencil for $\nabla^2$ which can be used on all levels and satisfies the Gal\"erkin condition, Eq.~(\ref{gal4}). 

The starting point of this process is to determine all of the linearly independent stencils for the $\nabla^2$ operator which are accurate for quadratic functions. This is done by constructing a ``test cube'' of three grid points on a side, where the center grid point of the cube is the grid point in which we will apply our various $\nabla^2$ operators. We now associate an unknown coefficient to each classification of grid point, much as we do with the stencils mentioned above. Specifically, associate the coefficient $A$ with the central point, the coefficient $B$ to those grid points once removed, $C$ for those twice removed, and $D$ for the grid points which are three times removed from the central grid point.

The condition that we wish our $\nabla^2$ to be accurate for quadratic functions means that it must yield the correct results for functions of the form $f(x) = 1$, $f(x) = x$, and $f(x) = x^2$. This means when we apply our $\nabla^2$ operator to our ``test cube'', we demand the expected results. For instance, if our test cube is filled with a constant value of $f(x) = 1$, then we know $\nabla^2 (1) = 0$ everywhere. This implies the application of the operator to our cube yields
\begin{equation}
\label{gal5}
  A + 6B + 12C + 8D = 0.
\end{equation}

We obviously need more equations to solve for the numerical coefficients, and this is achieved by applying the functions $f(x) = x$ and $f(x) = x^2$ to our test cube. The results of those operations are
\begin{equation}
\label{gal6}
  -B - 4C - 4D + B + 4C + 4D = 0,
\end{equation}
which is simply a tautology, and 
\begin{equation}
\label{gal7}
  2B + 8C + 8D = 2.
\end{equation}

We now use Eq.~(\ref{gal5}) and Eq.~(\ref{gal7}) to determine the coefficients. Noting we have some freedom in the determination of our coefficients, we first set $B$ to be zero---our ``traditional'' $\nabla^2$ stencil already depends on the nearest neighbors, so we would like to rule this stencil out of our remaining choices. Once this is done, we may choose $C$ to be zero, which yields the following stencil:
\begin{equation}
\label{gal8}
  [0,1,2,3]_b = \left( -2,0,0,\frac{1}{4} \right).
\end{equation}
If we were to choose $D$ to be zero, then we would generate a different stencil for $\nabla^2$:
\begin{equation}
\label{gal9}
  [0,1,2,3]_c = \left( -3,0,\frac{1}{4},0 \right).
\end{equation}

Armed with our three linearly independent stencils for $\nabla^2$, recall what the purpose of this exercise is: we wish to determine a stencil for $\nabla^2$ which has the same form on every level, and also satisfies the Gal\"erkin condition given by Eq.~(\ref{gal4}). Stated another way, assume we write our as-yet unknown Gal\"erkin operator as a linear combination of our three existing stencils for $\nabla^2$:
\begin{equation}
\label{gal10}
  \nabla^2_G = a (-6,1,0,0) + b \left( -2,0,0,\frac{1}{4} \right) + c \left(-3,0,\frac{1}{4},0 \right),
\end{equation}
where $\nabla^2_G$ denotes the Gal\"erkin operator, and $a$, $b$, and $c$ are unknown numerical coefficients. Our job is to determine these coefficients such that this Gal\"erkin operator has exactly the same stencil on every level---fine or coarse.

The method used to determine the unknown coefficients $a$, $b$, and $c$ relies on the definition of the Gal\"erkin condition, Eq.~(\ref{gal4}), along with some computer code written in C++. The computer code simply filled our test cube of 27 grid points with zeros everywhere, except at the center-most grid point. At this central point the value is set equal to unity. Then, the code simply applies the Gal\"erkin condition to each of our three stencils individually, acting on the test cube. For instance, recall our ``traditional'' stencil for $\nabla^2$ is simply $[0,1,2,3]_{a} = (-6,1,0,0)$. However, when our test cube is interpolated to a finer level, acted on by our traditional $\nabla^2$ operator, and in turn restricted back to the coarse level, the resulting stencil is $[0,1,2,3]_{Ga} = \left( -54,3,\frac{5}{2},\frac{3}{4} \right)$. The subscript $Ga$ indicates that the stencil is the result of the Gal\"erkin condition, and is associated with numerical coefficient $a$. Likewise, performing the same operations on the $b$ and $c$ stencils yield 
$[0,1,2,3]_{Gb} = \left( -38,-1,\frac{5}{2},\frac{7}{4} \right)$ and 
$[0,1,2,3]_{Gc} = \left( -45,\frac{1}{2},\frac{11}{4},\frac{9}{8} \right)$, respectively.

Likewise, when the three stencils are not interpolated and restricted, but simply applied to the test cube, the results are $[0,1,2,3]_a ^{2h} = (-96,16,0,0)$, 
$[0,1,2,3]_b ^{2h} = (-32,0,0,4)$ and 
$[0,1,2,3]_c ^{2h} = (-48,0,4,0)$, respectively. The superscript $2h$ is meant to denote the operators were determined on the coarse grid.

Because we are interested in a Gal\"erkin operator which has the same form on every level, we are interested in stencils which obey the following rule:
\begin{eqnarray}
\label{gal11}
  a [0,1,2,3]_a ^{2h} & + & b [0,1,2,3]_b ^{2h} + c [0,1,2,3]_c ^{2h} \nonumber \\
    & & = a [0,1,2,3]_{Ga} + b [0,1,2,3]_{Gb} + c [0,1,2,3]_{Gc}.
\end{eqnarray}

It is important to note this relationship must hold for all points in our test cube, so we must write down an equation for the central point, one for the grid points once removed, one for those twice removed, and finally an equation for the grid points three times removed. These equations, after some simplification, are:
\begin{eqnarray}
\label{gal12}
  2b - c & = & 14a,\\
  2b - c & = & -26a,\\
  2b - c & = &  -2a,\\
  2b - c & = & -\frac{2}{3}a.
\end{eqnarray}
These four equations indicate $a = 0$ and $c = 2b$. We may determine the value of $b$ by applying our Gal\"erkin operator (with $a = 0$ and $c = 2b$) to the function $f(x) = x^{2}$ in our test cube. Starting from Eqn.~(\ref{gal10}), we may write
\begin{eqnarray}
\label{gal12a}
  \nabla^{2}_{G} (x^{2}) &=& \left[b \nabla^{2}_{b} + c \nabla^{2}_{c}\right] (x^{2}) \nonumber \\
                       &=& \left[b \nabla^{2}_{b} + 2b \nabla^{2}_{c}\right] (x^{2}) \nonumber \\
		       &=& b\left[ \left(-2,0,0,\frac{1}{4}\right) + 2\left(-3,0,\frac{1}{4},0\right)\right] (x^{2}) \nonumber \\
		       &=& b \left(-8,0,\frac{1}{2},\frac{1}{4}\right) (x^{2}) \nonumber \\
		       &=& 6b
\end{eqnarray}
We know that $\nabla^2 x^2 = 2$, so we find that $b=\frac{1}{3}$. Hence, we may write the Gal\"erkin operator for $\nabla^2$ in one of many forms; abstractly it may be written as
\begin{equation}
\label{gal13}
  \nabla^2 _G = \frac{1}{3} [0,1,2,3]_b + \frac{2}{3} [0,1,2,3]_c,
\end{equation}
or with the coefficients of the stencils as
\begin{equation}
\label{gal14}
  \nabla^2 _G = \frac{1}{3} \left( -2,0,0,\frac{1}{4} \right) 
                          + \frac{2}{3} \left( -3,0,\frac{1}{4},0 \right),
\end{equation}
or in a convenient combined form:
\begin{equation}
\label{gal15}
  \nabla^2 _G = \left(-\frac{8}{3},0,\frac{1}{6},\frac{1}{12} \right) .
\end{equation}

Equation~(\ref{gal15}) is the main result of this section, and is the operator we employ in our nonlinear adaptive multigrid algorithm.

\subsection{The ADM Mass and Multipole Moments}
\label{ADM Mass}
It has long been known \cite{ADM,York1} when the Hamiltonian of General Relativity is written in the 3+1 formalism, certain terms may be re-expressed as a total divergence, and hence become a surface integral. Assuming the lapse function $N = 1$ at spatial infinity, the resulting surface integral may be written as
\begin{equation}
\label{adm1}
  16 \pi E_{\text{ADM}} = \oint_{\infty} \sqrt{\gamma} \gamma^{ij} \gamma^{kl} 
                           ( \gamma_{ik,j} - \gamma_{ij,k}  ) dS_{l},
\end{equation}
where $E_{\text{ADM}}$ is known as the ADM mass of the field and $\gamma_{ij}$ is the three metric of the geometry. The indices are purely spatial, and the integral is performed on a surface located at r = $\infty$ in a coordinate system which is asymptotically flat. If one were to calculate the ADM mass for the Schwarzschild metric, one would arrive at the expected result of $E_{\text{ADM}} = m$, where $m$ is the Schwarzschild mass parameter.

For situations in which the three metric is conformally flat, namely 
$\gamma_{ab} = \psi^4 g_{ab}$ where $g_{ab}$ is the flat three metric, it is relatively straightforward to derive an expression for the ADM mass which only depends on the conformal factor $\psi$. To do this, we simply substitute the conformally flat metric given above into Eq.~(\ref{adm1}). This results in
\begin{equation}
\label{adm2}
  16 \pi E_{\text{ADM}} = \oint_{\infty} \psi^{-2} \sqrt{g} g^{ij} g^{kl}
             [ ( g_{jk,l} - g_{kl,j} ) \psi^{4} 
	     + 4 \psi^{3} (\psi_{,l} g_{jk} - \psi_{,j} g_{kl} ) ] dS_{i},
\end{equation}
which in turn simplifies to
\begin{equation}
\label{adm3}
  16 \pi E_{\text{ADM}} = 4 \oint_{\infty} \sqrt{g} (1 - 3) \psi_{,l} g^{il}  dS_{i},
\end{equation}
which we may write in its final form as
\begin{equation}
\label{adm4}
  E_{\text{ADM}} = - \frac{1}{2 \pi} \oint_{\infty} \vec{\nabla} \psi \cdot d\vec{S}.
\end{equation}

One may note Gauss' law may be used to re-express Eq.~(\ref{adm4}) in terms of a volume integral, specifically
\begin{equation}
\label{adm5}
  E_{\text{ADM}} = - \frac{1}{2 \pi} \int_{V} \sqrt{g} \nabla^{2} \psi d^{3} x.
\end{equation}

In theory, if one could evaluate the integral of Eq.~(\ref{adm5}) over a volume bounded by a surface at r = $\infty$ then one could arrive at the ADM mass for the system. 

However, in the world of numerical modeling, this is typically not feasible. For starters, integrating over all of space is not possible due to memory limitations on the machine running the code---hence, it is necessary that we only integrate over a finite volume of space. Unfortunately, even this presents some difficulties. Imagine we choose to integrate a volume which consists of a cube with 65 grid points on a side. This translates to calculating a sum at nearly 300,000 points, which can be time consuming, to say the least! If we were instead to do a surface integral over the bounding surface, this would reduce the number of grid points involved in the calculation to approximately 13,000 points, which in turn would significantly reduce the amount of time spent computing the ADM mass.

However, if we wish to calculate the ADM mass via the surface integral, we discover another complication: we have a stencil describing $\nabla^2_{G}$, the Gal\"erkin operator associated with the volume integral, but we have no knowledge of the form of the stencil describing the 
$\vec{n} \cdot \vec{\nabla}$ operator in the surface integral. It would appear that if we are to succeed, we must determine the proper stencil for the surface integral, where the surface of integration consists of the faces of a cube.

The starting point for the determination of the surface integral operator is Eq.~(\ref{adm5}), the equation for the volume integral. One can imagine the volume of integration to be a cube with a small number of grid points on each side, say four. Then, the stencil associated with the Gal\"erkin $\nabla^2$ operator is applied to each point of our ``test cube'' in hopes of seeing how the volume integral manifests itself as a surface integral, and perhaps more importantly to determine the correct form of the stencil to be employed for the resulting surface integral.

With a little effort, it soon becomes apparent that all of the contributions from interior grid points of the cube cancel out, and the only contribution from the volume integral is due to grid points on the surface. In this way, the volume integral shows us directly that, after applying the Gal\"erkin stencil to our test cube, the volume integral given by Eq.~(\ref{adm5}) reduces to the surface integral given by Eq.~(\ref{adm4}). The surface integral, as mentioned above, allows for a relatively speedy and simple means of calculating the energy of the system. Perhaps more importantly, this method also generates the proper stencil associated with the surface integral.

In an effort to extract more information from the conformal factor, we have also determined a method to calculate the multipole moments of the field via a convenient surface integral. Roughly speaking, one can calculate the multipole moments of the field by integrating the spherical harmonics with indices $\ell$ and $m$ with $\nabla^{2} \psi$ over all space. However, for reasons mentioned above, volume integrals on a numerical grid are time consuming and costly---it would be to our benefit to determine an equivalent means to calculate multipole moments via surface integrals.

We begin by defining the function $\psi^{\dag}$ via
\begin{equation}
\label{mpole1}
  \psi^{\dag} \equiv r^{\ell} Y^{\ast}_{\ell m},
\end{equation}
where $r$ is the usual radial Cartesian coordinate and $Y^{\ast}_{\ell m}$ are the complex conjugates of the spherical harmonics. We claim the proper surface integral for the multipole moments $\Phi_{\ell m}$ with spherical harmonic indices $\ell$ and $m$ is of the form
\begin{equation}
\label{mpole2}
  \Phi_{\ell m} \equiv - \frac{1}{\sqrt{\pi} (2 \ell + 1)} 
          \oint \left[ \psi^{\dag} \nabla_{a} \psi - \psi \nabla_{a} \psi^{\dag} \right] d S^{a},
\end{equation}
where $d S^{a}$ denotes a unit of surface area with a unit vector normal to the surface of integration. 

Let us prove our assertion that Eq.~(\ref{mpole2}) is indeed the correct formula for the multipole moments by a simple example. Assume the conformal factor is a linear combination of terms proportional to the spherical harmonics. 
\begin{equation}
\label{mpole3}
  \psi = 1 + \sum_{\ell, m} \frac{Y_{\ell m}}{r^{\ell + 1}} \psi_{\ell m}
\end{equation}
\noindent where $\psi_{\ell m}$ is the multipole moment associated with a particular choice of $\ell$ and $m$. Note the assumed radial dependence of the conformal factor is determined by the asymptotic behavior $\psi \rightarrow 1$ as $r \rightarrow \infty$. After some simplifications, we evaluate $\Phi_{\ell m}$ to be
\begin{equation}
\label{mpole4}
  \Phi_{\ell m} = - \frac{1}{\sqrt{\pi} (2 \ell + 1)}
                     \oint \left[ r^{\ell} Y^{\ast}_{\ell m} \frac{\partial}{\partial r} 
		     \left(  \frac{Y_{\ell m}}{r^{\ell + 1}} \psi_{\ell m}\right) 
		     -  \frac{Y_{\ell m}}{r^{\ell + 1}} \psi_{\ell m}  
		     \frac{\partial}{\partial r} \left( r^{\ell} Y^{\ast}_{\ell m}\right) \right] r^{2} d \Omega.
\end{equation}
After evaluating the derivatives inside the integral, we find
\begin{equation}
\label{mpole5}
  \Phi_{\ell m} = - \frac{1}{\sqrt{\pi} (2 \ell + 1)} 
            \oint \left[ -(2 \ell + 1) \psi_{\ell m} Y_{\ell m} Y^{\ast}_{\ell m} \right] d \Omega.
\end{equation}
However, after usage of the orthogonality condition of the spherical harmonics, we ultimately find the value of the multipole moment $\Phi_{\ell m}$ to be
\begin{equation}
\label{mpole6}
  \Phi_{\ell m} =  \frac{1}{\sqrt{\pi}} \psi_{\ell m}.
\end{equation}
which is exactly the result we would expect. Imagine our field being spherically symmetric, and corresponding to $\psi = 1 / \alpha = M / (2r)$. For this case, the coefficient of $\psi$ would be $\psi_{\ell m} = \sqrt{\pi} M$, which would yield the lowest order multipole moment to be $\Phi_{00} = M$, the Newtonian mass of the system.

Now that our generalized form of the multipole moments has been determined via Eq.~(\ref{mpole2}), we may set about the task of determining a way in which to evaluate the integral numerically. By application of Gauss' law, we find we may write Eq.~(\ref{mpole2}) as a volume integral given by
\begin{equation}
\label{mpole7}
  \Phi_{\ell m} = - \frac{1}{\sqrt{\pi} (2 \ell + 1)} \int 
               \left[ \psi^{\dag} \nabla^{2} \psi - \psi \nabla^{2} \psi^{\dag} \right] \sqrt{f} d^{3} x.
\end{equation}

\noindent We can see from the above equation how we justify the statement that the multipole moments may be determined by the integral of the spherical harmonics with $\nabla^2 \psi$: $\psi^{\dag}$ is an eigenfunction of the Laplacian, and hence the second term in the above integral vanishes. With the above form of the integral, we may set about applying the Gal\"erkin stencil for $\nabla^2$ to the fields $\psi$ and $\psi^{\dag}$. On doing this, one will quickly see that all of the interior grid points will not contribute to the overall surface integral---just as was demonstrated above when examining the ADM mass. It is a tedious but straightforward task to determine the correct stencils for the multipole moments---one must be careful of the faces, edges, and corners of the numerical grid. The final result may be seen in the program \it{moment}\rm, which is listed in its entirety in the Appendix.

\subsection{Numerical Tests of the Adaptive Nonlinear Code}
\label{Numerical Tests}
Fortunately, we know of two cases in which we may test our numerical results to analytic expressions for the ADM mass. The first case is for a single boosted black hole; the second is for two black holes, separated by a ``small'' distance and boosted towards each other. Recall the expression for the ADM mass of the system, in the form of a volume integral given by Eq.~(\ref{adm5}):
\begin{equation}
\label{test1}
  E_{\text{ADM}} = - \frac{1}{2 \pi} \int_{V} \sqrt{g} \nabla^{2} \psi d^{3} x.
\end{equation}

Noting the conformal factor is given by $\psi = \frac{1}{\alpha} + 1 + u$, and also recalling the equation our adaptive nonlinear code is solving, given by
\begin{equation}
\label{test2}
  \nabla^2 u + \beta ( 1 + \alpha + \alpha u )^{-7} = 0,
\end{equation}

\noindent we realize that we may re-write the expression for the ADM mass as follows:
\begin{equation}
\label{test3}
  E_{\text{ADM}} = - \frac{1}{2 \pi} \int_{V} \nabla^{2} \left( \frac{1}{\alpha} \right) d^{3} x
              + \frac{1}{2 \pi} \int_{V} \beta ( 1 + \alpha + \alpha u )^{-7} d^{3} x,
\end{equation}

\noindent where we are evaluating the expression in regular Cartesian coordinates. Noting the first term simply yields the Newtonian masses of each individual puncture, we may write the ADM mass in the following form:
\begin{equation}
\label{test4}
  E_{\text{ADM}} = M_{1} + M_{2} 
                + \frac{1}{2 \pi} \int_{V} \beta ( 1 + \alpha + \alpha u )^{-7} d^{3} x.
\end{equation}

Let us examine Eq.~(\ref{test4}) for the situation of a single boosted black hole, where the linear momentum of the hole is small. In this situation, we assume the conformal factor correction is small $( u \ll 1)$ and can be ignored in the above integral. Hence, our ADM mass for a single boosted black hole becomes
\begin{equation}
\label{test5}
  E_{\text{ADM}} = M + \frac{1}{2 \pi} \int_{V} \beta ( 1 + \alpha )^{-7} d^{3} x.
\end{equation}

\noindent It is relatively straightforward to show that the integrand of Eq.~(\ref{test5}) is simply
\begin{equation}
\label{test6}
  \beta ( 1 + \alpha )^{-7} = 72 P^{2} \frac{r^{3}}{M^{7}} [ 1 + 2 \cos^{2} \theta ]
                                    \left( 1 + \frac{2 r}{M} \right) ^{-7}.
\end{equation}

\noindent When integrated using spherical coordinates, we find the ADM mass for a single boosted black hole is given by
\begin{equation}
\label{test7}
  E_{\text{ADM}} = M + \frac{5}{8} \frac{P^{2}}{M}
\end{equation}
for small momentum.

Tables \ref{n11test}, \ref{n19test} and \ref{n35test} show our numerical results for a single boosted black hole, evaluated at various distances. In each case, the momentum of the hole is given a value of $P = 0.10 M$. Each test is calculated using ten adaptive levels with a total of eight iterations, and the physical scale of the problem (given by the parameter $SCALE$ in the code) is equal to $4.0 M$. The only parameter which is varied from one table to the next is the number of grid points on the side of the smallest adaptive level in an octant of space, numbering $n=11$, $n=19$, and $n=35$ grid points respectively.\\
\renewcommand{\arraystretch}{2}
\begin{table}[]
\begin{center}
\begin{tabular}{ r  c } 
\multicolumn{2}{}{} \\ 
  \hline Distance & ADM Mass \\ \hline
   7~~~~ &    1.004478360538468\\
   14~~~~ &   1.005246092534632\\
   28~~~~ &   1.005687987991896\\
   56~~~~ &   1.005925410956947\\
   112~~~~ &  1.006048517746048\\
   224~~~~ &  1.006111206656130\\
   448~~~~ &  1.006142839716399\\
   896~~~~ &  1.006158728997733\\
   1792~~~~ & 1.006166691900793\\
   3584~~~~ & 1.006170677928971\\ \hline
\end{tabular}
\caption{Code test for a single boosted black hole: $n = 11$ grid points on a side, $ADAPT = 10$ adaptive levels, theoretical ADM mass $E_{\text{ADM}}= 1.00625 M$}
\label{n11test}
\end{center}
\end{table}
\renewcommand{\arraystretch}{2}
\begin{table}[]
\begin{center}
\begin{tabular}{ r  c } 
\multicolumn{2}{}{} \\ 
  \hline Distance & ADM Mass \\ \hline
   7~~~~ &    1.004456160293023\\
   14~~~~ &   1.005250846181358\\
   28~~~~ &   1.005710799727296\\
   56~~~~ &   1.005958644512548\\
   112~~~~ &  1.006087346697324\\
   224~~~~ &  1.006152934348016\\
   448~~~~ &  1.006186042680404\\
   896~~~~ &  1.006202676155737\\
   1792~~~~ & 1.006211012805630\\
   3584~~~~ & 1.006215186117088\\ \hline
\end{tabular}
\caption{Code test for a single boosted black hole: $n = 19$ grid points on a side, $ADAPT = 10$ adaptive levels, theoretical ADM mass $E_{\text{ADM}}= 1.00625 M$}
\label{n19test}
\end{center}
\end{table}
\renewcommand{\arraystretch}{2}
\begin{table}[]
\begin{center}
\begin{tabular}{ r  c } 
\multicolumn{2}{}{} \\ 
  \hline Distance & ADM Mass \\ \hline
   7~~~~ &    1.004433482500703\\
   14~~~~ &   1.005240400855877\\
   28~~~~ &   1.005708711698705\\
   56~~~~ &   1.005961421538742\\
   112~~~~ &  1.006092746446622\\
   224~~~~ &  1.006159695535913\\
   448~~~~ &  1.006193497436792\\
   896~~~~ &  1.006210480949185\\
   1792~~~~ & 1.006218993438650\\
   3584~~~~ & 1.006223254876639\\ \hline
\end{tabular}
\caption{Code test for a single boosted black hole: $n = 35$ grid points on a side, $ADAPT = 10$ adaptive levels, theoretical ADM mass $E_{\text{ADM}}= 1.00625 M$}
\label{n35test}
\end{center}
\end{table}
As can be seen from the tables, the ADM mass increases slightly as a function of the distance at which this quantity is evaluated---we attribute this to the nonlinear nature of the equation the code is solving. The most interesting point, however, is the amount our numerical results agree with the theoretical prediction for the mass correction. In Table~\ref{n11test} the discrepancy between the theoretical and numerical values of the mass correction are approximately 1.3\% when measured at the furthest grid point. Higher resolutions yield even more impressive results, showing deviations from the expected theoretical mass correction of 0.55\% and 0.42\%, respectively.

\begin{figure}[]
\begin{center}
\scalebox{0.8}[0.8]{
  \includegraphics{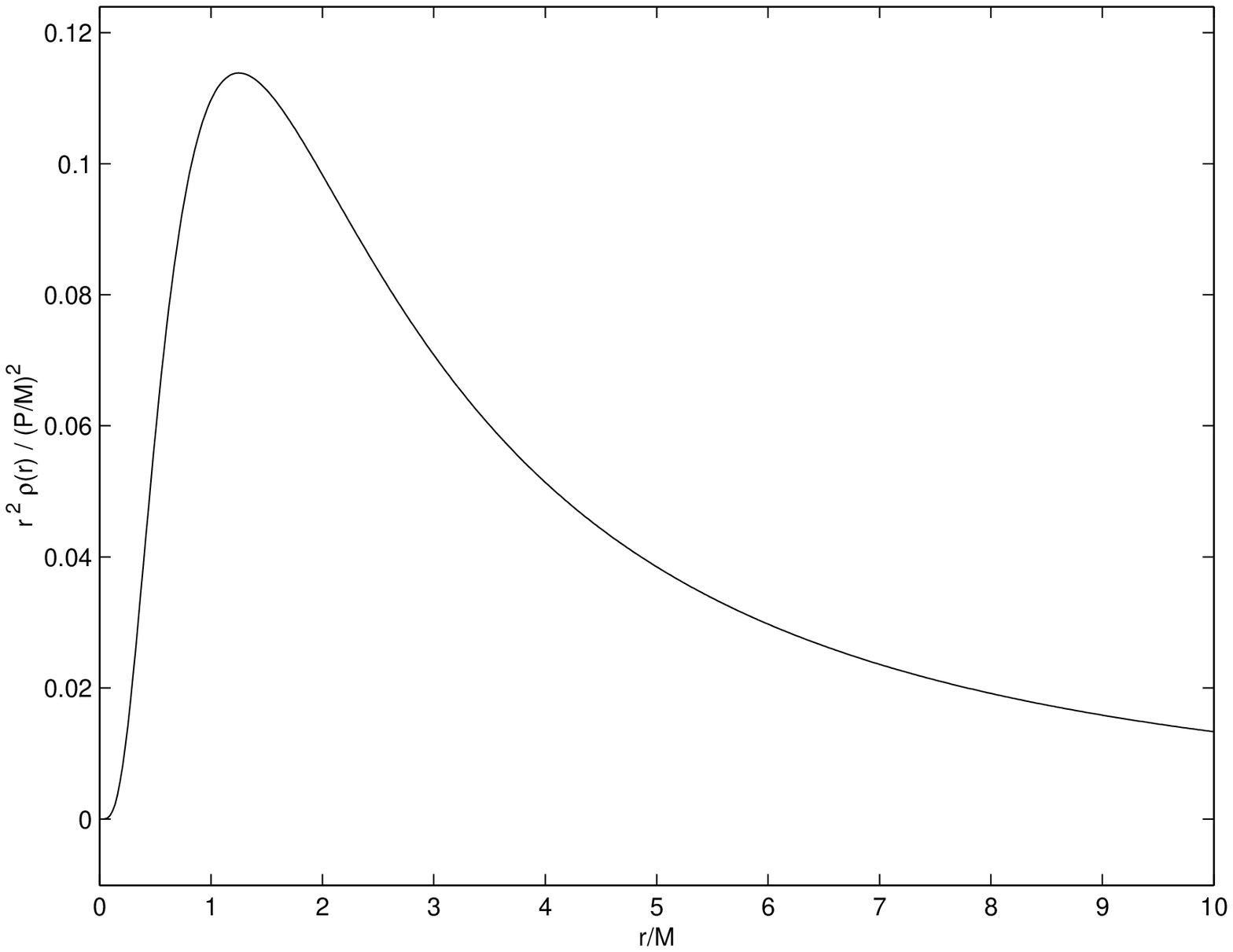}
}
\caption{The radial dependence of the normalized energy density $r^{2} \rho(r) / (P/M)^{2}$ versus the normalized distance $r/M$.}
\label{density3.eps}
\end{center}
\end{figure}

One may inquire why the ADM mass of a single boosted black hole yields a coefficient of $5/8$ in front of the kinetic energy term, rather than the expected Newtonian value of $1/2$. It would appear there is some additional energy present in the geometry. Figure \ref{density3.eps} shows the radial dependence of the normalized energy density of the field versus the normalized radial distance. If one were to integrate this curve from $r=0$ to $r=\infty$, the result would be $5/8$. The figure indicates it is difficult to localize the additional energy content of the geometry, which implies the additional energy is ``frozen'' into the geometry. It is not clear if this additional energy is due to the simplification introduced by a trace-free conformal extrinsic curvature, or if it is due to the conformal flatness conjecture. Additionally, it is not clear how this additional energy would effect an evolution of the geometry via the dynamical Einstein equations. We note the additional energy content will place a limitation on our ability to model astrophysically realistic binary black hole systems. The geometry will nonetheless provide a proving ground for the variational principle we employ.
 
We can test our numerical results against another situation---two black holes separated by a small distance, and boosted towards each other with relatively small momenta. The method of deriving an analytic result for the ADM mass is similar to that of above, but requires a little more effort. For this case, we used Mathematica to determine the form of the integrand for Eq.~(\ref{test4}), which is quite a bit more complicated than the single boosted black hole scenario. Once the integrand is determined, it is necessary to expand the integrand in powers of $ \frac{L}{M} $, where $L$ is the coordinate position of each hole along the $x$ axis. This limits our analysis to cases with $\frac{L}{M} \ll 1$, but still provides an adequate test bed for the code. When the expansion of the integrand is completed, we find the ADM mass for this situation is given by
\begin{equation}
\label{test8}
  E_{\text{ADM}} = 2M + M \left( \frac{P}{M} \right)^{2} \left[ \frac{11}{50} \left( \frac{L}{M} \right)^{2} 
                                       - \frac{24}{35} \left( \frac{L}{M} \right)^{4} \right], 
\end{equation}

\noindent where $M$ once again is the Newtonian mass of the individual punctures.

Tables \ref{n11twotest}, \ref{n19twotest} and \ref{n35twotest} show the results for two black holes located at coordinate distances $+0.1 M$ and $-0.1 M$ along the $x$ axis. The holes are boosted towards each other with a linear momentum of $P = 1.0 M$. The physical scale of the grids are the same as for the single black hole above, and again, each test was run with ten adaptive levels and eight complete iterations. The only parameter varied was once again the number of grid points. As shown in Table~\ref{n11twotest}, the discrepancy between the theoretical and numerical mass correction is 5.2\%. Going to higher grid densities results in mass correction discrepancies of 1.4\% and 0.6\%, respectively. 
\renewcommand{\arraystretch}{2}
\begin{table}[]
\begin{center}
\begin{tabular}{ r  c } 
\multicolumn{2}{}{} \\ 
  \hline Distance & ADM Mass\\ \hline
   7~~~~ &    2.002204960144522\\
   14~~~~ &   2.002235779237632\\
   28~~~~ &   2.002240987490862\\
   56~~~~ &   2.002241749371064\\
   112~~~~ &  2.002241852453771\\
   224~~~~ &  2.002241865652823\\
   448~~~~ &  2.002241867039063\\
   896~~~~ &  2.002241866886472\\
   1792~~~~ & 2.002241866678383\\
   3584~~~~ & 2.002241866670044\\ \hline
\end{tabular}
\caption{Code test for two boosted black holes: $n = 11$ grid points on a side, $ADAPT = 10$ adaptive levels, theoretical ADM mass $E_{\text{ADM}}= 2.0021314285 M$}
\label{n11twotest}
\end{center}
\end{table}
\renewcommand{\arraystretch}{2}
\begin{table}[]
\begin{center}
\begin{tabular}{ r  c } 
\multicolumn{2}{}{} \\ 
  \hline Distance & ADM Mass\\ \hline
   7~~~~ &    2.00212124423002\\
   14~~~~ &   2.00215537225156\\
   28~~~~ &   2.00216120866229\\
   56~~~~ &   2.00216206802759\\
   112~~~~ &  2.0021621847792\\
   224~~~~ &  2.00216219991037\\
   448~~~~ &  2.00216220173672\\
   896~~~~ &  2.00216220191223\\
   1792~~~~ & 2.00216220194748\\
   3584~~~~ & 2.00216220193432\\ \hline
\end{tabular}
\caption{Code test for two boosted black holes: $n = 19$ grid points on a side, $ADAPT = 10$ adaptive levels, theoretical ADM mass $E_{\text{ADM}}= 2.0021314285 M$}
\label{n19twotest}
\end{center}
\end{table}
\renewcommand{\arraystretch}{2}
\begin{table}[]
\begin{center}
\begin{tabular}{ r  c } 
\multicolumn{2}{}{} \\ 
  \hline Distance & ADM Mass\\ \hline
   7~~~~ &    2.00210214462189\\
  14~~~~ &   2.00213793576375\\
  28~~~~ &   2.00214409210936\\
  56~~~~ &   2.00214500147877\\
  112~~~~ &  2.00214512527282\\
  224~~~~ &  2.00214514140509\\
  448~~~~ &  2.00214514345247\\
  896~~~~ &  2.00214514371485\\
  1792~~~~ & 2.00214514374563\\
  3584~~~~ & 2.00214514374897\\ \hline
\end{tabular}
\caption{Code test for two boosted black holes: $n = 35$ grid points on a side, $ADAPT = 10$ adaptive levels, theoretical ADM mass $E_{\text{ADM}}= 2.0021314285 M$}
\label{n35twotest}
\end{center}
\end{table}

As these two test cases would indicate, we can be confident that our code is correctly solving the nonlinear partial differential equation for the correction to the conformal factor, given by Eq.~(\ref{test2}). We may now proceed to apply the code towards solving the Hamiltonian constraint for binary black holes in quasi-equilibrium circular orbits. 

\chapter{Numerical Results}

We present the numerical results from our adaptive multigrid computer code used to model binary black holes, as described in the previous chapter. The computer code employed may be found in its entirety in the Appendix.

The procedure for generating a sequence of quasi-stationary circular orbits which satisfy the constraint equations is, in theory, quite simple. The variational principle even guides us as to how to approach the problem: simply write some computer code which solves the Hamiltonian constraint for some fixed value of angular momentum $J$ and bare masses $\overline{m}$. Vary the separation distance of the black holes until you find a minimum in the value of the ADM mass. A minimum in the ADM mass ensures the sytem under study is a solution to the quasi-equilibrium Einstein equations for those particular values of $J$ and $\overline{m}$. The separation distance associated with this minimum corresponds to the coordinate separation distance of a quasi-equilibrium circular orbit for the system. The variational principle also indicates how to determine the angular frequency $\Omega$ for this circular orbit; simply determine the minimum in the ADM mass for a slightly different value of the angular momentum. One can then proceed in a similar fashion to generate a sequence of ``effective potential'' curves, where each minimum in the curve corresponds to a quasi-equilibrium circular orbit.

The resulting sequence of circular orbits can be viewed in the following fashion: a realistic binary system will radiate gravitational waves, causing their orbital separation distance to decrease slowly. Because of the circularizing nature on the orbits of gravitational radiation \cite{MTW}, the sequence of circular quasi-equilibrium orbits is a reasonable estimate to reality. The binary system will then ``evolve'' along the sequence, until the binary system reaches the innermost stable circular orbit (ISCO). It is at the ISCO that we can no longer employ the variational principle to generate information about circular orbits, and we can not determine the dynamics of the final plunge. Despite this shortcoming, the variational principle, and the sequence it generates, yields valuable information about the evolution of the system up to and including the ISCO. In particular, the orbital frequency at the ISCO is of vital interest to investigators in the field of gravitational wave detection.

One problem that arises is the question of mass. Specifically, what exactly does one mean when he speaks of the mass of a black hole? There is no general consensus on the answer to this question \cite{Cook2,Baumgarte2,Cook4}. Recall from \S(\ref{InitValue}) that we encountered two types of mass when describing the punctures of Brandt and Brugmann: the ``bare'' mass $\overline{m}$ and the ``Newtonian'' mass $M$, which are related via
\begin{equation}
\label{numresults1}
  \overline{m}_{i} = M_{i} \left( 1 + u_{i} + \frac{M_{j}}{ 2 D }  \right),
\end{equation} 
where $\overline{m}_{i}$ is the bare mass of the $i^{th}$ hole, $M_{i}$ is the corresponding Newtonian mass, $u_{i}$ is the value of the correction to the conformal factor at the $i^{th}$ hole, and $D$ is the coordinate separation distance between the holes.

However, these two masses are not the only two used to describe black holes. In Baumgarte's formalism \cite{Baumgarte2}, he chooses to use the mass associated with the area of the apparent horizon, described by Christodoulou's formula, to specify the system.

A different measure of the mass is the rest mass, which is given by the familiar special relativistic expression
\begin{equation}
\label{numresults2}
  E_{\text{rest}} = \sqrt{ E_{\text{ADM}}^{2} - P^{2} },
\end{equation}
where $E_{\text{ADM}}$ is the ADM mass of an isolated black hole with linear momentum $P$. We have opted to use the rest mass when comparing our numerical results to the expected post-Newtonian results, simply because it is a measure of an individual hole's energy without any contribution from the other hole.

We begin our numerical analysis by using a program named \it{VP}\rm, which is used to generate an ``effective potential'' curve for some fixed value of angular momentum $J$ and fixed bare mass $\overline{m}$. Let us describe, in detail, what the code does for a particular value of angular momentum. 

For a fixed value of $J$, the separation distance $D$ is initialized to some value. From these two quantities, the linear momentum $P$ of each hole is determined via $J = PD$. To ensure a solution to the quasi-equilibrium Einstein equations, recall that the variational principle dictates that we must hold the bare mass $\overline{m}$ fixed as we generate the sequence; the Newtonian mass $M$ may vary through the sequence. Therefore, for a fixed value of the bare mass, we initialize the Newtonian mass to an arbitrary value. We then solve the Hamiltonian constraint for the correction to the conformal factor $u$. We then calculate the new value of the Newtonian mass, which follows directly from Eq.~(\ref{numresults1}):
\begin{equation}
\label{numresults3}
  M = -D (1+u) + \sqrt{ D^{2} (1+u)^{2} + 2 D \overline{m} }.
\end{equation}
Using this new value of the Newtonian mass, we re-solve the Hamiltonian constraint. The old value of the conformal factor correction at the location of the hole is compared to the new value, and if the difference is below some tolerance level we conclude that we have reached convergence for that particular separation distance $D$. Once we have reached convergence, we calculate the ADM mass for the system and save the information to a file. Then, still using the unchanged values for the angular momentum $J$ and the bare mass $\overline{m}$, we decrement the separation distance $D$ and repeat the process to determine a new value of the ADM mass for that separation distance. This process continues until a single ``effective potential'' curve has been produced for a fixed value of $J$ and $\overline{m}$. Once this has been done, we decrement the value of the angular momentum $J$ and generate a new curve.

Once a series of curves have been generated, each curve corresponding to a different value of angular momentum, then we must determine the location of the local minima of each curve. The location of these minima correspond to a circular orbit for some value of angular momenta. This is achieved using the program \it{min}\rm. It is a minimization routine taken from Numerical Recipes which employs Brent's method \cite{numrecipes}. Once the minima of a curve has been determined, the information is saved to a file.

\begin{figure}[]
\begin{center}
\scalebox{0.8}[0.8]{
  \includegraphics[]{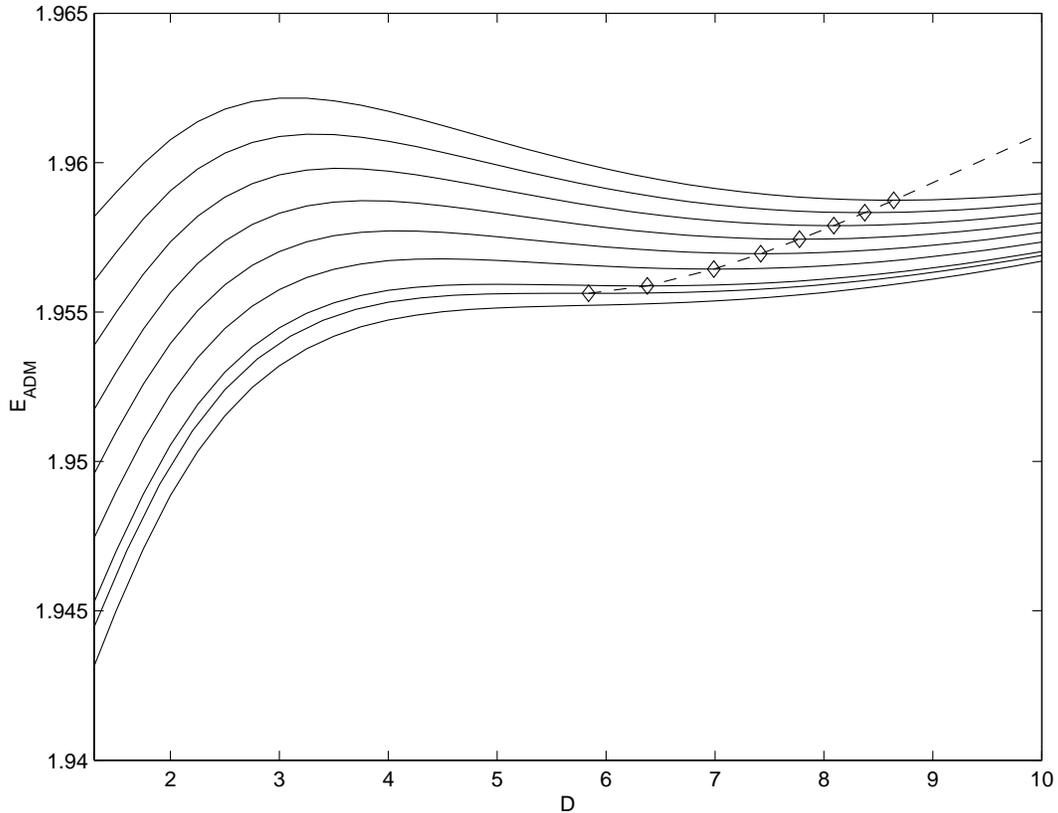}
}
\caption{ADM mass curves and the corresponding sequence of circular orbits for $n = 11$ grid points on a side and $ADAPT = 10$ adaptive levels. The values of $J$ range from $2.93 {\overline{m}}^{2}$ to $3.00 {\overline{m}}^{2}$. The sequence of circular orbits terminates at $J = 2.936 {\overline{m}}^{2}$, the innermost stable circular orbit.}
\label{MG3curves1.eps}
\end{center}
\end{figure}

Figure~\ref{MG3curves1.eps} is one example of the curves generated by this procedure, with the location of the circular orbits denoted by the dashed curve, with diamonds representing the locations of the circular orbits for particular values of $J$. This particular figure was generated for a grid with $n = 11$ grid points on a side, and ten adaptive levels. The vertical axis is the ADM mass of the system, and the horizontal axis is the coordinate separation distance between the two holes. The lowest curve corresponds to a value of the angular momentum $J = 2.93 \overline{m}^{2}$, and the topmost curve corresponds to a value of $J = 3.00 \overline{m}^{2}$. The sequence of circular orbits terminates at a separation distance of $D=5.8395 \overline{m}$, which corresponds to a value of $J = 2.936 \overline{m}^{2}$.

\begin{figure}[]
\begin{center}
\scalebox{0.8}[0.8]{
  \includegraphics[]{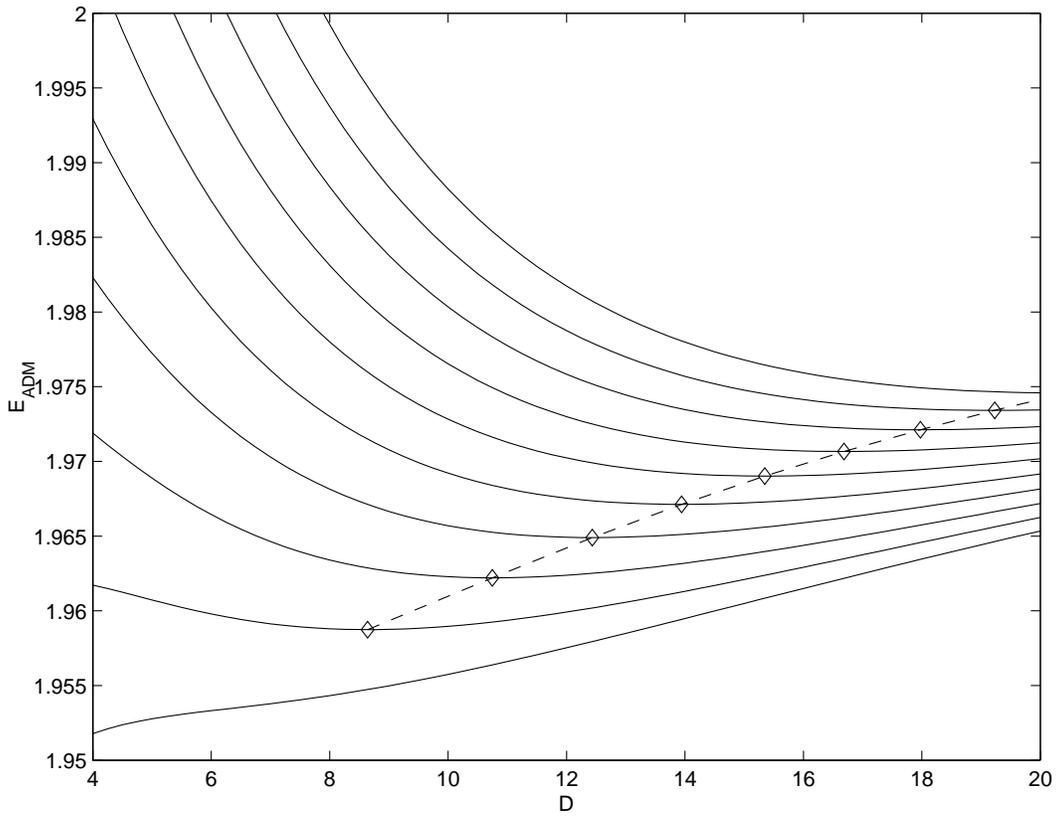}
}
\caption{ADM mass curves and the corresponding sequence of circular orbits for $n = 11$ grid points on a side  and $ADAPT= 10$ adaptive levels. The values of $J$ range from $2.90 {\overline{m}}^{2}$ to $3.80 {\overline{m}}^{2}$. The sequence of circular orbits terminates at $J = 3.00 {\overline{m}}^{2}$ only for illustrative purposes.}
\label{MG3curves2.eps}
\end{center}
\end{figure}

Likewise, Figure~\ref{MG3curves2.eps} is a similar figure of curves for values of angular momentum ranging from $J = 2.90 \overline{m}^{2}$ to $J = 3.80 \overline{m}^{2}$. In this figure, the the line representing the sequence of circular orbits terminates at a value of $J = 3.00 \overline{m}^{2}$ only for illustrative purposes---it is not the location of the innermost stable circular orbit.

Once we have determined the location of the circular orbits for various values of angular momentum, then we may apply the variational principle to determine the angular frequency $\Omega$. This is achieved by once again using the program \it{min}\rm. For example, imagine we are interested in determining the value of $\Omega$ for the circular orbit associated with $J = 3.00 \overline{m}^{2}$. We simply increase the angular momentum in \it{min}\rm~ to a value of $J = 3.001 \overline{m}^{2}$, and allow the code to search for a minimum value in the ADM mass. Once the program has converged to a solution and generated the minimum ADM mass associated with this new value of the angular momentum, then the angular frequency may be determined via
\begin{equation}
\label{numresults4}
  \Omega = \frac{ \Delta E_{\text{ADM}} }{ \Delta J }.
\end{equation}
For this particular case of interest, we find that $\Omega = 0.0407 \overline{m}^{-1}$. In this very fashion, the angular frequency can be determined for all circular orbits, up to and including the innermost stable circular orbit.

As was mentioned above, we choose to normalize all of our quantities by means of the rest mass of a single hole, given by Eq.~(\ref{numresults2}). This is achieved by calculating the ADM mass of a single, isolated, boosted black hole with linear momentum $P$. This value of the linear momentum is exactly the same linear momentum each hole in the binary system has for angular momentum $J$ and separation distance $D$. We believe this is a reasonable way to calculate the mass of a hole, as it is the mass due solely to the individual hole and its motion, and not due to the interaction with the other hole. Two quantities which we use to normalize the orbital parameters are the total rest mass of the system $m$, defined as
\begin{equation}
\label{numresults5}
  m \equiv 2 E_{\text{rest}},
\end{equation}
and the reduced mass of the system $\mu$, defined as
\begin{equation}
\label{numresults6}
  \mu \equiv \frac{1}{2} E_{\text{rest}}.
\end{equation}
It is important to note the above two definitions are valid only for equal mass black holes.

\section{Numerical Results of the Variational Principle}
We now present the data for three different resolutions in tabular form, and then discuss the agreement between the data and post-Newtonian expectations.

In Table \ref{n11vp}, the grid resolution was set to $n = 11$, with a total of ten adaptive levels. The variable $J$ denotes the angular momentum of the system, and is measured in units of $\overline{m}^{2}$, where $\overline{m}$ is the fixed bare mass of the holes. For all numerical simulations, we fix the value of $\overline{m}$ to be equal to 1.0. The variable $P$ is the linear momentum of each individual hole, and is measured in units of $\overline{m}$. The variable $D$ is the coordinate separation distance between the holes, and is also measured in units of $\overline{m}$. The Newtonian mass of the holes is denoted by $M$, is measured in units of $\overline{m}$, and is defined via Eq.~(\ref{numresults3}). The ADM mass of the binary system is denoted as $E_{\text{ADM}}$, and is measured in units of $\overline{m}$, and the angular frequency of the binary system is denoted as $\Omega$ and is measured in units of $\overline{m}^{-1}$.
\renewcommand{\arraystretch}{1.5}
\begin{table}[]
\begin{center}
\begin{tabular}{ c  c  c  c  c  c } 
\multicolumn{6}{}{} \\ 
  \hline J & D & P & $M$ & $E_{\text{ADM}}$ & $\Omega$ \\ \hline
   4.100 & 24.0155 & 0.1707 & 0.9791 & 1.9775 & 0.0085 \\ 
   4.000 & 22.8444 & 0.1751 & 0.9780 & 1.9766 & 0.0092 \\ 
   3.900 & 21.6597 & 0.1801 & 0.9768 & 1.9756 & 0.0101 \\
   3.800 & 20.4555 & 0.1858 & 0.9754 & 1.9746 & 0.0111 \\
   3.700 & 19.2296 & 0.1924 & 0.9738 & 1.9734 & 0.0123 \\
   3.600 & 17.9758 & 0.2003 & 0.9719 & 1.9721 & 0.0137 \\
   3.500 & 16.6858 & 0.2098 & 0.9697 & 1.9707 & 0.0154 \\
   3.400 & 15.3473 & 0.2215 & 0.9670 & 1.9690 & 0.0176 \\
   3.300 & 13.9428 & 0.2367 & 0.9635 & 1.9671 & 0.0204 \\
   3.200 & 12.4346 & 0.2573 & 0.9589 & 1.9649 & 0.0243 \\
   3.100 & 10.7475 & 0.2884 & 0.9522 & 1.9622 & 0.0300 \\
   3.000 & ~~8.6405 & 0.3472 & 0.9398 & 1.9587 & 0.0407 \\
   2.980 & ~~8.0898 & 0.3684 & 0.9355 & 1.9579 & 0.0444 \\
   2.960 & ~~7.4199 & 0.3989 & 0.9293 & 1.9570 & 0.0497 \\
   2.936 & ~~5.8395 & 0.5028 & 0.9087 & 1.9556 & 0.0666 \\ \hline
\end{tabular}
\caption{Orbital parameters for a binary black hole system: $n = 11$ grid points on a side, $ADAPT = 10$ adaptive levels.}
\label{n11vp}
\end{center}
\end{table}

We also tabulate the data used to determine the rest mass $E_{\text{rest}}$ of each individual hole for a particular value of angular momentum $J$ in Table \ref{n11rest}. The grid resolution and adaptive level number are the same as Table \ref{n11vp}. However, in Table \ref{n11rest}, $E_{\text{ADM}}$ is the ADM mass of the single boosted hole with linear momentum $P$, and $E_{\text{rest}}$ is the resulting rest mass, derived from Eq.~(\ref{numresults2}). Also note this data was calculated with only six adaptive levels. We found, experimentally, the ADM mass changed by only one part in $10^{6}$ when the number of adaptive levels was reduced from ten to six. This reduced the execution time of the code considerably.
\renewcommand{\arraystretch}{1.5}
\begin{table}[]
\begin{center}
\begin{tabular}{ c  c  c  c } 
\multicolumn{4}{}{} \\ 
  \hline J & P & $E_{\text{ADM}}$ & $E_{\text{rest}}$ \\ \hline
   4.100 & 0.1707 &  1.0144 & 0.9999509\\ 
   4.000 & 0.1751 &  1.0152 & 0.9999501\\ 
   3.900 & 0.1801 &  1.0160 & 0.9999495\\
   3.800 & 0.1858 &  1.0171 & 0.9999490\\
   3.700 & 0.1924 &  1.0183 & 0.9999489\\
   3.600 & 0.2003 &  1.0198 & 0.9999494\\
   3.500 & 0.2098 &  1.0217 & 0.9999510\\
   3.400 & 0.2215 &  1.0242 & 0.9999548\\
   3.300 & 0.2367 &  1.0276 & 0.9999632\\
   3.200 & 0.2573 &  1.0326 & 0.9999819\\
   3.100 & 0.2884 &  1.0408 & 1.0000297\\
   3.000 & 0.3472 &  1.0588 & 1.0002063\\
   2.980 & 0.3684 &  1.0660 & 1.0003051\\
   2.960 & 0.3989 &  1.0771 & 1.0004887\\
   2.936 & 0.5028 &  1.1207 & 1.0015630\\ \hline  
\end{tabular}
\caption{Single boosted black hole for the determination of the rest mass $E_{\text{rest}}$: $n = 11$ grid points on a side, $ADAPT = 6$ adaptive levels.}
\label{n11rest}
\end{center}
\end{table}

In anticipation of comparing our results to post-Newtonian predictions, we calculate the binding energy of the binary system via
\begin{equation}
\label{numresults7}
  E_{\text{b}} = E_{\text{ADM}} - 2 E_{\text{rest}}.
\end{equation}
We also normalize the binding energy, the angular momentum, the separation distance and the orbital angular frequency with $m$ and $\mu$ as defined by Eqs.~(\ref{numresults5}) and (\ref{numresults6}), respectively. This information is listed in Table \ref{n11normalized}.
\renewcommand{\arraystretch}{1.5}
\begin{table}[]
\begin{center}
\begin{tabular}{ c  c  c  c } 
\multicolumn{4}{}{} \\ 
  \hline $J / (m \mu)$ & $m \Omega$ & $E_{\text{b}} / \mu$ & $D / m$\\ \hline
   4.100 & 0.0170 & -0.04492 & 12.008\\
   4.000 & 0.0185 & -0.04659 & 11.423\\
   3.900 & 0.0202 & -0.04852 & 10.830\\
   3.800 & 0.0222 & -0.05063 & 10.228\\
   3.700 & 0.0246 & -0.05296 & ~~9.615\\
   3.600 & 0.0274 & -0.05556 & ~~8.988\\
   3.500 & 0.0309 & -0.05847 & ~~8.343\\
   3.400 & 0.0352 & -0.06178 & ~~7.674\\
   3.300 & 0.0408 & -0.06560 & ~~6.972\\
   3.200 & 0.0485 & -0.07012 & ~~6.217\\
   3.100 & 0.0601 & -0.07570 & ~~5.374\\
   2.999 & 0.0814 & -0.08333 & ~~4.319\\
   2.978 & 0.0889 & -0.08541 & ~~4.044\\
   2.957 & 0.0995 & -0.08800 & ~~3.708\\
   2.927 & 0.1334 & -0.09485 & ~~2.915\\ \hline
\end{tabular}
\caption{Normalized orbital parameters for a binary black hole system: $n = 11$ grid points on a side, $ADAPT = 10$ adaptive levels.}
\label{n11normalized}
\end{center}
\end{table}

We also list the corresponding data for a higher resolution, where $n = 19$, in Tables \ref{n19vp}, \ref{n19rest}, and \ref{n19normalized}.
\renewcommand{\arraystretch}{1.5}
\begin{table}[]
\begin{center}
\begin{tabular}{ c  c  c  c  c  c } 
\multicolumn{6}{}{} \\ 
  \hline J & D & P & $M$ & $E_{\text{ADM}}$ & $\Omega$ \\ \hline
   4.100 & 27.3284 & 0.1500 & 0.9810 & 1.9818 & 0.0083\\
   4.000 & 25.5797 & 0.1564 & 0.9797 & 1.9810 & 0.0093\\
   3.900 & 23.8083 & 0.1638 & 0.9781 & 1.9800 & 0.0104\\
   3.800 & 22.0133 & 0.1726 & 0.9762 & 1.9789 & 0.0117\\
   3.700 & 20.1909 & 0.1833 & 0.9740 & 1.9776 & 0.0133\\
   3.600 & 18.3405 & 0.1963 & 0.9712 & 1.9762 & 0.0154\\
   3.500 & 16.4578 & 0.2127 & 0.9677 & 1.9745 & 0.0180\\
   3.400 & 14.5414 & 0.2338 & 0.9632 & 1.9726 & 0.0215\\
   3.300 & 12.5857 & 0.2622 & 0.9571 & 1.9702 & 0.0262\\
   3.200 & 10.5755 & 0.3026 & 0.9484 & 1.9673 & 0.0330\\
   3.100 & ~~8.4456 & 0.3671 & 0.9346 & 1.9635 & 0.0437\\
   3.000 & ~~5.8004 & 0.5172 & 0.9036 & 1.9581 & 0.0675\\ 
   2.977 & ~~4.6086 & 0.6460 & 0.8784 & 1.9564 & 0.0860\\
   2.976 & ~~4.4902 & 0.6628 & 0.8753 & 1.9563 & 0.0882\\
   2.975 & ~~4.3060 & 0.6909 & 0.8700 & 1.9563 & 0.0920\\ \hline  
\end{tabular}
\caption{Orbital parameters for a binary black hole system: $n = 19$ grid points on a side, $ADAPT = 10$ adaptive levels.}
\label{n19vp}
\end{center}
\end{table}
\renewcommand{\arraystretch}{1.5}
\begin{table}[]
\begin{center}
\begin{tabular}{ c  c  c  c } 
\multicolumn{4}{}{} \\ 
  \hline J & P & $E_{\text{ADM}}$ & $E_{\text{rest}}$ \\ \hline
   4.100 & 0.1500 & 1.0111 & 0.9998793\\
   4.000 & 0.1564 & 1.0120 & 0.9998707\\
   3.900 & 0.1638 & 1.0132 & 0.9998607\\
   3.800 & 0.1726 & 1.0146 & 0.9998489\\
   3.700 & 0.1833 & 1.0165 & 0.9998348\\
   3.600 & 0.1963 & 1.0189 & 0.9998181\\
   3.500 & 0.2127 & 1.0222 & 0.9997985\\
   3.400 & 0.2338 & 1.0268 & 0.9997769\\
   3.300 & 0.2622 & 1.0336 & 0.9997573\\
   3.200 & 0.3026 & 1.0445 & 0.9997574\\
   3.100 & 0.3671 & 1.0651 & 0.9998619\\
   3.000 & 0.5172 & 1.1267 & 1.0009726\\
   2.977 & 0.6460 & 1.1934 & 1.0034880\\
   2.976 & 0.6628 & 1.2030 & 1.0039573\\
   2.975 & 0.6909 & 1.2194 & 1.0048253\\ \hline  
\end{tabular}  
\caption{Single boosted black hole for the determination of the rest mass $E_{\text{rest}}$: $n = 19$ grid points on a side, $ADAPT = 6$ adaptive levels.}
\label{n19rest}
\end{center}
\end{table}
\renewcommand{\arraystretch}{1.5}
\begin{table}[]
\begin{center}
\begin{tabular}{ c  c  c  c } 
\multicolumn{4}{}{} \\ 
  \hline $J / (m \mu)$ & $m \Omega$ & $E_{\text{b}} / \mu$ & $D / m$\\ \hline
   4.101 & 0.0167 & -0.03583 & 13.6658\\
   4.001 & 0.0185 & -0.03755 & 12.7915\\
   3.901 & 0.0207 & -0.03947 & 11.9058\\
   3.801 & 0.0234 & -0.04162 & 11.0083\\
   3.701 & 0.0267 & -0.04406 & 10.0971\\
   3.601 & 0.0308 & -0.04686 & ~~9.1719\\
   3.501 & 0.0360 & -0.05011 & ~~8.2306\\
   3.402 & 0.0430 & -0.05396 & ~~7.2723\\
   3.302 & 0.0524 & -0.05863 & ~~6.2944\\
   3.202 & 0.0659 & -0.06450 & ~~5.2890\\
   3.101 & 0.0875 & -0.07249 & ~~4.2234\\
   2.994 & 0.1352 & -0.08751 & ~~2.8974\\
   2.956 & 0.1726 & -0.10075 & ~~2.2963\\
   2.953 & 0.1772 & -0.10274 & ~~2.2363\\
   2.946 & 0.1848 & -0.10629 & ~~2.1427\\  \hline
\end{tabular} 
\caption{Normalized orbital parameters for a binary black hole system: $n = 19$ grid points on a side, $ADAPT = 10$ adaptive levels.}
\label{n19normalized}
\end{center}
\end{table}

Finally, we list the data for $n = 35$ in Tables \ref{n35vp}, \ref{n35rest}, and \ref{n35normalized}. For this data, we used six adaptive levels which, as mentioned above, yields the same energy content as ten adaptive levels.
\renewcommand{\arraystretch}{1.5}
\begin{table}[]
\begin{center}
\begin{tabular}{ c  c  c  c  c  c } 
\multicolumn{6}{}{} \\ 
  \hline J & D & P & $M$ & $E_{\text{ADM}}$ & $\Omega$ \\ \hline
   4.100 &  25.6240 & 0.1600 & 0.9788 & 1.9830 & 0.0100\\
   4.000 &  23.6889 & 0.1689 & 0.9769 & 1.9820 & 0.0112\\
   3.900 &  21.8122 & 0.1788 & 0.9749 & 1.9808 & 0.0125\\
   3.800 &  19.9945 & 0.1901 & 0.9725 & 1.9794 & 0.0142\\
   3.700 &  18.2343 & 0.2029 & 0.9697 & 1.9779 & 0.0161\\
   3.600 &  16.5273 & 0.2178 & 0.9664 & 1.9762 & 0.0183\\
   3.500 &  14.8654 & 0.2354 & 0.9626 & 1.9743 & 0.0211\\
   3.400 &  13.2360 & 0.2569 & 0.9578 & 1.9720 & 0.0246\\
   3.300 &  11.6187 & 0.2840 & 0.9518 & 1.9693 & 0.0290\\
   3.200 &  ~~9.9769 & 0.3207 & 0.9437 & 1.9661 & 0.0352\\
   3.100 &  ~~8.2322 & 0.3766 & 0.9316 & 1.9622 & 0.0445\\
   3.000 &  ~~6.0744 & 0.4939 & 0.9072 & 1.9569 & 0.0633\\
   2.970 &  ~~5.0079 & 0.5931 & 0.8875 & 1.9548 & 0.0781\\
   2.962 &  ~~4.4079 & 0.6720 & 0.8725 & 1.9542 & 0.0890\\ \hline
\end{tabular} 
\caption{Orbital parameters for a binary black hole system: $n = 35$ grid points on a side, $ADAPT = 6$ adaptive levels.}
\label{n35vp}
\end{center}
\end{table}
\renewcommand{\arraystretch}{1.5}
\begin{table}[]
\begin{center}
\begin{tabular}{ c  c  c  c } 
\multicolumn{4}{}{} \\ 
  \hline J & P & $E_{\text{ADM}}$ & $E_{\text{rest}}$ \\ \hline
   4.100 & 0.1600 & 1.0126 & 0.9998639\\
   4.000 & 0.1689 & 1.0140 & 0.9998518\\
   3.900 & 0.1788 & 1.0157 & 0.9998383\\
   3.800 & 0.1901 & 1.0177 & 0.9998233\\
   3.700 & 0.2029 & 1.0202 & 0.9998068\\
   3.600 & 0.2178 & 1.0232 & 0.9997892\\
   3.500 & 0.2354 & 1.0271 & 0.9997711\\
   3.400 & 0.2569 & 1.0322 & 0.9997548\\
   3.300 & 0.2840 & 1.0393 & 0.9997461\\
   3.200 & 0.3207 & 1.0500 & 0.9997633\\
   3.100 & 0.3766 & 1.0684 & 0.9998799\\
   3.000 & 0.4939 & 1.1159 & 1.0006720\\
   2.970 & 0.5931 & 1.1645 & 1.0022042\\
   2.962 & 0.6720 & 1.2083 & 1.0041878\\ \hline
\end{tabular}
\caption{Single boosted black hole for the determination of the rest mass $E_{\text{rest}}$: $n = 35$ grid points on a side, $ADAPT = 6$ adaptive levels.}
\label{n35rest}
\end{center}
\end{table}
\renewcommand{\arraystretch}{1.5}
\begin{table}[]
\begin{center}
\begin{tabular}{ c  c  c  c } 
\multicolumn{4}{}{} \\ 
  \hline $J / (m \mu)$ & $m \Omega$ & $E_{\text{b}} / \mu$ & $D / m$\\ \hline
   4.101 & 0.0200 & -0.03342 & 12.8138\\
   4.001 & 0.0223 & -0.03549 & 11.8462\\
   3.901 & 0.0251 & -0.03780 & 10.9079\\
   3.801 & 0.0283 & -0.04041 & ~~9.99901\\
   3.701 & 0.0321 & -0.04336 & ~~9.11892\\
   3.602 & 0.0366 & -0.04672 & ~~8.26540\\
   3.502 & 0.0422 & -0.05059 & ~~7.43439\\
   3.402 & 0.0491 & -0.05508 & ~~6.61964\\
   3.302 & 0.0580 & -0.06038 & ~~5.81082\\
   3.202 & 0.0703 & -0.06683 & ~~4.98963\\
   3.101 & 0.0890 & -0.07518 & ~~4.11662\\
   2.996 & 0.1267 & -0.08876 & ~~3.03515\\
   2.957 & 0.1565 & -0.09890 & ~~2.49844\\
   2.937 & 0.1788 & -0.10792 & ~~2.19476\\ \hline
\end{tabular}  
\caption{Normalized orbital parameters for a binary black hole system: $n = 35$ grid points on a side, $ADAPT = 6$ adaptive levels.}
\label{n35normalized}
\end{center}
\end{table}
\section{Post-Newtonian Comparisons}
At this point, we will compare our numerical results to theoretical predictions from Newtonian and post-Newtonian calculations. In particular, we compare our results to the $($post$)^{2}$-Newtonian predictions of Kidder et al. \cite{Kidder}, which are written in a convenient form in Cook \cite{Cook4}. In brief, the expressions derived in the post-Newtonian approximation are expansions in $v/c$. The $($post$)^{2}$-Newtonian expressions are expanded out to powers of $(v/c)^{4}$. 

Cook rewrites the expressions of Kidder et al. to relate the normalized binding energy $E_{\text{b}}/\mu$, the normalized angular momentum $J/\mu m$, and the normalized angular frequency $m \Omega$. To  $($post$)^{2}$ order, these equations for an equal mass binary system are
\begin{equation}
\label{pn1}
  \frac{E_{\text{b}}}{\mu} = - \frac{1}{2} \left( \frac{\mu m}{J} \right)^{2}
                       \left[ 1 + \frac{37}{16} \left( \frac{\mu m}{J} \right)^{2}
		       + \frac{1269}{128} \left( \frac{\mu m}{J} \right)^{4} + \ldots \right],
\end{equation}

\begin{equation}
\label{pn2}
  \frac{E_{\text{b}}}{\mu} = - \frac{1}{2} (m \Omega)^{2/3}
                       \left[ 1 - \frac{37}{48} (m \Omega)^{2/3}
		       - \frac{1069}{384} (m \Omega)^{4/3} + \ldots \right],
\end{equation}
and finally
\begin{equation}
\label{pn3}
  \left( \frac{J}{\mu m} \right)^{2} = (m \Omega)^{-2/3}
                                        \left[ 1 + \frac{37}{12} (m \Omega)^{2/3}
					+ \frac{143}{18} (m \Omega)^{4/3} + \ldots  \right].
\end{equation}
In the above three equations, the first term in the square brackets is the Newtonian result, the second term is the $($post$)^{3/2}$-Newtonian result, and the third term corresponds to the $($post$)^{2}$-Newtonian result.

We can also compare our results to a purely Newtonian equation relating the normalized binding energy $E_{\text{b}}/\mu$ to the normalized coordinate separation distance $D/m$. Although the range of validity of this equation is severely limited, it acts as a nice check of the asymptotic behavior of our numerical data as the separation distance of the holes becomes large. The Newtonian equation is
\begin{equation}
\label{pn4}
  \frac{E_{\text{b}}}{\mu} = -\frac{1}{2} \frac{m}{D}.
\end{equation}

In the following figures, we present our numerical results, compared to Newtonian and post-Newtonian expectations.

\begin{figure}[]
\begin{center}
\scalebox{0.8}[0.8]{
  \includegraphics[]{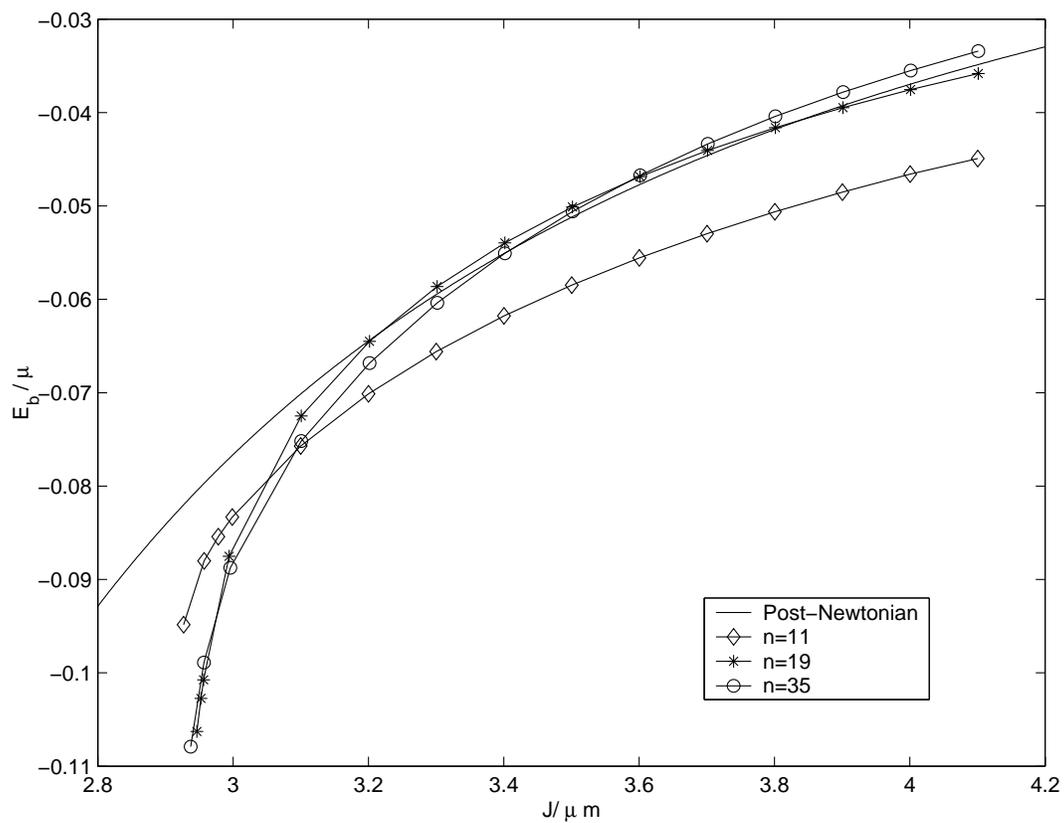}
}
\caption{The normalized binding energy $E_{\text{b}}/ \mu$ versus the normalized angular momentum $J /\mu m$ for various grid resolutions. }
\label{EvJ.eps}
\end{center}
\end{figure}

\begin{figure}[]
\begin{center}
\scalebox{0.8}[0.8]{
  \includegraphics[]{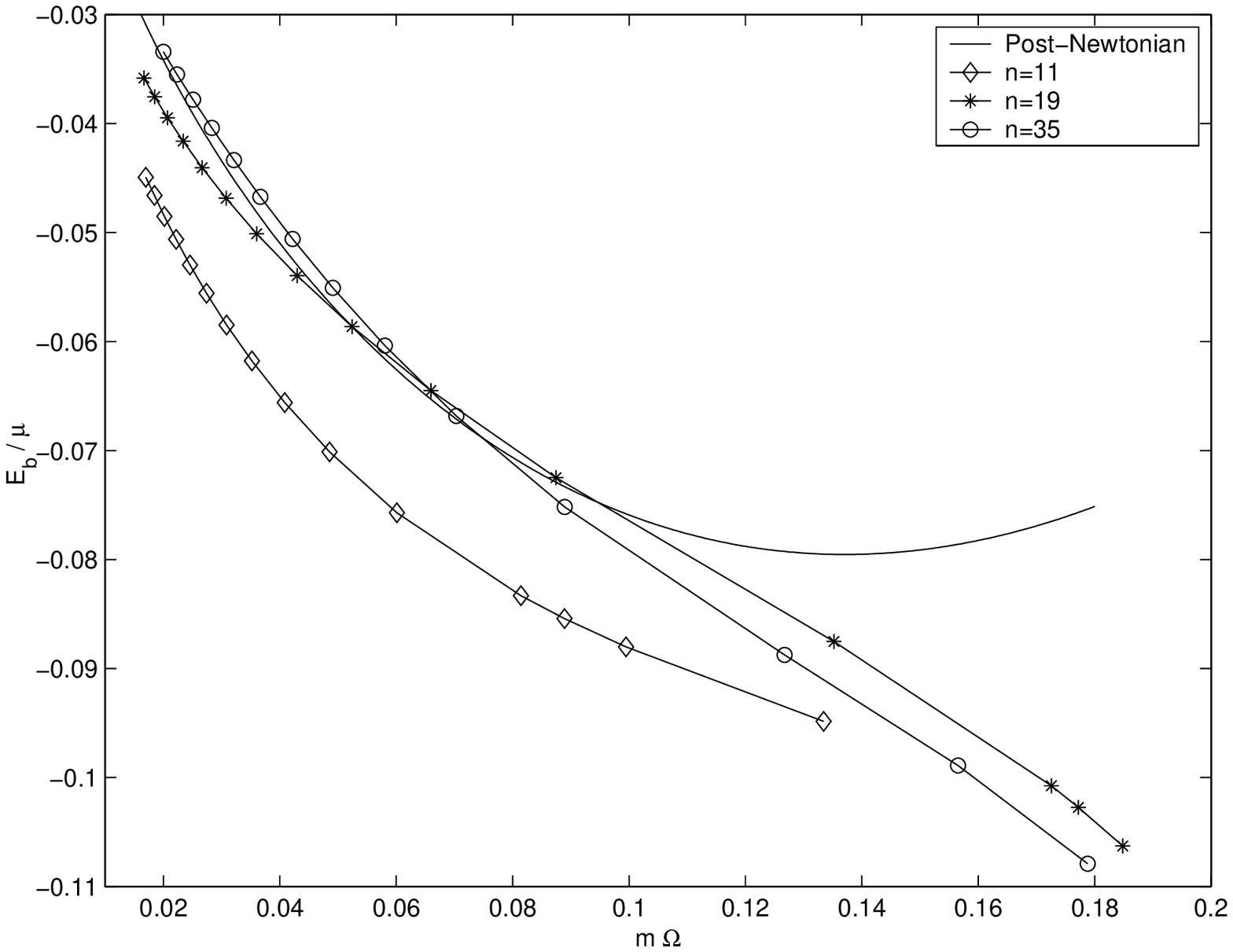}
}
\caption{The normalized binding energy $E_{\text{b}}/ \mu$ versus the normalized angular frequency $m \Omega$ for various grid resolutions.}
\label{EvOmega.eps}
\end{center}
\end{figure}

\begin{figure}[]
\begin{center}
\scalebox{0.8}[0.8]{
  \includegraphics[]{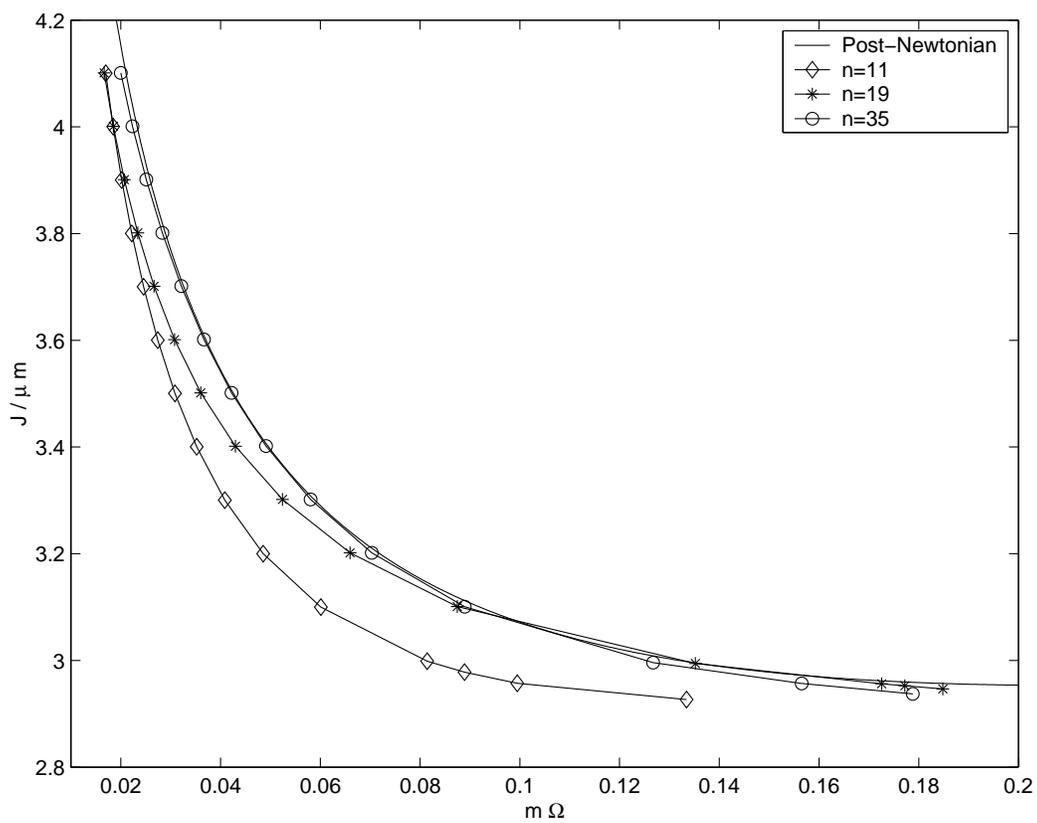}
}
\caption{The normalized angular momentum $J/ \mu m$ versus the normalized angular frequency $m \Omega$ for various grid resolutions.}
\label{JvOmega.eps}
\end{center}
\end{figure}

\begin{figure}[]
\begin{center}
\scalebox{0.8}[0.8]{
  \includegraphics[]{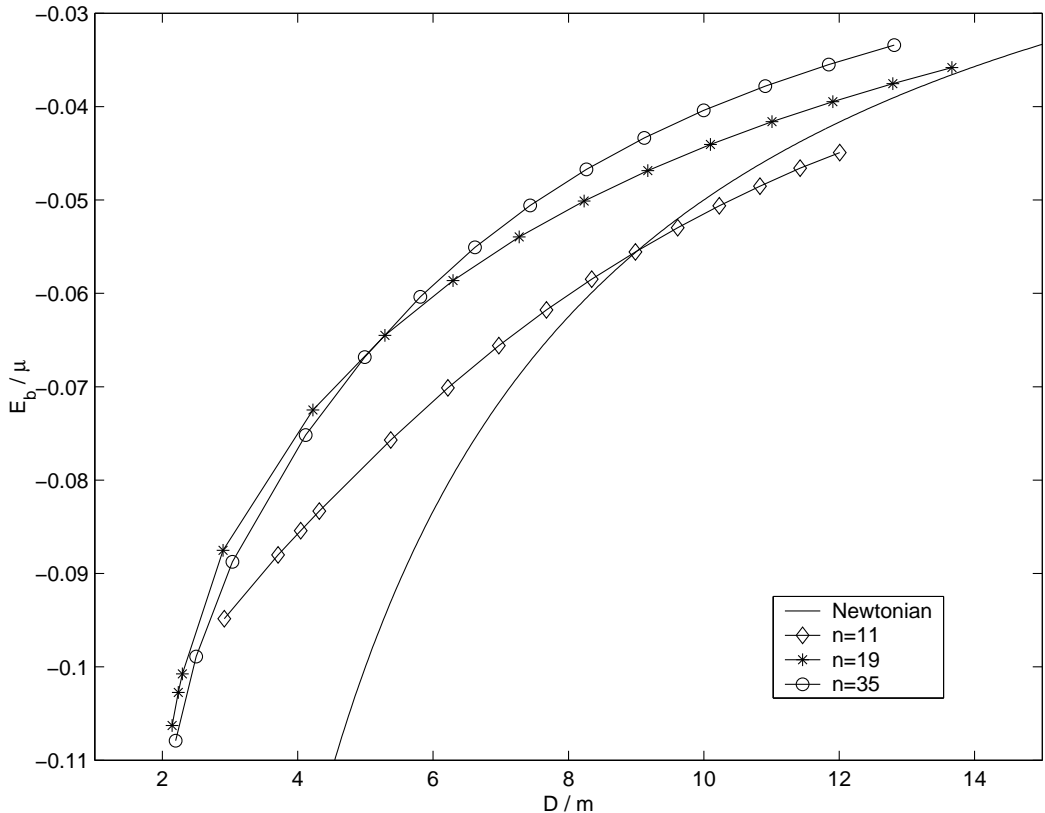}
}
\caption{The normalized binding energy $E_{\text{b}}/ \mu$ versus the normalized coordinate separation distance $D / m$ for various grid resolutions.}
\label{EvD.eps}
\end{center}
\end{figure}

Figure~\ref{EvJ.eps} displays the normalized binding energy $E_{\text{b}}/\mu$ versus the normalized angular momentum $J/\mu m$. The solid line is the theoretical prediction given by Eq.~(\ref{pn1}). One notes the curves corresponding to $n=19$ and $n=35$ actually cross the theoretical prediction, as opposed to asymptotically approaching the solid line. We believe this is due to the limited grid resolution of the data; if the grid density could be increased, we would expect closer asymptotic agreement with post-Newtonian results.\\
\indent Figure~\ref{EvOmega.eps} displays the normalized binding energy $E_{\text{b}}/\mu$ versus the normalized angular frequency $m \Omega$. The solid line is the theoretical prediction given by Eq.~(\ref{pn2}). There are several interesting points to this figure. First, note the Newtonian regime of the numerical curves, corresponding to smaller values of $m \Omega$, appear to agree with the post-Newtonian prediction given by the solid line. There appears to be good agreement between the theoretical prediction and the two higher resolution curves up until $m \Omega \approx 0.1$. At this point, the curvature of the two higher resolution curves deviates from the theoretical prediction, and tends to more negative values of the normalized binding energy. We attribute this to the fact that the system is becoming highly relativistic at this point, and higher order post-Newtonian effects would have to be considered to accurately predict the normalized binding energy.\\
\indent Figure~\ref{JvOmega.eps} displays the normalized angular momentum $J/\mu m$ versus the normalized angular frequency $m \Omega$. The post-Newtonian prediction is given by Eq.~(\ref{pn3}), and is denoted in the figure by the solid line. The curves indicate convergence at both the Newtonian and relativistic ends of the figure.\\
\indent Finally, Fig.~\ref{EvD.eps} displays the normalized binding energy $E_{\text{b}}/\mu$ versus the normalized coordinate separation distance $D/m$. The solid line is the theoretical prediction dictated by Newtonian theory only, as given by Eq.~(\ref{pn4}). Again, this figure's value lies only in large values of $D/m$, where the system is approaching Newtonian speeds. We can see from the figure that as we increase our grid resolution, our results do indeed asymptotically approach Newtonian theory.

Before we compare our data to that of Cook \cite{Cook4} and Baumgarte \cite{Baumgarte2}, we feel it is important once again to stress that there is some ambiguity as to what the ``mass'' of a black hole in a binary system means. Both Cook and Baumgarte associate the area of the apparent horizon to the mass of an individual hole, and hold the area of the apparent horizon fixed as they generate their sequence of circular orbits. We, on the other hand, chose to use the rest mass to describe the individual hole, and held the bare mass of each hole fixed as we generated our sequence of orbits.

Despite the fact that we determine the normalized separation distance for our data, we must keep in mind that this is a coordinate separation distance, and not a proper separation distance. Hence, we can not compare our $D/m$ to the proper separation distances of Cook and Baumgarte, primarily because of the fact that we do not know the locations of the apparent horizons in our geometries. However, we can get an approximate value of the proper separation distance of the innermost stable circular orbit by examining the quadrupole nature of the field. Roughly speaking, the amplitude of the quadrupole moment of the geometry measured at $r=\infty$ may be approximated to be
\begin{equation}
\label{quaddist1}
  {\cal I}_{xx} \approx \frac{1}{8} E_{\text{ADM}} d^{2}.
\end{equation} 
The variable $d$ may be interpreted as the proper separation distance of two objects of mass $\frac{1}{2} E_{\text{ADM}}$. In reality, the amplitude of the quadrupole moment consists of two parts: a portion due to the two objects of mass $E_{\text{rest}}$, and a portion due to the correction in the conformal factor. We may write this quadrupole moment amplitude as
\begin{equation}
\label{quaddist2}
  {\cal I}_{xx} = \frac{1}{4} D^{2} E_{\text{rest}} + {\cal I}_{u}.
\end{equation}
The computer code \it{moment}\rm~ calculates the contribution from the conformal factor correction ${\cal I}_{u}$. 

We may equate Eq.~(\ref{quaddist1}) and Eq.~(\ref{quaddist2}) to generate an expression for the approximate separation distance of the innermost circular orbit, as measured at infinity. We choose to normalize the separation distance $d$ by the ADM mass of the system:
\begin{equation}
\label{quaddist3}
  \frac{d}{E_{\text{ADM}}} = \sqrt{ \frac{2 D^{2} E_{\text{rest}} + 8 {\cal I}_{u} }{E_{\text{ADM}}^{3}} }.
\end{equation}
We present the normalized coordinate distance and the normalized separation distance as measured at infinity in Table \ref{quaddisttable}.
\renewcommand{\arraystretch}{1.5}
\begin{table}[]
\begin{center}
\begin{tabular}{ c  c  c }
\multicolumn{3}{}{} \\  
  \hline Resolution  & $D/m$ & $d/E_{\text{ADM}}$ \\ \hline
   $n = 11$      & 2.915 & 3.078\\
   $n = 19$      & 2.143 & 2.290\\
   $n = 35$      & 2.195 & 2.349\\ \hline
\end{tabular} 
\caption{The normalized coordinate separation distance $D/m$ and the normalized separation distance (as measured at infinity) $d/E_{\text{ADM}}$ for three grid resolutions.}
\label{quaddisttable}
\end{center}
\end{table}

Table \ref{datacompare} lists the various parameters of interest corresponding to the ISCO, as listed in Baumgarte \cite{Baumgarte2}. It is important to note Cook \cite{Cook4} employed the conformal imaging approach with a two-sheeted geometry, while Baumgarte employed the three-sheeted puncture method. Both held the areas of the apparent horizons fixed as they generated their sequence of circular orbits.
\renewcommand{\arraystretch}{1.5}
\begin{table}[]
\begin{center}
\begin{tabular}{ c  l  l  l }
\multicolumn{4}{}{} \\  
  \hline Data  & $~~E_{\text{b}}/\mu$ & $J/\mu m$ & $~~m \Omega$ \\ \hline
   Cook      & -0.09030 & 2.976 & ~0.172\\
   Baumgarte       & -0.092 & 2.95 & ~0.18\\
   This work       & -0.10792 & 2.937 & ~0.1788\\ \hline
\end{tabular} 
\caption{Comparison of the orbital parameters of Cook, Baumgarte, and the $n=35$ results of this work.}
\label{datacompare}
\end{center}
\end{table}

The agreement between our data and that of Cook and Baumgarte may lend some credence both to their method of approach, as well to the variational principle we employed. On one hand, our method is based on a mathematically derived variational principle for binary black holes, in which the variational principle dictates to us that we must hold the bare mass fixed, and not the area of the apparent horizon. This greatly simplifies the analysis, as one does not need to be concerned with the location of the apparent horizon, which can be a daunting numerical task \cite{Cook2,Baumgarte1}. On the other hand, the close agreement of Cook and Baumgarte's results with ours may indicate that there is some interesting physics yet to be discovered concerning the relationship between the variational principle we employed and the methods employed by Cook and Baumgarte.

\section{Other Numerical Results}
\subsection{The Lapse Function}
As is indicated by the variational principle of Blackburn and Detweiler \cite{Det1} given by Eq.~(\ref{stevevp1}) in \S(\ref{InitValue}), we may also generate approximate values for the ratio of the lapse functions on the upper and lower sheets of our three-sheeted geometry. Recall the lapse function is a measure of the proper time elapsed as one moves from one hypersurface to the next. In particular, the variable $N$ in Eq.~(\ref{stevevp1}) is defined as
\begin{equation}
\label{lapseresults1}
  N \equiv \frac{n_{\text{lower sheet}}}{n_{\text{upper sheet}}},
\end{equation}
where $n$ is the respective lapse on an upper or lower sheet of the geometry. If we choose the lapse on the upper sheet to be normalized to $n_{\rm upper~sheet} \equiv 1$, then we may interpret $N$ to be a measure of the lapse on the lower sheets.

Determination of $N$ numerically is not particularly straight forward. Recall the variational principle dictates $N$ is determined via
\begin{equation}
\label{lapseresults2}
  N = - \frac{1}{2} \frac{\Delta E_{\text{ADM}}}{\Delta \overline{m}},
\end{equation}
for fixed values of the angular momentum $J$. It is not a simple task to carry out the variation indicated by Eq.~(\ref{lapseresults2}), simply because the bare mass $\overline{m}$ is in some sense the fundamental ``unit'' associated with the physical quantities in the computer code.

This problem is side stepped via the following procedure. Assume a particular choice for the bare mass $\overline{m}$ has a value of $m_{B} L$, where $m_{B}$ is simply a number, and $L$ is the unit associated with the mass. Likewise, assume a particular choice of the angular momentum to be $J L^{2}$, where again $J$ is simply a pure number. These two choices result in an ADM mass with a value of $E L$, where once again $E$ is a pure number.

Now, imagine incrementing the angular momentum to a new value, given by $(J + \delta J) L^{2}$, while holding the value of the bare mass fixed. This in turn yields a new value for the ADM mass, given as $(E + \delta E) L$. This is not quite in line with the method the variational principle indicates---to determine the lapse, we should be holding the angular momentum fixed while varying the bare mass. In order to achieve this, we imagine a change in the units used to describe the quantities mentioned above. In particular, to hold the angular momentum fixed we demand $ J L^{2} = (J + \delta J) {L'}^{\; 2}$, where $L'$ is the new unit system. This indicates a rescaling of the units will take place, namely $L = \sqrt{J / (J + \delta J)} L' \equiv \lambda L'$, where $\lambda$ is the scale factor.

With this scale change in our units, this implies our initial bare mass $m_{B} L$ becomes $m_{B} \lambda L'$. Likewise, the ADM mass associated with the change in the angular momentum becomes $(E + \delta E) \lambda L'$. Now that the rescaling has taken place, we drop the prime superscript on our units. Once this is done, we may write an expression for the lapse function.

The bare mass was normalized to a value of 1.0 in all computer code runs. If we denote the initial angular momentum and it's corresponding ADM mass by $(J,E_{\text{ADM}})$, and the incremented angular momentum and it's corresponding ADM mass by $(J',E'_{\text{ADM}})$, then the lapse function is given by
\begin{equation}
\label{lapseresults3}
  N = - \frac{1}{2} \frac{ \lambda E'_{\text{ADM}} - E_{\text{ADM}} }{ 1 - \lambda },
\end{equation}
where $\lambda$ is the scale factor defined as
\begin{equation}
\label{lapseresults4}
  \lambda \equiv \sqrt{\frac{J}{J'}}.
\end{equation}

We now present the numerical data for the lapse function, which is a measure of the flow of time on the lower sheets, keeping in mind we normalize the lapse on the upper sheet to take a value of 1.0. We list the dimensionless coordinate separation distance $D/m$ and the lapse $N$ for the three different resolutions in Tables \ref{11Lapse}, \ref{19Lapse}, and \ref{35Lapse}.
\renewcommand{\arraystretch}{1.5}
\begin{table}[]
\begin{center}
\begin{tabular}{ c  c } 
\multicolumn{2}{}{} \\ 
  \hline $D / m$ & $N$ \\ \hline
    12.0084 & 0.9539\\
    11.4228 & 0.9514\\
    10.8304 & 0.9485\\
    10.2283 & 0.9451\\
    ~~9.6153 & 0.9413\\
    ~~8.9884 & 0.9367\\
    ~~8.3433 & 0.9313\\
    ~~7.6740 & 0.9247\\
    ~~6.9717 & 0.9162\\
    ~~6.2174 & 0.9048\\
    ~~5.3736 & 0.8879\\
    ~~4.3194 & 0.8572\\
    ~~4.0437 & 0.8465\\
    ~~3.7081 & 0.8313\\
    ~~2.9152 & 0.7822\\ \hline
\end{tabular}  
\caption{The normalized coordinate separation distance $D/m$ and the lapse function $N$ for $n = 11$ grid points on a side, with $ADAPT = 10$ adaptive levels.}
\label{11Lapse}
\end{center}
\end{table}
\renewcommand{\arraystretch}{1.5}
\begin{table}[]
\begin{center}
\begin{tabular}{ c  c } 
\multicolumn{2}{}{} \\ 
  \hline $D / m$ & $N$ \\ \hline
   13.6658 & 0.9568\\
   12.7915 & 0.9535\\
   11.9058 & 0.9496\\
   11.0083 & 0.9450\\
   10.0971 & 0.9395\\
   ~~9.1719 & 0.9327\\
   ~~8.2306 & 0.9242\\
   ~~7.2723 & 0.9132\\
   ~~6.2944 & 0.8986\\
   ~~5.2890 & 0.8781\\
   ~~4.2234 & 0.8461\\
   ~~2.8974 & 0.7765\\
   ~~2.2963 & 0.7222\\
   ~~2.2363 & 0.7155\\
   ~~2.1427 & 0.7045\\ \hline
\end{tabular}  
\caption{The normalized coordinate separation distance $D/m$ and the lapse function $N$ for $n = 19$ grid points on a side, with $ADAPT = 10$ adaptive levels.}
\label{19Lapse}
\end{center}
\end{table}
\renewcommand{\arraystretch}{1.5}
\begin{table}[]
\begin{center}
\begin{tabular}{ c  c } 
\multicolumn{2}{}{} \\
  \hline$D / m$ & $N$ \\ \hline
    12.8138 & 0.9506\\
    11.8462 & 0.9463\\
    10.9079 & 0.9415\\
    ~~9.9990 & 0.9359\\
    ~~9.1189 & 0.9296\\
    ~~8.2654 & 0.9221\\
    ~~7.4344 & 0.9133\\
    ~~6.6196 & 0.9025\\
    ~~5.8108 & 0.8889\\
    ~~4.9896 & 0.8705\\
    ~~4.1166 & 0.8432\\
    ~~3.0352 & 0.7885\\
    ~~2.4984 & 0.7455\\
    ~~2.1948 & 0.7134\\ \hline 
\end{tabular}  
\caption{The normalized coordinate separation distance $D/m$ and the lapse function $N$ for $n = 35$ grid points on a side, with $ADAPT = 6$ adaptive levels.}
\label{35Lapse}
\end{center}
\end{table}

\begin{figure}[]
\begin{center}
\scalebox{0.8}[0.8]{
  \includegraphics[]{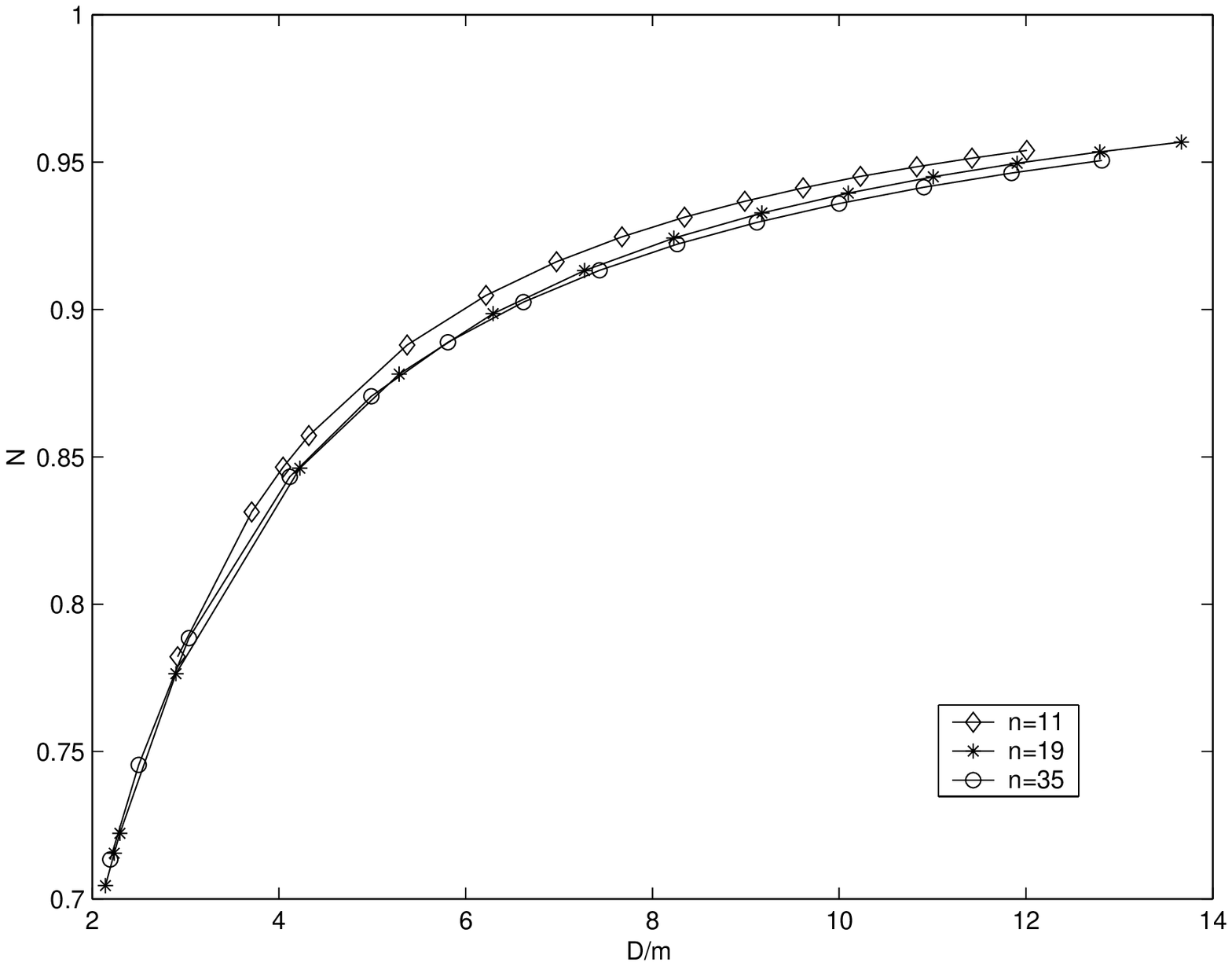}
}
\caption{The lapse function $N$ versus the normalized coordinate separation distance $D/m$ for three resolutions. }
\label{lapse.eps}
\end{center}
\end{figure}

The data from Tables \ref{11Lapse}, \ref{19Lapse}, and \ref{35Lapse} are displayed in Fig.~\ref{lapse.eps}. From inspection of the figure, the data are all in close agreement with one another. The interpretation of the data is as follows: When the holes are separated by large distances, then the geometry has approximately the same lapse as a completely isolated hole. This corresponds to an asymptotic limit of $N \rightarrow 1$ as $D / m \rightarrow \infty$, which the figure demonstrates. However, as the two holes come closer together, the gravitational interaction increases, causing the relativistic effects to increase as well. However, on the lower sheets, the isolated holes remain unchanged (the bare masses are fixed, and the holes on the lower sheet have no linear or angular momentum \cite{BB}). Because the geometry on the lower sheets is unchanging, the lapse function on the lower sheets is unchanging as well. Hence, as the value of $N$ decreases with a decreasing separation distance, this corresponds to the proper time on the upper sheet lagging the proper time on the lower sheet. For instance, when the holes are at the innermost stable circular orbit, for a single tick of the clock on the upper sheet, the clock on the lower sheet has undergone approximately 1.4 ticks.
\subsection{Gravitational Waves}

As mentioned in \S(\ref{ADM Mass}), we developed a method for determining the multipole moments of the gravitational field, denoted as $\Phi_{\ell m}$. It is possible to relate these multipole moments to the reduced quadrupole moments ${\cal I}_{i j}$, which in turn yields information about the gravitational wave content of the system. First, let us briefly review some definitions. Recall from Eq.~(\ref{mpole2}) we defined the multipole moments as
\begin{equation}
\label{GW1}
  \Phi_{\ell m} \equiv - \frac{1}{\sqrt{\pi} (2 \ell + 1)} 
          \oint \left[ \psi^{\dag} \nabla_{a} \psi - \psi \nabla_{a} \psi^{\dag} \right] d S^{a},
\end{equation}
where $\psi^{\dag}$ are polynomials related to the spherical harmonics, and the standard definition of the reduced quadrupole moment is
\begin{equation}
\label{GW2}
  {\cal I}_{i j} \equiv \int \rho \left[ x_{i} x_{j} - \frac{1}{3} \delta_{i j} r^{2} \right] d^{3} x,
\end{equation}
where the volume integral is performed over all space \cite{MTW}.

We may rewrite Eq.~(\ref{GW1}) as a volume integral, which in turn allows us to relate the multipole moments $\Phi_{\ell m}$ and the reduced quadrupole moments ${\cal I}_{i j}$. Specifically, the only non-zero radiating modes are of the form
\begin{equation}
\label{GW3}
  {\cal I}_{x x} = - {\cal I}_{y y} = 2 \sqrt{\frac{2 \pi}{15}} \Phi_{2 2}.
\end{equation}

The computer code \it{moment}\rm~ calculates the multipole moment $\Phi_{2 2}$, which allows us to calculate the gravitational wave luminosity via the standard quadrupole moment formula \cite{MTW} given by
\begin{equation}
\label{GW4}
  L \equiv \frac{dE_{\text{ADM}}}{dt} = \frac{1}{5} \left\langle \dddot{{\cal I}}_{i j} \dddot{{\cal I}}_{i j} \right\rangle,
\end{equation}
where the three dots above the reduced quadrupole moments indicate three time derivatives, and the angled brackets indicate averaging over several orbital periods. Summation is implied over all spatial indices $i$ and $j$.

Because of the quadrupole nature of the radiation, we may assume a time dependence of the form $\cos(2 \Omega t)$, which in turn implies the three time derivatives on each of the reduced quadrupole moments may be replaced by $(2 \Omega)^{3}$ in the above equation for the luminosity.

We calculate the reduced quadrupole moments by assuming they consist of two parts: one part is due to the existence of the point masses (which correspond to the black holes of mass $E_{\text{rest}}$), and a second part which is due to the existence of the nonlinear correction to the conformal factor (which corresponds to the numerical value of $\Phi_{2 2}$). It is then straightforward to write an expression for the gravitational wave luminosity in terms of the rest masses of the black holes $E_{\text{rest}}$, the coordinate separation distance between the holes $D$, the orbital angular frequency $\Omega$, and the multipole moment $\Phi_{2 2}$:
\begin{equation}
\label{GW5}
  L = \frac{128}{5} \Omega^{6} \left[ \frac{1}{8} D^{4} E_{\text{rest}}^{2} 
                                     + \sqrt{\frac{2 \pi}{15}} D^{2} E_{\text{rest}} \Phi_{2 2}
				     + \frac{8 \pi}{15} \left( \Phi_{2 2} \right)^{2} \right].
\end{equation}
The first term of Eq.~(\ref{GW5}) is the expected Newtonian result, and the remaining terms are the relativistic corrections.

Tables \ref{MG_3-GW}, \ref{MG_4-GW}, and \ref{MG_5-GW} display various quantities derived from the gravitational wave luminosity for the three different grid resolutions we have been studying. All three tables display the normalized coordinate separation distance $D/m$, the gravitational wave luminosity $L$ as given by Eq.~(\ref{GW5}), the expected Newtonian gravitational wave luminosity $L_{\text{N}}$ given by only the first term of Eq.~(\ref{GW5}), the energy radiated per orbit $E/\text{orbit}$, and the time scale $t$ the binary system will spend at a particular separation distance. A few important things to note before we display our numerical results is the Newtonian expectation of the luminosity is determined strictly via Kepler's laws, and not from the variational principle for the orbital angular frequency. Also, the energy radiated per orbit is calculated by multiplying the luminosity and the orbital period: $E/\text{orbit} = \frac{2 \pi}{\Omega} L$. Finally, the time scale of the orbit was determined by dividing the absolute value of the binding energy by the luminosity: $t = \frac{\vert E_{\text{b}} \vert}{L}$.
\renewcommand{\arraystretch}{1.5}
\begin{table}[]
\begin{center}
\begin{tabular}{ c  c  c  c  c } 
\multicolumn{5}{}{} \\
  \hline$D / m$ & $L$ & $L_{\text{N}}$ & $E/\text{orbit}$ & $t$ \\ \hline
    12.0084& 4.00E-07& 1.60E-06& 2.96E-04& 5.60E+04\\
    11.4228& 5.43E-07& 2.06E-06& 3.69E-04& 4.29E+04\\ 
    10.8304& 7.48E-07& 2.68E-06& 4.66E-04& 3.24E+04\\ 
    10.2283& 1.05E-06& 3.57E-06& 5.96E-04& 2.41E+04\\ 
    ~~9.6153& 1.51E-06& 4.87E-06& 7.72E-04& 1.75E+04\\ 
    ~~8.9884& 2.22E-06& 6.82E-06& 1.02E-03& 1.25E+04\\ 
    ~~8.3433& 3.38E-06& 9.89E-06& 1.37E-03& 8.66E+03\\ 
    ~~7.6740& 5.33E-06& 1.50E-05& 1.90E-03& 5.79E+03\\ 
    ~~6.9717& 8.87E-06& 2.43E-05& 2.73E-03& 3.70E+03\\ 
    ~~6.2174& 1.59E-05& 4.31E-05& 4.11E-03& 2.21E+03\\ 
    ~~5.3736& 3.19E-05& 8.93E-05& 6.68E-03& 1.19E+03\\ 
    ~~4.3194& 8.30E-05& 2.66E-04& 1.28E-02& 5.02E+02\\ 
    ~~4.0437& 1.08E-04& 3.70E-04& 1.53E-02& 3.95E+02\\ 
    ~~3.7081& 1.51E-04& 5.71E-04& 1.90E-02& 2.92E+02\\ 
    ~~2.9152& 3.38E-04& 1.90E-03& 3.19E-02& 1.40E+02\\ \hline 
\end{tabular}  
\caption{Gravitational wave data for a black hole binary system with $n = 11$ grid points on a side and $ADAPT = 10$ adaptive levels.}
\label{MG_3-GW}
\end{center}
\end{table}
\renewcommand{\arraystretch}{1.5}
\begin{table}[]
\begin{center}
\begin{tabular}{ c  c  c  c  c } 
\multicolumn{5}{}{} \\
  \hline$D / m$ & $L$ & $L_{\text{N}}$ & $E/\text{orbit}$ & $t$ \\ \hline
   13.6658& 5.99E-07& 8.39E-07& 4.52E-04& 2.99E+04\\
   12.7915& 8.64E-07& 1.17E-06& 5.87E-04& 2.17E+04\\
   11.9058& 1.28E-06& 1.67E-06& 7.74E-04& 1.55E+04\\
   11.0083& 1.93E-06& 2.47E-06& 1.04E-03& 1.08E+04\\
   10.0971& 3.01E-06& 3.81E-06& 1.42E-03& 7.32E+03\\
   ~~9.1719& 4.85E-06& 6.16E-06& 1.98E-03& 4.83E+03\\
   ~~8.2306& 8.13E-06& 1.06E-05& 2.84E-03& 3.08E+03\\
   ~~7.2723& 1.43E-05& 1.97E-05& 4.17E-03& 1.89E+03\\
   ~~6.2944& 2.64E-05& 4.05E-05& 6.34E-03& 1.11E+03\\
   ~~5.2890& 5.26E-05& 9.66E-05& 1.00E-02& 6.13E+02\\
   ~~4.2234& 1.17E-04& 2.98E-04& 1.68E-02& 3.09E+02\\
   ~~2.8974& 3.59E-04& 1.96E-03& 3.34E-02& 1.22E+02\\
   ~~2.2963& 6.19E-04& 6.27E-03& 4.52E-02& 8.17E+01\\
   ~~2.2363& 6.54E-04& 7.15E-03& 4.65E-02& 7.89E+01\\
   ~~2.1427& 7.11E-04& 8.86E-03& 4.86E-02& 7.51E+01\\ \hline 
\end{tabular}  
\caption{Gravitational wave data for a black hole binary system with $n = 19$ grid points on a side and $ADAPT = 10$ adaptive levels.}
\label{MG_4-GW}
\end{center}
\end{table}
\renewcommand{\arraystretch}{1.5}
\begin{table}[]
\begin{center}
\begin{tabular}{ c  c  c  c  c } 
\multicolumn{5}{}{} \\
  \hline $D / m$ & $L$ & $L_{\text{N}}$ & $E/\text{orbit}$ & $t$ \\ \hline 
  12.8138& 1.38E-06& 1.16E-06& 8.68E-04& 1.21E+04\\
   11.8462& 1.98E-06& 1.71E-06& 1.11E-03& 8.98E+03\\
   10.9079& 2.85E-06& 2.59E-06& 1.43E-03& 6.63E+03\\
   ~~9.9990& 4.16E-06& 4.00E-06& 1.85E-03& 4.86E+03\\
   ~~9.1189& 6.13E-06& 6.34E-06& 2.40E-03& 3.54E+03\\
   ~~8.2654& 9.17E-06& 1.04E-05& 3.14E-03& 2.55E+03\\
   ~~7.4344& 1.40E-05& 1.76E-05& 4.16E-03& 1.81E+03\\
   ~~6.6196& 2.19E-05& 3.15E-05& 5.60E-03& 1.26E+03\\
   ~~5.8108& 3.56E-05& 6.04E-05& 7.70E-03& 8.48E+02\\
   ~~4.9896& 6.13E-05& 1.29E-04& 1.10E-02& 5.45E+02\\
   ~~4.1166& 1.17E-04& 3.38E-04& 1.65E-02& 3.21E+02\\
   ~~3.0352& 2.93E-04& 1.55E-03& 2.90E-02& 1.52E+02\\
   ~~2.4984& 4.81E-04& 4.11E-03& 3.87E-02& 1.03E+02\\
   ~~2.1948& 6.41E-04& 7.85E-03& 4.53E-02& 8.45E+01\\ \hline  
\end{tabular}  
\caption{Gravitational wave data for a black hole binary system with $n = 35$ grid points on a side and $ADAPT = 6$ adaptive levels.}
\label{MG_5-GW}
\end{center}
\end{table}

\begin{figure}[]
\begin{center}
\scalebox{0.8}[0.8]{
  \includegraphics[]{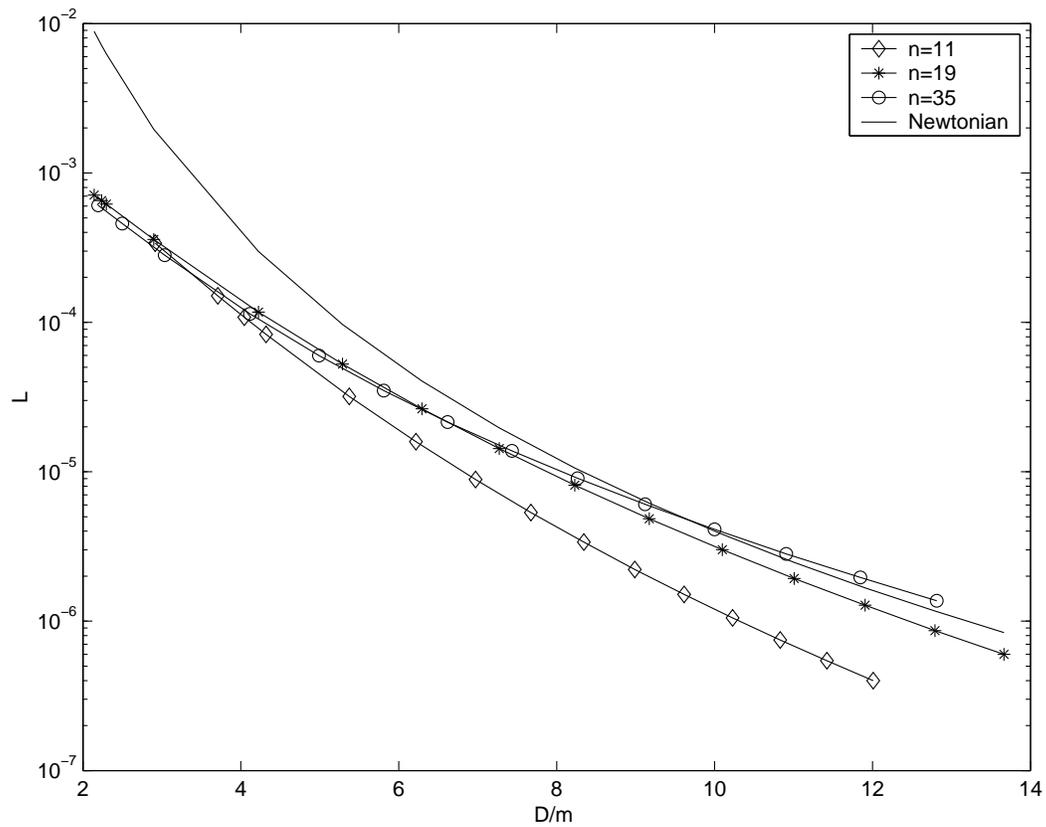}
}
\caption{The gravitational wave luminosity $L$ as a function of the normalized separation distance $D/m$.}
\label{LvD.eps}
\end{center}
\end{figure}

We also present the data from Tables \ref{MG_3-GW}, \ref{MG_4-GW} and \ref{MG_5-GW} in graphical form. First, Fig.~\ref{LvD.eps} displays the luminosity $L$ as a function of the normalized separation distance $D/m$. The three different resolutions are displayed, as well as the Newtonian expectation for the gravitational wave luminosity. Figure~\ref{LvD.eps} indicates relatively good agreement in both the Newtonian domain as well as the highly relativistic. In particular, the numerical results appear to be converging to a limit which is in line with the Newtonian expectation. Also, the numerical results are in excellent agreement with each other as one examines the curves as the separation distance closes to the innermost stable circular orbit.

\begin{figure}[]
\begin{center}
\scalebox{0.8}[0.8]{
  \includegraphics[]{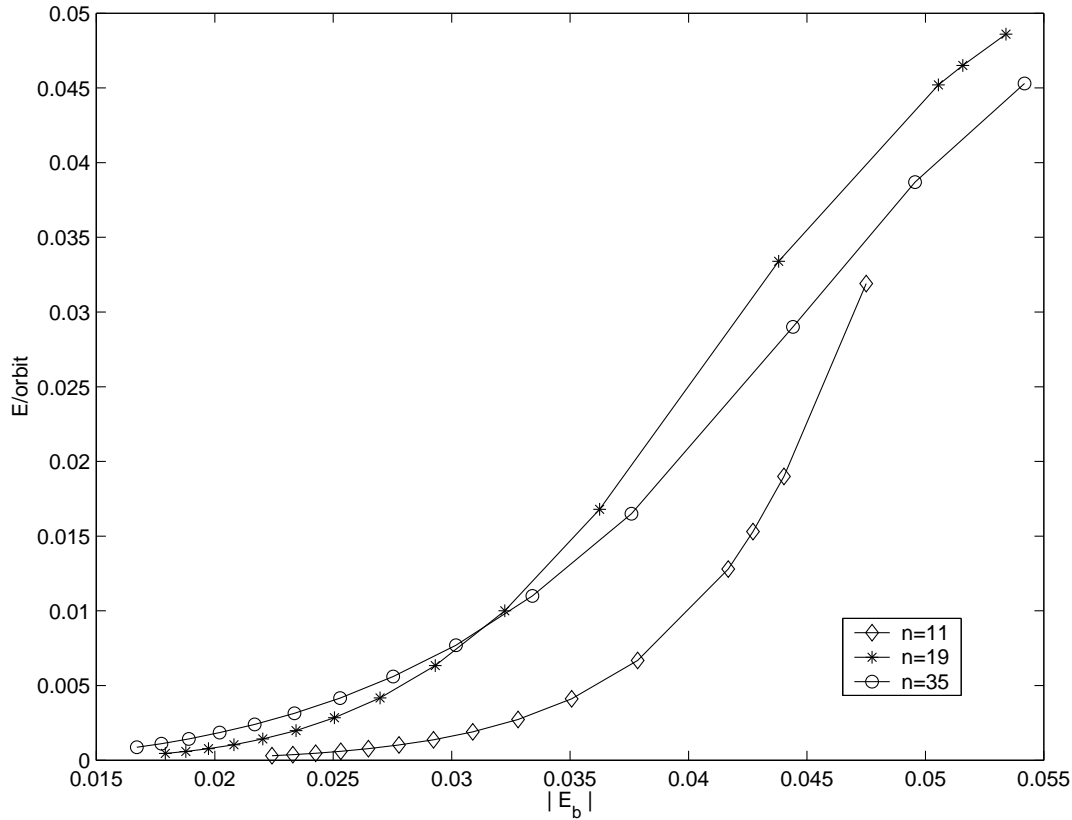}
}
\caption{The energy radiated per orbit $E/\text{orbit}$ as a function of the absolute value of the non-normalized binding energy $E_{\text{b}}$.}
\label{EvE.eps}
\end{center}
\end{figure}

We now present Fig.~\ref{EvE.eps}, the energy radiated per orbit $E/\text{orbit}$ versus the absolute value of the non-normalized binding energy $E_{\text{b}}$. Once again, all three resolutions are displayed. The figure shows the numerical results are converging, and the data gives an approximate view of the number of orbits the system will complete at a particular separation distance. For instance, examining the $n=35$ curve, when the separation distance is $D/m = 12.8138$, the ratio of the binding energy to the energy radiated per orbit is on the order of 20, which indicates the system will complete approximately 20 orbits at this separation distance. Likewise, when the separation distance is $D/m =2.1948$, the ratio of the binding energy to the energy radiated per orbit is on the order of 1, which indicates the system will complete approximately one orbit at this separation distance before the final plunge.

\begin{figure}[]
\begin{center}
\scalebox{0.8}[0.8]{
  \includegraphics[]{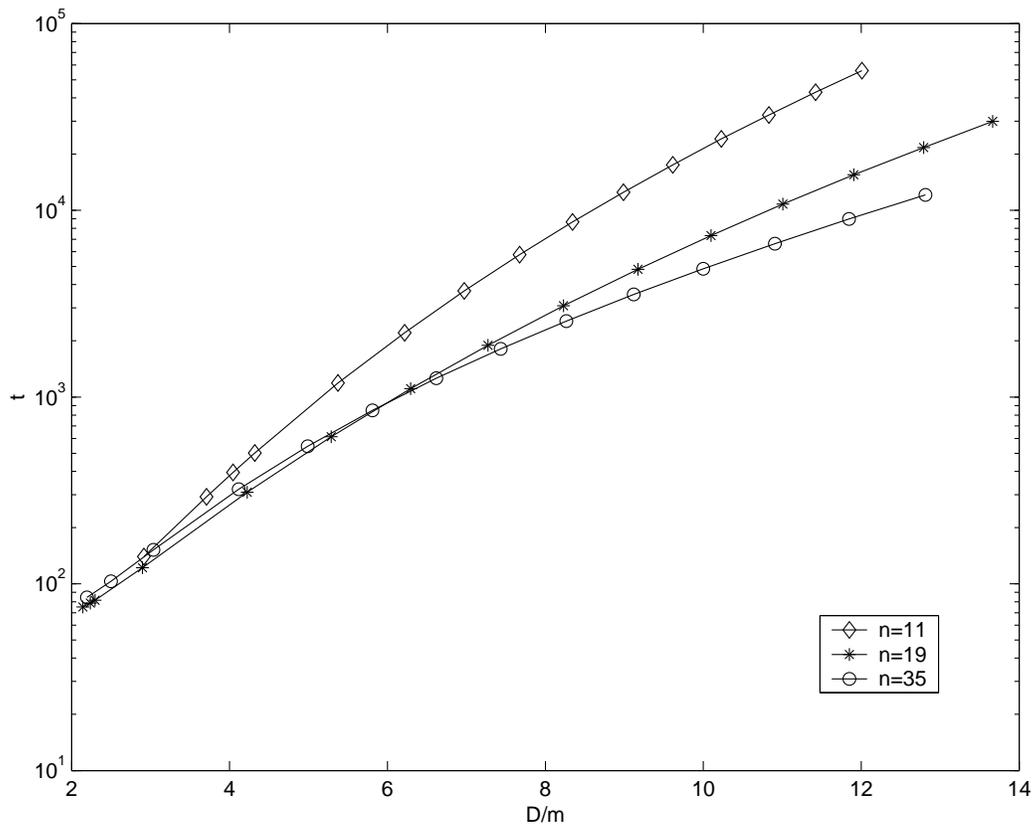}
}
\caption{The time scale of radiation reaction as a function of the normalized separation distance $D/m$.}
\label{tvD.eps}
\end{center}
\end{figure}

We may get an approximate idea of the time scale involved with the evolution of our system by examining Fig.~\ref{tvD.eps}, which displays the time scale of radiation reaction versus the normalized separation distance $D/m$. The figure shows the data is converging as the grid resolution is increased, and all three resolutions agree relatively well with each other as the holes approach the innermost stable circular orbit. As indicated in the figure, the holes will spend approximately $t \approx 100 \overline{m}$ at the innermost stable circular orbit before the final plunge and merger.

\section{Discussion}
As demonstrated, our variational principle for describing binary black hole systems in the framework of the three sheeted Brill geometry greatly simplifies numerical implementation. Due to the necessity of holding the bare masses of the holes fixed, we did not concern ourselves with calculations involving the apparent horizon. Comparisons of our numerical results with that of Cook \cite{Cook4} and Baumgarte \cite{Baumgarte2} would seem to point to some interesting mathematical physics: a potentially interesting project would be to find a transformation law for variational principles which hold the bare mass fixed to variational principles in which the apparent horizon area is held fixed.

Another advantage our analysis has over previous work is the extraction of information pertaining to the gravitational wave content of the geometry. This information yields estimates pertaining to the dynamics of the system, which is of vital interest to the community involved with gravitational wave detectors like LIGO.

\chapter{Conclusion}
In an era in which numerical relativity is attempting to solve the full set of  nonlinear Einstein equations, we feel numerical simulations based upon mathematically sound variational principles is a step in the right direction. In general, they greatly simplify the algorithms involved in calculations: rather than attempting to solve the full set of Einstein equations one simply need to minimize a single function after satisfying the constraint equations. Another pleasing property of variational principles is they provide a road map for deriving accurate estimates of orbital parameters for binary systems, which may in turn be used to generate waveform templates for gravitational wave observatories.

Assuming the constraint equations are satisfied, our method requires only the minimization of the effective mass to ensure an approximate solution to the quasi-stationary Einstein equations, as well as the equations governing the irrotational motion of the neutron stars. Our treatment of irrotational binary neutron stars yields accurate estimates of the mass, gravitational radiation, and orbital frequency of the system---quantities necessary to construct viable waveform templates for gravitational wave detectors \cite{FlanaganHughes}. The variational principle also serves as a relatively simple way to test many of the current conformally flat treatments of binary neutron stars investigated by other authors \cite{Bonna3,Bonna4,Shibata2,Uryu3,Wilson4}. 

With the power of nonlinear adaptive multigrid techniques, we demonstrated the use of a variational principle for the ADM mass for binary black holes. The variational principle ensures solutions to the quasi-equilibrium Einstein equations for fixed values of the angular momentum and the black hole bare masses. We took advantage of the simplifying assumptions of a conformally flat metric, in which the conformal extrinsic curvature is trace-free. This introduced some additional energy into the system, which limits this geometry's ability to model astrophysically realistic black hole systems. Despite this, the geometry acted as a proving ground for the variational principle. We employed the ``puncture'' method to describe the black holes. This greatly simplified the numerical analysis by allowing the three-sheeted geometry to be compactified into $R^{3}$, covered by a regular Cartesian grid. 

It is important to reiterate our treatment holds the bare masses of the holes and the angular momentum of the system fixed as we generate our sequence of circular orbits. This method is based upon a mathematically derived variational principle, and is algorithmically more desirable than other methods in which the apparent horizon area is held fixed \cite{Baumgarte2,Cook4}. Agreement between our results and the fixed horizon area results of Baumgarte and Cook may indicate a correspondence between the two different approaches.

We demonstrated agreement between our numerical results and post-Newtonian approximations for various orbital parameters, as well as generated lapse function and gravitational wave information for the system. The gravitational wave data may in turn facilitate gravitational wave observatories in their identification of astrophysical binary systems.

\backmatter

\chapter[Computer Code]{Computer Code for the Black Hole Variational Principle}
In this appendix, we list all of the computer code we developed in C++ to solve the Hamiltonian constraint and subsequently determine sequences of quasi-stationary circular orbits by means of employing the variational principle for the ADM mass. Effort will be made to list the program functions in order of relative importance, with as many comments as possible to clarify the purpose of the individual functions. We also note that the header files included with each of these individual programs are the standard files assert.h, fstream.h, iostream.h, math.h and iomanip.h. Two additional header files, written by Steven Detweiler and which will be listed later, are grid.h and adapt.h. Any additional header files will be listed in the individual programs.

It is important to note the parameter $ADAPT$, mentioned in \S(\ref{Nonlinear Adaptive Multigrid}), is called $NMG$ (for non-multigrid) in the code that follows.

We begin by listing the program \it{min}\rm, which finds a minimum in the ADM mass:
\begin{verbatim}
// min.cxx

// Determines the minimum value of the ADM mass

// Designed for Variational Principle
// Code decreases the spacing between punctures with fixed
// angular momentum J
// Calculates ADM mass, and saves it to a file 

// Galerkin operator to be used on *ALL* levels
// symmetries built into relaxation and residual code!
// This is the "true" adaptive multigrid code for the non-linear 
// equation describing the conformal factor of a black hole
// binary system, using the puncture method described in 
// Brandt and Brugmann. 

// Be careful of the largest grid you have to force the physical
// size of the largest grid to a certain value in order to
// maintain consistency (smaller grids have n+2 grid points,
// where as the largest has n grid points)

// Relax on interior points only on each non-mg level
// Use Robin conditions on the mg levels

// This code solves for u, where Phi = 1/alpha + 1 + u. This
// allows the Robin boundary conditions to be employed

// Be sure to run on wombat!

#include <assert.h>
#include <fstream.h>
#include <iostream.h>
#include <iomanip.h>
#include <math.h>
#include "grid.h"
#include "adapt.h"	// contains all function prototypes

// Coefficients of Relaxation schemes 

//Regular Del^2 Operator
extern const double DEL0 = 6.0;  // coeff for relaxation point
extern const double DEL1 = 1.0;  // coeff for relaxation point
                                 // once removed
extern const double DEL2 = 0.0;  // coeff for relaxation point
                                 // twice removed
extern const double DEL3 = 0.0;  // coeff for relaxation point
                                 // thrice removed

//Galerkin Operator for all points of the coarser grid
extern const double GAL0 = -8.0/3;  // coeff for relaxation
                                    // point
extern const double GAL1 = 0.0;     // coeff for relaxation
                                    // point once removed
extern const double GAL2 = 1.0/6;   // coeff for relaxation
                                    // point twice removed
extern const double GAL3 = 1.0/12;  // coeff for relaxation
                                    // point thrice removed
				    
extern const double BAREM1 = 1.00;   // bare mass of hole #1
extern const double BAREM2 = 1.00;   // bare mass of hole #2

extern const double PI = 4.0*atan(1.0);

extern const double Y1 = 0.0*BAREM1;  // Note that X,Y,and Z 
                                      // are all >0
extern const double Y2 = 0.0*BAREM1; 
extern const double Z1 = 0.0*BAREM1; 
extern const double Z2 = 0.0*BAREM1; 

extern const double PX1 = 0.00*BAREM1;         
extern const double PX2 = 0.00*BAREM1; 
extern const double PZ1 = 0.0*BAREM1; 
extern const double PZ2 = 0.0*BAREM1;

extern const double SX1 = 0.0*BAREM1;         
extern const double SX2 = 0.0*BAREM1; 
extern const double SY1 = 0.0*BAREM1;         
extern const double SY2 = 0.0*BAREM1;         
extern const double SZ1 = 0.0*BAREM1; 
extern const double SZ2 = 0.0*BAREM1; 

extern const int NMG = 6;	// # of non-mg levels
extern const int MG = 5;        // # of mg levels

extern const int PMAX = 8;      //sets the number of v-cycles

extern const double j = 3.00/(BAREM1*BAREM1); // dimensionless
                                              // # to hold J
					      // fixed
//extern double J = 1.0;
extern const double J = j*BAREM1*BAREM2;    // Fixed angular
                                            // momentum of
                                            // system

//----------------------------------------------------------

int main()
{
  using namespace Parameters;
  using namespace Energy;

  // MG, NMG, j, ax, bx, and cx must be specified prior!!
  
  double ax = 4.1*BAREM1;              // left most location on
                                       // curve
  double bx = 4.4*BAREM1;              // center location on
                                       // curve
  double cx = 4.7*BAREM1;              // right most location on
                                       // curve
  
  const double TOL = 1.0e-6;   // TOL ~ sqrt(machine error)
  
  double fmin, xmin;
  
  for(J=j*BAREM1*BAREM2;J<=3.0*BAREM1*BAREM2;
      J=J+0.01*BAREM1*BAREM2) {
    
    // see Numerical Recipes for brent.c code
    fmin = brent(ax,bx,cx,newADMmass,TOL,&xmin);  // minimum ADM
                                                  // mass
  
    cout << "Circular orbit for J = " << J << " at D = " 
         << xmin << " with E(adm) = " << fmin << endl;
       
    ofstream output;
    output.open("minimum-orbits.dat", ios::app|ios::ate);
//appends to end of file
    
// prints out J, sep. dist., Newtonian mass, Newtonian binding
// energy, and ADM mass
    //double m = 2*M1;
    //double mu = 1.0/2*M1;
    //double Jdivmmu = J/(m*mu);
    //double Eb = -(fmin - 2*M1);
    //double Ebdivmu = Eb/mu;
    double Eadm = fmin;
     
    output << "# J, D, P, M, Eadm" << endl;
    output << J <<  setprecision(12) << "   " << xmin 
           << "   " << PY1 << "   " << M1 << "   " << Eadm
           << endl;
       
  } // end for - J 
} // end main
\end{verbatim}

Another useful program is \it{VP}\rm, which allows us to generate ``effective potential'' curves, plotting the ADM mass as a function of puncture separation distance for fixed values of angular momentum:
\begin{verbatim}
// VP.cxx

// Designed for Variational Principle
// Code decreases the spacing between punctures with fixed
// angular momentum J
// Calculates ADM mass, and saves it to a file 

// Galerkin operator to be used on *ALL* levels
// symmetries built into relaxation and residual code!
// This is the "true" adaptive multigrid code for the non-linear
// equation describing the conformal factor of a black hole
// binary system, using the puncture method described in
// Brandt and Brugmann. 

// Be careful of the largest grid you have to force the 
// physical size of the largest grid to a certain
// value in order to maintain consistency (smaller grids have
// n+2 grid points, where as the largest has n grid points)

// Relax on interior points only on each non-mg level
// Use Robin conditions on the mg levels

// This code solves for u, where Phi = 1/alpha + 1 + u. This
// allows the Robin 
// boundary conditions to be employed

// Be sure to run on wombat!

#include <assert.h>
#include <fstream.h>
#include <iostream.h>
#include <iomanip.h>
#include <math.h>
#include "grid.h"
#include "adapt.h"	// contains all function prototypes


// Coefficients of Relaxation schemes 

//Regular Del^2 Operator
extern const double DEL0 = 6.0;  // coeff for relaxation point
extern const double DEL1 = 1.0;  // coeff for relaxation point
                                 // once removed
extern const double DEL2 = 0.0;  // coeff for relaxation point
                                 // twice removed
extern const double DEL3 = 0.0;  // coeff for relaxation point
                                 // thrice removed

//Galerkin Operator for all points of the coarser grid
extern const double GAL0 = -8.0/3;  // coeff for relaxation
                                    // point
extern const double GAL1 = 0.0;  // coeff for relaxation point
                                 // once removed
extern const double GAL2 = 1.0/6;  // coeff for relaxation point
                                   // twice removed
extern const double GAL3 = 1.0/12;  // coeff for relaxation
                                    // point thrice removed

extern const double BAREM1 = 1.00;   // bare mass of hole #1
extern const double BAREM2 = 1.00;   // bare mass of hole #2

extern const double PI = 4.0*atan(1.0);

extern const double Y1 = 0.0*BAREM1;      // Note that X,Y,and Z
                                          // are all >0
extern const double Y2 = 0.0*BAREM1; 
extern const double Z1 = 0.0*BAREM1; 
extern const double Z2 = 0.0*BAREM1; 

extern const double PX1 = 0.00*BAREM1;         
extern const double PX2 = 0.00*BAREM1; 
extern const double PZ1 = 0.0*BAREM1; 
extern const double PZ2 = 0.0*BAREM1;

extern const double SX1 = 0.0*BAREM1;         
extern const double SX2 = 0.0*BAREM1; 
extern const double SY1 = 0.0*BAREM1;         
extern const double SY2 = 0.0*BAREM1;         
extern const double SZ1 = 0.0*BAREM1; 
extern const double SZ2 = 0.0*BAREM1; 

extern const int NMG = 6;	// # of non-mg levels
//extern int MG = 3;            // # of mg levels
extern const int MG = 5;        // # of mg levels

extern const int PMAX = 8;        //sets the number of v-cycles

extern const double j = 2.962/(BAREM1*BAREM1);  //dimensionless
                                                // # to hold J
                                                // fixed
//extern double J=1.0;    // Fixed angular momentum of system,
                          // but give initial arb. value
extern const double J = j*BAREM1*BAREM2;    // Fixed angular
                                            // momentum of
                                            // system

//-----------------------------------------------------------

int main()
{
  using namespace Parameters;
  
  for(MG = 3; MG <= 3; MG++) {
    for(J = j*BAREM1*BAREM2;J >= 2.961*BAREM1*BAREM1;
          J = J-0.001*BAREM1*BAREM2) {  
      
      double sepdist = 8.5*BAREM1;  // separation distance
  
      while(sepdist>=3.6*BAREM1) {  // puncture shouldn't
                                    // merge...
        setup(sepdist);  // determines parameter values
        checkparam(sepdist);   // checks parameters for
                               // consistency
        newADMmass();       // iterates to find M1,M2, then
                            // solves, writes to file...
        checkparam(sepdist);   // checks parameters for
                               // consistency
    
        sepdist = sepdist - 0.3*BAREM1; //decrement puncture
                                        // positions  
  
      } // end - while sepdist

   } // end - for J
  } // end - for MG
} // end main
\end{verbatim}

The following program, \it{newADMmass}\rm, is the ``workhorse'' program, and contains most of the functions used to solve the Hamiltonian constraint:
\begin{verbatim}
// newADMmass.cxx - portion of code which calculates correction
// to conformal factor and the corresponding ADM mass

#include <assert.h>
#include <fstream.h>
#include <stdio.h>   // C header file for sprintf
#include <iostream.h>
#include <iomanip.h>
#include <math.h>
#include "grid.h"
#include "adapt.h"


void newADMmass(void) {

  extern double DEL0,DEL1,DEL2,DEL3;
  extern double GAL0,GAL1,GAL2,GAL3;
  extern int NMG, MG, PMAX;
  extern double BAREM1, BAREM2, PI;
  extern double J;
  extern double Y1,Y2,Z1,Z2;
  extern double PX1,PX2,PZ1,PZ2;
  extern double SX1,SX2,SY1,SY2,SZ1,SZ2;
  
  using namespace Parameters;
  
  // note that i = 0 corresponds to the finest grid

  AGrid3D v[NMG+1];	// v[x] goes from 0 to x-1
  AGrid3D vmg[MG];	// v denotes a previous best solution,
                        // except at smallest level
  AGrid3D r[NMG];	// r denotes residual
  AGrid3D rmg[MG];
  AGrid3D s[NMG+1];	// s denotes source
  AGrid3D smg[MG];
  AGrid3D a[NMG];	// a denotes a dummy array associated
                        // with corrections
  AGrid3D amg[MG];
  AGrid3D e[NMG+1];     // e denotes an error
  AGrid3D emg[MG];
  AGrid3D u[NMG+1];	// u denotes the best solution 
  AGrid3D umg[MG];
  AGrid3D alpha[NMG+1];  // alpha and beta are geometrical
                         // quantities
  AGrid3D alphamg[MG];
  AGrid3D beta[NMG+1];
  AGrid3D betamg[MG];
  AGrid3D psi[NMG+1];	// psi = 1 +  u, u = var code solves
                        // for, psi = correction to conformal
                        // factor
  AGrid3D uold[1];	// dummy arrays for error analysis
  AGrid3D unew[1];
  
  double maxsize = 1.0*pow(2,NMG)*SCALE;   // ensures largest
                                           // grid matches with
                                           // smallest grid
  
  
  for(int i=0;i<MG;i++) {               // allocate memory for
                                        // mg arrays
    int n=(int)(pow(2,MG-i)+1);		// must be defined for
                                        // i<MG	
    vmg[i].reset(n,maxsize);                
    rmg[i].reset(n,maxsize);                
    smg[i].reset(n,maxsize);                
    amg[i].reset(n,maxsize);
    umg[i].reset(n,maxsize);
    emg[i].reset(n,maxsize);
    alphamg[i].reset(n,maxsize);
    betamg[i].reset(n,maxsize);
  }

  for(int i=0;i<NMG;i++) {
    int n=(int)(pow(2,MG)+3);	// adds 2 grid points to 
                                // non-mg arrays
    v[i].reset(n,pow(2,i)*LEN);	// must be defined for i<NMG
    u[i].reset(n,pow(2,i)*LEN);
    r[i].reset(n,pow(2,i)*LEN);
    s[i].reset(n,pow(2,i)*LEN);
    a[i].reset(n,pow(2,i)*LEN);
    e[i].reset(n,pow(2,i)*LEN);
    alpha[i].reset(n,pow(2,i)*LEN);
    beta[i].reset(n,pow(2,i)*LEN);
    psi[i].reset(n,pow(2,i)*LEN); 
    if(i==0) {
      uold[0].reset(n,LEN);
      unew[0].reset(n,LEN);
    }
  }
  
  int n=(int)(pow(2,MG)+1);	
  v[NMG].reset(n,maxsize);	// v[NMG] must be handled
                                // carefully
  u[NMG].reset(n,maxsize);	// u[NMG] must be handled
                                // carefully
  e[NMG].reset(n,maxsize);      // due to different # of grid
                                // points
  s[NMG].reset(n,maxsize);	// from one level to the next
  alpha[NMG].reset(n,maxsize);	
  beta[NMG].reset(n,maxsize);
  psi[NMG].reset(n,maxsize);
 
  double uoldvalue, unewvalue;  // old and new values of u at
                                // puncture location
  double uerror = 1.0;          // initilize to arbit. value
  int counter = 0;              // initilize to zero
  
  // initialize M1
  M1 = 2.0*X1*( -1.0 + sqrt( 1.0 + 1.0/X1*BAREM1 ) );  
  // initialize M2
  M2 = 2.0*X1*( -1.0 + sqrt( 1.0 + 1.0/X1*BAREM2 ) );  
  
  while(uerror>=10E-6) {

    for(int i=0;i<=NMG;i++) {
      newbeta(alpha[i],beta[i]);      // use for 2 BH's, boosted
                                      // to each other, with
                                      // spin, etc...
    }

    fillsource(s[0],v[0],alpha[0],beta[0]);    // source info
                                               // on finest grid

    for(int p=0;p<PMAX;p++) {                // p is the # of
                                             // total sweeps
      cout << "Sweep #" << p+1 << endl;
      int i;

      for(i=0; i<NMG; i++) {
        if(i!=0) {		
          zerointerior(u[i]);	// this takes restricted
                                // interior solution
	  add(u[i],v[i],u[i]);	// and adds it to the exterior
                                // solution
	  copy(v[i],u[i]);	// copies solution into v
        }      

        /****going to larger levels************************/
        // relax on interior n-3 grid pts only
        newrelax(u[i],v[i],s[i],alpha[i],beta[i]);  
        newresidual(r[i],u[i],v[i],s[i],alpha[i],beta[i]);
        // find residual everywhere 
        symadaptrstrct(u[i+1],u[i]);   // restrict solution to
                                       // larger level      
        s[i+1].setZero();              // zero out source on
                                       // larger level
        // must take into account beta*(1+alpha)^-6 term 
        // from v...                                     
        fillsource(s[i+1],v[i+1],alpha[i+1],beta[i+1]);  
        psi[i+1].setZero();            // zero out temp array
        newresidual(psi[i+1],u[i+1],v[i+1],s[i+1],
                      alpha[i+1],beta[i+1]);  // find source 
        s[i+1].setZero();          // ensure source is zeroed 
        copy(s[i+1],psi[i+1]);     // copy source from temp
                                   // array in to s[i+1]
        psi[i+1].setZero();        // zero out temp array   
        symresrstrct(s[i+1],r[i]);   // fill interior dummy
                                     // source with restricted
                                     // residual    
        symadaptrstrct(v[i+1],u[i]);
        /********end of larger levels*******************/    
    
      }

      copy(smg[0],s[NMG]);         // copy info into mg arrays
      copy(alphamg[0],alpha[NMG]);
      copy(betamg[0],beta[NMG]);
      copy(umg[0],u[NMG]);
      copy(v[NMG],u[NMG]);
      copy(vmg[0],v[NMG]);

      /******************/
      // Now do usual multigrid on largest,finest using Robin
      // conditions   
      /******************/       
      for(int k=0;k<16;k++) { 	
      for(i=0; i<MG;i++) {
        if(i!=0) copy(vmg[i],umg[i]);   // copy best solution
                                        // into v
        if(i<MG-1) {
          newrelax(umg[i],vmg[i],smg[i],alphamg[i],betamg[i]);
          newresidual(rmg[i],umg[i],vmg[i],smg[i],
                        alphamg[i],betamg[i]);
          symrstrct(umg[i+1],umg[i]);   // rstrct solution to
                                        // larger, coarser
          symrstrct(smg[i+1],rmg[i]);   // rstct residual to
                                        // largest, coarser
          symrstrct(alphamg[i+1],alphamg[i]);
          symrstrct(betamg[i+1],betamg[i]);	
        }
        else {
          copy(vmg[i],umg[i]);      // copy best solution to v
          for(int j=1; j<8; j++) {
            newrelax(umg[i],vmg[i],smg[i],alphamg[i],betamg[i]);
            newresidual(rmg[i],umg[i],vmg[i],smg[i],
                               alphamg[i],betamg[i]);
	  }
        } 
      }
    
      // reconstruct the solution on finer, largest grid
      for (i=MG-1; i>0; i--) {
        subtract(emg[i],umg[i],vmg[i]);
        interp(amg[i-1],emg[i]);      // interp to finer, save
                                      // to amg
        update(umg[i-1],amg[i-1]);    // update solution
        newrelax(umg[i-1],vmg[i-1],smg[i-1],
                       alphamg[i-1],betamg[i-1]);
        newresidual(rmg[i-1],umg[i-1],vmg[i-1],smg[i-1],
                       alphamg[i-1],betamg[i-1]);
      }    
    } //end for-k

    /******************/  
    // At this point umg[0] contains the best soln. on largest,
    // finest grid
    /******************/  

    /******* method tacks on best solution to exterior,
    /******* yields same results as correction method
    copy(u[NMG],umg[0]);		// copy solution
    subtract(e[NMG],u[NMG],v[NMG]);     // determine error on
                                        // largest grid  
    newresidual(r[NMG],u[NMG],v[NMG],s[NMG],
                         alpha[NMG],beta[NMG]);
  
    for(i=NMG;i>0;i--) {
      if(p!=PMAX-1) zeroexterror(v[i]);
      adaptinterp(a[i-1],e[i]);	// adapt-interpolate correction
                                // to smaller grid
      update(u[i-1],a[i-1]);	// update correction on smaller
                                // grid
      psi[i-1].setZero();       // zero dummy array
      adaptinterp(psi[i-1],u[i]); // adaptinterp large solution
                                  // into small array
      zeroexterior(u[i-1]);       // zero the exterior of small
                                  // solution
      zerointeriorsoln(psi[i-1]);   // zero the interior of
                                    // large solution
      update(u[i-1],psi[i-1]);    // update small solution with
                                  // exterior solution
      psi[i-1].setZero();
      newrelax(u[i-1],v[i-1],s[i-1],alpha[i-1],beta[i-1]);    
      newresidual(r[i-1],u[i-1],v[i-1],s[i-1],
                       alpha[i-1],beta[i-1]);
      subtract(e[i-1],u[i-1],v[i-1]);   // determine the error
    }
    /***********end of exterior solution method**************/

    /******calculate errors************/  
    for(int i= 0;i>=0;i--) {  
      cout << endl << "Calculating error for interior n-3 points
               on level " << i<< " ..." << endl << endl;
      if(p==0) {
        cout << "residual error = "<< norm(r[i]) << " for
                 v-cycle " << p+1 << endl << endl;
        copy(uold[i],u[i]);
      }
      else {
        cout << "residual error = "<< norm(r[i]) << " for
                 v-cycle " << p+1 << endl;
        subtract(unew[i],uold[i],u[i]);
        cout << "truncation error = " << norm(unew[i]) 
             << " for v-cycle " << p+1 << endl << endl;
        copy(uold[i],u[i]);  
      }
    } // end for i
     
    /*******end of errors***********/  
    
    }
  
  if(counter==0) {
    uoldvalue = getu(u[0]);  // value of u at puncture 
    counter++;
  }
  else {
    unewvalue = getu(u[0]);   // value of u at puncture
    if(uoldvalue>unewvalue) uerror = uoldvalue - unewvalue;  
    // the above ensures error is > 0
    else uerror = unewvalue - uoldvalue;
    uoldvalue = unewvalue;   // stores new value of u at
                             // puncture
    counter++;
  }
  
  double D = 2.0*X1;
  M1 = D*( -(1.0+uoldvalue) 
      + sqrt( (1.0 + uoldvalue)*(1.0 + uoldvalue) 
      + 2.0*BAREM1/D ) );  // calculates new mass
  M2 = M1;
  
  cout << "M1 = M2 = " << setprecision(12) << M1 
       << "     uerror = " << uerror <<  "    counter = " 
       << counter << endl;
  
  } // end while - uerror
    
    for(int i=0;i<=NMG;i++) {     // determines the conformal
                                  // factor on the surfaces
      fillpsi(psi[i],alpha[i],u[i]);
    }
 
    for(int i=0;i<=NMG;i++) {
      cout << "Calculating Mass for Level " << i << " << endl;
      n = psi[i].sideX();
      double h = psi[i].hX();
      if(psi[i].LenX()<maxsize) {
        for(int k = n/2;k<n-2;k++) {
          cout << k << "  " << h*(k-1) << "  " 
          << setprecision(16) << M1+M2+moment(psi[i],0,0,k)  
          << endl;
        }
      }
      else{
        for(int k = n/2+1;k<n;k++) {
          cout << "J = " << J << "     d = " << 2.0*X1 
               << "     ADM Mass =  " << setprecision(16) 
               << M1+M2+moment(psi[i],0,0,n-1)  << endl;
	  cout << k << "  " << h*(k-1) << "  " 
          << setprecision(16) << M1+M2+moment(psi[i],0,0,k)  
          << endl;
        }
      }  end else - i
    }   end for - i
    
    char filename[50];
    sprintf(filename,"MG_%1d-J_%1f.dat",MG,J); 
    save(filename,psi[NMG],u[0]);
}

\end{verbatim}

The program \it{setup }\rm determines all of the other system parameters, such as the hole separation distance:
\begin{verbatim}
//setup.cxx - changes parameters of system

#include <assert.h>
#include <fstream.h>
#include <iostream.h>
#include <iomanip.h>
#include <math.h>
#include "grid.h"
#include "adapt.h"

namespace Parameters {   
  
  double M1;      
  double M2;
  double X1;      
  double X2;      
  double SCALE;   
  double LEN;     
  double PY1;
  double PY2;     

}

void setup(double sepdist) {
  
  using namespace Parameters;
  
  extern int MG;
  extern double J;
  
  X1 = sepdist/2;
  X2 = X1;            // punctures are equal distances away
  SCALE = 2.0*X1;   // SETS SCALE OF ARRAYS
  LEN = 1.0*(pow(2,MG)+2)/(pow(2,MG))*SCALE;  // SETS SIZE OF
                                              // NMG ARRAYS
  PY1 = 1.0*J/(2.0*X1);
  PY2 = -1.0*J/(2.0*X1);  // J = j1+j2=d1*P1+d2*P2 = 2*d*P...
  cout << endl << "MG = " << MG << "  J = " << J << endl;
  cout << "X1 = " << X1 << "  X2 = " << X2 << endl;
  cout << "SCALE = " << SCALE << "  LEN = " << LEN << endl;
  cout << "P1 = -P2 = " << PY1 << endl << endl;
}

\end{verbatim}

The program \it{newbeta }\rm calculates the geometrical quantities $\alpha$ and $\beta$, used to solve the Hamiltonian constraint:
\begin{verbatim}
// newbetacxx - Fills alpha and beta arrays for binary 
// black hole system
#include <assert.h>
#include <fstream.h>
#include <iostream.h>
#include <iomanip.h>
#include <math.h>
#include "grid.h"
#include "adapt.h"

extern const double Y1;         // Y1 = Y2 = 0
extern const double Y2; 
extern const double Z1;         // Z1 = Z2 = 0
extern const double Z2; 
extern const double PX1; 
extern const double PX2; 
extern const double PZ1;        // PZ1 = PZ2 = 0 
extern const double PZ2; 
extern const double SX1;        // SX1 = SX2 = SY1 = SY2 = 0 
extern const double SX2; 
extern const double SY1;         
extern const double SY2;         
extern const double SZ1; 
extern const double SZ2; 


void newbeta(AGrid3D& alpha, AGrid3D& beta){

  using namespace Parameters;  // looks up parameter values
  
  int i,j,k;
  int n=alpha.sideX();
  double h = alpha.hX(); 
  double x1,y1,z1,r1;
  double x2,y2,z2,r2;
  double A;
//cout << "calculating for binary black holes!!!" << endl;  

  for(i=1;i<=n;i++) {        // calculate for equatorial plane
    for(j=1;j<=n;j++) {
      k = 1;
      
      x1=(h*(i-1)-X1);
      y1=(h*(j-1)-Y1);
      z1=(h*(k-1)-Z1);
      x2=(h*(i-1)+X2);
      y2=(h*(j-1)+Y2);
      z2=(h*(k-1)+Z2);
     
     
      r1=sqrt(x1*x1+y1*y1+z1*z1);
      r2=sqrt(x2*x2+y2*y2+z2*z2);
      
      A = M1*r2+M2*r1;
     
      alpha(i,j,1) = 2.0*r1*r2/(M1*r2+M2*r1);
      
      // this expression for beta is the same as below, but 
      // z1 and z2 = 0 in two important places
      // along with some minor simplification to get rid of
      // diverging terms
  
beta(i,j,1) =  
  pow(A,-7)*(72*pow(r2,7)*r1*(2*pow(y1,2)+pow(r1,2))*pow(PY1,2)
  +(72*pow(r2,2)*pow(r1,2)*(2*x2*x1*pow(r2,2)*pow(r1,2)
  +2*x2*x1*pow(y2,2)*pow(r1,2)+2*x2*x1*pow(r2,2)*pow(y1,2)
  +2*x2*x1*pow(y2,2)*pow(y1,2)+2*pow(y2,3)*pow(y1,3)
  +y2*pow(y1,3)*pow(r2,2)+pow(y1,3)*y2*pow(z2,2)
  +2*z2*z1*pow(y2,2)*pow(y1,2)+2*z2*z1*pow(r2,2)*pow(y1,2)
  +y1*pow(z1,2)*pow(y2,3)-y2*y1*pow(r2,2)*pow(z1,2)
  +2*y2*y1*pow(z2,2)*pow(z1,2)+pow(y2,3)*y1*pow(r1,2)
  +2*y2*y1*pow(r2,2)*pow(r1,2)-y2*y1*pow(z2,2)*pow(r1,2)
  +2*z2*z1*pow(y2,2)*pow(r1,2)+2*z2*z1*pow(r2,2)*pow(r1,2)
  +2*y2*z2*x2*y1*z1*x1)*PY2+288*SZ1*x1*pow(r2,7)*r1
  +288*pow(r2,2)*pow(r1,2)*(-2*x1*pow(y2,2)*pow(r1,2)
  -x1*pow(z2,2)*pow(r1,2)+x1*pow(r2,2)*pow(r1,2)
  -2*x1*pow(y2,2)*pow(y1,2)-x1*pow(y1,2)*pow(z2,2)
  +x1*pow(y1,2)*pow(r2,2)+x2*z2*z1*pow(r1,2)
  +x2*z2*z1*pow(y1,2)+y2*x2*y1*pow(r1,2)
  +y2*x2*y1*pow(z1,2)+2*y2*x2*pow(y1,3)-y2*z2*y1*z1*x1)*SZ2
  +72*pow(r2,2)*r1*(-3*PX2*r1*pow(y1,3)*x2*pow(r2,2)
  +4*PX2*pow(r1,3)*y2*x1*pow(r2,2)-2*PX2*pow(r1,3)*pow(y2,3)*x1
  -3*PX2*r1*y1*x2*pow(r2,2)*pow(z1,2)
  +4*PX2*r1*y2*x1*pow(r2,2)*pow(y1,2)+4*PX1*pow(r2,5)*x1*y1
  -2*PX2*r1*z2*y1*z1*x1*pow(y2,2)+PX2*r1*pow(y1,3)*x2*pow(z2,2)
  +2*PX2*pow(r1,3)*y2*z2*x2*z1+2*PX2*r1*y2*z2*x2*z1*pow(y1,2)
  +PX2*r1*y1*pow(z1,2)*x2*pow(y2,2)
  -2*PX2*r1*pow(y2,3)*x1*pow(y1,2)
  -2*PX2*pow(r1,3)*y2*x1*pow(z2,2)-2*PX2*r1*z1*x1*y1*pow(z2,3)
  -PX2*pow(r1,3)*y1*x2*pow(z2,2)+PX2*pow(r1,3)*y1*x2*pow(y2,2)
  -2*PX2*r1*x1*pow(y1,2)*y2*pow(z2,2)
  +4*PX2*r1*z2*y1*z1*x1*pow(r2,2)
  +2*PX2*r1*y1*x2*pow(z2,2)*pow(z1,2)
  +2*PX2*r1*pow(y1,3)*x2*pow(y2,2)))*PY1
  +72*pow(r1,7)*r2*(2*pow(y2,2)+pow(r2,2))*pow(PY2,2)
  +(288*pow(r2,2)*pow(r1,2)*(2*pow(y2,3)*x1*y1
  +y2*x1*y1*pow(r2,2)
  +y2*x1*y1*pow(z2,2)+x1*z1*z2*pow(r2,2)+x1*z1*z2*pow(y2,2)
  -2*x2*pow(y2,2)*pow(y1,2)-2*x2*pow(r2,2)*pow(y1,2)
  -x2*pow(z1,2)*pow(r2,2)-pow(z1,2)*x2*pow(y2,2)-y2*z2*x2*y1*z1
  +x2*pow(r2,2)*pow(r1,2)+x2*pow(y2,2)*pow(r1,2))*SZ1
  +288*SZ2*x2*pow(r1,7)*r2-72*r2*pow(r1,2)*(
  -PX1*pow(r2,3)*y2*x1*pow(y1,2)+2*PX1*r2*pow(y1,3)*x2*pow(y2,2)
  +2*PX1*r2*y2*z2*x2*z1*pow(y1,2)-2*PX1*r2*z2*y1*z1*x1*pow(y2,2)
  -2*PX1*r2*y2*x1*pow(z2,2)*pow(z1,2)
  -4*PX1*r2*pow(r1,2)*y1*x2*pow(y2,2)
  -4*PX1*r2*pow(r1,2)*y2*z2*x2*z1
  +3*PX1*r2*pow(r1,2)*pow(y2,3)*x1
  +3*PX1*r2*pow(r1,2)*y2*x1*pow(z2,2)
  -PX1*r2*x1*pow(z1,2)*pow(y2,3)-2*PX1*r2*pow(y2,3)*x1*pow(y1,2)
  -4*PX2*pow(r1,5)*y2*x2+PX1*pow(r2,3)*y2*x1*pow(z1,2)
  +2*PX1*pow(r2,3)*pow(y1,3)*x2+2*PX1*pow(r2,3)*y1*x2*pow(z1,2)
  -PX1*r2*x1*pow(y1,2)*y2*pow(z2,2)-2*PX1*pow(r2,3)*z2*y1*z1*x1
  -4*PX1*pow(r2,3)*pow(r1,2)*y1*x2+2*PX1*r2*y2*z2*x2*pow(z1,3)
  +2*PX1*r2*y1*pow(z1,2)*x2*pow(y2,2)))*PY2
  +288*pow(r2,7)*r1*pow(SZ1,2)
  +(576*pow(r2,2)*pow(r1,2)*(4*pow(y2,2)*pow(y1,2)
  +2*pow(y1,2)*pow(z2,2)-2*pow(y1,2)*pow(r2,2)
  +2*pow(y2,2)*pow(z1,2)+pow(z1,2)*pow(z2,2)-pow(z1,2)*pow(r2,2)
  -2*pow(y2,2)*pow(r1,2)-pow(z2,2)*pow(r1,2)+pow(r2,2)*pow(r1,2)
  +4*y2*x2*x1*y1+z2*x2*x1*z1+y2*z2*y1*z1)*SZ2
  -288*pow(r2,2)*r1*(-2*PX2*r1*y2*pow(y1,2)*pow(z2,2)
  -PX2*r1*x2*y1*x1*pow(z2,2)+4*PX2*pow(r2,2)*r1*y2*pow(y1,2)
  -PX2*r1*pow(z1,2)*pow(y2,3)+pow(y2,3)*pow(r1,3)*PX2
  +2*PX2*pow(r2,2)*r1*y2*pow(z1,2)-2*PX2*pow(r2,2)*pow(r1,3)*y2
  -PX2*r1*x1*z1*y2*z2*x2-2*PX2*r1*pow(y1,2)*pow(y2,3)
  +PX1*pow(r2,5)*y1-PX2*r1*y2*pow(z1,2)*pow(z2,2)
  +2*PX2*pow(r2,2)*r1*y1*z1*z2-PX2*r1*y1*z1*z2*pow(y2,2)
  -PX2*r1*y1*z1*pow(z2,3)-2*PX2*r1*x2*y1*x1*pow(y2,2)
  +3*PX2*pow(r2,2)*r1*x2*y1*x1+pow(z2,2)*y2*pow(r1,3)*PX2))*SZ1
  +288*pow(r1,7)*r2*pow(SZ2,2)-288*r2*pow(r1,2)*(
  -2*PX1*r2*y2*x2*x1*pow(y1,2)-2*PX1*pow(r2,3)*pow(r1,2)*y1
  -2*PX1*r2*pow(y1,3)*pow(y2,2)-PX1*r2*y2*x2*x1*pow(z1,2)
  -PX1*r2*y2*z2*z1*pow(y1,2)+PX1*pow(r2,3)*y1*pow(z1,2)
  +2*PX1*r2*pow(r1,2)*z1*z2*y2-PX1*r2*pow(y1,3)*pow(z2,2)
  +PX2*pow(r1,5)*y2+3*PX1*r2*pow(r1,2)*y2*x2*x1
  -PX1*r2*y2*z2*pow(z1,3)-PX1*r2*x2*z2*y1*z1*x1
  +PX1*pow(r2,3)*pow(y1,3)+4*PX1*r2*pow(r1,2)*y1*pow(y2,2)
  +2*PX1*r2*pow(r1,2)*y1*pow(z2,2)
  -2*PX1*r2*y1*pow(y2,2)*pow(z1,2)
  -PX1*r2*y1*pow(z2,2)*pow(z1,2))*SZ2
  +72*r2*r1*(-4*PX1*pow(r2,3)*r1*PX2*z2*z1*pow(y1,2)
  -4*PX1*pow(r2,3)*r1*PX2*y2*y1*pow(z1,2)
  -3*PX1*pow(r2,3)*r1*PX2*x2*x1*pow(y1,2)
  -4*PX1*r2*pow(r1,3)*PX2*y2*y1*pow(z2,2)
  +2*PX1*r2*r1*PX2*pow(z2,3)*pow(z1,3)
  -3*PX1*r2*pow(r1,3)*PX2*x2*x1*pow(z2,2)
  -3*PX1*r2*pow(r1,3)*PX2*x2*x1*pow(y2,2)
  -2*pow(PX1,2)*pow(r2,6)*pow(z1,2)
  +PX1*r2*r1*PX2*x1*pow(z1,2)*x2*pow(y2,2)
  +2*PX1*r2*r1*PX2*x2*x1*pow(z2,2)*pow(z1,2)
  +2*PX1*r2*r1*PX2*pow(y1,3)*y2*pow(z2,2)
  +8*PX1*pow(r2,3)*pow(r1,3)*PX2*y2*y1
  -4*PX1*pow(r2,3)*r1*PX2*z2*pow(z1,3)
  -4*PX1*pow(r2,3)*r1*PX2*y2*pow(y1,3)
  +8*PX1*pow(r2,3)*pow(r1,3)*PX2*z2*z1
  +PX1*r2*r1*PX2*x1*pow(y1,2)*x2*pow(z2,2)
  +2*PX1*r2*r1*PX2*pow(z1,3)*z2*pow(y2,2)
  -4*PX1*r2*pow(r1,3)*PX2*z2*z1*pow(y2,2)
  +2*PX1*r2*r1*PX2*x2*x1*pow(y2,2)*pow(y1,2)
  +2*PX1*r2*r1*PX2*y1*pow(z1,2)*pow(y2,3)
  +2*PX1*r2*r1*PX2*z1*pow(y1,2)*pow(z2,3)
  -3*PX1*pow(r2,3)*r1*PX2*x2*x1*pow(z1,2)
  +2*PX1*r2*r1*PX2*y2*y1*pow(z2,2)*pow(z1,2)
  -4*PX1*r2*pow(r1,3)*PX2*pow(z2,3)*z1
  +6*PX1*pow(r2,3)*pow(r1,3)*PX2*x2*x1
  -2*pow(PX1,2)*pow(r2,6)*pow(y1,2)
  +2*PX1*r2*r1*PX2*z2*z1*pow(y2,2)*pow(y1,2)
  +2*PX1*r2*r1*PX2*pow(y2,3)*pow(y1,3)
  +2*PX1*r2*r1*PX2*y2*z2*x2*y1*z1*x1
  -4*PX1*r2*pow(r1,3)*PX2*pow(y2,3)*y1
  +3*pow(PX2,2)*pow(r1,6)*pow(r2,2)
  +3*pow(PX1,2)*pow(r2,6)*pow(r1,2)
  -2*pow(PX2,2)*pow(r1,6)*pow(z2,2)
  -2*pow(PX2,2)*pow(r1,6)*pow(y2,2))); 
    }
  }

  
  for(i=1;i<=n;i++) {        // calculate for everywhere else
    for(j=1;j<=n;j++) {
      for(k=2;k<=n;k++) {
      
      x1=(h*(i-1)-X1);
      y1=(h*(j-1)-Y1);
      z1=(h*(k-1)-Z1);
      x2=(h*(i-1)+X2);
      y2=(h*(j-1)+Y2);
      z2=(h*(k-1)+Z2);
     
     
      r1=sqrt(x1*x1+y1*y1+z1*z1);
      r2=sqrt(x2*x2+y2*y2+z2*z2);
      
      A = M1*r2+M2*r1;
     
      alpha(i,j,k) = 2.0*r1*r2/(M1*r2+M2*r1);
      
beta(i,j,k) =
  pow(A,-7)*(72*pow(r2,7)*r1*(2*pow(y1,2)+pow(r1,2))*pow(PY1,2)
  +(72*pow(r2,2)*pow(r1,2)*(2*x2*x1*pow(r2,2)*pow(r1,2)
  +2*x2*x1*pow(y2,2)*pow(r1,2)+2*x2*x1*pow(r2,2)*pow(y1,2)
  +2*x2*x1*pow(y2,2)*pow(y1,2)+2*pow(y2,3)*pow(y1,3)
  +y2*pow(y1,3)*pow(r2,2)+pow(y1,3)*y2*pow(z2,2)
  +2*z2*z1*pow(y2,2)*pow(y1,2)+2*z2*z1*pow(r2,2)*pow(y1,2)
  +y1*pow(z1,2)*pow(y2,3)-y2*y1*pow(r2,2)*pow(z1,2)
  +2*y2*y1*pow(z2,2)*pow(z1,2)+pow(y2,3)*y1*pow(r1,2)
  +2*y2*y1*pow(r2,2)*pow(r1,2)-y2*y1*pow(z2,2)*pow(r1,2)
  +2*z2*z1*pow(y2,2)*pow(r1,2)+2*z2*z1*pow(r2,2)*pow(r1,2)
  +2*y2*z2*x2*y1*z1*x1)*PY2+288*SZ1*x1*pow(r2,7)*r1
  +288*pow(r2,2)*pow(r1,2)*(-2*x1*pow(y2,2)*pow(r1,2)
  -x1*pow(z2,2)*pow(r1,2)+x1*pow(r2,2)*pow(r1,2)
  -2*x1*pow(y2,2)*pow(y1,2)-x1*pow(y1,2)*pow(z2,2)
  +x1*pow(y1,2)*pow(r2,2)+x2*z2*z1*pow(r1,2)+x2*z2*z1*pow(y1,2)
  +y2*x2*y1*pow(r1,2)+y2*x2*y1*pow(z1,2)+2*y2*x2*pow(y1,3)
  -y2*z2*y1*z1*x1)*SZ2+72*pow(r2,2)*r1*(
  -3*PX2*r1*pow(y1,3)*x2*pow(r2,2)
  +4*PX2*pow(r1,3)*y2*x1*pow(r2,2)-2*PX2*pow(r1,3)*pow(y2,3)*x1
  -3*PX2*r1*y1*x2*pow(r2,2)*pow(z1,2)
  +4*PX2*r1*y2*x1*pow(r2,2)*pow(y1,2)+4*PX1*pow(r2,5)*x1*y1
  -2*PX2*r1*z2*y1*z1*x1*pow(y2,2)+PX2*r1*pow(y1,3)*x2*pow(z2,2)
  +2*PX2*pow(r1,3)*y2*z2*x2*z1+2*PX2*r1*y2*z2*x2*z1*pow(y1,2)
  +PX2*r1*y1*pow(z1,2)*x2*pow(y2,2)
  -2*PX2*r1*pow(y2,3)*x1*pow(y1,2)
  -2*PX2*pow(r1,3)*y2*x1*pow(z2,2)-2*PX2*r1*z1*x1*y1*pow(z2,3)
  -PX2*pow(r1,3)*y1*x2*pow(z2,2)+PX2*pow(r1,3)*y1*x2*pow(y2,2)
  -2*PX2*r1*x1*pow(y1,2)*y2*pow(z2,2)
  +4*PX2*r1*z2*y1*z1*x1*pow(r2,2)
  +2*PX2*r1*y1*x2*pow(z2,2)*pow(z1,2)
  +2*PX2*r1*pow(y1,3)*x2*pow(y2,2)))*PY1
  +72*pow(r1,7)*r2*(2*pow(y2,2)+pow(r2,2))*pow(PY2,2)
  +(288*pow(r2,2)*pow(r1,2)*(2*pow(y2,3)*x1*y1
  +y2*x1*y1*pow(r2,2)+y2*x1*y1*pow(z2,2)+x1*z1*z2*pow(r2,2)
  +x1*z1*z2*pow(y2,2)-2*x2*pow(y2,2)*pow(y1,2)
  -2*x2*pow(r2,2)*pow(y1,2)-x2*pow(z1,2)*pow(r2,2)
  -pow(z1,2)*x2*pow(y2,2)-y2*z2*x2*y1*z1+x2*pow(r2,2)*pow(r1,2)
  +x2*pow(y2,2)*pow(r1,2))*SZ1+288*SZ2*x2*pow(r1,7)*r2
  -72*r2*pow(r1,2)*(-PX1*pow(r2,3)*y2*x1*pow(y1,2)
  +2*PX1*r2*pow(y1,3)*x2*pow(y2,2)
  +2*PX1*r2*y2*z2*x2*z1*pow(y1,2)
  -2*PX1*r2*z2*y1*z1*x1*pow(y2,2)
  -2*PX1*r2*y2*x1*pow(z2,2)*pow(z1,2)
  -4*PX1*r2*pow(r1,2)*y1*x2*pow(y2,2)
  -4*PX1*r2*pow(r1,2)*y2*z2*x2*z1
  +3*PX1*r2*pow(r1,2)*pow(y2,3)*x1
  +3*PX1*r2*pow(r1,2)*y2*x1*pow(z2,2)
  -PX1*r2*x1*pow(z1,2)*pow(y2,3)
  -2*PX1*r2*pow(y2,3)*x1*pow(y1,2)-4*PX2*pow(r1,5)*y2*x2
  +PX1*pow(r2,3)*y2*x1*pow(z1,2)+2*PX1*pow(r2,3)*pow(y1,3)*x2
  +2*PX1*pow(r2,3)*y1*x2*pow(z1,2)
  -PX1*r2*x1*pow(y1,2)*y2*pow(z2,2)
  -2*PX1*pow(r2,3)*z2*y1*z1*x1-4*PX1*pow(r2,3)*pow(r1,2)*y1*x2
  +2*PX1*r2*y2*z2*x2*pow(z1,3)
  +2*PX1*r2*y1*pow(z1,2)*x2*pow(y2,2)))*PY2
  +288*pow(r2,7)*(r1-z1)*(r1+z1)/r1*pow(SZ1,2)
  +(576*pow(r2,2)*pow(r1,2)*(4*pow(y2,2)*pow(y1,2)
  +2*pow(y1,2)*pow(z2,2)-2*pow(y1,2)*pow(r2,2)
  +2*pow(y2,2)*pow(z1,2)+pow(z1,2)*pow(z2,2)
  -pow(z1,2)*pow(r2,2)-2*pow(y2,2)*pow(r1,2)
  -pow(z2,2)*pow(r1,2)+pow(r2,2)*pow(r1,2)+4*y2*x2*x1*y1
  +z2*x2*x1*z1+y2*z2*y1*z1)*SZ2
  -288*pow(r2,2)*r1*(-2*PX2*r1*y2*pow(y1,2)*pow(z2,2)
  -PX2*r1*x2*y1*x1*pow(z2,2)+4*PX2*pow(r2,2)*r1*y2*pow(y1,2)
  -PX2*r1*pow(z1,2)*pow(y2,3)+pow(y2,3)*pow(r1,3)*PX2
  +2*PX2*pow(r2,2)*r1*y2*pow(z1,2)
  -2*PX2*pow(r2,2)*pow(r1,3)*y2-PX2*r1*x1*z1*y2*z2*x2
  -2*PX2*r1*pow(y1,2)*pow(y2,3)+PX1*pow(r2,5)*y1
  -PX2*r1*y2*pow(z1,2)*pow(z2,2)+2*PX2*pow(r2,2)*r1*y1*z1*z2
  -PX2*r1*y1*z1*z2*pow(y2,2)-PX2*r1*y1*z1*pow(z2,3)
  -2*PX2*r1*x2*y1*x1*pow(y2,2)+3*PX2*pow(r2,2)*r1*x2*y1*x1
  +pow(z2,2)*y2*pow(r1,3)*PX2))*SZ1
  +288*pow(r1,7)*(r2-z2)*(r2+z2)/r2*pow(SZ2,2)
  -288*r2*pow(r1,2)*(-2*PX1*r2*y2*x2*x1*pow(y1,2)
  -2*PX1*pow(r2,3)*pow(r1,2)*y1-2*PX1*r2*pow(y1,3)*pow(y2,2)
  -PX1*r2*y2*x2*x1*pow(z1,2)-PX1*r2*y2*z2*z1*pow(y1,2)
  +PX1*pow(r2,3)*y1*pow(z1,2)+2*PX1*r2*pow(r1,2)*z1*z2*y2
  -PX1*r2*pow(y1,3)*pow(z2,2)+PX2*pow(r1,5)*y2
  +3*PX1*r2*pow(r1,2)*y2*x2*x1-PX1*r2*y2*z2*pow(z1,3)
  -PX1*r2*x2*z2*y1*z1*x1+PX1*pow(r2,3)*pow(y1,3)
  +4*PX1*r2*pow(r1,2)*y1*pow(y2,2)
  +2*PX1*r2*pow(r1,2)*y1*pow(z2,2)
  -2*PX1*r2*y1*pow(y2,2)*pow(z1,2)
  -PX1*r2*y1*pow(z2,2)*pow(z1,2))*SZ2
  +72*r2*r1*(-4*PX1*pow(r2,3)*r1*PX2*z2*z1*pow(y1,2)
  -4*PX1*pow(r2,3)*r1*PX2*y2*y1*pow(z1,2)
  -3*PX1*pow(r2,3)*r1*PX2*x2*x1*pow(y1,2)
  -4*PX1*r2*pow(r1,3)*PX2*y2*y1*pow(z2,2)
  +2*PX1*r2*r1*PX2*pow(z2,3)*pow(z1,3)
  -3*PX1*r2*pow(r1,3)*PX2*x2*x1*pow(z2,2)
  -3*PX1*r2*pow(r1,3)*PX2*x2*x1*pow(y2,2)
  -2*pow(PX1,2)*pow(r2,6)*pow(z1,2)
  +PX1*r2*r1*PX2*x1*pow(z1,2)*x2*pow(y2,2)
  +2*PX1*r2*r1*PX2*x2*x1*pow(z2,2)*pow(z1,2)
  +2*PX1*r2*r1*PX2*pow(y1,3)*y2*pow(z2,2)
  +8*PX1*pow(r2,3)*pow(r1,3)*PX2*y2*y1
  -4*PX1*pow(r2,3)*r1*PX2*z2*pow(z1,3)
  -4*PX1*pow(r2,3)*r1*PX2*y2*pow(y1,3)
  +8*PX1*pow(r2,3)*pow(r1,3)*PX2*z2*z1
  +PX1*r2*r1*PX2*x1*pow(y1,2)*x2*pow(z2,2)
  +2*PX1*r2*r1*PX2*pow(z1,3)*z2*pow(y2,2)
  -4*PX1*r2*pow(r1,3)*PX2*z2*z1*pow(y2,2)
  +2*PX1*r2*r1*PX2*x2*x1*pow(y2,2)*pow(y1,2)
  +2*PX1*r2*r1*PX2*y1*pow(z1,2)*pow(y2,3)
  +2*PX1*r2*r1*PX2*z1*pow(y1,2)*pow(z2,3)
  -3*PX1*pow(r2,3)*r1*PX2*x2*x1*pow(z1,2)
  +2*PX1*r2*r1*PX2*y2*y1*pow(z2,2)*pow(z1,2)
  -4*PX1*r2*pow(r1,3)*PX2*pow(z2,3)*z1
  +6*PX1*pow(r2,3)*pow(r1,3)*PX2*x2*x1
  -2*pow(PX1,2)*pow(r2,6)*pow(y1,2)
  +2*PX1*r2*r1*PX2*z2*z1*pow(y2,2)*pow(y1,2)
  +2*PX1*r2*r1*PX2*pow(y2,3)*pow(y1,3)
  +2*PX1*r2*r1*PX2*y2*z2*x2*y1*z1*x1
  -4*PX1*r2*pow(r1,3)*PX2*pow(y2,3)*y1
  +3*pow(PX2,2)*pow(r1,6)*pow(r2,2)
  +3*pow(PX1,2)*pow(r2,6)*pow(r1,2)
  -2*pow(PX2,2)*pow(r1,6)*pow(z2,2)
  -2*pow(PX2,2)*pow(r1,6)*pow(y2,2)));
      }
    }
  }
}
\end{verbatim}

The program \it{newrelax }\rm is the main relaxation scheme for the nonlinear Hamiltonian constraint. The code only relaxes on the interior points of adaptive levels, except the largest adaptive and all multigrid levels, in which the Robin boundary condition is employed.
\begin{verbatim}
// newrelax - galerkin (with symmetry) relaxation scheme for
// nlamg code
// using galerkin operator on *ALL* levels

// makes no assumption of solution u being small

#include <assert.h>
#include <fstream.h>
#include <iostream.h>
#include <iomanip.h>
#include <math.h>

#include "grid.h"
#include "adapt.h"

#define SHOW(a)   " "<<#a<<" = "<<a<<" " 

void newrelax(AGrid3D& u, AGrid3D& v, AGrid3D& s,
                  AGrid3D& alpha, AGrid3D& beta)
{

  extern double DEL0, DEL1, DEL2, DEL3;
  extern double GAL0, GAL1, GAL2, GAL3;
  extern int MG;
  extern double LEN;
   
  int p;
  int n = u.sideX();
  double h=u.hX();
  double D = DEL0+2.0/(n-1);
  double drdu,ures,vres,nl;
  int pmaxsweep = 4;	// sets # of relaxation sweeps
  
    int N;
    
    if(n==pow(2,MG)+3) {    // for non-mg levels
      N = n-2;
    }
    else {    // for mg levels
      N = n;
    }

    for(p=1;p<=pmaxsweep;p++) {
// note: pass = (i+j+k)%2  if pass=1 -> red; if pass=0 -> black
      for(int pass =1; pass >= 0; pass--) { 
        for(int i=1;i<N;i++) {
          for(int j=1; j<N;j++) {
            for(int k=1;k<N;k++) {
              if((i+j+k)%2!=pass) continue;
              drdu = -GAL0/(h*h)
                +7*alpha(i,j,k)*beta(i,j,k)*pow(1+alpha(i,j,k)
                  +alpha(i,j,k)*u(i,j,k),-8);
              ures = 1.0/(h*h)*(GAL1*(u.sy(i+1,j,k)
                +u.sy(i-1,j,k)+u.sy(i,j+1,k)+u.sy(i,j-1,k)
                  +u.sy(i,j,k+1)+u.sy(i,j,k-1))
                +GAL2*(u.sy(i+1,j+1,k)+u.sy(i+1,j-1,k)
                  +u.sy(i-1,j+1,k)+u.sy(i-1,j-1,k)
                    +u.sy(i,j+1,k+1)+u.sy(i,j-1,k+1)
                      +u.sy(i+1,j,k+1)+u.sy(i-1,j,k+1)
                        +u.sy(i,j+1,k-1)+u.sy(i,j-1,k-1)
                          +u.sy(i+1,j,k-1)+u.sy(i-1,j,k-1))
                +GAL3*(u.sy(i+1,j+1,k+1)+u.sy(i-1,j+1,k+1)
                  +u.sy(i+1,j-1,k+1)+u.sy(i-1,j-1,k+1)
                    +u.sy(i+1,j+1,k-1)+u.sy(i-1,j+1,k-1)
                      +u.sy(i+1,j-1,k-1)+u.sy(i-1,j-1,k-1)));
              vres = 1.0/(h*h)*(GAL0*v.sy(i,j,k)
                +GAL1*(v.sy(i+1,j,k)+v.sy(i-1,j,k)
                  +v.sy(i,j+1,k)+v.sy(i,j-1,k)
                    +v.sy(i,j,k+1)+v.sy(i,j,k-1))
                +GAL2*(v.sy(i+1,j+1,k)+v.sy(i+1,j-1,k)
                  +v.sy(i-1,j+1,k)+v.sy(i-1,j-1,k)
                    +v.sy(i,j+1,k+1)+v.sy(i,j-1,k+1)
                      +v.sy(i+1,j,k+1)+v.sy(i-1,j,k+1)
                        +v.sy(i,j+1,k-1)+v.sy(i,j-1,k-1)
                          +v.sy(i+1,j,k-1)+v.sy(i-1,j,k-1))
               +GAL3*(v.sy(i+1,j+1,k+1)+v.sy(i-1,j+1,k+1)
                 +v.sy(i+1,j-1,k+1)+v.sy(i-1,j-1,k+1)
                   +v.sy(i+1,j+1,k-1)+v.sy(i-1,j+1,k-1)
                     +v.sy(i+1,j-1,k-1)+v.sy(i-1,j-1,k-1)))
                +beta(i,j,k)*pow(1+alpha(i,j,k)
                  +alpha(i,j,k)*v.sy(i,j,k),-7);
	      nl = -beta(i,j,k)*pow(1+alpha(i,j,k)
                     +alpha(i,j,k)*u(i,j,k),-7)
	            *(1.0+7*alpha(i,j,k)*u.sy(i,j,k)/(1
                      +alpha(i,j,k)+alpha(i,j,k)*u(i,j,k)));
              u(i,j,k) = 1.0/drdu*(-s(i,j,k) + ures 
                           - vres - nl);      
            } // end for-k 
          } // end for-j 
        } // end for-i 
	
	// Outermost corner, edges, and faces
	if(n<=pow(2,MG)+1) {
	  // Outermost corner
	  if(pass==1) {
          drdu = D/(h*h)
              +7*alpha(n,n,n)*beta(n,n,n)*pow(1+alpha(n,n,n)
              +alpha(n,n,n)*u(n,n,n),-8);
            ures = 1.0/(h*h)*(2*u.sy(n-1,n,n)
                     +2*u.sy(n,n-1,n)+2*u.sy(n,n,n-1));
            vres = 1.0/(h*h)*(2*v.sy(n-1,n,n)+2*v.sy(n,n-1,n)
              +2*v.sy(n,n,n-1)-D*v.sy(n,n,n))
              +beta(n,n,n)*pow(1+alpha(n,n,n)
                +alpha(n,n,n)*v.sy(n,n,n),-7);
            nl = -beta(n,n,n)*pow(1+alpha(n,n,n)
                +alpha(n,n,n)*u(n,n,n),-7)
                *(1.0+7*alpha(n,n,n)*u.sy(n,n,n)/(1
                  +alpha(n,n,n)+alpha(n,n,n)*u(n,n,n)));
            u(n,n,n) = 1.0/drdu*(-s(n,n,n) + ures - vres - nl);
	    
          } //end if corner
	  
	  // Edges
	  for(int i=1;i<n;i++) {
	    if((n+n+i)%2!=pass) continue;
	    
            double S = (1-1.0*(i-1)/(n-1));  // S for subtract
            double A = (1+1.0*(i-1)/(n-1));  // A for add
	    
            drdu = D/(h*h)
             +7*alpha(i,n,n)*beta(i,n,n)*pow(1+alpha(i,n,n)
               +alpha(i,n,n)*u(i,n,n),-8);
            ures = 1.0/(h*h)*(2*u.sy(i,n-1,n)
              +2*u.sy(i,n,n-1)+S*u.sy(i+1,n,n)+A*u.sy(i-1,n,n));
            vres = 1.0/(h*h)*(2*v.sy(i,n-1,n)
              +2*v.sy(i,n,n-1)+S*v.sy(i+1,n,n)+A*v.sy(i-1,n,n)
               -D*v.sy(i,n,n))
             +beta(i,n,n)*pow(1+alpha(i,n,n)
               +alpha(i,n,n)*v.sy(i,n,n),-7);
            nl = -beta(i,n,n)*pow(1+alpha(i,n,n)
             +alpha(i,n,n)*u(i,n,n),-7)
              *(1.0+7*alpha(i,n,n)*u.sy(i,n,n)/(1
                +alpha(i,n,n)+alpha(i,n,n)*u(i,n,n)));
            u(i,n,n) = 1.0/drdu*(-s(i,n,n) + ures - vres - nl);
	    
            drdu = D/(h*h)
              +7*alpha(n,i,n)*beta(n,i,n)*pow(1+alpha(n,i,n)
                +alpha(n,i,n)*u(n,i,n),-8);
            ures = 1.0/(h*h)*(2*u.sy(n-1,i,n)
              +2*u.sy(n,i,n-1)+S*u.sy(n,i+1,n)+A*u.sy(n,i-1,n));
            vres = 1.0/(h*h)*(2*v.sy(n-1,i,n)
              +2*v.sy(n,i,n-1)+S*v.sy(n,i+1,n)+A*v.sy(n,i-1,n)
                -D*v.sy(n,i,n))
              +beta(n,i,n)*pow(1+alpha(n,i,n)
                +alpha(n,i,n)*v.sy(n,i,n),-7);
            nl = -beta(n,i,n)*pow(1+alpha(n,i,n)
              +alpha(n,i,n)*u(n,i,n),-7)
                *(1.0+7*alpha(n,i,n)*u.sy(n,i,n)/(1
                  +alpha(n,i,n)+alpha(n,i,n)*u(n,i,n)));
            u(n,i,n) = 1.0/drdu*(-s(n,i,n) + ures - vres - nl);
	    
            drdu = D/(h*h)
              +7*alpha(n,n,i)*beta(n,n,i)*pow(1+alpha(n,n,i)
                +alpha(n,n,i)*u(n,n,i),-8);
            ures = 1.0/(h*h)*(2*u.sy(n-1,n,i)
              +2*u.sy(n,n-1,i)+S*u.sy(n,n,i+1)+A*u.sy(n,n,i-1));
            vres = 1.0/(h*h)*(2*v.sy(n-1,n,i)
              +2*v.sy(n,n-1,i)+S*v.sy(n,n,i+1)+A*v.sy(n,n,i-1)
                -D*v.sy(n,n,i))
              +beta(n,n,i)*pow(1+alpha(n,n,i)
                +alpha(n,n,i)*v.sy(n,n,i),-7);
            nl = -beta(n,n,i)*pow(1+alpha(n,n,i)
              +alpha(n,n,i)*u(n,n,i),-7)
                *(1.0+7*alpha(n,n,i)*u.sy(n,n,i)/(1
                  +alpha(n,n,i)+alpha(n,n,i)*u(n,n,i)));
            u(n,n,i) = 1.0/drdu*(-s(n,n,i) + ures - vres - nl);
          } // end for-i edges
	
	  // Faces
          for(int i=1;i<n;i++) {
            for(int j=1;j<n;j++) {
              if((i+j+n)%2!=pass) continue;
              // A for add, S for subtract 
              double Aj = (1+1.0*(j-1)/(n-1));
              double Sj = (1-1.0*(j-1)/(n-1));
              double Ai = (1+1.0*(i-1)/(n-1));
              double Si = (1-1.0*(i-1)/(n-1));
	      
              drdu = D/(h*h)
                +7*alpha(i,j,n)*beta(i,j,n)*pow(1+alpha(i,j,n)
                  +alpha(i,j,n)*u(i,j,n),-8);
              ures = 1.0/(h*h)*(2*u.sy(i,j,n-1)
                +Si*u.sy(i+1,j,n)+Ai*u.sy(i-1,j,n)
                  +Sj*u.sy(i,j+1,n)+Aj*u.sy(i,j-1,n));
              vres = 1.0/(h*h)*(2*v.sy(i,j,n-1)
                +Si*v.sy(i+1,j,n)+Ai*v.sy(i-1,j,n)
                  +Sj*v.sy(i,j+1,n)+Aj*v.sy(i,j-1,n)
                    -D*v.sy(i,j,n))
                      +beta(i,j,n)*pow(1+alpha(i,j,n)
                        +alpha(i,j,n)*v.sy(i,j,n),-7);
              nl = -beta(i,j,n)*pow(1+alpha(i,j,n)
                +alpha(i,j,n)*u(i,j,n),-7)
                  *(1.0+7*alpha(i,j,n)*u.sy(i,j,n)/(1
                    +alpha(i,j,n))+alpha(i,j,n)*u(i,j,n));
              u(i,j,n) = 1.0/drdu*(-s(i,j,n) + ures 
                           - vres - nl);
	      
              drdu = D/(h*h)
                +7*alpha(i,n,j)*beta(i,n,j)*pow(1+alpha(i,n,j)
                  +alpha(i,n,j)*u(i,n,j),-8);
              ures = 1.0/(h*h)*(2*u.sy(i,n-1,j)
                +Si*u.sy(i+1,n,j)+Ai*u.sy(i-1,n,j)
                  +Sj*u.sy(i,n,j+1)+Aj*u.sy(i,n,j-1));
              vres = 1.0/(h*h)*(2*v.sy(i,n-1,j)
               +Si*v.sy(i+1,n,j)+Ai*v.sy(i-1,n,j)
                 +Sj*v.sy(i,n,j+1)+Aj*v.sy(i,n,j-1)
                   -D*v.sy(i,n,j))
                     +beta(i,n,j)*pow(1+alpha(i,n,j)
                       +alpha(i,n,j)*v.sy(i,n,j),-7);
              nl = -beta(i,n,j)*pow(1+alpha(i,n,j)
                +alpha(i,n,j)*u(i,n,j),-7)
                  *(1.0+7*alpha(i,n,j)*u.sy(i,n,j)/(1
                    +alpha(i,n,j)+alpha(i,n,j)*u(i,n,j)));
              u(i,n,j) = 1.0/drdu*(-s(i,n,j) + ures 
                            - vres - nl);
	      
              drdu = D/(h*h)
                +7*alpha(n,j,i)*beta(n,j,i)*pow(1+alpha(n,j,i)
                  +alpha(n,j,i)*u(n,j,i),-8);
              ures = 1.0/(h*h)*(2*u.sy(n-1,j,i)
                +Sj*u.sy(n,j+1,i)+Aj*u.sy(n,j-1,i)
                  +Si*u.sy(n,j,i+1)+Ai*u.sy(n,j,i-1));
              vres = 1.0/(h*h)*(2*v.sy(n-1,j,i)
                +Sj*v.sy(n,j+1,i)+Aj*v.sy(n,j-1,i)
                  +Si*v.sy(n,j,i+1)+Ai*v.sy(n,j,i-1)
                    -D*v.sy(n,j,i))
                      +beta(n,j,i)*pow(1+alpha(n,j,i)
                        +alpha(n,j,i)*v.sy(n,j,i),-7);
              nl = -beta(n,j,i)*pow(1+alpha(n,j,i)
                +alpha(n,j,i)*u(n,j,i),-7)
                  *(1.0+7*alpha(n,j,i)*u.sy(n,j,i)/(1
                    +alpha(n,j,i)+alpha(n,j,i)*u(n,j,i)));
              u(n,j,i) = 1.0/drdu*(-s(n,j,i) + ures 
                            - vres - nl);  
            }  // end for-j faces
          } // end for-i faces
        } // end if-n      
      } // end for-pass
    } // end for-p
} // end newrelax

\end{verbatim}

The program \it{newresidual }\rm calculates the residual, based upon the formula for the nonlinear relaxation scheme above:\\
\begin{verbatim}
// newresidual.cxx - Galerkin (with symmetry) residual for nl
// adaptive mg code
// using galerkin operator on *ALL* levels

// makes no assumption about solution u being small

#include <assert.h>
#include <fstream.h>
#include <iostream.h>
#include <iomanip.h>
#include <math.h>

#include "grid.h"
#include "adapt.h"


// Calculates the residual from r = A(u)-A(v)
void newresidual(AGrid3D& r, AGrid3D& u, AGrid3D& v, AGrid3D& s,
                       AGrid3D& alpha, AGrid3D& beta)
{
  extern  double DEL0, DEL1, DEL2, DEL3;
  extern double GAL0, GAL1, GAL2, GAL3;
  extern int MG;
  extern double LEN;
  
  int n = r.sideX();
  double h = r.hX();
  int i, j, k;
  double D = DEL0+2.0/(n-1);
  
     for(i=1;i<n;i++) {              
        for(j=1;j<n;j++) {
          for(k=1;k<n;k++) {
              
              double drdu = -GAL0/(h*h)
                +7*alpha(i,j,k)*beta(i,j,k)*pow(1+alpha(i,j,k)
                  +alpha(i,j,k)*u(i,j,k),-8);
              double ures = 1.0/(h*h)*(GAL1*(u.sy(i+1,j,k)
                +u.sy(i-1,j,k)+u.sy(i,j+1,k)+u.sy(i,j-1,k)
                  +u.sy(i,j,k+1)+u.sy(i,j,k-1))
                +GAL2*(u.sy(i+1,j+1,k)+u.sy(i+1,j-1,k)
                  +u.sy(i-1,j+1,k)+u.sy(i-1,j-1,k)
                    +u.sy(i,j+1,k+1)+u.sy(i,j-1,k+1)
                      +u.sy(i+1,j,k+1)+u.sy(i-1,j,k+1)
                        +u.sy(i,j+1,k-1)+u.sy(i,j-1,k-1)
                          +u.sy(i+1,j,k-1)+u.sy(i-1,j,k-1))
                +GAL3*(u.sy(i+1,j+1,k+1)+u.sy(i-1,j+1,k+1)
                  +u.sy(i+1,j-1,k+1)+u.sy(i-1,j-1,k+1)
                    +u.sy(i+1,j+1,k-1)+u.sy(i-1,j+1,k-1)
                      +u.sy(i+1,j-1,k-1)+u.sy(i-1,j-1,k-1)));
              double vres = 1.0/(h*h)*(GAL0*v.sy(i,j,k)
                +GAL1*(v.sy(i+1,j,k)+v.sy(i-1,j,k)
                  +v.sy(i,j+1,k)+v.sy(i,j-1,k)
                    +v.sy(i,j,k+1)+v.sy(i,j,k-1))
                +GAL2*(v.sy(i+1,j+1,k)+v.sy(i+1,j-1,k)
                  +v.sy(i-1,j+1,k)+v.sy(i-1,j-1,k)
                    +v.sy(i,j+1,k+1)+v.sy(i,j-1,k+1)
                      +v.sy(i+1,j,k+1)+v.sy(i-1,j,k+1)
                        +v.sy(i,j+1,k-1)+v.sy(i,j-1,k-1)
                          +v.sy(i+1,j,k-1)+v.sy(i-1,j,k-1))
                +GAL3*(v.sy(i+1,j+1,k+1)+v.sy(i-1,j+1,k+1)
                  +v.sy(i+1,j-1,k+1)+v.sy(i-1,j-1,k+1)
                    +v.sy(i+1,j+1,k-1)+v.sy(i-1,j+1,k-1)
                      +v.sy(i+1,j-1,k-1)+v.sy(i-1,j-1,k-1)))
                +beta(i,j,k)*pow(1+alpha(i,j,k)
                  +alpha(i,j,k)*v.sy(i,j,k),-7);
              double nl = -beta(i,j,k)*pow(1+alpha(i,j,k)
                +alpha(i,j,k)*u(i,j,k),-7)
                  *(1.0+7*alpha(i,j,k)*u.sy(i,j,k)/(1
                    +alpha(i,j,k)+alpha(i,j,k)*u(i,j,k)));
              r(i,j,k) = s(i,j,k) - ures + vres 
                           + nl + drdu*u(i,j,k);
          } // end for k 
        } // end for-j
      } // end for-i 
   
    if(n<=pow(2,MG)+1) {  // if on finest mg level, use Robin 
                          // AND Galerkin condition
       double drdu = D/(h*h)
         +7*alpha(n,n,n)*beta(n,n,n)*pow(1+alpha(n,n,n)
           +alpha(n,n,n)*u(n,n,n),-8);
       double ures = 1.0/(h*h)*(2*u.sy(n-1,n,n)
         +2*u.sy(n,n-1,n)+2*u.sy(n,n,n-1));
       double vres = 1.0/(h*h)*(2*v.sy(n-1,n,n)
         +2*v.sy(n,n-1,n)+2*v.sy(n,n,n-1)-D*v.sy(n,n,n))
           +beta(n,n,n)*pow(1+alpha(n,n,n)
             +alpha(n,n,n)*v.sy(n,n,n),-7);
       double nl = -beta(n,n,n)*pow(1+alpha(n,n,n)
         +alpha(n,n,n)*u(n,n,n),-7)
           *(1.0+7*alpha(n,n,n)*u.sy(n,n,n)/(1
             +alpha(n,n,n)+alpha(n,n,n)*u(n,n,n)));
       r(n,n,n) = s(n,n,n) - ures + vres + nl + drdu*u(n,n,n);
       
      // Edges
  
      for(i=1;i<n;i++) {
  
        double S = (1-1.0*(i-1)/(n-1));      // S for subtract
        double A = (1+1.0*(i-1)/(n-1));      // A for add
    
        drdu = D/(h*h)
          +7*alpha(i,n,n)*beta(i,n,n)*pow(1+alpha(i,n,n)
            +alpha(i,n,n)*u(i,n,n),-8);
        ures = 1.0/(h*h)*(2*u.sy(i,n-1,n)
          +2*u.sy(i,n,n-1)+S*u.sy(i+1,n,n)+A*u.sy(i-1,n,n));
        vres = 1.0/(h*h)*(2*v.sy(i,n-1,n)+2*v.sy(i,n,n-1)
          +S*v.sy(i+1,n,n)+A*v.sy(i-1,n,n)-D*v.sy(i,n,n))
            +beta(i,n,n)*pow(1+alpha(i,n,n)
              +alpha(i,n,n)*v.sy(i,n,n),-7);
        nl = -beta(i,n,n)*pow(1+alpha(i,n,n)
          +alpha(i,n,n)*u(i,n,n),-7)
            *(1.0+7*alpha(i,n,n)*u.sy(i,n,n)/(1
              +alpha(i,n,n)+alpha(i,n,n)*u(i,n,n)));
        r(i,n,n) = s(i,n,n) - ures + vres + nl + drdu*u(i,n,n);    
	    
        drdu = D/(h*h)
          +7*alpha(n,i,n)*beta(n,i,n)*pow(1+alpha(n,i,n)
            +alpha(n,i,n)*u(n,i,n),-8);
        ures =1.0/(h*h)*(2*u.sy(n-1,i,n)
          +2*u.sy(n,i,n-1)+S*u.sy(n,i+1,n)+A*u.sy(n,i-1,n));
        vres = 1.0/(h*h)*(2*v.sy(n-1,i,n)+2*v.sy(n,i,n-1)
          +S*v.sy(n,i+1,n)+A*v.sy(n,i-1,n)-D*v.sy(n,i,n))
            +beta(n,i,n)*pow(1+alpha(n,i,n)
              +alpha(n,i,n)*v.sy(n,i,n),-7);
        nl = -beta(n,i,n)*pow(1+alpha(n,i,n)
          +alpha(n,i,n)*u(n,i,n),-7)
            *(1.0+7*alpha(n,i,n)*u.sy(n,i,n)/(1
              +alpha(n,i,n)+alpha(n,i,n)*u(n,i,n)));
        r(n,i,n) = s(n,i,n) - ures + vres + nl + drdu*u(n,i,n);    
	
        drdu = D/(h*h)
          +7*alpha(n,n,i)*beta(n,n,i)*pow(1+alpha(n,n,i)
            +alpha(n,n,i)*u(n,n,i),-8);
        ures = 1.0/(h*h)*(2*u.sy(n-1,n,i)
          +2*u.sy(n,n-1,i)+S*u.sy(n,n,i+1)+A*u.sy(n,n,i-1));
        vres = 1.0/(h*h)*(2*v.sy(n-1,n,i)+2*v.sy(n,n-1,i)
          +S*v.sy(n,n,i+1)+A*v.sy(n,n,i-1)-D*v.sy(n,n,i))
            +beta(n,n,i)*pow(1+alpha(n,n,i)
              +alpha(n,n,i)*v.sy(n,n,i),-7);
        nl = -beta(n,n,i)*pow(1+alpha(n,n,i)
          +alpha(n,n,i)*u(n,n,i),-7)
            *(1.0+7*alpha(n,n,i)*u.sy(n,n,i)/(1
              +alpha(n,n,i)+alpha(n,n,i)*u(n,n,i)));
        r(n,n,i) = s(n,n,i) - ures + vres + nl + drdu*u(n,n,i); 
      }  
      // Faces

      for(i=1;i<n;i++) {
        for(j=1;j<n;j++) {
          // A for add, S for subtract
          double Aj = (1+1.0*(j-1)/(n-1)); 
          double Sj = (1-1.0*(j-1)/(n-1));
          double Ai = (1+1.0*(i-1)/(n-1));
          double Si = (1-1.0*(i-1)/(n-1));
    
          drdu = D/(h*h)
            +7*alpha(i,j,n)*beta(i,j,n)*pow(1+alpha(i,j,n)
              +alpha(i,j,n)*u(i,j,n),-8);
          ures = 1.0/(h*h)*(2*u.sy(i,j,n-1)
            +Si*u.sy(i+1,j,n)+Ai*u.sy(i-1,j,n)
              +Sj*u.sy(i,j+1,n)+Aj*u.sy(i,j-1,n));
          vres = 1.0/(h*h)*(2*v.sy(i,j,n-1)
            +Si*v.sy(i+1,j,n)+Ai*v.sy(i-1,j,n)
              +Sj*v.sy(i,j+1,n)+Aj*v.sy(i,j-1,n)-D*v.sy(i,j,n))
                +beta(i,j,n)*pow(1+alpha(i,j,n)
                  +alpha(i,j,n)*v.sy(i,j,n),-7);
          nl = -beta(i,j,n)*pow(1+alpha(i,j,n)
            +alpha(i,j,n)*u(i,j,n),-7)
              *(1.0+7*alpha(i,j,n)*u.sy(i,j,n)/(1
                +alpha(i,j,n))+alpha(i,j,n)*u(i,j,n));
          r(i,j,n) = s(i,j,n) - ures + vres 
                        + nl + drdu*u(i,j,n); 
	      
          drdu = D/(h*h)
            +7*alpha(i,n,j)*beta(i,n,j)*pow(1+alpha(i,n,j)
              +alpha(i,n,j)*u(i,n,j),-8);
          ures = 1.0/(h*h)*(2*u.sy(i,n-1,j)
            +Si*u.sy(i+1,n,j)+Ai*u.sy(i-1,n,j)
              +Sj*u.sy(i,n,j+1)+Aj*u.sy(i,n,j-1));
          vres = 1.0/(h*h)*(2*v.sy(i,n-1,j)
            +Si*v.sy(i+1,n,j)+Ai*v.sy(i-1,n,j)
              +Sj*v.sy(i,n,j+1)+Aj*v.sy(i,n,j-1)-D*v.sy(i,n,j))
                +beta(i,n,j)*pow(1+alpha(i,n,j)
                  +alpha(i,n,j)*v.sy(i,n,j),-7);
          nl = -beta(i,n,j)*pow(1+alpha(i,n,j)
            +alpha(i,n,j)*u(i,n,j),-7)
              *(1.0+7*alpha(i,n,j)*u.sy(i,n,j)/(1
                +alpha(i,n,j)+alpha(i,n,j)*u(i,n,j)));
          r(i,n,j) = s(i,n,j) - ures + vres 
                        + nl + drdu*u(i,n,j); 
	      
          drdu = D/(h*h)
            +7*alpha(n,j,i)*beta(n,j,i)*pow(1+alpha(n,j,i)
              +alpha(n,j,i)*u(n,j,i),-8);
          ures = 1.0/(h*h)*(2*u.sy(n-1,j,i)
            +Sj*u.sy(n,j+1,i)+Aj*u.sy(n,j-1,i)
              +Si*u.sy(n,j,i+1)+Ai*u.sy(n,j,i-1));
          vres = 1.0/(h*h)*(2*v.sy(n-1,j,i)
            +Sj*v.sy(n,j+1,i)+Aj*v.sy(n,j-1,i)
              +Si*v.sy(n,j,i+1)+Ai*v.sy(n,j,i-1)-D*v.sy(n,j,i))
                +beta(n,j,i)*pow(1+alpha(n,j,i)
                  +alpha(n,j,i)*v.sy(n,j,i),-7);
          nl = -beta(n,j,i)*pow(1+alpha(n,j,i)
            +alpha(n,j,i)*u(n,j,i),-7)
              *(1.0+7*alpha(n,j,i)*u.sy(n,j,i)/(1
                +alpha(n,j,i)+alpha(n,j,i)*u(n,j,i)));
          r(n,j,i) = s(n,j,i) - ures + vres 
                        + nl + drdu*u(n,j,i);         
        }// end for-j
      } // end for-i   
    } // end if-n      
} // end newresidual
\end{verbatim}

The program \it{moment }\rm allows for the calculation of multipole moments for any value of $\ell$ and $m$, using the appropriate Gal\"erkin operator. For our purposes, we only calculate the monopole moment, which is equivalent to the mass correction to the ADM mass, and the $\ell = m = 2$ multipole moment, which is related to the quadrupole moment.
\begin{verbatim}
// moment.cxx - calculates quadrupole moments of field
// for some l,m

#include <assert.h>
#include <fstream.h>
#include <iostream.h>
#include <iomanip.h>
#include <math.h>

#include "grid.h"
#include "adapt.h"

// k denotes the grid point of the surface
double moment(AGrid3D& u, int l, int m, int k) {

  extern const double PI;
  extern const int MG;
  
  int n = u.sideX();
  double len = u.LenX();
  double h = u.hX();
  
  AGrid3D psib;       // psib = r^l*Y(l,m) in polynomial form
  psib.reset(n,len);  // must have one extra grid point in psib
  
  psibar(psib,l,m);  // determine psibar
  
    double sum = 0.0;

    //Faces
    for(int i = 2;i<k;i++) {
      for(int j = 2;j<k;j++) {
        sum = sum + h*((1.0/6*(psib.sy(k,i+1,j)+psib.sy(k,i-1,j)
            +psib.sy(k,i,j+1)+psib.sy(k,i,j-1))
          +1.0/12*(psib.sy(k,i+1,j+1)+psib.sy(k,i+1,j-1)
            +psib.sy(k,i-1,j+1)
            +psib.sy(k,i-1,j-1)))*u.sy(k+1,i,j)
          -(1.0/6*(psib.sy(k+1,i+1,j)+psib.sy(k+1,i-1,j)
            +psib.sy(k+1,i,j+1)+psib.sy(k+1,i,j-1))
          +1.0/12*(psib.sy(k+1,i+1,j+1)+psib.sy(k+1,i+1,j-1)
            +psib.sy(k+1,i-1,j+1)
            +psib.sy(k+1,i-1,j-1)))*u.sy(k,i,j));
		       
        sum = sum + h*((1.0/6*(psib.sy(i+1,k,j)+psib.sy(i-1,k,j)
            +psib.sy(i,k,j+1)+psib.sy(i,k,j-1))
          +1.0/12*(psib.sy(i+1,k,j+1)+psib.sy(i+1,k,j-1)
            +psib.sy(i-1,k,j+1)
            +psib.sy(i-1,k,j-1)))*u.sy(i,k+1,j)
          -(1.0/6*(psib.sy(i+1,k+1,j)+psib.sy(i-1,k+1,j)
            +psib.sy(i,k+1,j+1)+psib.sy(i,k+1,j-1))
          +1.0/12*(psib.sy(i+1,k+1,j+1)+psib.sy(i+1,k+1,j-1)
            +psib.sy(i-1,k+1,j+1)
            +psib.sy(i-1,k+1,j-1)))*u.sy(i,k,j));
		       
        sum = sum + h*((1.0/6*(psib.sy(i+1,j,k)+psib.sy(i-1,j,k)
            +psib.sy(i,j+1,k)+psib.sy(i,j-1,k))
          +1.0/12*(psib.sy(i+1,j+1,k)+psib.sy(i+1,j-1,k)
            +psib.sy(i-1,j+1,k)
            +psib.sy(i-1,j-1,k)))*u.sy(i,j,k+1)
          -(1.0/6*(psib.sy(i+1,j,k+1)+psib.sy(i-1,j,k+1)
            +psib.sy(i,j+1,k+1)+psib.sy(i,j-1,k+1))
          +1.0/12*(psib.sy(i+1,j+1,k+1)+psib.sy(i+1,j-1,k+1)
            +psib.sy(i-1,j+1,k+1)
            +psib.sy(i-1,j-1,k+1)))*u.sy(i,j,k));
      } // end for-j
    } // end for-i
    
    // symmetry edges
    for(int i =2;i<k;i++) {
      // (i,1,k) edge
      sum = sum + 1.0/2*h*((1.0/6*(psib.sy(i+1,1,k)
          +psib.sy(i-1,1,k)+psib.sy(i,2,k)+psib.sy(i,0,k))
        +1.0/12*(psib.sy(i+1,2,k)+psib.sy(i+1,0,k)
          +psib.sy(i-1,2,k)+psib.sy(i-1,0,k)))*u.sy(i,1,k+1)
        -(1.0/6*(psib.sy(i+1,1,k+1)+psib.sy(i-1,1,k+1)
          +psib.sy(i,2,k+1)+psib.sy(i,0,k+1))
        +1.0/12*(psib.sy(i+1,2,k+1)+psib.sy(i+1,0,k+1)
          +psib.sy(i-1,2,k+1)+psib.sy(i-1,0,k+1)))*u.sy(i,1,k));
      // (1,i,k) edge
      sum = sum + 1.0/2*h*((1.0/6*(psib.sy(1,i+1,k)
          +psib.sy(1,i-1,k)+psib.sy(2,i,k)+psib.sy(0,i,k))
        +1.0/12*(psib.sy(2,i+1,k)+psib.sy(0,i+1,k)
          +psib.sy(2,i-1,k)+psib.sy(0,i-1,k)))*u.sy(1,i,k+1)
        -(1.0/6*(psib.sy(1,i+1,k+1)+psib.sy(1,i-1,k+1)
          +psib.sy(2,i,k+1)+psib.sy(0,i,k+1))
        +1.0/12*(psib.sy(2,i+1,k+1)+psib.sy(0,i+1,k+1)
          +psib.sy(2,i-1,k+1)+psib.sy(0,i-1,k+1)))*u.sy(1,i,k));
      
      // (i,k,1) edge
      sum = sum + 1.0/2*h*((1.0/6*(psib.sy(i+1,k,1)
          +psib.sy(i-1,k,1)+psib.sy(i,k,2)+psib.sy(i,k,0))
        +1.0/12*(psib.sy(i+1,k,2)+psib.sy(i+1,k,0)
          +psib.sy(i-1,k,2)+psib.sy(i-1,k,0)))*u.sy(i,k+1,1)
        -(1.0/6*(psib.sy(i+1,k+1,1)+psib.sy(i-1,k+1,1)
          +psib.sy(i,k+1,2)+psib.sy(i,k+1,0))
        +1.0/12*(psib.sy(i+1,k+1,2)+psib.sy(i+1,k+1,0)
          +psib.sy(i-1,k+1,2)+psib.sy(i-1,k+1,0)))*u.sy(i,k,1));
      
      // (1,k,i) edge
      sum = sum + 1.0/2*h*((1.0/6*(psib.sy(1,k,i+1)
          +psib.sy(1,k,i-1)+psib.sy(2,k,i)+psib.sy(0,k,i))
        +1.0/12*(psib.sy(2,k,i+1)+psib.sy(0,k,i+1)
          +psib.sy(2,k,i-1)+psib.sy(0,k,i-1)))*u.sy(1,k+1,i)
        -(1.0/6*(psib.sy(1,k+1,i+1)+psib.sy(1,k+1,i-1)
          +psib.sy(2,k+1,i)+psib.sy(0,k+1,i))
        +1.0/12*(psib.sy(2,k+1,i+1)+psib.sy(0,k+1,i+1)
          +psib.sy(2,k+1,i-1)+psib.sy(0,k+1,i-1)))*u.sy(1,k,i));
      
      // (k,i,1) edge
      sum = sum + 1.0/2*h*((1.0/6*(psib.sy(k,i+1,1)
          +psib.sy(k,i-1,1)+psib.sy(k,i,2)+psib.sy(k,i,0))
        +1.0/12*(psib.sy(k,i+1,2)+psib.sy(k,i+1,0)
          +psib.sy(k,i-1,2)+psib.sy(k,i-1,0)))*u.sy(k+1,i,1)
        -(1.0/6*(psib.sy(k+1,i+1,1)+psib.sy(k+1,i-1,1)
          +psib.sy(k+1,i,2)+psib.sy(k+1,i,0))
        +1.0/12*(psib.sy(k+1,i+1,2)+psib.sy(k+1,i+1,0)
          +psib.sy(k+1,i-1,2)+psib.sy(k+1,i-1,0)))*u.sy(k,i,1));
		       
      // (k,1,i) edge
      sum = sum + 1.0/2*h*((1.0/6*(psib.sy(k,2,i)
          +psib.sy(k,0,i)+psib.sy(k,1,i+1)+psib.sy(k,1,i-1))
        +1.0/12*(psib.sy(k,2,i+1)+psib.sy(k,2,i-1)
          +psib.sy(k,0,i+1)+psib.sy(k,0,i-1)))*u.sy(k+1,1,i)
        -(1.0/6*(psib.sy(k+1,2,i)+psib.sy(k+1,0,i)
          +psib.sy(k+1,1,i+1)+psib.sy(k+1,1,i-1))
        +1.0/12*(psib.sy(k+1,2,i+1)+psib.sy(k+1,2,i-1)
          +psib.sy(k+1,0,i+1)+psib.sy(k+1,0,i-1)))*u.sy(k,1,i));
     } //end for-i
    
    // interior symmetry corners
    int i = 1;
    int j = 1;
    // (1,1,k), (k,1,1) and (1,k,1)
    sum = sum + 1.0/4*h*((1.0/6*(psib.sy(i+1,j,k)
        +psib.sy(i-1,j,k)+psib.sy(i,j+1,k)+psib.sy(i,j-1,k))
      +1.0/12*(psib.sy(i+1,j+1,k)+psib.sy(i+1,j-1,k)
        +psib.sy(i-1,j+1,k)+psib.sy(i-1,j-1,k)))*u.sy(i,j,k+1)
      -(1.0/6*(psib.sy(i+1,j,k+1)+psib.sy(i-1,j,k+1)
        +psib.sy(i,j+1,k+1)+psib.sy(i,j-1,k+1))
      +1.0/12*(psib.sy(i+1,j+1,k+1)+psib.sy(i+1,j-1,k+1)
        +psib.sy(i-1,j+1,k+1)
        +psib.sy(i-1,j-1,k+1)))*u.sy(i,j,k))
      +1.0/4*h*((1.0/6*(psib.sy(k,i+1,j)+psib.sy(k,i-1,j)
        +psib.sy(k,i,j+1)+psib.sy(k,i,j-1))
      +1.0/12*(psib.sy(k,i+1,j+1)+psib.sy(k,i+1,j-1)
        +psib.sy(k,i-1,j+1)+psib.sy(k,i-1,j-1)))*u.sy(k+1,i,j)
      -(1.0/6*(psib.sy(k+1,i+1,j)+psib.sy(k+1,i-1,j)
        +psib.sy(k+1,i,j+1)+psib.sy(k+1,i,j-1))
      +1.0/12*(psib.sy(k+1,i+1,j+1)+psib.sy(k+1,i+1,j-1)
        +psib.sy(k+1,i-1,j+1)
        +psib.sy(k+1,i-1,j-1)))*u.sy(k,i,j))
      +1.0/4*h*((1.0/6*(psib.sy(i+1,k,j)+psib.sy(i-1,k,j)
        +psib.sy(i,k,j+1)+psib.sy(i,k,j-1))
      +1.0/12*(psib.sy(i+1,k,j+1)+psib.sy(i+1,k,j-1)
        +psib.sy(i-1,k,j+1)+psib.sy(i-1,k,j-1)))*u.sy(i,k+1,j)
      -(1.0/6*(psib.sy(i+1,k+1,j)+psib.sy(i-1,k+1,j)
        +psib.sy(i,k+1,j+1)+psib.sy(i,k+1,j-1))
      +1.0/12*(psib.sy(i+1,k+1,j+1)+psib.sy(i+1,k+1,j-1)
        +psib.sy(i-1,k+1,j+1)
        +psib.sy(i-1,k+1,j-1)))*u.sy(i,k,j));
   
    // boundary edges
    for(int i = 2;i<k;i++) {
      
      // (k,k,i) edge		  
      sum = sum + h*((1.0/6*(psib.sy(k,k,i+1)+psib.sy(k,k-1,i)
          +psib.sy(k,k,i-1))
        +1.0/12*(psib.sy(k,k-1,i-1)
          +psib.sy(k,k-1,i+1)))*u.sy(k+1,k,i)
        +(1.0/6*(psib.sy(k,k,i+1)+psib.sy(k-1,k,i)
          +psib.sy(k,k,i-1))
        +1.0/12*(psib.sy(k-1,k,i-1)
          +psib.sy(k-1,k,i+1)))*u.sy(k,k+1,i)
        +(1.0/6*(psib.sy(k,k,i))
        +1.0/12*(psib.sy(k,k,i+1)
          +psib.sy(k,k,i-1)))*u.sy(k+1,k+1,i)
        -(1.0/6*(psib.sy(k,k+1,i+1)+psib.sy(k-1,k+1,i)
          +psib.sy(k,k+1,i-1)+psib.sy(k+1,k,i-1)
          +psib.sy(k+1,k,i+1)+psib.sy(k+1,k+1,i)
          +psib.sy(k+1,k-1,i))
        +1.0/12*(psib.sy(k+1,k+1,i-1)+psib.sy(k-1,k+1,i-1)
          +psib.sy(k-1,k+1,i+1)+psib.sy(k+1,k-1,i-1)
          +psib.sy(k+1,k-1,i+1)
          +psib.sy(k+1,k+1,i+1)))*u.sy(k,k,i));
			  
      // (k,i,k) edge
      sum = sum + h*((1.0/6*(psib.sy(k,i+1,k)+psib.sy(k,i,k-1)
          +psib.sy(k,i-1,k))
        +1.0/12*(psib.sy(k,i-1,k-1)
          +psib.sy(k,i+1,k-1)))*u.sy(k+1,i,k)
        +(1.0/6*(psib.sy(k,i+1,k)+psib.sy(k-1,i,k)
          +psib.sy(k,i-1,k))
        +1.0/12*(psib.sy(k-1,i-1,k)
          +psib.sy(k-1,i+1,k)))*u.sy(k,i,k+1)
        +(1.0/6*(psib.sy(k,i,k))
        +1.0/12*(psib.sy(k,i+1,k)
          +psib.sy(k,i-1,k)))*u.sy(k+1,i,k+1)
        -(1.0/6*(psib.sy(k,i+1,k+1)+psib.sy(k-1,i,k+1)
          +psib.sy(k,i-1,k+1)+psib.sy(k+1,i-1,k)
          +psib.sy(k+1,i+1,k)+psib.sy(k+1,i,k+1)
          +psib.sy(k+1,i,k-1))
        +1.0/12*(psib.sy(k+1,i-1,k+1)+psib.sy(k-1,i-1,k+1)
          +psib.sy(k-1,i+1,k+1)+psib.sy(k+1,i-1,k-1)
          +psib.sy(k+1,i+1,k-1)
          +psib.sy(k+1,i+1,k+1)))*u.sy(k,i,k));
    
      // (i,k,k) edge
      sum = sum + h*((1.0/6*(psib.sy(i+1,k,k)+psib.sy(i,k,k-1)
          +psib.sy(i-1,k,k))
        +1.0/12*(psib.sy(i-1,k,k-1)
          +psib.sy(i+1,k,k-1)))*u.sy(i,k+1,k)
        +(1.0/6*(psib.sy(i+1,k,k)+psib.sy(i,k-1,k)
          +psib.sy(i-1,k,k))
        +1.0/12*(psib.sy(i-1,k-1,k)
          +psib.sy(i+1,k-1,k)))*u.sy(i,k,k+1)
        +(1.0/6*(psib.sy(i,k,k))
        +1.0/12*(psib.sy(i+1,k,k)
          +psib.sy(i-1,k,k)))*u.sy(i,k+1,k+1)
        -(1.0/6*(psib.sy(i+1,k,k+1)+psib.sy(i,k-1,k+1)
          +psib.sy(i-1,k,k+1)+psib.sy(i-1,k+1,k)
          +psib.sy(i+1,k+1,k)+psib.sy(i,k+1,k+1)
          +psib.sy(i,k+1,k-1))
        +1.0/12*(psib.sy(i-1,k+1,k+1)+psib.sy(i-1,k-1,k+1)
          +psib.sy(i+1,k-1,k+1)+psib.sy(i-1,k+1,k-1)
          +psib.sy(i+1,k+1,k-1)
          +psib.sy(i+1,k+1,k+1)))*u.sy(i,k,k));  
    } // end for-i
    
    // boundary corners
    
    // (k,k,1) corner
    sum = sum + 0.5*h*((1.0/6*(psib.sy(k,k,2)+psib.sy(k,k-1,1)
        +psib.sy(k,k,0))
      +1.0/12*(psib.sy(k,k-1,0)
        +psib.sy(k,k-1,2)))*u.sy(k+1,k,1)
      +(1.0/6*(psib.sy(k,k,2)+psib.sy(k-1,k,1)
        +psib.sy(k,k,0))
      +1.0/12*(psib.sy(k-1,k,0)
        +psib.sy(k-1,k,2)))*u.sy(k,k+1,1)
      +(1.0/6*(psib.sy(k,k,1))
      +1.0/12*(psib.sy(k,k,2)
        +psib.sy(k,k,0)))*u.sy(k+1,k+1,1)
      -(1.0/6*(psib.sy(k,k+1,2)+psib.sy(k-1,k+1,1)
        +psib.sy(k,k+1,0)+psib.sy(k+1,k,0)
        +psib.sy(k+1,k,2)+psib.sy(k+1,k+1,1)
        +psib.sy(k+1,k-1,1))
      +1.0/12*(psib.sy(k+1,k+1,0)+psib.sy(k-1,k+1,0)
        +psib.sy(k-1,k+1,2)+psib.sy(k+1,k-1,0)
        +psib.sy(k+1,k-1,2)
        +psib.sy(k+1,k+1,2)))*u.sy(k,k,1));
    
    // (k,1,k) corner
    sum = sum + 0.5*h*((1.0/6*(psib.sy(k,2,k)+psib.sy(k,1,k-1)
        +psib.sy(k,0,k))
      +1.0/12*(psib.sy(k,0,k-1)
        +psib.sy(k,2,k-1)))*u.sy(k+1,1,k)
      +(1.0/6*(psib.sy(k,2,k)+psib.sy(k-1,1,k)
        +psib.sy(k,0,k))
      +1.0/12*(psib.sy(k-1,0,k)
        +psib.sy(k-1,2,k)))*u.sy(k,1,k+1)
      +(1.0/6*(psib.sy(k,1,k))
      +1.0/12*(psib.sy(k,2,k)
        +psib.sy(k,0,k)))*u.sy(k+1,1,k+1)
      -(1.0/6*(psib.sy(k,2,k+1)+psib.sy(k-1,1,k+1)
        +psib.sy(k,0,k+1)+psib.sy(k+1,0,k)
        +psib.sy(k+1,2,k)+psib.sy(k+1,1,k+1)
        +psib.sy(k+1,1,k-1))
      +1.0/12*(psib.sy(k+1,0,k+1)+psib.sy(k-1,0,k+1)
        +psib.sy(k-1,2,k+1)+psib.sy(k+1,0,k-1)
        +psib.sy(k+1,2,k-1)
        +psib.sy(k+1,2,k+1)))*u.sy(k,1,k));
    
    // (1,k,k) corner
    sum = sum + 0.5*h*((1.0/6*(psib.sy(2,k,k)+psib.sy(1,k,k-1)
        +psib.sy(0,k,k))
      +1.0/12*(psib.sy(0,k,k-1)
        +psib.sy(2,k,k-1)))*u.sy(1,k+1,k)
      +(1.0/6*(psib.sy(2,k,k)+psib.sy(1,k-1,k)
        +psib.sy(0,k,k))
      +1.0/12*(psib.sy(0,k-1,k)
        +psib.sy(2,k-1,k)))*u.sy(1,k,k+1)
      +(1.0/6*(psib.sy(1,k,k))
      +1.0/12*(psib.sy(2,k,k)
        +psib.sy(0,k,k)))*u.sy(1,k+1,k+1)
      -(1.0/6*(psib.sy(2,k,k+1)+psib.sy(1,k-1,k+1)
        +psib.sy(0,k,k+1)+psib.sy(0,k+1,k)
        +psib.sy(2,k+1,k)+psib.sy(1,k+1,k+1)
        +psib.sy(1,k+1,k-1))
      +1.0/12*(psib.sy(0,k+1,k+1)+psib.sy(0,k-1,k+1)
        +psib.sy(2,k-1,k+1)+psib.sy(0,k+1,k-1)
        +psib.sy(2,k+1,k-1)
        +psib.sy(2,k+1,k+1)))*u.sy(1,k,k)); 
    
    // outermost corner
    sum = sum + h*((1.0/6*(psib.sy(k,k,k-1)+psib.sy(k,k-1,k))
      +1.0/12*psib.sy(k,k-1,k-1))*u.sy(k+1,k,k)
      +(1.0/6*(psib.sy(k,k,k-1)+psib.sy(k-1,k,k))
      +1.0/12*psib.sy(k-1,k,k-1))*u.sy(k,k+1,k)
      +(1.0/6*(psib.sy(k,k-1,k)+psib.sy(k-1,k,k))
      +1.0/12*psib.sy(k-1,k-1,k))*u.sy(k,k,k+1)
      +(1.0/6*psib.sy(k,k,k)
      +1.0/12*psib.sy(k,k,k-1))*u.sy(k+1,k+1,k)
      +(1.0/6*psib.sy(k,k,k)
      +1.0/12*psib.sy(k,k-1,k))*u.sy(k+1,k,k+1)
      +(1.0/6*psib.sy(k,k,k)
      +1.0/12*psib.sy(k-1,k,k))*u.sy(k,k+1,k+1)
      +(1.0/12*psib.sy(k,k,k))*u.sy(k+1,k+1,k+1)
      -(1.0/6*(psib.sy(k-1,k,k+1)+psib.sy(k-1,k+1,k)
        +psib.sy(k,k-1,k+1)+psib.sy(k,k+1,k-1)
        +psib.sy(k,k+1,k+1)+psib.sy(k+1,k-1,k)
        +psib.sy(k+1,k,k-1)+psib.sy(k+1,k,k+1)
        +psib.sy(k+1,k+1,k))
      +1.0/12*(psib.sy(k-1,k-1,k+1)+psib.sy(k-1,k+1,k-1)
        +psib.sy(k-1,k+1,k+1)+psib.sy(k+1,k-1,k+1)
        +psib.sy(k+1,k+1,k+1)+psib.sy(k+1,k+1,k-1)
        +psib.sy(k+1,k-1,k-1)))*u.sy(k,k,k));
    
    // Now, add all contributions together to determine the 
    // multipole moment...
    // Don't forget to multiply by 8 for the octant!
    
    double norm = -1.0/sqrt(PI);   // Ensures phi(0,0) for bh's
    
    double mom = 0.0;    // ensures moment is zero
    
    mom = 8.0*norm/(2*l+1)*sum;

    return(mom);  
}
\end{verbatim}

The function \it{psibar }\rm simply fills an array with radial polynomials which are dependent upon spherical harmonics, depending upon the values of $\ell$ and $m$:
\begin{verbatim}
// psibar.cxx - code that determines the polynomial form 
// of r^l*Y(l,m)

#include <assert.h>
#include <fstream.h>
#include <iostream.h>
#include <iomanip.h>
#include <math.h>

#include "grid.h"
#include "adapt.h"


void psibar(AGrid3D& psib, int l, int m) 
{

  extern const double PI;
  int n = psib.sideX();
  double h = psib.hX();

  for(int i=1;i<=n;i++) {
    for(int j=1;j<=n;j++) {
      for(int k=1;k<=n;k++) {
  
 // Note "x" is the symmetry axis!!
  
        double x = h*(i-1);  
        double y = h*(j-1);  
        double z = h*(k-1);
  
        if(l==0) {
          psib(i,j,k) = 1.0/sqrt(4*PI);
        }
  
        else if(l==1) {
          
          if(m==0) {
            psib(i,j,k) = 1.0*sqrt(3.0/(4*PI))*z;
          } // end if m=0
          else if(m==1) {
            psib(i,j,k) = 1.0*sqrt(3.0/(8*PI))*x;
          }// end if m=1
          else {
            psib(i,j,k) = -1.0*sqrt(3.0/(8*PI))*x;
          }// end if m=-1
        
        } // end if l=1
  
        else if(l==2) {
          
          if(m==0) {
            psib(i,j,k) = -0.5*sqrt(5.0/(4*PI))*(x*x+y*y-2*z*z);
          } // end if m=0
         
          else if(m==1) {
            psib(i,j,k) = -1.0*sqrt(15.0/(8*PI))*x*y;
          } // end if m=1
          else if(m==-1) {
            psib(i,j,k) = 1.0*sqrt(15.0/(8*PI))*x*y;
          }// end if m=-1
        
          else if(m==2) {
            psib(i,j,k) = 1.0/4*sqrt(15.0/(2*PI))*(x*x-y*y);
          } // end if m=2
          else {
            psib(i,j,k) = 1.0/4*sqrt(15.0/(2*PI))*(x*x-y*y);
          } // end if m=-2
       
        } // end if-l=2
      } //end for-k
    } // end for-j
  } // end for-i
} // end psibar
\end{verbatim}

The function \it{symadaptrstrct }\rm is a adaptive grid restriction routine, designed to restrict only some of the arrays to larger adaptive levels:
\begin{verbatim}
#include <assert.h>
#include <fstream.h>
#include <iostream.h>
#include <iomanip.h>
#include <math.h>
#include "grid.h"
#include "adapt.h"

extern int MG;

// Half-weighting restriction for non-mg levels
void symadaptrstrct(AGrid3D& vl, AGrid3D& vs)
{
  int il, is, jl, js, kl, ks;
  int nl = vl.sideX();
  int N = nl/2-1;
  
  if(nl==pow(2,MG)+3) N = nl/2-1;
  else N = nl/2;

  // Calculate the interior
  
  for(ks=1,kl=1; kl<=N; kl++,ks+=2) {
    for(js=1,jl=1; jl<=N; jl++,js+=2) {			
      for(is=1,il=1; il<=N; il++,is+=2) {
        vl(il,jl,kl) = 0.125*(vs.sy(is,js,ks)
          +0.5*(vs.sy(is+1,js,ks)+vs.sy(is-1,js,ks)
            +vs.sy(is,js+1,ks)+vs.sy(is,js-1,ks)
            +vs.sy(is,js,ks+1)+vs.sy(is,js,ks-1))
          +0.25*(vs.sy(is,js+1,ks+1)+vs.sy(is,js+1,ks-1)
            +vs.sy(is,js-1,ks+1)+vs.sy(is,js-1,ks-1)
            +vs.sy(is+1,js,ks+1)+vs.sy(is+1,js,ks-1)
            +vs.sy(is-1,js,ks+1)+vs.sy(is-1,js,ks-1)
            +vs.sy(is+1,js+1,ks)+vs.sy(is+1,js-1,ks)
            +vs.sy(is-1,js+1,ks)+vs.sy(is-1,js-1,ks))
          +0.125*(vs.sy(is+1,js+1,ks+1)+vs.sy(is+1,js+1,ks-1)
            +vs.sy(is+1,js-1,ks+1)+vs.sy(is+1,js-1,ks-1)
            +vs.sy(is-1,js+1,ks+1)+vs.sy(is-1,js+1,ks-1)
            +vs.sy(is-1,js-1,ks+1)+vs.sy(is-1,js-1,ks-1)));
      } // end for-is
    } // end for-js
  } // end for-ks
} // end symadaptrstrct
\end{verbatim}

Likewise, the program \it{symresrstrct }\rm is a adaptive multigrid restriction routine designed specifically to restrict residuals to larger adaptive levels:
\begin{verbatim}
#include <assert.h>
#include <fstream.h>
#include <iostream.h>
#include <iomanip.h>
#include <math.h>
#include "grid.h"
#include "adapt.h"

// Full-weighting restriltion of residuals
// for use with trueamg.cxx

void symresrstrct(AGrid3D& vl, AGrid3D& vs)
{
  int il, iis, jl, js, kl, ks;
  int ns = vs.sideX();

  // Calculate the interior
  for(ks=1,kl=1; ks<=ns-2; kl++,ks+=2) {
    for(js=1,jl=1; js<=ns-2; jl++,js+=2) {	
      for(iis=1,il=1; iis<=ns-2; il++,iis+=2) {
        vl(il,jl,kl) = 0.125*(vs.sy(iis,js,ks)
          +0.5*(vs.sy(iis+1,js,ks)+vs.sy(iis-1,js,ks)
            +vs.sy(iis,js+1,ks)+vs.sy(iis,js-1,ks)
            +vs.sy(iis,js,ks+1)+vs.sy(iis,js,ks-1))
          +0.25*(vs.sy(iis,js+1,ks+1)+vs.sy(iis,js+1,ks-1)
            +vs.sy(iis,js-1,ks+1)+vs.sy(iis,js-1,ks-1)
            +vs.sy(iis+1,js,ks+1)+vs.sy(iis+1,js,ks-1)
            +vs.sy(iis-1,js,ks+1)+vs.sy(iis-1,js,ks-1)
            +vs.sy(iis+1,js+1,ks)+vs.sy(iis+1,js-1,ks)
            +vs.sy(iis-1,js+1,ks)+vs.sy(iis-1,js-1,ks))
          +0.125*(vs.sy(iis+1,js+1,ks+1)+vs.sy(iis+1,js+1,ks-1)
            +vs.sy(iis+1,js-1,ks+1)+vs.sy(iis+1,js-1,ks-1)
            +vs.sy(iis-1,js+1,ks+1)+vs.sy(iis-1,js+1,ks-1)
            +vs.sy(iis-1,js-1,ks+1)+vs.sy(iis-1,js-1,ks-1)));
      } // end for-iis
    } // end for-js
  } // end for-ks
} // end symresrstrct
\end{verbatim}

The program \it{symrstrct }\rm is a restriction routine designed specifically for multigrid levels only:
\begin{verbatim}
#include <assert.h>
#include <fstream.h>
#include <iostream.h>
#include <iomanip.h>
#include <math.h>
#include "grid.h"
#include "adapt.h"

// Half-weighting restriction
void symrstrct(AGrid3D& vc, AGrid3D& vf)
{
  int ic, iif, jc, jf, kc, kf;
  int nc = vc.sideX();
  int nf = vf.sideX();
  
  // Calculate the interior
  
  for(kf=1,kc=1; kc<nc; kc++,kf+=2) {
    for(jf=1,jc=1; jc<nc; jc++,jf+=2) {			
      for(iif=1,ic=1; ic<nc; ic++,iif+=2) {
        vc(ic,jc,kc) = 0.125*(vf.sy(iif,jf,kf)
          +0.5*(vf.sy(iif+1,jf,kf)+vf.sy(iif-1,jf,kf)
            +vf.sy(iif,jf+1,kf)+vf.sy(iif,jf-1,kf)
            +vf.sy(iif,jf,kf+1)+vf.sy(iif,jf,kf-1))
          +0.25*(vf.sy(iif,jf+1,kf+1)+vf.sy(iif,jf+1,kf-1)
            +vf.sy(iif,jf-1,kf+1)+vf.sy(iif,jf-1,kf-1)
            +vf.sy(iif+1,jf,kf+1)+vf.sy(iif+1,jf,kf-1)
            +vf.sy(iif-1,jf,kf+1)+vf.sy(iif-1,jf,kf-1)
            +vf.sy(iif+1,jf+1,kf)+vf.sy(iif+1,jf-1,kf)
            +vf.sy(iif-1,jf+1,kf)+vf.sy(iif-1,jf-1,kf))
          +0.125*(vf.sy(iif+1,jf+1,kf+1)+vf.sy(iif+1,jf+1,kf-1)
            +vf.sy(iif+1,jf-1,kf+1)+vf.sy(iif+1,jf-1,kf-1)
            +vf.sy(iif-1,jf+1,kf+1)+vf.sy(iif-1,jf+1,kf-1)
            +vf.sy(iif-1,jf-1,kf+1)+vf.sy(iif-1,jf-1,kf-1)));
      } // end for-iif
    } // end for-jf
  } // end for-kf
  
// Calculate the corners
  
  vc(nc,nc,nc)= 8.0/27.0*(vf.sy(nf,nf,nf)
    +0.5*(vf.sy(nf-1,nf,nf)+vf.sy(nf,nf-1,nf)
      +vf.sy(nf,nf,nf-1))
    +0.25*(vf.sy(nf,nf-1,nf-1)+vf.sy(nf-1,nf,nf-1)
      +vf.sy(nf-1,nf-1,nf))
    +0.125*(vf.sy(nf-1,nf-1,nf-1)));

// Calculate the edges

  for(kc=1,kf=1;kc<nc;kc++,kf+=2) {
  
    vc(kc,nc,nc)= 2.0/9.0*(vf.sy(kf,nf,nf)
      +0.5*(vf.sy(kf+1,nf,nf)+vf.sy(kf-1,nf,nf)
        +vf.sy(kf,nf-1,nf)+vf.sy(kf,nf,nf-1))
      +0.25*(vf.sy(kf,nf-1,nf-1)+vf.sy(kf+1,nf,nf-1)
        +vf.sy(kf-1,nf,nf-1)+vf.sy(kf+1,nf-1,nf)
        +vf.sy(kf-1,nf-1,nf))
      +0.125*(vf.sy(kf+1,nf-1,nf-1)+vf.sy(kf-1,nf-1,nf-1)));
  
    vc(nc,kc,nc)= 2.0/9.0*(vf.sy(nf,kf,nf)
      +0.5*(vf.sy(nf-1,kf,nf)+vf.sy(nf,kf+1,nf)
        +vf.sy(nf,kf-1,nf)+vf.sy(nf,kf,nf-1))
      +0.25*(vf.sy(nf,kf+1,nf-1)+vf.sy(nf,kf-1,nf-1)
        +vf.sy(nf-1,kf,nf-1)+vf.sy(nf-1,kf+1,nf)
        +vf.sy(nf-1,kf-1,nf))
      +0.125*(vf.sy(nf-1,kf+1,nf-1)+vf.sy(nf-1,kf-1,nf-1)));
 
    vc(nc,nc,kc)= 2.0/9.0*(vf.sy(nf,nf,kf)
      +0.5*(vf.sy(nf-1,nf,kf)+vf.sy(nf,nf-1,kf)
        +vf.sy(nf,nf,kf+1)+vf.sy(nf,nf,kf-1))
      +0.25*(vf.sy(nf,nf-1,kf+1)+vf.sy(nf,nf-1,kf-1)
        +vf.sy(nf-1,nf,kf+1)+vf.sy(nf-1,nf,kf-1)
        +vf.sy(nf-1,nf-1,kf))
      +0.125*(vf.sy(nf-1,nf-1,kf+1)+vf.sy(nf-1,nf-1,kf-1)));  
  } // end for-kc

  // Calculate the faces
  
  for(jc=1,jf=1;jc<nc;jc++,jf+=2) {
    for(kc=1,kf=1;kc<nc;kc++,kf+=2) {
      
      vc(jc,nc,kc)= 1.0/6.0*(vf.sy(jf,nf,kf)
        +0.5*(vf.sy(jf+1,nf,kf)+vf.sy(jf-1,nf,kf)
          +vf.sy(jf,nf-1,kf)+vf.sy(jf,nf,kf+1)
          +vf.sy(jf,nf,kf-1))
        +0.25*(vf.sy(jf,nf-1,kf+1)+vf.sy(jf,nf-1,kf-1)
          +vf.sy(jf+1,nf,kf+1)+vf.sy(jf+1,nf,kf-1)
          +vf.sy(jf-1,nf,kf+1)+vf.sy(jf-1,nf,kf-1)
          +vf.sy(jf+1,nf-1,kf)+vf.sy(jf-1,nf-1,kf))
        +0.125*(vf.sy(jf+1,nf-1,kf+1)+vf.sy(jf+1,nf-1,kf-1)
          +vf.sy(jf-1,nf-1,kf+1)+vf.sy(jf-1,nf-1,kf-1)));
    
      vc(nc,jc,kc)= 1.0/6.0*(vf.sy(nf,jf,kf)
        +0.5*(vf.sy(nf-1,jf,kf)+vf.sy(nf,jf+1,kf)
          +vf.sy(nf,jf-1,kf)+vf.sy(nf,jf,kf+1)
          +vf.sy(nf,jf,kf-1))
        +0.25*(vf.sy(nf,jf+1,kf+1)+vf.sy(nf,jf+1,kf-1)
          +vf.sy(nf,jf-1,kf+1)+vf.sy(nf,jf-1,kf-1)
          +vf.sy(nf-1,jf,kf+1)+vf.sy(nf-1,jf,kf-1)
          +vf.sy(nf-1,jf+1,kf)+vf.sy(nf-1,jf-1,kf))
        +0.125*(vf.sy(nf-1,jf+1,kf+1)+vf.sy(nf-1,jf+1,kf-1)
          +vf.sy(nf-1,jf-1,kf+1)+vf.sy(nf-1,jf-1,kf-1)));
    
      vc(kc,jc,nc)= 1.0/6.0*(vf.sy(kf,jf,nf)
        +0.5*(vf.sy(kf+1,jf,nf)+vf.sy(kf-1,jf,nf)
          +vf.sy(kf,jf+1,nf)+vf.sy(kf,jf-1,nf)
          +vf.sy(kf,jf,nf-1))
        +0.25*(vf.sy(kf,jf+1,nf-1)+vf.sy(kf,jf-1,nf-1)
          +vf.sy(kf+1,jf,nf-1)+vf.sy(kf-1,jf,nf-1)
          +vf.sy(kf+1,jf+1,nf)+vf.sy(kf+1,jf-1,nf)
          +vf.sy(kf-1,jf+1,nf)+vf.sy(kf-1,jf-1,nf))
        +0.125*(vf.sy(kf+1,jf+1,nf-1)+vf.sy(kf+1,jf-1,nf-1)
          +vf.sy(kf-1,jf+1,nf-1)+vf.sy(kf-1,jf-1,nf-1)));
    } // end for-kc
  } // end for-jc
} // end symrstrct
\end{verbatim}

The following program, \it{adaptinterp}\rm, is designed to interpolate data from one adaptive level to a smaller adaptive level:
\begin{verbatim}
#include <assert.h>
#include <fstream.h>
#include <iostream.h>
#include <iomanip.h>
#include <math.h>
#include "grid.h"
#include "adapt.h"

// Coarse-to-fine prolongation by bilinear interpolation for
// adaptive grid
void adaptinterp(AGrid3D& a, AGrid3D& v)
{
  
  extern const int MG;
  
  int i, j, k, x, y, z;
  int n=a.sideX();
  
  for(k=1;k<=n;k+=2) {		// we do hold the nc faces fixed
    for(j=1;j<=n;j+=2) {	// identifies large to small
                                // grid points
      for(i=1;i<=n;i+=2) {
        x = (i+1)/2;
        y = (j+1)/2;
        z = (k+1)/2;
        a(i,j,k) = v(x,y,z);
      } // end for-i
    } // end for-j
  } // end for-k
   
  for(k=1;k<=n;k+=2) {
    for(j=1;j<=n;j+=2) {		// averages horizontally
      for(i=2;i<n;i+=2) {
        a(i,j,k) = 0.5*(a(i+1,j,k)+a(i-1,j,k));
      } // end for-i
    } // end for-j
    for(j=2;j<n;j+=2){		// averages vertically		
      for(i=1;i<=n;i++){
        a(i,j,k) = 0.5*(a(i,j+1,k)+a(i,j-1,k));
      } // end for-i
    } // end for-j
  } // end for-k
  
  for(k=2;k<n;k+=2) {		// average "up" and "down"
    for(i=1;i<=n;i++) {
      for(j=1;j<=n;j++) {
        a(i,j,k) = 0.5*(a(i,j,k+1)+a(i,j,k-1));
      } // end for-j
    } // end for-i
  } // end for-k 
} // end adaptinterp
\end{verbatim}

The program \it{interp }\rm interpolates from one multigrid level to a smaller multigrid level:
\begin{verbatim}
#include <assert.h>
#include <fstream.h>
#include <iostream.h>
#include <iomanip.h>
#include <math.h>
#include "grid.h"
#include "adapt.h"

// Coarse-to-fine prolongation by bilinear interpolation
void interp(AGrid3D& af, AGrid3D& vc)
{
  int ic, iif, jc, jf, kc, kf;
  int nf = af.sideX();
  int nc = vc.sideX();
  
  for(kc=1;kc<=nc;kc++) {	// we do not hold the nc faces fixed
    for(jc=1;jc<=nc;jc++) {	// identifies course to fine
                                // grid points
      for(ic=1;ic<=nc;ic++) {
        af(2*ic-1,2*jc-1,2*kc-1) = vc(ic,jc,kc);
      } // end for-ic
    } // end for-jc
    kf = 2*kc-1;
    for(jf=1;jf<=nf;jf+=2) {		// averages horizontally
      for(iif=2;iif<nf;iif+=2) {
        af(iif,jf,kf) = 0.5*(af(iif+1,jf,kf)+af(iif-1,jf,kf));
      } // end for-iif
    } // end for-jf
    for(jf=2;jf<nf;jf+=2){		// averages vertically		
      for(iif=1;iif<=nf;iif++){
        af(iif,jf,kf) = 0.5*(af(iif,jf+1,kf)+af(iif,jf-1,kf));
      } // end for-iif
    } // end for-jf
  } // end for-kc
  
  for(kf=2;kf<nf;kf+=2) {	// average "up" and "down"
    for(iif=1;iif<=nf;iif++) {
      for(jf=1;jf<=nf;jf++) {
        af(iif,jf,kf) = 0.5*(af(iif,jf,kf+1)+af(iif,jf,kf-1));
      } // end for-jf
    } // end for-iif
  } // end for-kf 
} // end interp
\end{verbatim}

The program \it{gluegrid }\rm ``pastes'' together several adaptive levels, relegated to the equitorial plane, allowing the solution to be viewed over a region larger than the smallest adaptive level. It is not a necessary component in the application of the variational principle, but it is sometimes convenient to view the behavior of the field graphically:
\begin{verbatim}
// gluegrid.cxx - function to "glue" several adaptive multigrid
// levels together

#include <assert.h>
#include <fstream.h>
#include <iostream.h>
#include <iomanip.h>
#include <math.h>

#include "grid.h"
#include "adapt.h"

// sm is the small grid, lg is the large grid, lev is the
// smallest level number

void gluegrid(AGrid3D& usm, AGrid3D& ulg, int lev, int count)
{
  // ONLY WRITE DATA FOR EQUATORIAL PLANE!!!!
  
  // first read in the large grid data, and interpolate it to
  // a finer grid

  int jc,ic,iif,jf;
  int nc=ulg.sideX();
  int nf = 2*nc-1;
  double len = ulg.LenX();
  
  AGrid3D alg;          // alg is a dummy array for gluing
                        // solutions together
  alg.reset(nf,len);
    
    
    for(jc=1;jc<=nc;jc++) {	// identifies course to fine
                                // grid points
      for(ic=1;ic<=nc;ic++) {
        alg(2*ic-1,2*jc-1,1) = ulg(ic,jc,1);
      } // end for-ic
    } // end for-jc
    for(jf=1;jf<=nf;jf+=2) {		// averages horizontally
      for(iif=2;iif<nf;iif+=2) {
        alg(iif,jf,1) = 0.5*(alg(iif+1,jf,1)+alg(iif-1,jf,1));
      } // end for-iif
    } // end for-jf
    for(jf=2;jf<nf;jf+=2){		// averages vertically	
      for(iif=1;iif<=nf;iif++){
        alg(iif,jf,1) = 0.5*(alg(iif,jf+1,1)+alg(iif,jf-1,1));
      } // end for-iif
    } // end for-jf
  
  // Now alg contains the interpolated larger grid solution
  // Zero interior of larger solution, just to be "safe"
  
  int i,j;
  double h = usm.hX();
  
  
  for(i=1;i<=nf/(pow(2,count+2))+1;i++) {    // zeroing interior
                                             // depends upon
                                             // the level...
    for(j=1;j<=nf/(pow(2,count+2))+1;j++) {
      alg(i,j,1) = 0.0;
    } // end for-j
  } // end for-i

  // Now glue together solutions, save into usm and write to a
  // file if lev == 1
 
  AGrid3D bsm;        // bsm is a dummy array
  bsm.reset(nf,len);
  
  for(i=1;i<=nf/(pow(2,count+2))+1;i++) {  // size of interior
                                           // depends upon level
    for(j=1;j<=nf;j++) {
      if(j<=nf/(pow(2,count+2))+1) bsm(i,j,1) = usm(i,j,1);
      else bsm(i,j,1) = alg(i,j,1);
    } // end for-j
  } // end for-i
  for(i=nf/(pow(2,count+2))+2;i<=nf;i++) {
    for(j=1;j<=nf;j++) {
      bsm(i,j,1) = alg(i,j,1);
    } // end for-j
  } // end for-i

  usm.reset(nf,len);     // reset usm to copy bsm into it...
  copy(usm,bsm);
  
  if(lev==1) {  // if lev = 1, then write glued 
                // solution to a file

    h = usm.hX();
    cout << "Set dgrid3d in gnuplot to " << nf << endl;
    ofstream output;
    output.open("gluesoln.dat", ios::app|ios::trunc);

    for(i=1;i<=nf;i++) {
      for(j=1;j<=nf;j++) {
        output << i << "   " << j << "  " << h*(i-1) 
               << "   "  << h*(j-1) << "   " << setprecision(10)
               << usm(i,j,1) << endl;
      } // end for-j
    } // end for-i

    for(i=-nf;i<=-1;i++) {
      for(j=1;j<=nf;j++) {
        output << i << "   " << j << "  " << h*(i+1) 
               << "   "  << h*(j-1) << "   " << setprecision(10)
               << usm(-i,j,1) << endl;
      } // end for-j
    } // end for-i

    for(i=1;i<=nf;i++) {
      for(j=1;j<=nf;j++) {
        output << i << "   " << -j << "  " << h*(i-1) 
               << "   " << -h*(j-1) << "   " << setprecision(10)
               << usm(i,j,1) << endl;
      } // end for-j
    } // end for-i

    for(i=1;i<=nf;i++) {
      for(j=1;j<=nf;j++) {
        output << -i << "   " << -j << "  " << -h*(i-1) 
        << "   "  << -h*(j-1) << "   " << setprecision(10) 
        << usm(i,j,1) << endl;
      } // end for-j
    } // end for-i  
  } // end if-lev=1  
} // end gluegrid
\end{verbatim}

The program \it{save }\rm does exactly as advertised, saving data into a file:
\begin{verbatim}
//save.cxx - writes J, separation distance, and mass correction

#include <assert.h>
#include <fstream.h>
#include <iostream.h>
#include <iomanip.h>
#include <math.h>
#include "grid.h"
#include "adapt.h"

void save(char* filename, AGrid3D& psi, AGrid3D& u) 
{
  
  extern double J;
  extern double BAREM1;
  
  using namespace Parameters;  // looks up parameter values
    
  int largen = psi.sideX();
  double smallh = u.hX();
  double D = 2*X1; // coordinate separation
  int xpunct = X1/smallh + 1;  // grid point location of
                               // puncture
  double upunct = u(xpunct,1,1);   // value of u at puncture
  double admmass = 2*M1 + moment(psi,0,0,largen-1);
  
  double q22 = moment(psi,2,2,largen-1);
  cout << " Phi(l = 2, m = 2) = "  << q22 << endl;
 
 ofstream output;
  output.open(filename, ios::app|ios::ate);  //appends to 
                                             // end of file
  cout << setprecision(6) << "J = " << J 
       << "     coordinate separation = " 
       << D << "     ADM = "<< setprecision(16) 
       << admmass << "     Newtonian mass = " << M1 << endl;
} // end save
\end{verbatim}

The following programs are ``helper'' programs, in the sense that they are relatively small programs that do simple tasks, such as filling arrays with source information or adding and subtracting arrays. 
\begin{verbatim}
// Fills source array on finest level such that we are relaxing
// only to find the best approximation, not the error
void fillsource(AGrid3D& s, AGrid3D& v, AGrid3D& alpha, 
                   AGrid3D& beta)
{
  int i, j, k;
  int n = s.sideX();
  double h=s.hX();
  
  for (i=1;i<n;i++) {
    for(j=1;j<n;j++) {
      for(k=1;k<n;k++) {
        double vres = 1.0/(h*h)*(GAL0*v.sy(i,j,k)
          +GAL1*(v.sy(i+1,j,k)+v.sy(i-1,j,k)
            +v.sy(i,j+1,k)+v.sy(i,j-1,k)
            +v.sy(i,j,k+1)+v.sy(i,j,k-1))
          +GAL2*(v.sy(i+1,j+1,k)+v.sy(i+1,j-1,k)
            +v.sy(i-1,j+1,k)+v.sy(i-1,j-1,k)
            +v.sy(i,j+1,k+1)+v.sy(i,j-1,k+1)
            +v.sy(i+1,j,k+1)+v.sy(i-1,j,k+1)
            +v.sy(i,j+1,k-1)+v.sy(i,j-1,k-1)
            +v.sy(i+1,j,k-1)+v.sy(i-1,j,k-1))
          +GAL3*(v.sy(i+1,j+1,k+1)+v.sy(i-1,j+1,k+1)
            +v.sy(i+1,j-1,k+1)+v.sy(i-1,j-1,k+1)
            +v.sy(i+1,j+1,k-1)+v.sy(i-1,j+1,k-1)
            +v.sy(i+1,j-1,k-1)+v.sy(i-1,j-1,k-1)))
          +beta(i,j,k)*pow(1+alpha(i,j,k)
            +alpha(i,j,k)*v.sy(i,j,k),-7);
	
	s(i,j,k) = -vres;
      }
    }
  }
}
\end{verbatim}

\begin{verbatim}
//Ecalc.cxx - calculates the ADM mass

#include <assert.h>
#include <fstream.h>
#include <iostream.h>
#include <iomanip.h>
#include <math.h>
#include "grid.h"
#include "adapt.h"

namespace Energy {
  double ADM;
}

void Ecalc(AGrid3D& psi) 
{
  
  using namespace Parameters;  // looks up parameter values
  using namespace Energy;      // allows for modification 
                               // of ADM mass
 
  int n = psi.sideX();
  
  ADM = 2*M1 + moment(psi,0,0,n-1);
  
  cout << "ADM mass = " << ADM << " for a D = " 
       << 2.0*X1 << endl;
}
\end{verbatim}

\begin{verbatim}
// getu.cxx - returns value of u at puncture

#include <assert.h>
#include <fstream.h>
#include <iostream.h>
#include <iomanip.h>
#include <math.h>
#include "grid.h"
#include "adapt.h"

double getu(AGrid3D& u) {

  using namespace Parameters;
  
  double h = u.hX();
  int x = X1/h + 1;  // grid point location of puncture
  
  return(u(x,1,1));
}

\end{verbatim}

\begin{verbatim}
extern int MG,NMG;
extern double GAL0,GAL1,GAL2,GAL3;
\\checks to make sure system parameters are unchanged
void checkparam(double sepdist) {

  using namespace Parameters;
  extern double J;

  assert( X1 == sepdist/2 );
  assert( X2 == X1 );
  assert( SCALE == 2.0*X1 );
  assert( LEN == 1.0*(pow(2,MG)+2)/(pow(2,MG))*SCALE );
  assert( PY1 == 1.0*J/(2.0*X1) );
  assert( PY2 == -1.0*J/(2.0*X1) );

}
\end{verbatim}

\begin{verbatim}
//determines the conformal factor on the surfaces in order to
// calculate a mass
void fillpsi(AGrid3D& psi, AGrid3D& alpha, AGrid3D& u) 
{
  int i,j,k;
  int n = u.sideX();
  
  // Calculating the mass from 1/alpha introduces
  // discontinuities...
  // Just calculate correction from 1 + u, and then add bare
  // mass in main...
  
  for(i = 1;i<=n;i++) {
    for(j = 1;j<=n;j++) {
      for(k=1;k<=n;k++) {
        psi(i,j,k) = 1.0 + u(i,j,k);
      }
    }
  }   
}
\end{verbatim}

\begin{verbatim}
// zeros interior points of grid
void zerointerior(AGrid3D& v)
{
  int m;
  int n=v.sideX();
  
  if(n==pow(2,MG)+3) m = n/2-1;
  else  m = n/2;

  for(int i=1;i<=m;i++)  
    for(int j=1;j<=m;j++) {
      for(int k=1;k<=m;k++) {
        v(i,j,k) = 0.0;
      }
    }
  }
}
\end{verbatim}

\begin{verbatim}
// zeros interior points of grid
void zerointeriorsoln(AGrid3D& v)
{
  int m;
  int n=v.sideX();
  
  m = n-2;
  
  for(int i=1;i<m;i++) {
    for(int j=1;j<m;j++) {
      for(int k=1;k<m;k++) {
        v(i,j,k) = 0.0;
      }
    }
  }
}
\end{verbatim}

\begin{verbatim}
// zeros everything but fine grid "relaxed" interior
void zeroexterror(AGrid3D& v)
{
  int i,j,k,m;
  int n = v.sideX();
  if(n==pow(2,MG)+3) m = n/2;
  else m = n/2+1;

  for(i=m;i<=n;i++) {
    for(j=1;j<=n;j++) {
      for(k=1;k<=n;k++) {
        v(i,j,k) = 0.0;
	v(j,i,k) = 0.0;
	v(j,k,i) = 0.0;
      }
    }
  }
}
\end{verbatim}

\begin{verbatim}
// adds two arrays together
void add(AGrid3D& u, AGrid3D& a, AGrid3D& b)
{
  int i, j, k;
  int n = u.sideX();
  
  for (i=1;i<=n;i++) {
    for(j=1;j<=n;j++) {
      for(k=1;k<=n;k++) {
        u(i,j,k) = a(i,j,k) + b(i,j,k);
      }
    }
  }
}
\end{verbatim}

\begin{verbatim}
// copies array
void copy(AGrid3D& u, AGrid3D& v)
{
  int i,j,k;
  int n=u.sideX();
  
  for (i=1;i<=n;i++) {
    for(j=1;j<=n;j++) {
      for(k=1;k<=n;k++) {
        u(i,j,k) = v(i,j,k);
      }
    }
  }
}
\end{verbatim}

\begin{verbatim}
// finds difference of two arrays
void subtract(AGrid3D& u, AGrid3D& v, AGrid3D& w)
{
  int i,j,k;
  int n = u.sideX();

  for (i=1;i<=n;i++) {
    for(j=1;j<=n;j++) {
      for(k=1;k<=n;k++) {
        u(i,j,k) = v(i,j,k) - w(i,j,k);
      }
    }
  }

}
\end{verbatim}

\begin{verbatim}
// Find the norm of an array
double norm(AGrid3D& v)
{
  double err,sum = 0.0;
  int i,j, k;
  int N;
  int n = v.sideX();
  if(n==pow(2,MG)+3) N = n-3;
  else N = n;

  for(i=1;i<=N;i++) {
    for(j=1;j<=N;j++) {
      for(k=1;k<=N;k++) {
        sum = sum + v(i,j,k)*v(i,j,k);
      }
    }
  }
  err = sqrt(1.0/(N*N*N)*sum);
  return(err);
}
\end{verbatim}

\begin{verbatim}
// Adds the correction to the "old" solution
void update(AGrid3D& v, AGrid3D& a)
{
  int i, j, k, n;
  n = v.sideX();
  
  for(i=1;i<=n;i++) {
    for(j=1;j<=n;j++) {
      for(k=1;k<=n;k++) {
        v(i,j,k) = v(i,j,k) + a(i,j,k);
      }
    }
  }
}
\end{verbatim}

The following header files, \it{grid.h }\rm and \it{adapt.h}\rm, were written by Steven Detweiler and allow convenient manipulation of arrays:
\begin{verbatim}
// grid.h

#ifndef GRID__H
#define GRID__H

// a Vect goes either from 0 to n-1,  for Vect v(n)
//                 or from nl to nh   for Vect(nl,nh).
// Inbounds-checking is enabled for v[i],
// but not for v.val(i)

class Vect {
private:
   int nl;
   int nh;
   int len;
   double *v;
public:

// Constructors:
   Vect(): nl(0), nh(0), len(1){	// changed 1's to 0's
                                        // for nl, nh
     v = new double[len];
     setZero();
   };
   Vect(int n): nl(0), nh(n-1), len(n){
     v = new double[len];
     setZero();
   };
   Vect(int nlow, int nhi): nl(nlow), nh(nhi), len(nh-nl+1){
     v = new double[len];
     setZero();
   };

// Destructors:
   ~Vect(){
     delete [] v;
   };

// initialize (done automatically on construction)
   void setZero(){
     for(int i = 0; i < len; i++) {
       v[i]=0;
     }
   };

// information
   int size() const {return len;}
   int high() const {return nh;};
   int low() const {return nl;};
   int index(int i) const {return i-nl;};

// value includes out-of-bounds check
   double& operator[](int n) {
     if(n < nl) {cout << n <<" "; 
       VectError("Vect: Out of bounds too low");};
     if(n > nh) {cout << n <<" "; 
       VectError("Vect: Out of bounds too high");};
     return v[n-nl];
   };
// value with no out-of-bounds check
   double& val(int n) {
     return v[index(n)];
   };

// Reset size
   void reset(int n){
     if (n != len){
       nl = 0; nh = n-1; len = nh-nl+1;
       delete [] v;
       v = new double[len];
     }
     setZero();
   };
   void reset(int nlow, int nhi){
     nl=nlow; nh = nhi;
     if (len != nh-nl+1){
       len = nh-nl+1;
       delete [] v;
       v = new double[len];
     }
     setZero();
   };

private:
   void VectError(char* str) const {
     cout.flush();
     cout << str << endl;
     cout << "press 'c' and return to continue ..." << endl;
     //getch();
     char kb;
     cin >> kb;
     assert(0);
   };
};

// Grid2D is a two dimensional grid of n x n for Grid2D u(n)
//                                or nx x ny for Grid2D u(nx,ny)
// access with u(i,j) with 1 <= i <= nx and 1 <= j <= ny
// Inbonds-checking is enabled, but could be removed.

class Grid2D {
private:
   int ngx;
   int ngy;
   Vect v;
public:
// Constructors
   Grid2D(): ngx(1), ngy(1), v(ngx*ngy){
   };
   Grid2D(int N): ngx(N), ngy(N), v(ngx*ngy){
     if (N<=0) GridError("side smaller than 1");
   };
   Grid2D(int Nx, int Ny): ngx(Nx), ngy(Ny), v(ngx*ngy){
     if (Nx<=0) GridError("X-side smaller than 1");
     if (Ny<=0) GridError("Y-side smaller than 1");
   };
// Destructors
   ~Grid2D(){};

// Initialize--done automatically through Vector::v on construction
   void setZero() {v.setZero();};

// Information
   int sideX() const {return ngx;};
   int sideY() const {return ngy;};
   int index(int i, int j) const {return ((i-1)*ngx + j - 1);};

// Value with out-of-bounds check
   double& operator()(int i, int j) {
     if(i < 1) {cout << i <<" ";
       GridError("Grid2D: i too low");}
     if(j < 1) {cout << j <<" "; 
       GridError("Grid2D: j too low");}
     if(i > ngx) {cout << i <<" "; 
       GridError("Grid2D: i too big");}
     if(j > ngy) {cout << j <<" "; 
       GridError("Grid2D: j too big");}
     return v[index(i,j)];
   };

// Reset sides
   void reset(int N){
     if (N<=0) GridError("side smaller than 1");
     ngx = N;
     ngy = N;
     v.reset(ngx*ngy);
   };
   void reset(int Nx, int Ny){
     if (Nx<=0) GridError("X-side smaller than 1");
     if (Ny<=0) GridError("Y-side smaller than 1");
     ngx = Nx;
     ngy = Ny;
     v.reset(ngx*ngy);
   };

private:
   void GridError(char* str) const {
     cout.flush();
     cout << str << endl;
     cout << "press 'c' and return to continue ..." << endl;
     //getch();
     char kb;
     cin >> kb;
     assert(0);
   };
};


// Grid2D is a three dimensional grid of n x n x n for 
// Grid2D u(n) or nx x ny x nz for Grid2D u(nx,ny,nz)
// access with u(i,j,k) with 1 <= i <= nx; 1 <= j <= ny; 
// 1 <= k <= nz;
// Inbonds-checking is enabled, but could be removed.

class Grid3D {
private:
   int ngx;
   int ngy;
   int ngz;
   Vect v;
public:
// Constructors
   Grid3D(): ngx(1), ngy(1), ngz(1), v(ngx*ngy*ngz){
   };
   Grid3D(int n): ngx(n), ngy(n), ngz(n), v(ngx*ngy*ngz){
     if (n<=0) GridError("side smaller than 1");
   };
   Grid3D(int Nx, int Ny, int Nz): ngx(Nx), ngy(Ny), ngz(Nz),
   v(ngx*ngy*ngz){
     if (Nx<=0) GridError("X-side smaller than 1");
     if (Ny<=0) GridError("Y-side smaller than 1");
     if (Nz<=0) GridError("Z-side smaller than 1");
   };

// Destructor
   ~Grid3D(){};

//Initialze -- done automaticall on constructoin through Vect.v
   void setZero() {v.setZero();};

// Information
   int sideX() const {return ngx;}
   int sideY() const {return ngy;}
   int sideZ() const {return ngz;}
   int index(int i, int j, int k)
   const {return (((i-1)*ngy + j-1)*ngz + k-1);}

// Value with out-of-bounds check
   double& operator()(int i, int j, int k) {
     if(i < 1) {cout << i <<" "; 
       GridError("Grid3D: i too low");}
     if(j < 1) {cout << j <<" "; 
       GridError("Grid3D: j too low");}
     if(k < 1) {cout << k <<" "; 
       GridError("Grid3D: k too low");}
     if(i > ngx) {cout << i <<" "; 
       GridError("Grid3D: i too big");}
     if(j > ngy) {cout << j <<" "; 
       GridError("Grid3D: j too big");}
     if(k > ngz) {cout << k <<" "; 
       GridError("Grid3D: k too big");}
     return v[index(i,j,k)];
   };
   
// Value with symmetry imposed at i=1, j=1, k=1
   double& sy(int i, int j, int k) {
     if(i < 1) {i = 2 - i;}
     if(j < 1) {j = 2 - j;}
     if(k < 1) {k = 2 - k;}
     if(i > ngx) {cout << i <<" "; 
       GridError("Grid3D: i too big");}
     if(j > ngy) {cout << j <<" "; 
       GridError("Grid3D: j too big");}
     if(k > ngz) {cout << k <<" "; 
       GridError("Grid3D: k too big");}
     return v[index(i,j,k)];
   };

// reset size
   void reset(int n){
     if (n<=0) GridError("side smaller than 1");
     ngx = n;
     ngy = n;
     ngz = n;
     v.reset(ngx*ngy*ngz);
   };
   void reset(int Nx, int Ny, int Nz) {
     if (Nx<=0) GridError("X-side smaller than 1");
     if (Ny<=0) GridError("Y-side smaller than 1");
     if (Nz<=0) GridError("Z-side smaller than 1");
     ngx = Nx;
     ngy = Ny;
     ngz = Nz;
     v.reset(ngx*ngy*ngz);
   };

private:
   void GridError (char* str) const{
     cout.flush();
     cout << str << endl;
     cout << "press 'c' and return to continue ..." << endl;
     //getch();
     char kb;
     cin >> kb;
     assert(0);
   };
};

#endif

\end{verbatim}

\begin{verbatim}
// adapt.h

#ifndef ADAPT__H
#define ADAPT__H

#include "grid.h"

class AGrid3D : public Grid3D {
private:
  double Lx;
  double Ly;
  double Lz;
  double ghx;
  double ghy;
  double ghz;
public:
// Constructors
  AGrid3D(): Grid3D(), Lx(1),Ly(1),Lz(1) { setsize();};
  AGrid3D(int n): Grid3D(n), Lx(1),Ly(1),Lz(1) { setsize();};
  AGrid3D(int Nx, int Ny, int Nz): Grid3D(Nx,Ny,Nz),
    Lx(1),Ly(1),Lz(1) {
    setsize();
  }
  AGrid3D(int n, double L): Grid3D(n), Lx(L),Ly(L),Lz(L) {
    setsize();};
  AGrid3D(int Nx, double LX, int Ny, double LY, int Nz,
    double LZ):
    Grid3D(Nx,Ny,Nz), Lx(LX),Ly(LY),Lz(LZ) {
    setsize();
  }
// Destructor
  ~AGrid3D(){;}

  void reset(int n){
    Grid3D::reset(n);
    setsize();
  }
  void reset(int n, double L){
    Grid3D::reset(n);
    Lx = L;
    Ly = L;
    Lz = L;
    setsize();
  }

  void reset(int Nx, int Ny, int Nz) {
    Grid3D::reset(Nx,Ny,Nz);
    setsize();
  }
  void reset(int Nx, double LX, int Ny, double LY, int Nz,
    double LZ) {
    Grid3D::reset(Nx,Ny,Nz);
    Lx = LX;
    Ly = LY;
    Lz = LZ;
    setsize();
  }

  double LenX() { return Lx;}
  double LenY() { return Ly;}
  double LenZ() { return Lz;}

  double hX() { return ghx;}
  double hY() { return ghy;}
  double hZ() { return ghz;}

private:
  void setsize() {
    if (sideX()<=1) ghx = 0; else ghx = Lx/(sideX()-1);
    if (sideY()<=1) ghy = 0; else ghy = Ly/(sideY()-1);
    if (sideZ()<=1) ghz = 0; else ghz = Lz/(sideZ()-1);
  }
};


namespace Parameters {   
  
  extern double M1;      
  extern double M2;
  extern double X1;      
  extern double X2;      
  extern double SCALE;   
  extern double LEN;     
  extern double PY1;
  extern double PY2;     

}

double getu(AGrid3D& u);
void checkparam(double dist);
void newADMmass(void);
void setup(double dist);
void save(char* filename, AGrid3D& psi, AGrid3D& u);
void newbeta(AGrid3D& alpha, AGrid3D& beta);
void psibar(AGrid3D& psib, int l, int m);
double moment(AGrid3D& u, int l, int m, int k);
void adaptinterp(AGrid3D& u, AGrid3D& v);
void symadaptrstrct(AGrid3D& u, AGrid3D& v);
void adaptrstrct(AGrid3D& u, AGrid3D& v);
void add(AGrid3D& u, AGrid3D& a, AGrid3D& b);
void copy(AGrid3D& u, AGrid3D& v);
void subtract(AGrid3D& u, AGrid3D& v, AGrid3D& w);
void fillpsi(AGrid3D& psi, AGrid3D& alpha, AGrid3D& u);
void fillsource(AGrid3D& s, AGrid3D& v, AGrid3D& alpha, 
                    AGrid3D& beta);
void gluegrid(AGrid3D& usm, AGrid3D& ulg, int lev, int count);
void interp(AGrid3D& af, AGrid3D& vc);
double norm(AGrid3D& v);
void symresrstrct(AGrid3D& u, AGrid3D& v);
void resrstrct(AGrid3D& u, AGrid3D& v);
void newrelax(AGrid3D& u, AGrid3D& v, AGrid3D& s, 
                 AGrid3D& alpha, AGrid3D& beta);
void newresidual(AGrid3D& r, AGrid3D& u, AGrid3D& v, 
                   AGrid3D& s, AGrid3D& alpha, AGrid3D& beta);
void symrstrct(AGrid3D& vc, AGrid3D& vf);
void rstrct(AGrid3D& vc, AGrid3D& vf);
void update(AGrid3D& v, AGrid3D& a);
void write(AGrid3D& a, AGrid3D& b,AGrid3D& c, AGrid3D& d,
               AGrid3D& e, AGrid3D& f);
void zeroexterior(AGrid3D& v);
void zeroexterror(AGrid3D& v);
void zerointerior(AGrid3D& v);
void zerointeriorsoln(AGrid3D& v);

#endif
\end{verbatim}


\chapter{Biographical Sketch}
Brian Douglas Baker was born on August 10, 1972 in Renton, Washington. He is the only son of Phillip and Cherie Baker, and the brother of Carrie Klippenstein.

Brian's childhood was a relatively normal existence, except for his paralyzing fear of kindergarten at the age of 5 years. It is somewhat ironic Brian later went on to pursue a field of study which would require formal schooling until he was nearly 30 years old.

The year 1979 was an important year for the Baker family. After living in various cities and towns in Washington and Oregon most of Brian's life, the family decided to move to Whitetail, Montana---the home town of Brian's father, and the location of a 4,000 acre family farm. Until that time, the farm had been managed by Brian's grandparents, Hugh and Susie Baker. At this time Brian's interest in the natural world blossomed. A large farm with a variety of wildlife to study and farm equipment to dismantle is an ideal environment for an inquisitive young mind.

Brian was educated at Flaxville Public School, and was given all the opportunities that a small, rural public education system can offer. He demonstrated an interest in science at an early age, often reading from the set of encyclopedias his parents owned, learning as much as he could about a variety of topics. 

In high school, Brian was fortunate enough to have teachers who were excited about teaching and who demanded excellence. They helped to develop Brian's inquisitive nature, and taught him the value of doing something that most take for granted---thinking. Brian graduated from Flaxville High School in 1991, finishing at the top of his class. Of course, that is a simple task when the class consists of four people---the Flaxville class of 1991 was the smallest graduating class in the state of Montana for that year.

After graduation, Brian enrolled at Montana State University in Bozeman, Montana, where he majored in physics. This was a wonderful time for Brian, as it was the first time in his life he was exposed to the amazing culture of physics. The people he met from around the state and from around the world enriched his experience and gave him a deep appreciation for diversity and differences. During his stay at Montana State, Brian gained valuable experience in a laser optics lab. He studied issues involving electronic noise suppression in laser diodes and frequency locking of laser diodes. Eventually he designed and built a version of a low-noise laser diode mount. The highlights of Brian's brief career as an experimentalist include presenting his designs to the engineers of ILX Lightwave, and on a separate occasion presenting his work to the Governor of Montana, Marc Rasicot. Brian was also featured in the University's annual research report for the College of Letters and Science.

Brian earned a B.S. in physics from Montana State in 1995. He was then employed by Lattice Materials Inc.~in Bozeman, Montana, as a crystal growth engineer. After tolerating a year of laughable pay and questionable safety precautions at the workplace, Brian decided to pursue graduate study at the University of Florida in Gainesville, Florida.

At the University of Florida, Brian was fortunate enough to work with Steven Detweiler. Over the years, their work covered a variety of topics including variational principles, neutron stars, the Newman-Penrose formalism, perturbation theory, and numerical methods using C++. Aside from being an excellent advisor and physicist, Steve became a trusted friend and confidant. His words of encouragement and understanding were a welcome change from some of the aloof attitudes prevalent in the department at the time.

Aside from research activities, Brian was also active in educating literally thousands of undergraduates who attended the University of Florida. Everything from labs for introductory and advanced physics, to discussion sections for introductory physics, and eventually to being the main instructor for an introductory physics course for two consecutive summer terms. At times, teaching was very enjoyable for Brian, and the physics department and University recognized him for his efforts by naming Brian one of the top twenty teaching assistants on the University of Florida campus in 1999.

In his spare time, Brian enjoys playing with his cat Cleopatra, drinking wine, listening to music, cooking, playing his guitar, reading poetry, visiting with friends, and sleeping.


\begin{thebibliography}{10}
\addcontentsline{toc}{chapter}{Bibliography}

\bibitem{Weber}
J.~Weber.
\newblock Detection and generation of gravitational waves.
\newblock {\em Phys. Rev.}, 117:306, 1960.

\bibitem{Abramovici}
{A. Abramovici, W. Althouse, R. Drever, Y. G\"ursel, S. Kawamura, F. Raab, D.
  Shoemaker, L. Sievers, R. Spero, K. Thorne, R. Vogt, R. Weiss, S. Whitcomb
  and M. Zucker}.
\newblock {LIGO: The Laser Interferometer Gravitational-Wave Observatory}.
\newblock {\em Science}, 256:325, 1992.

\bibitem{FlanaganHughes}
E.~E. Flanagan and S.~A. Hughes.
\newblock Measuring gravitational waves from binary black hole coalescences:
  \uppercase{II}. \uppercase{t}he waves' information and its extraction, with
  and without templates.
\newblock {\em Phys. Rev. D}, 57:4566, 1998.

\bibitem{BB}
S.~Brandt and B.~Brugmann.
\newblock A simple construction of initial data for multiple black holes.
\newblock {\em Phys. Rev. Lett.}, 78(19):3606, May 1997.

\bibitem{Kochanek}
C.~Kochanek.
\newblock Coalescing binary neutron stars.
\newblock {\em Astrophys. J.}, 398:234, 1992.

\bibitem{Bildsten}
L.~Bildsten and C.~Cutler.
\newblock Tidal interactions of inspiraling compact binaries.
\newblock {\em Astrophys. J.}, 400:175, 1992.

\bibitem{Teuk1}
S.~A. Teukolsky.
\newblock Irrotational binary neutron stars in quasiequilibrium in
  \uppercase{G}eneral \uppercase{R}elativity.
\newblock {\em Astrophys. J.}, 504:442, 1998.

\bibitem{Regge}
T.~Regge and C.~Teitelboim.
\newblock Role of surface integrals in the \uppercase{H}amiltonian formulation
  of \uppercase{G}eneral \uppercase{R}elativity.
\newblock {\em Ann. Phys.}, 88:286, 1974.

\bibitem{ADM}
{R. Arnowitt, S. Deser and C. W. Misner}.
\newblock {\em Gravitation: an Introduction to Current Research}.
\newblock edited by L. Witten (John Wiley and Sons), New York, 1962.

\bibitem{MTW}
{C. W. Misner, K. S. Thorne and J. A. Wheeler}.
\newblock {\em Gravitation}.
\newblock W. H. Freeman and Company, New York, 1970.

\bibitem{York3+1}
J.~W. York.
\newblock {\em Sources of Gravitational Radiation}.
\newblock edited by L. Smarr (Cambridge University Press), Cambridge, 1979.

\bibitem{Det2}
S.~Detweiler.
\newblock {Periodic solutions of the Einstein equations for binary systems}.
\newblock {\em Phys. Rev. D}, 50(8):4929, October 1994.

\bibitem{Shibata1}
M.~Shibata.
\newblock A relativistic formalism for computation of irrotational binary stars
  in quasi-equilibrium states.
\newblock {\em Phys. Rev. D}, 58:024012, 1998.

\bibitem{Det3}
B.~D. Baker and S.~Detweiler.
\newblock Variational principles for binary neutron stars and black holes in
  \uppercase{g}eneral \uppercase{r}elativity.
\newblock {\em In preparation}, 2002.

\bibitem{Uryu3}
{K. Ury\=u, M. Shibata and Y. Eriguchi}.
\newblock Properties of general relativistic, irrotational binary neutron stars
  in close quasiequilibrium orbits: Polytropic equations of state.
\newblock {\em Phys. Rev. D}, 62:104015, 2000.

\bibitem{Shibata2}
M.~Shibata and K.~Ury\=u.
\newblock Simulation of merging binary neutron stars in full
  \uppercase{g}eneral \uppercase{r}elativity: $\gamma = 2$ case.
\newblock {\em Phys. Rev. D}, 61:064001, 2000.

\bibitem{GibbonsStewart}
G.~W. Gibbons and J.~M. Stewart.
\newblock Classical \uppercase{G}eneral \uppercase{R}elativity.
\newblock In W.~B. Bonner, J.~N. Islam, and M.~A.~H. MacCallum, editors, {\em
  Classical {G}eneral {R}elativity}, pages 77--94. Cambridge University Press,
  Cambridge, 1984.

\bibitem{Wald}
R.~M. Wald.
\newblock {\em General Relativity}.
\newblock The University of Chicago Press, Chicago, 1984.

\bibitem{Hayward}
G.~Hayward.
\newblock Gravitational action for spacetimes with nonsmooth boundaries.
\newblock {\em Phys. Rev. D}, 47(8):3275, April 1993.

\bibitem{Thorne1}
K.~S. Thorne.
\newblock Multipole expansions of gravitational radiation.
\newblock {\em Rev. Mod. Phys.}, 52(2):299, April 1980.

\bibitem{Bonna3}
{S. Bonazzola, E. Gourgoulhon and J.-A. Marck}.
\newblock Numerical models of irrotational binary neutron stars in
  \uppercase{g}eneral \uppercase{r}elativity.
\newblock {\em Phys. Rev. Lett.}, 82:892, 1999.

\bibitem{Bonna4}
{S. Bonazzola, E. Gourgoulhon and J.-A. Marck}.
\newblock Evolutionary sequences of irrotational binary neutron stars.
\newblock In {\em Proceedings of the 19th Texas Symposium on Relativistic
  Astrophysics}. World Scientific, 2000.
\newblock ed. Aubourg, Montmerle, Paul and Peter.

\bibitem{Baumgarte7}
{M. D. Duez, T. W. Baumgarte, S. L. Shapiro, M. Shibata and K. Ury\=u}.
\newblock Comparing the inspiral of irrotational and corotational binary
  neutron stars.
\newblock {\em Phys. Rev. D}, 65:024016, 2002.

\bibitem{Wilson4}
{P. Marronetti, G. J. Mathews and J. R. Wilson}.
\newblock Binary neutron stars systems: Irrotational quasi-equilibrium
  sequences.
\newblock In {\em Proceedings of the 19th Texas Symposium on Relativistic
  Astrophysics}. World Scientific, 2000.
\newblock ed Aubourg, Montmerle, Paul and Peter.

\bibitem{Bowen}
J.~M. Bowen and J.~W. York.
\newblock Time-asymmetric initial data for black holes and black-hole
  collisions.
\newblock {\em Phys. Rev. D}, 21(8):2047, April 1980.

\bibitem{Wheeler1}
C.~Misner and J.~Wheeler.
\newblock Classical physics as geometry: gravitation, electromagnetism,
  unquantized charge, and mass as properties of curved empty space.
\newblock {\em Ann. Phys.}, 2:525, 1957.

\bibitem{Misner1}
C.~W. Misner.
\newblock Wormhole initial conditions.
\newblock {\em Phys. Rev.}, 118(4):1110, May 1960.

\bibitem{Brill1}
D.~R. Brill and R.~W. Lindquist.
\newblock Interaction energy in geometrostatics.
\newblock {\em Phys. Rev.}, 131(1):471, July 1963.

\bibitem{Cook1}
{G. B. Cook, M. W. Choptuik, M. R. Dubal, S. Klasky, R. Matzner and S. R.
  Oliveira}.
\newblock Three-dimensional initial data for the collision of two black holes.
\newblock {\em Phys. Rev. D}, 47(4):1471, February 1993.

\bibitem{Cook4}
G.~B. Cook.
\newblock Three-dimensional initial data for the collision of two black holes.
  \uppercase{II}. \uppercase{Q}uasicircular orbits for equal-mass black holes.
\newblock {\em Phys. Rev. D}, 50(8):5025, October 1994.

\bibitem{Cook3}
G.~B. Cook.
\newblock Initial data for axisymmetric black-hole collisions.
\newblock {\em Phys. Rev. D}, 44(10):2983, November 1991.

\bibitem{Bonna1}
{E. Gourgoulhon, P. Grandclement and S. Bonazzola}.
\newblock Binary black holes in circular orbits. \uppercase{I}. \uppercase{A}
  global spacetime approach.
\newblock {\em Phys. Rev. D}, 65:044020, 2002.

\bibitem{Friedman1}
{J. L. Friedman, K. Ury\=u and M. Shibata}.
\newblock Thermodynamics of binary black holes and neutron stars.
\newblock {\em Phys. Rev. D}, 65:064035, 2002.

\bibitem{Baumgarte2}
T.~W. Baumgarte.
\newblock Innermost stable circular orbit of binary black holes.
\newblock {\em Phys. Rev. D}, 62:024018, 2000.

\bibitem{Christo}
D.~Christodoulou.
\newblock Reversible and irreversible transformations in black-hole physics.
\newblock {\em Phys. Rev. Lett.}, 25(22):1596, November 1970.

\bibitem{Baumgarte1}
{T. W. Baumgarte, G. B. Cook, M. A. Scheel, S. Shapiro and S. A. Teukolsky}.
\newblock Implementing an apparent-horizon finder in three dimensions.
\newblock {\em Phys. Rev. D}, 54(8):4849, October 1996.

\bibitem{Det1}
J.~K. Blackburn and S.~Detweiler.
\newblock Close black-hole binary systems.
\newblock {\em Phys. Rev. D}, 46(6):2318, September 1992.

\bibitem{multigrid1}
W.~L. Briggs.
\newblock {\em A Multigrid Tutorial}.
\newblock Society for Industrial and Applied Mathematics, Philadelphia,
  Pennsylvania, 1987.

\bibitem{multigrid2}
{W. L. Briggs, V. E. Henson and S. F. McCormick}.
\newblock {\em A Multigrid Tutorial, Second Edition}.
\newblock Society for Industrial and Applied Mathematics, Philadelphia,
  Pennsylvania, 2000.

\bibitem{multigrid3}
P.~Wesseling.
\newblock {\em An Introduction to Multigrid Methods}.
\newblock John Wiley and Sons, New York, 1992.

\bibitem{numrecipes}
{W. H. Press, S. A. Teukolsky, W. T. Vetterling and B. P. Flannery}.
\newblock {\em Numerical Recipes in C: The Art of Scientific Computing}.
\newblock Cambridge University Press, New York, 1988.

\bibitem{York1}
N.~O' Murchadha and J.~W. York.
\newblock Gravitational energy.
\newblock {\em Phys. Rev. D}, 10(8):2345, October 1974.

\bibitem{Cook2}
G.~B. Cook and J.~W. York.
\newblock Apparent horizons for boosted or spinning black holes.
\newblock {\em Phys. Rev. D}, 41(4):1077, February 1990.

\bibitem{Kidder}
{L. E. Kidder, C. M. Will and A. G. Wiseman}.
\newblock Coalescing binary systems of compact objects to post 5/2 -
  \uppercase{N}ewtonian order. \uppercase{III}. \uppercase{T}ransition from
  inspiral to plunge.
\newblock {\em Phys. Rev. D}, 47(8):3281, April 1993.

\end{thebibliography}
\end{document}